\documentclass[twocolumn,showpacs,preprintnumbers,amsmath,amssymb,rmp]{revtex4}

\usepackage{graphicx}
\usepackage{dcolumn}
\usepackage{bm}
\usepackage{longtable}
\newcommand{\ket}[1] {\left| #1 \right\rangle}

\begin{document}

\preprint{APS/123-QED}

\title{Quantum Simulation}

\author{I. M. Georgescu\footnote{I. M. Georgescu was the
lead author of the first version of the manuscript.
Subsequent revisions were made mainly by S. Ashhab and
F. Nori.}}
\affiliation{ CEMS, RIKEN,  Saitama, 351-0198, Japan}
\author{S. Ashhab}
\email{sashhab@qf.org.qa}
\affiliation{ CEMS, RIKEN,  Saitama, 351-0198, Japan
\\
Qatar Environment and Energy Research Institute, Doha, Qatar}
\author{Franco Nori}
\email{fnori@riken.jp}
\affiliation{
CEMS, RIKEN, Saitama, 351-0198, Japan
\\
Department of Physics, University of Michigan, Ann Arbor, Michigan 48109-1040, USA
\\
Department of Physics, Korea University, Seoul 136-713, Korea
}

\date{\today}

\begin{abstract}
Simulating quantum mechanics is known to be a difficult
computational problem, especially when dealing with large systems.
However, this difficulty may be overcome by using some
controllable quantum system to study another less controllable or
accessible quantum system, i.e., quantum simulation. Quantum
simulation promises to have applications in the study of many
problems in, e.g., condensed-matter physics, high-energy physics,
atomic physics, quantum chemistry and cosmology. Quantum
simulation could be implemented using quantum computers, but also
with simpler, analog devices that would require less control, and
therefore, would be easier to construct. A number of quantum
systems such as neutral atoms, ions, polar molecules, electrons
in semiconductors, superconducting circuits, nuclear spins
and photons have been proposed as quantum simulators. This review
outlines the main theoretical and experimental aspects of quantum
simulation and emphasizes some of the challenges and promises of
this fast-growing field.
\end{abstract}

\pacs{03.65.-w, 03.67.Ac}
\maketitle
\tableofcontents

\section{Introduction}

 \begin{quote}
\emph{``Let the computer itself be built of quantum mechanical
elements which obey quantum mechanical laws.''} \cite{feynman82}.
 \end{quote}

Thirty years ago, in the early 1980s, it had become clear that
simulating quantum mechanics is a very challenging problem
\cite{Man80,feynman82}. One obvious difficulty is the huge amount
of computer memory needed to store the quantum state of a large
physical system. This state is described by a number of parameters
that grows exponentially with the system size (which is generally
defined as the number of particles or degrees of freedom in the
system). In particular, one needs to keep track of the probability
amplitudes for all the possible classical configurations of the
system. Furthermore, simulating the temporal evolution of the
system requires a number of operations that also increases
exponentially with the size of the system. This
\emph{exponential explosion} is unavoidable,
unless approximation methods (e.g.~Monte Carlo methods) are used.
However, depending on the specifics of the problem under study,
good approximations are not always available or they also face
some limitations. Therefore, the simulation of quantum systems in
general remains a hard task even for today's supercomputers.
\par
A proposed solution to this problem came in the new type of
computer envisaged by Richard Feynman \cite{feynman82} --- the
quantum computer. In fact, as has become clear over the past three
decades, a quantum computer promises to do much more than
simulating quantum mechanics, and today quantum computation and
quantum information theory are very active research fields (see
e.g.~the books \cite{NC,SS08,SW08}). Feynman realized at the time
that a quantum machine would itself experience an exponential
explosion, but with good consequences. The machine would have the
capacity to contain an exponentially large amount of information
without using an exponentially large amount of physical resources,
thus making it a natural tool to perform quantum simulation.
Despite the great importance of his insight, however, he was not
very specific about how his proposed quantum mechanical computer
was supposed to function or how the simulation itself would be
realized, as can be seen from the quote given at the beginning of
this section: ``\emph{Let the computer itself be built of quantum
mechanical elements which obey quantum mechanical laws.} ''
\cite{feynman82}.
\par
More than a decade later, it was shown \cite{lloyd96} that a
quantum computer (i.e., an ensemble of well-defined qubits that
can be initialized, measured and on which universal quantum gates
can be performed) can indeed act as a universal quantum simulator.
Here, the word universal refers to the fact that, except for
changes in the programs that it runs, the same machine is capable
of tackling vastly different problems. However, a quantum computer
(as defined above) is not necessarily required for implementing
quantum simulation. Simpler quantum devices that mimic the
evolution of other quantum systems in an analog manner could be
used for this task (note that these are not universal simulators,
but rather problem-specific machines). It is therefore expected
that practical quantum simulation will become a reality well
before full-fledged quantum computers.
\par
In recent years, the interest in quantum simulation has been
growing rapidly, and the reason for this is twofold. First, there
are a large number of potential applications of quantum simulation
in physics, chemistry and even biology. Second, the technologies
required for the coherent control of quantum systems have matured
enough to allow for the physical implementation of practical
quantum simulation in the very near future. In fact, some
proof-of-principle experiments on quantum simulation have already
been realized (see e.g.~\cite{Gre02,Lei02,Lan09,Fri08,Nee09,Ger10,Kim10}).
\par
Quantum simulation will provide a valuable tool that researchers
from numerous fields will wish to add to their toolbox of research
methods. For instance, in condensed-matter physics, quantum
simulation would allow the study of many difficult problems, such
as quantum phase transitions, quantum magnetism or high-$T_c$
superconductivity. Other potential application areas include
high-energy physics, quantum chemistry, cosmology and nuclear
physics.
\par
With the latest advances in the coherent manipulation of quantum
systems \cite{Bul10,Lad10}, such as atoms in optical lattices,
trapped ions, nuclear spins, superconducting circuits, or spins in
semiconductors, practical applications of quantum simulation can
be expected in the coming years. Several research groups are now
actively aiming at the experimental realization of quantum
simulators with tens of qubits, which would be the first practical
applications in which quantum computers outperform their classical
counterparts.
\par
There is by now a sizable literature on quantum simulation,
especially papers published in the past decade. However, besides
the brief overview of quantum simulators in \cite{Bul09} and the
specialized reviews focused on cold atoms
\cite{Blo12,Lew07,Jak05}, ions \cite{Sch12,Bla12}, photons
\cite{Asp12}, superconducting circuits \cite{Hou12} and quantum
chemistry \cite{Kas10,Lu12}, a global review of the field is
missing. Moreover, a comprehensive, pedagogic introduction to the
subject would benefit researchers just entering the field, as well
as those already working on quantum simulation and looking for a
quick reference guide. Since quantum simulation is a subject of
interest to a broad audience, this review attempts to provide a
self-contained description of the current status of theoretical
and experimental research on the subject. However, given the
breadth of the topics touched by quantum simulation, not all
technical details can be provided here, and the reader is directed
to the relevant references instead.
\par
The remainder of this article is organized as follows. Sections
II--V discuss in some detail the basic theory. Readers interested
only in the physical implementations of quantum simulation can
concentrate on Section VI, while those interested in the
applications of quantum simulation can concentrate on Section VII.
Tables \ref{tcomp}, \ref{tapp} and \ref{tapp1} provide
quick reference guides for the content of Sections VI and VII. In
Section VIII we discuss the challenges and prospects of quantum
simulation.

\section{The problem}
 \begin{quote}
 \emph{``The rule of simulation that I would like to have is that the number of computer elements required to simulate a large physical system is only to be proportional to the space-time volume of the physical system. I don't want to have an explosion.''} \cite{feynman82}
 \end{quote}
Let us consider a rather general quantum simulation problem,
namely that of finding the state of a quantum system described by
the wavefunction $\ket{\phi}$ at some time $t$ and computing the
value of some physical quantity of interest. Focusing for
simplicity on time-independent Hamiltonians (denoted $H$), the
solution of the Schr\"odinger equation:
\begin{equation}
i\hbar \frac{d}{dt}\ket{\phi} = H \ket{\phi} \label{sch}
\end{equation}
is given by $\ket{\phi(t)} = \exp\{-i\hbar Ht\}\ket{\phi(0)}$. In
order to compute $\ket{\phi (t)}$ numerically, it is first
necessary to discretize the problem such that it can be encoded in
the computer memory. As mentioned earlier, the amount of memory
required for representing quantum systems grows exponentially with
the system size, and so does the number of operations required to
simulate the time evolution. For instance, representing the state
of $N$ spin-1/2 particles requires $2^N$ numbers (namely the
complex probability amplitudes for the different spin
configurations), and this is without including the particles'
motional degrees of freedom. Calculating the time evolution of
this system requires exponentiating a $2^N \times 2^N$ matrix. Let
us take the standard ``threshold'' $N=40$ frequently cited in the
literature (\cite{lloyd96,Cir03,Rae07,Fri08}). This implies
storing $2^{40} \approx 10^{12}$ numbers for $\ket{\phi}$ (For the
moment, we will not worry about the Hamiltonian with its $2^{40}
\times 2^{40} \approx 10^{24}$ matrix elements, because for
realistic physical problems the Hamiltonian has a very regular
structure and just encoding it in the computer memory does not
suffer form the exponential-explosion problem). Assuming single
precision, about $\sim 3.2 \times 10^{13}$ bits, that is 4 TB
(terabytes), are required to represent the spin state of 40
particles in a computer memory. In order to put this in
perspective, the US Library of Congress has almost 160 TB of data.
Double the number of spins, and $\sim 3.8 \times 10^{25}$ bits (or
$5 \times 10^{12}$ TB) would be required. This is roughly ten
thousand times more than the amount of information stored by
humankind in 2007, which was estimated to be $2.4 \times 10^{21}$
bits \cite{Hil11}.
\par
Classical stochastic methods, namely Monte Carlo algorithms
\cite{Suz93}, have been developed as a way of tackling the
difficult problem of simulating large quantum systems. These
methods allow the evaluation of phase space integrals for
many-body problems in a time that scales polynomially with the
number of particles. Such stochastic methods generally work well
when the functions being integrated do not change sign (and
ideally vary slowly with respect to the relevant variables), such
that sampling the function at a relatively small number of points
gives a good approximation to the integral of the function. For
some quantum systems, especially fermionic and frustrated systems,
the numerical evaluation of the integrals encounters the problem
of sampling with non-positive-semidefinite weight functions, which
is the so-called {\it sign problem} [see, e.g.~\cite{Tro05}]. This
results in an exponential growth of the statistical error, and
hence the required simulation time, with the number of particles,
which cancels the advantage of the Monte Carlo methods. Other
methods of solving quantum many-body problems such as density
functional theory, mean-field theories, many-body perturbation
theories or Green's function-based methods, coupled clusters,
etc., (see \cite{FetWal,Thou,Zago}) have similar validity criteria
that restrict their applicability to well-behaved systems.

\section{Definitions}

The alternative simulation method initially proposed by Feynman,
i.e., \textit{quantum simulation}, can be loosely defined as
simulating a quantum system by quantum mechanical means.
This very general definition allows us to include three types of
simulation:
\begin{itemize}
\item{Digital quantum simulation}

\item{Analog quantum simulation}

\item{Quantum-information-inspired algorithms for the classical
simulation of quantum systems}
\end{itemize}
These will be discussed in some detail in the following sections.
\par
By \textit{quantum simulator}, we understand \textit{a
controllable quantum system used to simulate/emulate other quantum
systems} (see, e.g., \cite{Bul09}).
\par
\begin{figure}
\includegraphics[width=0.5\textwidth ]{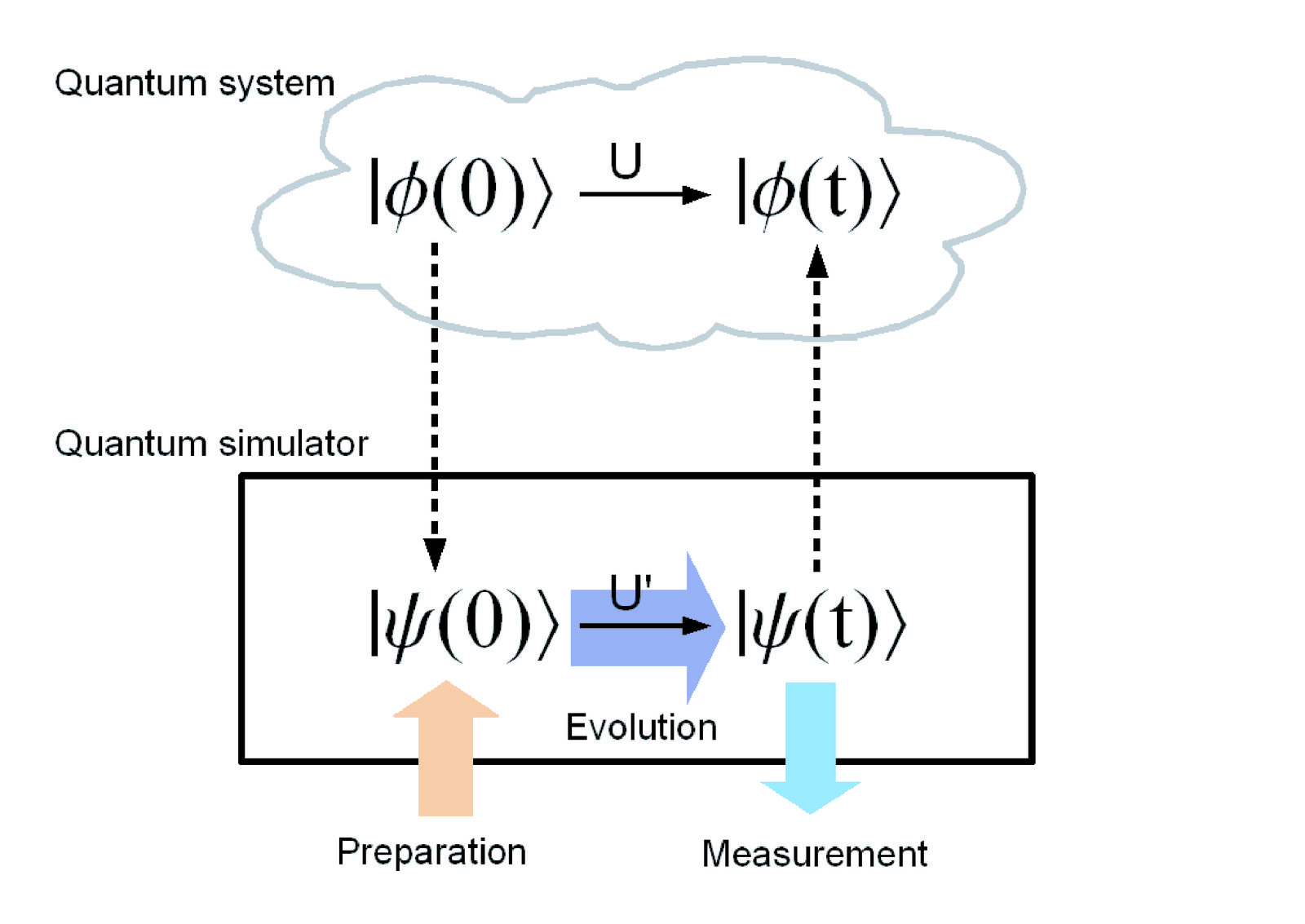}
\caption{\label{qs} (color online) Schematic representation of a
quantum system and a corresponding quantum simulator. The quantum state
$\ket{\phi (0)}$ evolves to $\ket{\phi (t)}$ via the unitary
transformation $U=\exp\{-i\hbar H_{\rm{sys}} t\}$. The quantum
simulator evolves from the state $\ket{\psi (0)}$ to $\ket{\psi (t)}$
via $U'=\exp\{-i\hbar H_{\rm{sim}} t\}$. The simulator is designed
such that there is a mapping between the simulator and the
simulated system, in particular the mappings $\ket{\phi (0)}
\leftrightarrow \ket{\psi (0)}$, $\ket{\phi (t)}\leftrightarrow
\ket{\psi (t)}$ and $U \leftrightarrow U'$. While the simulated
system may not be controllable (or not experimentally accessible
in some cases), the quantum simulator is. Namely, the initial state
$\ket{\psi (0)}$ can be prepared, the unitary evolution $U'$ can
be engineered, and the final state $\ket{\psi (t)}$ can be measured.
The result of this measurement provides information about the simulated
system. The color arrows denote the controllable operations. The
solid black arrows describe the time evolution of the system and
the simulator. The dashed arrows indicate the correspondence
between the quantum states of the simulator and the simulated system.}
\end{figure}

Let us denote the state of the simulated system by $\ket{\phi}$.
The system evolves from the initial state $\ket{\phi (0)}$ to
$\ket{\phi (t)}$ via the unitary transformation $U=\exp\{-i\hbar
H_{\rm{sys}} t\}$, where $H_{\rm{sys}}$ is the Hamiltonian of the
system. The quantum simulator is a controllable system: the
initial state $\ket{\psi (0)}$ can be prepared, the desired
unitary evolution $U'=\exp\{-i\hbar H_{\rm{sim}} t\}$, with
$H_{\rm{sim}}$ being the controllable Hamiltonian of the simulator,
can be engineered, and the final state $\ket{\psi (t)}$ can be
measured. If a mapping between the system and the simulator (i.e.,
between $\ket{\phi (0)}$ and $\ket{\psi (0)}$, and between
$\ket{\phi (t)}$ and $\ket{\psi (t)}$, exists, then the system can
be simulated. The basic idea of quantum simulation is represented
schematically in Figure \ref{qs}.

\section{Digital and analog quantum simulation}

The advantage of quantum simulators over classical devices is
that, being quantum systems themselves, they are capable of
storing large amounts of information in a relatively small amount
of physical space. For example, the storage capacity of $N$ {\it
qubits} is exponentially larger than that of $N$ classical {\it
bits}. Going back to the example given in the previous section,
the quantum state of $N=40$ spin-1/2 particles, which would
require a 4TB classical memory register, can be represented by a
$40$-qubit (i.e.~5-quantum-byte) register. If the time evolution
of the simulator reproduces the time evolution of the simulated
system, the desired final state can be obtained \textit{without}
the need for numerically exponentiating a $2^N \times 2^N$ matrix.
This sounds very promising, but the quantum simulation problem is
not really solved unless the initial state preparation, the
implementation of the time evolution and the measurement are
realized using only polynomial resources. The importance of
measurement must be stressed because the success of quantum
simulation ultimately depends on the ability to extract useful
information from the simulator. As will be discussed later, these
are not easy tasks, even for quantum simulators.

\subsection{\label{DQS}Digital quantum simulation (DQS)}

We consider the well-known circuit model for quantum computation
\cite{NC}. First, the wavefunction $\ket{\phi}$ has to be encoded
using the computational basis, i.e., as a superposition of binary
bit strings. A very simple example is the simulation of spin-1/2
particles. Each particle is represented by a qubit: the spin-up
state $\ket{\uparrow}$ is encoded as the qubit state
$\ket{1}$, and the spin-down state $\ket{\downarrow}$ as
$\ket{0}$. For example, the three-spin state $\ket{\phi} = \ket{\uparrow
\uparrow \downarrow}$ is represented in the simulator by
$\ket{\psi} = \ket{110}$.
\par
In order to obtain $\ket{\psi(t)} = \exp\{-i\hbar
Ht\}\ket{\psi(0)}$, $U=\exp\{-i\hbar Ht\}$ must be applied to the
initial state. The complicated many-qubit unitary transformation
$U$ is implemented through the application of a sequence of
single- and two-qubit gates (we will come back and discuss the
decomposition of $U$ into these simple gates shortly). Such a
circuit-based quantum simulation recreating the evolution
$\ket{\psi(0)} \rightarrow \ket{\psi(t)}$ is usually referred to
as digital quantum simulation (DQS). Some of the representative
studies on DQS are
\cite{lloyd96,Wis96,Abr97,Lid97,Zal98,Za98,Ter00,Ort01,Mar02,Som02,Ver08,Rae12}.
\par
Since any unitary operation can be written in terms of universal
quantum gates, it follows that in principle ``anything'' can be
simulated, i.e.~DQS is universal \cite{lloyd96}. However, it
must be noted that not any unitary operation can be {\it
efficiently} simulated (that is with polynomial resources) and,
therefore, there are Hamiltonians that cannot be efficiently
simulated in this way. Nevertheless, it is possible to efficiently
simulate any finite-dimensional local Hamiltonian. This is
particularly important since all local spin systems, and all
Hamiltonians that can be efficiently mapped to such systems, are
included in this class. In other words, although not all
mathematically allowed Hamiltonians can be simulated efficiently,
those that appear in most physical theories can be simulated
efficiently. Note that finding an efficient decomposition in terms
of universal gates can in itself be a difficult problem
\cite{Das11}. Furthermore, it must be stressed that the
implemented unitary operation (i.e.~that obtained from the
decomposition into single- and two-qubit gates) is generally an
approximation of the desired unitary evolution. In principle, this
approximation can be made arbitrarily accurate (by refining the
decomposition), but this comes at the cost of an ever-increasing
number of gates.
\par
Although DQS algorithms rely on applying a time-ordered sequence
of gates, thus implementing a unitary evolution of the simulator,
DQS is not restricted to recreating the temporal evolution of the
simulated system. Applications of DQS also include obtaining
certain properties of a given quantum system (e.g., phase
estimation for computing eigenvalues of operators, particularly
the Hamiltonian \cite{Abr99,Guz05,WanWu10}, or computing partition
functions \cite{Lid97}). Moreover, according to \cite{Mey01} it
should also be possible to use quantum computers to simulate
classical physics more efficiently (see also \cite{Sin10,Yun10}).
\par
In general, DQS consists of three steps: initial state preparation
$\ket{\psi(0)}$, unitary evolution $U$ and the final measurement.
These steps will be discussed in detail in the remainder of this
subsections (see also \cite{Bro10}).

\textbf{Initial state preparation} --- The first step of the
simulation is to initialize the quantum register to the state
$\ket{\psi(0)}$.
\begin{figure}
\includegraphics[width=0.5\textwidth ]{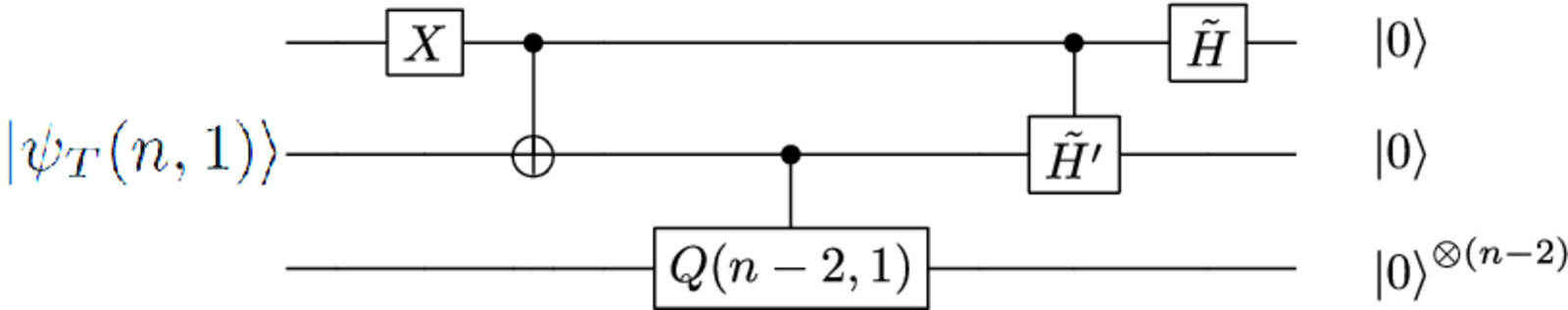}
\caption{\label{prep}Initial state preparation. Quantum circuit
for the recursive procedure used to find an efficient gate
sequence for preparing a given target state $\ket{\psi_T(n,1)}$ of
one electron occupying $n$ possible orbitals. The procedure uses
reverse engineering, where one considers the problem of
transforming the target state to the initial state $\ket{0}
^{\otimes n}$. This reverse problem allows an intuitive,
systematic solution. Once the solution of this inverse problem is
found, it can be inverted in order to prepare the target state
$\ket{\psi_T(n,1)}$ from the initial state $\ket{0}^{\otimes n}$.
The unitary operations $\tilde{H}$ and $\tilde{H}'$ can be calculated
easily from the given target state: each one of them transforms the
known state of the corresponding qubit to $\ket{0}$. The unitary
operation $Q(n-2,1)$ transforms $\ket{\psi_T(n-2,1)}$ into
$\ket{0}^{\otimes (n-2)}$ (adapted from \cite{Wan09}).}
\end{figure}
In many cases the preparation of the initial state is difficult
and an efficient algorithm may not be available. Fortunately, for
particular cases of interest efficient state preparation is
possible. For example, a method for generating a state that
encodes the antisymmetrized many-particle state of fermions
(including all the possible permutations), starting from an
unsymmetrized state (e.g.~$\ket{000 \cdots 0}$), with polynomial
resources was given in \cite{Abr97}. The preparation of
$N$-particle fermionic states of the form:
\begin{equation}
\ket{\psi(0)} = \prod_{j=1}^{N} b^{\dagger} _j \ket{\rm vac},
\end{equation}
where $\ket{\rm vac}$ is the vacuum state and $b^{\dagger}_j$ and
$b_j$ are the fermionic creation and annihilation operators, was
discussed in \cite{Ort01,Ort01a,Som02,Som03}. A quantum algorithm
for the efficient preparation of physically realistic quantum
states on a lattice (arbitrary pure or mixed many-particle states
with an arbitrary number of particles) was proposed in
\cite{War08}, while in \cite{Kas08} it was shown that most
commonly used chemical wave functions can be efficiently prepared.
A quantum algorithm for preparing a pure state of a molecular
system with a given number $m$ of electrons occupying a given
number $n$ of spin orbitals that exhibits polynomial scaling in
$m$ (regardless of $n$) was proposed in \cite{Wan09} (see Figure
\ref{prep}). In \cite{Wang11} a state-preparation algorithm that
incorporates quantum simulation was proposed: the time evolution
of the quantum system is simulated including the interaction with
ancilla, i.e.~auxiliary, qubits that can inject or absorb any
specified amount of energy from the system, thus preparing any
desired energy eigenstate.

\textbf{Unitary evolution} --- Let us now discuss in some more
detail how to obtain $U$. We assume that the Hamiltonian can be
written as a sum of many terms that describe local interactions:
\begin{equation}
H= \sum _{l=1}^{M} H_l.
\end{equation}
Examples of Hamiltonians of this form include the Hubbard and
Ising Hamiltonians. If $[H_l,H_{l'}]=0$ for all $l$ and $l'$, then
\begin{equation}
U =  \prod _l \exp\{-i\hbar H_l t\}.
\end{equation}
In this case, the decomposition of $U$ into a sequence of local
gates is straightforward. Unfortunately, in most cases of
practical interest $[H_l,H_{l'}]\neq 0$ in general. As a result,
when taken as a whole, the decomposition of $U$ cannot be obtained
efficiently using classical methods. An important step in this
regard is breaking up the evolution time into a large number of
small time steps of duration $\Delta t$ each:
\begin{equation}
U = \left( \exp\{-i\hbar H \Delta t\} \right)^{t/\Delta t}.
\end{equation}
There are approximations available for decomposing $\exp\{-i\hbar
H \Delta t\}$ into local gates. For example, the first-order
Trotter formula (see, e.g., \cite{NC,Ort01,Som02}) gives
\begin{equation}
U(\Delta t)=e^{-i\hbar \sum _l H_l \Delta t}=\prod _l e^{-i\hbar
H_l\Delta t}+{\cal{O}} ((\Delta t)^2).
\end{equation}
As a result, when $\Delta t \rightarrow 0$,
\begin{equation}
U(\Delta t) \approx \prod _l \exp\{-i\hbar H_l\Delta t\}.
\end{equation}
The drawback of this approach is that high accuracy comes at the
cost of very small $\Delta t$ and therefore a very large number of
quantum gates. Recent results have re-emphasized the shortcomings
of using this first-order Trotter formula \cite{Bro06,Cla09,Whi10},
showing that higher-order decompositions can be more efficient
(see, e.g., \cite{Dur08}). Recently quantum algorithms for
simulating time-dependent Hamiltonian evolutions on a quantum
computer have also been investigated \cite{Wie11}. The topic was
further discussed in \cite{Pou11} where it was shown that using
randomness, it is possible to efficiently simulate local bounded
Hamiltonians with arbitrary time dependence.
\par
\begin{figure}
\includegraphics[width=0.45\textwidth ]{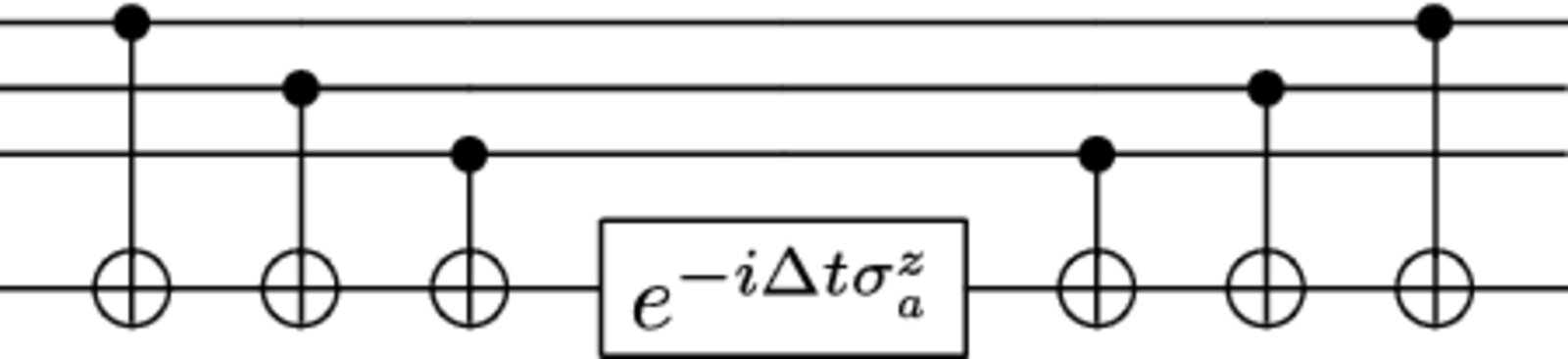}
\caption{\label{circuit1} Quantum circuit for simulating the
three-body Hamiltonian $H= \sigma_1 ^z \otimes \sigma_2 ^z \otimes
\sigma_3 ^z$. The circuit contains six CNOT gates and utilizes a
fourth, ancilla qubit (bottom line) in order to achieve the
desired effective Hamiltonian \cite{NC}.}
\end{figure}

Let us now consider an example of constructing rather complex
operations from simple quantum gates. Take the Hamiltonian
\begin{equation}
H= \sigma_1 ^z \otimes \sigma_2 ^z \otimes \cdots \otimes \sigma_N
^z,
\end{equation}
where $\sigma_i ^z$ is the Pauli matrix acting on spin (qubit)
$i$. Throughout the paper we denote by $\sigma_i ^\alpha$, with
$\alpha=x,y,z$, the corresponding Pauli matrix acting on spin
(qubit) $i$. The quantum circuit in Figure \ref{circuit1} realizes
the unitary transformation $U=\exp\{-i\hbar H t\}$ for $N=3$ \cite{NC}.
It is composed of six two-qubit (CNOT) gates and one single-qubit
gate. Note that an ancilla qubit is used. Similar quantum circuits
can be written for any product of Pauli matrices
\begin{equation}\label{Hpauli}
H= \otimes _{l=1} ^N \,\sigma_l ^\alpha.
\end{equation}
\par
Although the example above might look simple, the efficient simulation
of a general many-body interaction Hamiltonian using two-body
interactions is by no means easy \cite{Ben02,Nie02}. This question has
been thoroughly studied, and several methods have been developed (see, e.g.,
\cite{Dod02,Wan02,Woc02,Wocj02,Wo02,Bre05,Has06,Berr07,Bra08,Dur08,Bro11}),
but it still remains a difficult problem. Moreover, note that
ancilla qubits are required, which adds to the resource
requirements (see Section V. B).
\par
\begin{figure}
\includegraphics[width=0.45\textwidth ]{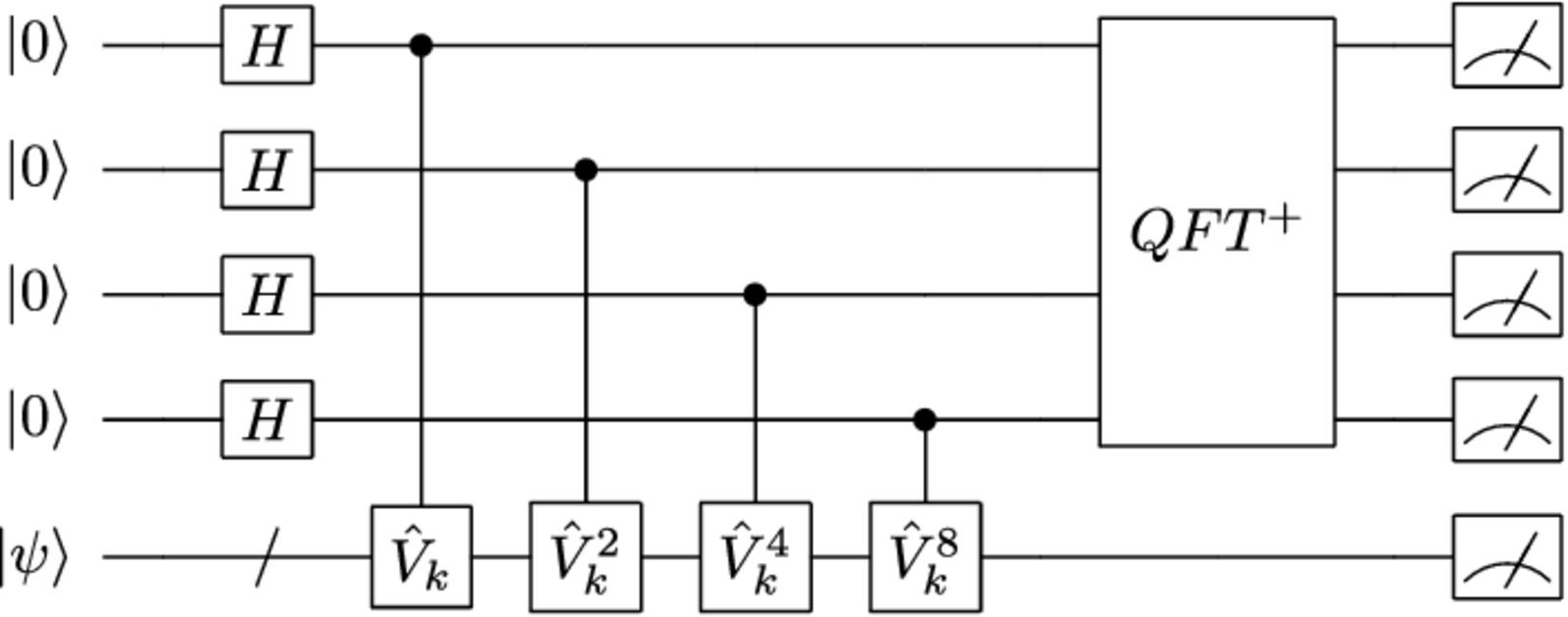}
\caption{\label{molecenerg}Quantum circuit for the calculation of
molecular energies in \cite{Guz05}. The circuit implements
the recursive phase estimation algorithm. The first iteration
gives the phase $\phi$ (which represents the molecular energy) to
four bits of accuracy. Each subsequent iteration incorporates the
previous estimate and increases the accuracy by one bit,
i.e.~reduces the uncertainty by a factor of 2. Here $H$ denotes the
Hadamard gate, $QFT^+$ is the inverse quantum Fourier transform
and $V_k=[\exp (-i2\pi\phi _{k-1})\hat{V}_{k-1}]^2$ (adapted from
\cite{Guz05}).}
\end{figure}

Let us now take a look at another example: the algorithm given in
\cite{Guz05} for the calculation of molecular energies using a
recursive phase-estimation algorithm. The quantum circuit is shown
in Figure \ref{molecenerg}. This procedure provides an arbitrarily
accurate estimate of the energy, with the accuracy increasing with
increasing number of iterations. The first iteration gives a rough
estimate for the energy. This estimate is then used as a reference
point for the next iteration, which yields a better estimate. The
procedure is repeated until the desired precision is obtained.
\par
So far, the literature has generally focused on the discrete
evolution of a quantum system, but recently, continuous evolution
has also been discussed \cite{Mck09,Bia11}. Furthermore, it is
usually assumed that there is no restriction in applying one- and
two-qubit gates and that all qubits of the simulator can be
individually addressed and measured. An interesting question is
what Hamiltonians can be simulated under certain control constraints.
For example, \cite{Kra07} discussed the class of Hamiltonians that can be
simulated when one is restricted to applying translationally invariant
Hamiltonians. The authors showed that if both local and
nearest-neighbor interactions are controllable, then the
simulation of interactions in quadratic fermionic and bosonic
systems is possible. However, for spins this is still an open
problem.

\textbf{Measurement} --- After obtaining $\ket{\psi(t)} = U
\ket{\psi(0)}$ via the unitary evolution, we need to perform the
final measurement in order to extract the desired information. In
general, for characterizing a quantum state, quantum state
tomography (QST) (e.g., \cite{Dar03}) can be used. However, QST
requires resources that grow exponentially with the size of the
system, making it inefficient for large quantum systems. In order
to avoid this problem, the direct estimation of certain physical
quantities such as correlation functions or spectra of operators
is more desirable than taking the long route through QST. A
detailed discussion is given in \cite{Ort01,Som02}.
\par
\begin{figure}
\includegraphics[width=0.3\textwidth ]{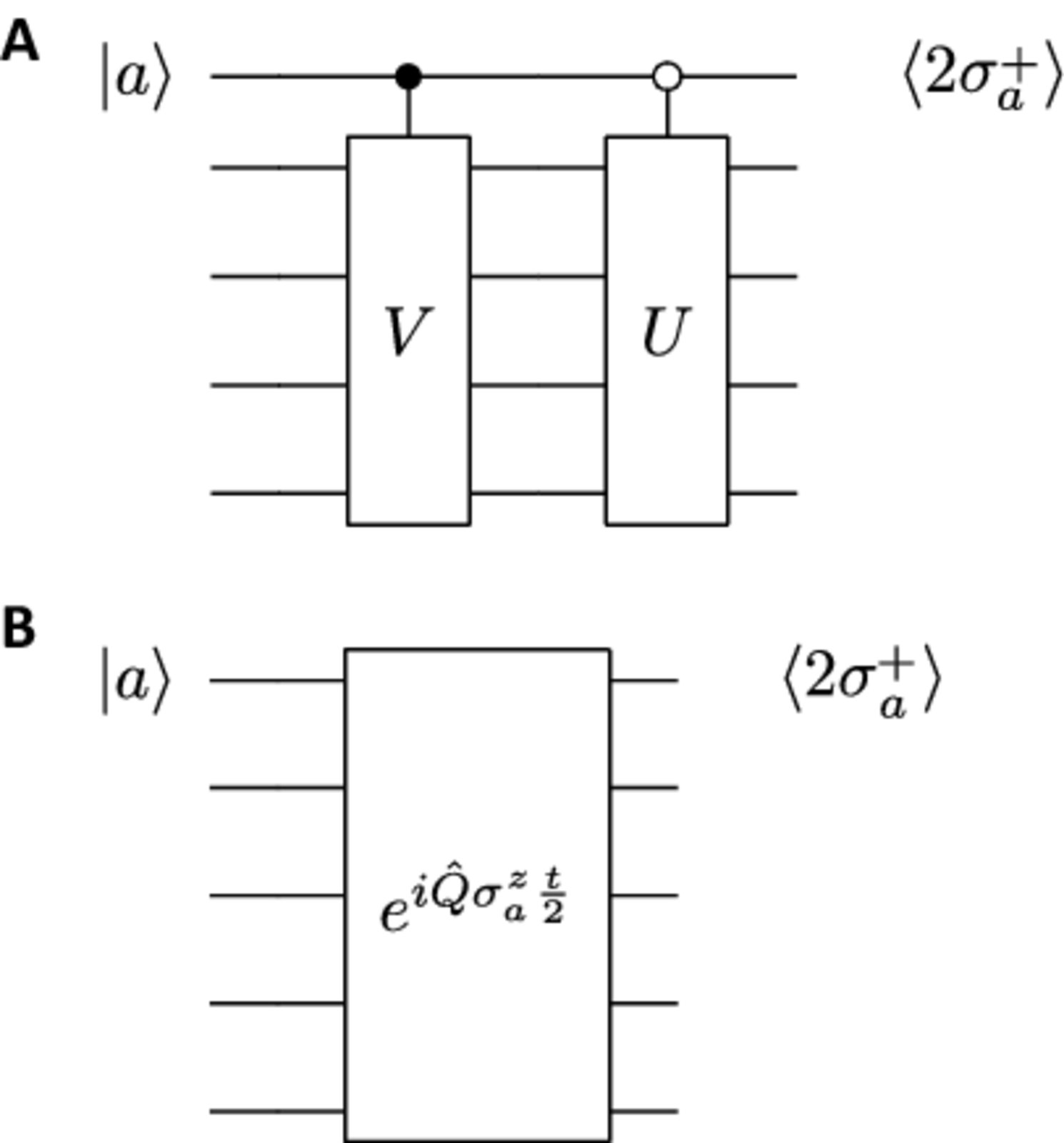}
\caption{\label{FigSomma}Quantum circuits (A) for the measurement of
the quantity $\langle U^{\dagger}V \rangle$ for two unitary
operators $U$ and $V$, and (B) for the measurement of the spectrum
of a Hermitian operator $\hat Q$. Both algorithms use one
ancilla qubit, which is initially prepared in the state $\ket{+} =
(\ket{0}+\ket{1})/\sqrt{2}$. The black dot represents a
$\ket{1}$-controlled gate and the white dot a $\ket{0}$-controlled
gate (adapted from \cite{Som02}).}
\end{figure}

We consider two examples. The first one refers to measurements
of quantities that can be written in the form $\langle
U^{\dagger}V \rangle$, where $U$ and $V$ are unitary operators.
The measurement circuit is shown in Figure \ref{FigSomma}. One
ancilla qubit that is initially in the state $\ket{+} =
(\ket{0}+\ket{1})/\sqrt{2}$ is needed. The desired quantity,
i.e.~$\langle U^{\dagger}V \rangle$, is given by the expectation
value $\langle 2\sigma_{+}^a\rangle$ of the ancilla at the end
of the simulation (Here $2\sigma_{+}^a = \sigma_{x}^a +i\sigma_{y}^a$).
The second example pertains to measuring the spectrum of
a Hermitian operator $\hat Q$. Again, one ancilla qubit that is
initially in the state $\ket{+} = (\ket{0}+\ket{1})/\sqrt{2}$ is
needed, and the desired spectrum is obtained by analyzing
the time dependence of $\langle 2\sigma_{+}^a\rangle$. The
measurement circuit is shown in Figure \ref{FigSomma}.

\vspace{15pt}

\subsection{Analog quantum simulation (AQS)}
 \begin{quote}
 \emph{``...there is to be an exact simulation, that the computer will do exactly the same as nature.''} \cite{feynman82}
 \end{quote}
Another approach to simulating quantum systems by quantum
mechanical means is analog quantum simulation (AQS), in which one
quantum system mimics (emulates) another (see, e.g.,
\cite{Wei97,Man02,Fis04,Por04,Zag07,Smi07}). The Hamiltonian of
the system to be simulated, $H_{\rm{sys}}$, is directly mapped
onto the Hamiltonian of the simulator, $H_{\rm{sim}}$, which can
be controlled at least to some extent:
\begin{equation}
\label{eqAQS} H_{\rm{sys}} \longleftrightarrow H_{\rm{sim}}.
\end{equation}
This can be done if there is a mapping between the system and the
simulator \cite{Som99}. Then $\ket{\phi (0)}$ can be mapped to
$\ket{\psi (0)}$ via an operator $f$
[$\ket{\psi (0)} = f \ket{\phi (0)}$], and $\ket{\psi (t)}$ can be
mapped back to $\ket{\phi (t)}$ via $f^{-1}$. For Hamiltonians
$H_{\rm{sim}} = fH_{\rm{sys}}f^{-1}$. Note that the simulator may
only partly reproduce the dynamics of the system. The choice of
the mapping depends on what needs to be simulated and on the
capabilities of the simulator. In AQS one is usually emulating an
effective many-body model of the simulated system. A controllable
``toy-model'' of the system is used to reproduce the property of
interest, e.g.~the dynamics or ground state.
\par
An important advantage of AQS is that it could be useful even in
the presence of errors, up to a certain tolerance level. For
example, one is sometimes interested in knowing whether a certain
set of physical conditions leads to a given quantum phase
transition. Even without having the full quantitative details, a
qualitative answer can be quite valuable in this context. If the
quantum simulator suffers from uncertainties in the control
parameters, the phase transition under study could still be
observed, hence providing the answer to the question of interest.
\par
Finding the mapping in an AQS might, at first glance, look
simpler than obtaining the most efficient gate decomposition for
a given Hamiltonian in DQS. Sometimes the mapping is
indeed straightforward, but this is not always the case, and quite
often clever mappings have to be devised, sometimes involving
additional externally applied fields or ancillary systems to
mediate various interactions.
\par
Let us now look at two examples of mappings between quantum
systems and the corresponding simulators. The first is the
Hamiltonian describing a gas of interacting bosonic atoms in
a periodic potential
\begin{equation}
\label{eqH} H_{\rm{sim}}=-J\sum_{i,j}\hat{a}_i^{\dagger} \hat{a}_j
+\sum_{i}\epsilon_i \hat{n}_i +
\frac{1}{2}U\sum_{i}\hat{n}_i(\hat{n}_i-1),
\end{equation}
where $\hat{a}_i^{\dagger}$ and $\hat{a}_i$ correspond to the
bosonic creation and  annihilation operators of atoms on the $i$th
lattice site, $ \hat{n}_i =\hat{a}_i^{\dagger} \hat{a}_i$ is the
atomic number operator counting the number of atoms on the $i$th
lattice site, and $\epsilon_i$ denotes the energy offset of the
$i$th lattice site due to an external confining potential. The
coefficient $J$ quantifies the hopping strength between lattice
sites, and $U$ quantifies the interaction strength between atoms
occupying the same lattice site. This Hamiltonian has a similar
form to the Bose-Hubbard Hamiltonian
\begin{equation}
\label{eqBH} H_{\rm{BH}}=-J\sum_{i,j}\hat{b}_i^{\dagger}
\hat{b}_j + \frac{1}{2}U\sum_{i}\hat{n}_i(\hat{n}_i-1)-\mu
\sum _i \hat{n}_i,
\end{equation}
where $J$ and $U$ are the same as above, and $\mu$ is the chemical
potential. The analog simulation of the Bose-Hubbard model using
atoms in optical lattices is therefore straightforward. However,
in other situations one must rewrite $H_{\rm{sim}}$ in order to
obtain the mapping $H_{\rm{sys}} \leftrightarrow H_{\rm{sim}}$.
For example, in the case of an array of Josephson junctions as
in \cite{Oud96} the system is described by the quantum phase model,
which can be connected to the Bose-Hubbard model via a mapping where
the field operators $\hat{a}_i$ are reformulated in terms of the
amplitude and phase of the superconducting order parameter at
different points in the circuit.
\par
The second example of a mapping between a quantum system and its
simulator is the trapped-ion simulation of the Dirac equation
\cite{Lam07,Ger10}. The Dirac equation in (1 + 1) dimensions for a
spin-1/2 particle with rest mass $m$ is
\begin{equation}
i\hbar\frac{\partial \phi}{\partial t}=H_D\phi=(c\hat{p}\sigma_x +
mc^2 \sigma_z)\phi,
\end{equation}
where $c$ is the speed of light, $\hat{p}$ is the momentum
operator, $\sigma_x$ and $\sigma_z$ are the Pauli matrices. The
Hamiltonian of a single trapped ion interacting with a bichromatic
light field can be written as
\begin{equation}
H_I=2\eta \Delta\bar{\Omega} \sigma_x \hat{p} +  \hbar \Omega
\sigma_z,
\end{equation}
where $\eta$ is the Lamb-Dicke parameter, $\Delta$ is the spatial
size of the ground-state wavefunction, and $\bar{\Omega}$ is
controlled via the intensity of the bichromatic light field. With
the identifications $c\equiv 2\eta \hat{\omega} \Delta$ and $mc^2
\equiv \hbar \Omega$, $H_I$ has the same form as $H_D$.
With this analogy, effects such as Zitterbewegung and the Klein
paradox can be studied in a non-relativistic quantum system
\cite{Ger10,Ger11}.
\par
In the following sections the Hamiltonians of several proposed
quantum simulators and those of the systems to be simulated will
be discussed in more detail and the relation in equation
\ref{eqAQS} will become clearer for each particular case.
\par
The initial-state preparation and measurement in AQS have not been
thoroughly discussed in the literature. Because the system and
simulator are presumed to be very similar, it is expected that the
preparation of the initial state can occur naturally in processes
mimicking the natural relaxation of the simulated system to an
equilibrium state. Moreover, directly measuring some physical
quantity of the simulator would yield information about its
analogue in the simulated system. In this sense, AQS has the
additional advantage that physical quantities can be measured
directly, without the need for computational manipulation of
measurement results as in DQS. Nevertheless, both the initial
state preparation and measurement process in AQS will need to be
studied in more detail as AQS becomes a widely used research tool.

\subsection{Quantum-information-inspired algorithms for the classical simulation of quantum systems}

In an interesting recent development, \textit{classical} numerical
algorithms for the simulation of quantum many-body systems came
out of research on quantum information theory (For detailed
studies on the subject, see e.g.~\cite{Ver04,Ver05,Vid08}).

As discussed in the previous sections, fully characterizing a
quantum system requires an exponentially large number of
parameters. It would be very useful if many-particle states could
be represented in such a way that some physical quantities could
be classically calculated in a more efficient way. In order to
achieve this goal, some techniques from quantum information theory
have been used in recently developed algorithms. The first steps
in this direction were taken in \cite{Ver04,Ver05} and thereafter
a significant effort has been made to explore this idea. Using
matrix product states (MPS) and projected entangled-pair states
(PEPS) one can simulate more efficiently infinite-size quantum
lattice systems in one and two dimensions. This new class of
algorithms makes it possible to simulate spin systems for longer
times, and to study physical phenomena which would have been
inaccessible with previous methods. Moreover, these methods can be
combined with Monte Carlo techniques. For more details we direct
the reader to the two reviews \cite{VMC08,Cir09} and the
references therein. Another widely used stochastic method is the
Metropolis algorithm. Its quantum version allows for the direct
sampling from the eigenstates of the Hamiltonian, overcoming in
this way the sign problem \cite{Tem11}.

\section{Resource estimation and fault tolerance}
\subsection{Resource estimation}

Using a quantum simulator instead of a classical computer does not
necessarily provide an efficient solution to the problem of
simulating quantum systems. This is because it is not always easy
to prepare the initial state, evolve it, and measure it with
polynomial resources. The amount of physical resources (i.e.,
number of qubits, number of operations, number of steps, etc.)
needed for the quantum simulation in the case of an $N$-body
problem strongly depends on the type of problem and the
particularities of the simulator. In this subsection, we review
some results on the estimation of the required resources for some
particular cases.
\par
How many particles or qubits are needed to realize useful quantum
simulations? The answer to this question depends on the type of
simulation one wishes to implement. As a general
rule of thumb, it is sometimes said that in order to outperform
classical computers quantum simulators require somewhere between
40 and 100 qubits \cite{Bul09}. However, there are some
interesting applications that could be realized with fewer qubits.
For instance, with ten or fewer qubits one could perform
proof-of-principle simulations, including the simulation of
frustrated spin systems \cite{Por06,Kim10,Ma11,Ma12}, quantum
chaos \cite{Wei02,How00}, some simple chemical reactions
\cite{Smi07}, Dirac particles \cite{Lam07,Ber07,Ger10}, the Unruh
effect \cite{Als05,Nat11}, or anyons \cite{Lu09,You08} (Note that
these few-qubit simulations can in principle be readily performed
on a present-day classical computer). With a few tens of qubits,
one could perform frustrated-spin simulations or molecular-energy
calculations at the limits of present-day supercomputers.
\par
\begin{figure}
\includegraphics[width=0.5\textwidth ]{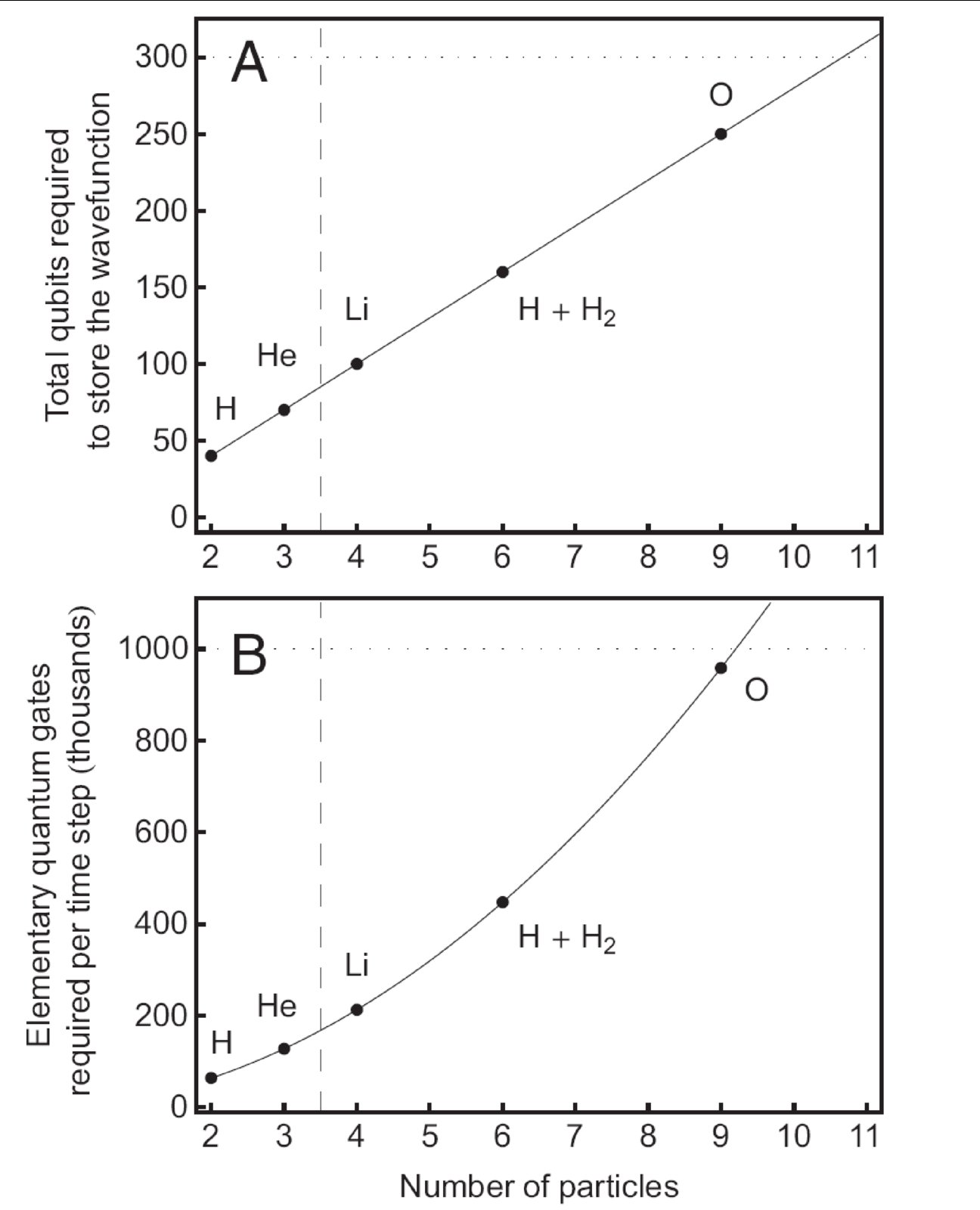}
\caption{\label{estim} Resource requirements for the quantum
simulation of the dynamics of $N$ particles interacting through a
pairwise potential, maintaining a relatively small error level.
The chemical symbols are a guide to show what type of problem can
be simulated with a given computation size. The vertical dashed
line represents the current limit of numerically-exact quantum
simulations on classical computers on a grid. (reproduced from
\cite{Kas08}; Copyright (2008) National Academy of Sciences, U.S.A.).}
\end{figure}

There have been rather extensive studies on the resource estimation
for DQS. The estimation of the requirements for simulating $N$ particles
interacting through a pairwise potential has been performed in
\cite{Kas08}. The results are reproduced in Figure \ref{estim}. A
discrete-variable representation of the wave function in an $n$
qubit basis is used. Furthermore, a number $m$ of ancilla qubits
is required to represent the desired range of potential values
with a certain precision, four of which give a reasonably high
accuracy for the Coulomb potential. This gives a total of $n(3N -
6) + 4m$ qubits. The Coulomb potential can be evaluated in
${\cal{O} }{(N^2m^2)}$ steps, so chemical dynamics could be
simulated on a quantum computer in ${\cal{O} }{(N^2m^2)}$ steps,
which is exponentially faster than known classical algorithms.
However, from the graph in Figure \ref{estim}, it follows that in
order to outperform current classical computers at least 100
qubits and over 200,000 quantum gates per step would be required.
\par
Compared with the studies on scaling with the system size, less
attention has been paid in the literature to the scaling of
required resources with desired accuracy. For example, if one
considers a case where increasing the accuracy of the answer
(i.e., the desired number of bits in the final answer) leads to an
exponential increase in the number of quantum gates, it is not
obvious that the quantum simulation can be called efficient.
Indeed, it was pointed out in \cite{Bro06} that several current
algorithms for quantum simulation exhibit poor scaling as a
function of desired accuracy, even if they seem efficient based on
the scaling with system size. One should also note here that, at
first sight, one might think that making the step size for the
Trotter decomposition in DQS smaller does not affect the total run
time of the algorithm, because the gates can be implemented more
quickly for small time steps. However, there is typically an
overhead that is proportional to the number of gates that need to
be implemented, and this number can increase rapidly with
decreasing step size. The precision requirements in a given
quantum simulation is therefore an important question for purposes
of resource estimation.
\par
Recently, the resource requirements (total number of physical
qubits and computation time) for computing the ground state
energy of the one-dimensional quantum transverse Ising model with
$N$ spin-1/2 particles, as functions of the system size and the
numerical precision, were investigated in \cite{Cla09}. The
quantum circuit was decomposed into fault-tolerant operations, and
the total number of qubits and the total number of steps were
estimated as functions of the desired precision. The authors found
that the computation time grows exponentially with desired
precision. In order to obtain polynomial scaling, new quantum
simulation algorithms are needed. Alternatively, systems where the
phase estimation algorithm can be implemented without the Trotter
formula \cite{Cla09} could be used.
\par
A related recent study \cite{You13} compared the resource
requirements for two alternatives that could be used in
fault-tolerant DQS: topologically protected surface codes and
circuit models with quantum error correction. By analyzing the
Ising model as a representative example, and using parameters
that are relevant to present-day experiments, the authors
concluded that surface codes are superior for quantum simulation.
\begin{figure}
\includegraphics[width=0.5\textwidth ]{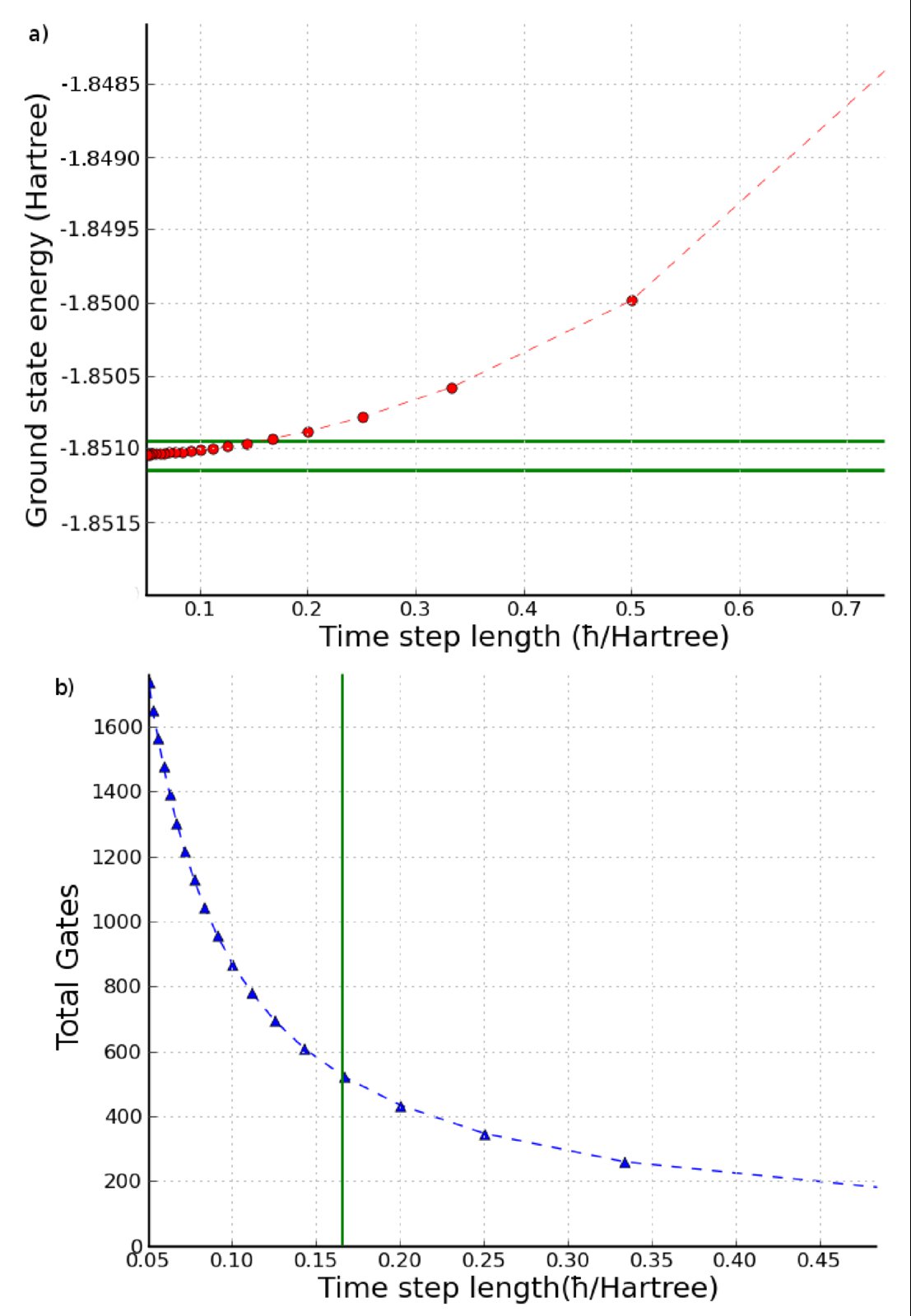}
\caption{\label{mol} (color online) Trotter error analysis and
gate count for a simulation of the hydrogen molecule using a DQS
algorithm. a) The calculated ground state energy of the
hydrogen molecule as a function of the time step duration
$\Delta t$.  The green horizontal lines indicate the bounds for
$\pm10^{-4}E_h$ precision. (b) Total number of gates for a single
construction of the approximate unitary as a function of $\Delta t$.
(reproduced from \cite{Lan09} supplementary material)}
\end{figure}
\par
Another example of resource estimation for a DQS implementation is
given in \cite{Lan09}. The results are reproduced in Figure
\ref{mol}, where the error in the ground state energy as a
function of the time step duration $\Delta t$ is shown. The ground state
energies were obtained via direct diagonalization on a classical
computer. A precision of $\pm10^{-4}E_h$, where $E_h\approx 27.21$
eV, is achieved for about 522 gates. The gate count includes both
one- and two-qubit operations and the estimate does not take into
consideration error correction for the qubits. An extended
discussion of the resource estimation for such molecular energy
simulations is provided in \cite{Whi10}.

There has not been much work in the literature on resource
estimation for AQS. However, it is sometimes said that AQS has
less stringent resource requirements in order to produce useful
results that are intractable for classical computational methods.
This statement does not necessarily imply that few particles are
sufficient in order to obtain results in AQS, but rather that
large numbers of particles could be collectively manipulated in
AQS using a small number of controls. For example, in \cite{Gre02}
hundreds of thousands of atoms were trapped using three laser
beams.

\subsection{\label{deco}Decoherence and errors}

Although quantum simulators are affected by the interactions with
the environment in the same way as quantum computers, it is
generally believed that the effects of decoherence are less
dramatic. This is most clearly seen with AQS, where only limited
precision (or just a qualitative answer) might be required. As a
result, a few imperfections in an ensemble of particles performing
AQS might not affect the overall behavior of the ensemble, such
that the AQS might still produce useful results even in the
presence of these imperfections. Moreover, it has been suggested
that the decoherence of the simulator might be useful
\cite{lloyd96} as it could serve as a rough way of modeling the
decoherence of the simulated system. A simple argument could go as
follows: if the noise level that is naturally present in the
simulator is lower than the noise level in the simulated system,
then it is rather straightforward to artificially supplement noise
in the simulator so that the combined noise in the simulator
faithfully mimics that present in the simulated system. This idea
has in fact been demonstrated recently in experiment \cite{Li13}.
More sophisticated methods of dealing with noise are also possible.
In \cite{Tse00} it was demonstrated, through calculations and a nuclear
magnetic resonance (NMR) experiment, that in the quantum simulation
of open systems it is possible to exploit the natural decoherence of
the simulator by varying the choice of mapping between the simulated
system and the simulator. In principle, one could characterize how
decoherence affects a simulation. Then, by an appropriate choice of
the mapping between the system and simulator, one may take advantage
of the natural symmetries in order to modify the effective
decoherence of the simulator. It was also suggested \cite{Tse00}
that it should be possible to simplify decoherence effects in a
simulation within certain subspaces. Decoherence may also provide
a  useful tool for extracting information about a critical system
(spectral structure or critical point of its quantum phase
transition) as suggested in \cite{Cuc07}. This idea has been
investigated in an NMR setting \cite{Zhan08} with the simulation
of the Ising Hamiltonian.
\par
Unfortunately, there are certain limitations and the inclusion of
the simulator's decoherence in the simulation must be carefully
considered. The interaction between the system and the environment
could be qualitatively different from that between the simulator
and its environment \cite{Bro07}. For example, when simulating
spin Hamiltonians with degenerate ground states using trapped
ions, spontaneous emission of the ions drives the system to states
outside the Hilbert space used in the system-simulator mapping
\cite{Bro07}. This shows that one should be cautious when trying
to include decoherence in the simulation. First, one needs to
understand how decoherence will affect the simulation and,
whenever possible, find clever mappings in order to take advantage
of the uncontrollable properties of the simulator. Second, one
needs to pay attention to the way the system and simulator are
described. It is therefore necessary to pay more attention to the
role of errors in AQS than has been done so far in the literature.
Note that the simulation of open quantum systems does not
necessarily require the inclusion of the decoherence of the
simulator \cite{Sch01,Pii06}. The ideal situation therefore
remains that uncontrollable errors should be minimized as much as
possible.
\par
In \cite{Bro06} a detailed study of the algorithm for finding the
low-lying spectrum of a pairing Hamiltonian was conducted in an
NMR implementation. Such simulations were found to be sensitive to
systematic errors in the applied Hamiltonian and fault-tolerant
implementations to be inefficient with respect to precision in the
current Trotter approximation methods. Other studies have shown
that for simulating the Schr\"odinger equation the minimization
of amplitude errors would be required \cite{Str02}. In
\cite{Mon04} the two-qubit entanglement in a simulation of a
dynamically localized system was found to be exponentially
sensitive both to small changes of the Hamiltonian and to the
locations of the chosen qubits. This sensitivity is due to the
natural ordering introduced on the qubit by the coding of the
simulated system. More recently, there have been studies
\cite{Dur08} on the effect of noise (timing errors in pairwise
interactions and noisy pairwise interactions described by master
equations of Lindblad form) in two-body interactions and local
control operations used for the simulation of many-body
interaction Hamiltonians.
\par
Further problems may arise for each physical implementation. In
this context, the specific limitations of each system should be
considered in more detail. So far there are few studies
investigating how the simulator's imperfections affect the quantum
simulation (e.g., in trapped ions \cite{Por04,Por06,Bul08}).
\par
Recently, the reliability, complexity and efficiency of analog
quantum simulations have been considered in more detail than in
past studies \cite{Hau12}. Reliability refers to the need to
ensure that the results of the simulation faithfully reflect the
simulated system. Cross-validation over a number of different
physical systems could be used, and in this way the particular
imperfections of each implementation could be ruled out as
possible sources of error. However, this approach is limited as
implementations in different systems are not always available. The
quantum simulation results can also be validated against
analytical and numerical predictions, but this is possible only
for small systems. Complexity/efficiency refers to the requirement
that the quantum simulator is able to solve problems that cannot
be solved on a classical computer in polynomial time (i.e., the
simulator is more efficient than a classical computer). Note that
in the case of the quantum simulation of \textit{experimentally
challenging} problems (see Section VII) this is not a necessary
requirement. Disorder, noise and other imperfections might affect
the reliability of the quantum simulation \cite{Hau12}. This issue
is illustrated in the case of a disordered quantum spin chain
where strong disorder introduces large errors.

\section{Physical realizations}

\begin{figure}
\includegraphics[width=0.51\textwidth ]{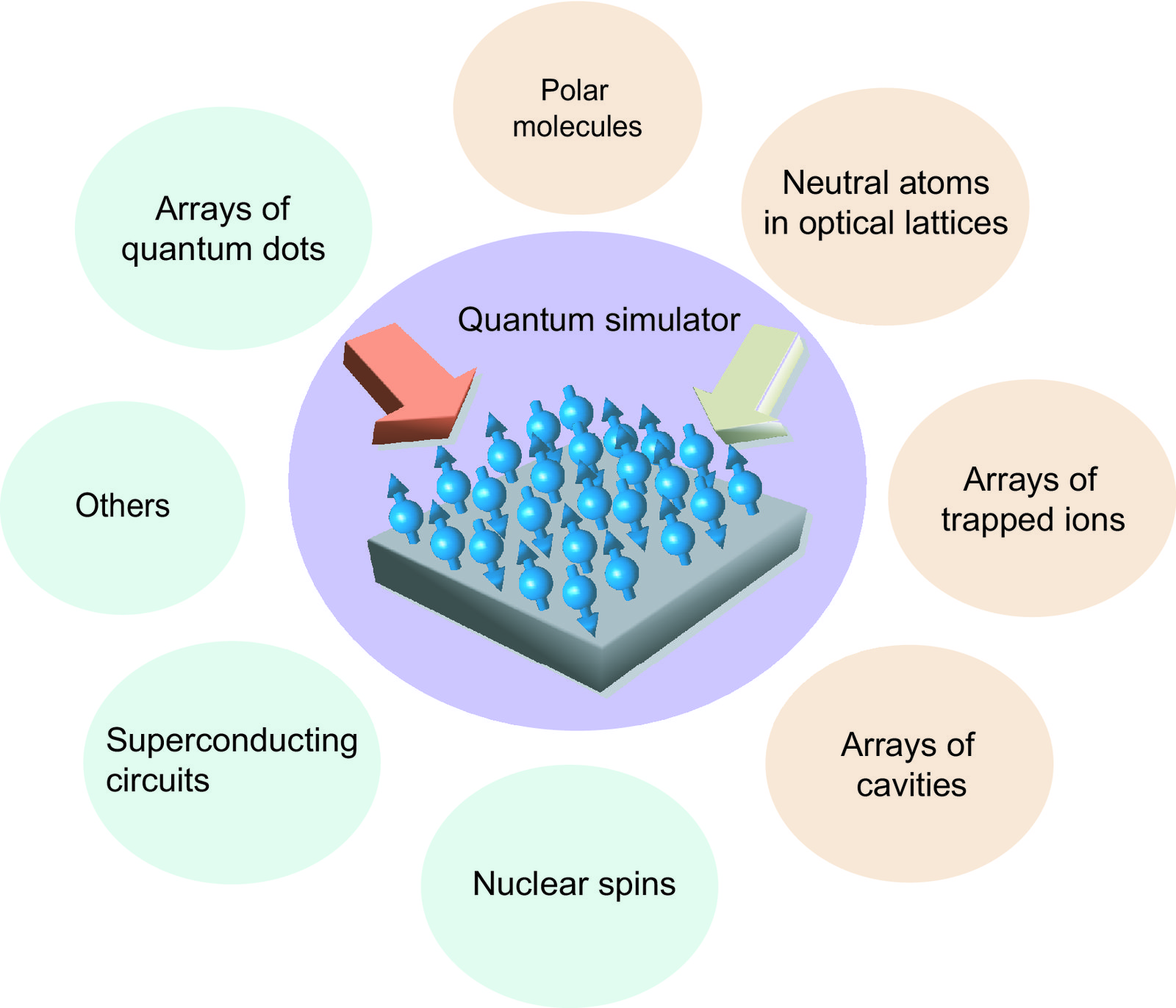}
\caption{\label{simulators} (color online) Different systems that
could implement a specialized quantum simulator for the study of
problems in condensed-matter physics. Examples of such analog
quantum simulators include: atoms, ions, photons, nuclear and
electronic spins, as well as superconducting circuits. These
systems could be designed such that they form one- or
two-dimensional arrays of qubits that can be manipulated in
different manners. They can be thought of as toy-models of the
magnified lattice structure of a ``solid'', with a magnification
factor of a few orders of magnitude.}
\end{figure}

The physical implementation of a quantum simulator requires a
controllable quantum mechanical system. Any physical system that
can be used as a quantum computer would also be a universal machine for DQS.
Possible routes and experimental progress towards building a
quantum computer have been thoroughly discussed in the last decade
(see, e.g., \cite{Bul10,Lad10,QCD,SS08,SW08} and references
therein). However, a quantum system that is not a potential
quantum computer could still implement AQS. For instance, the
propagation of sound waves in a two-component BEC was proposed for
the study of cosmic inflation \cite{Fis04}, and a rotating Fermi
gas could be used to understand nuclear physics phenomena
\cite{Nuclear}. We will not discuss the physical realization of
such highly specialized quantum simulators, but focus on a more
widely studied design of a quantum simulator: quantum simulators
for various models in condensed matter physics. For many problems
in this class, an array of qubits plus their controls (see Figure
\ref{simulators}) would make an ideal quantum simulator because it
can be thought of as the simplified, magnified lattice structure
of a ``solid'', that can be manipulated in a number of different
ways in order to test various models. Each qubit resides in its
own potential energy well and is used to encode a spin 1/2
particle. The array is configurable in the sense that its
dimensionality and geometry can be changed. Such an array could be
realized, for example, with atoms in optical lattices
\cite{Gre08}, atoms in arrays of cavities
\cite{Ang07,Har06,Gre06,Bra07}, ions either in microtrap arrays
\cite{Chi08,Cla08,Sch09} or in two-dimensional crystals
\cite{Por06}, electrons in arrays of quantum dots
\cite{Man02,Byr07,Byr08}, and so on. The desired evolution of the
system would be induced by the simulator's control fields.
This can either directly realize the desired Hamiltonian (AQS) or
reconstruct it out of elementary one- and two-qubit gates (DQS).
The control can be applied individually or to the entire array. In
this section, we will look at different physical systems and describe
how the array and controls can be realized experimentally. For a
recent review of the state-of-the-art capabilities of the physical
systems that we consider here see \cite{Bul10,Lad10}.

\subsection{\label{AtomsAndIons}Atoms and ions}

\textbf{Neutral atoms} in optical lattices are very well suited
for mimicking solid-state systems. Indeed, optical lattices
provide the highly desirable properties of being easily tunable
and essentially defect-free. The optical potentials can be
adjusted to allow the change of the geometry and dimensionality of
the lattice (e.g., triangular lattice \cite{Stu11} and Kagome lattice
\cite{Liu10}, etc.). Interestingly, the optical potential can be
tuned \textit{in situ} rather easily via the intensity, frequency
or phase of the applied lasers.
\par
Since the first experiment on the simulation of the quantum phase
transition from a superfluid to a Mott insulator using a cold
atomic gas in an optical lattice \cite{Gre02}, there has been
increasing interest in the study of condensed matter physics with
atoms in optical lattices. A theoretical review \cite{Lew07}
discusses in detail atoms in optical lattices as potential quantum
simulators, providing various examples of quantum systems that
could be simulated. Other reviews \cite{Blo08,Blo12} describe
recent experimental progress.
\par
Atoms in optical lattices are flexible systems with several
controllable parameters: tunneling strength, on-site,
nearest-neighbor, long-range and multiparticle interactions,
nonuniform potentials and coupling between internal quantum
states. Furthermore, there are both bosonic and fermionic elements
that can be used for quantum simulation with atoms in optical
lattices. A rather general type of Hubbard Hamiltonian that can be
realized in these systems is:
\begin{equation}
H=H_{\rm{hop}}+H_{\rm{interaction}}+H_{\rm{pot}}+H_{\rm{internal}},
\end{equation}
where $H_{\rm{hop}}$ describes the tunneling of atoms from one
lattice site to another, $H_{\rm{interaction}}$ is the interaction part,
$H_{\rm{pot}}$ combines all the effects of the nonuniform
potentials felt by the atoms, and $H_{\rm{internal}}$ describes
coherent on-site transitions between the internal levels
of the atoms.
\par
A quantum simulation of the Mott insulator-superfluid phase
transition in the Hubbard model can be realized by adjusting the
depth of the optical lattice, which mainly modifies the tunneling
strength and to a lesser extent modifies the on-site interaction
strength, or by controlling the on-site interactions via Feschbach
resonances \cite{Lew07}(see Section VII.A.1 on the simulation of
the Hubbard model).
\par
By tuning interatomic interactions using Feshbach resonances, it
was possible to investigate the crossover from a
Bardeen-Cooper-Schrieffer (BCS) state of weakly attractive
fermions to a Bose-Einstein condensate (BEC) of tightly bound
fermion pairs \cite{Reg04,Zwi05}. The continuous tunability of the
interaction strength also allowed access to the so-called
unitarity regime \cite{Oha02}, where the interaction strength is
comparable to the Fermi energy, meaning that there is a single
energy scale in the problem. This regime was previously
inaccessible and served as one more example demonstrating the
power of atoms as quantum simulators.
\par
Using laser-assisted tunneling and lattice tilting, \cite{Sim11}
achieved the simulation of an antiferromagnetically coupled spin
chain in an external magnetic field. In this simulation the
occupation of lattice sites was mapped onto the spin states of a
quantum magnet. In particular, a pair of neighboring lattice
sites sharing a single particle are mapped onto a (static) spin-1/2
particle in the quantum magnet. It should also be possible to
utilize the intrinsic spins of atoms in optical lattices for this
purpose. However, no such simulations have been performed to date.
\par
Another recently emerging direction for quantum simulation using
atomic gases is the simulation of artificial gauge fields
\cite{dal11}. At a basic level, an overall rotation of the
trapping potential can be used to simulate a magnetic field
(for the orbital degree of freedom). More intricate techniques
that rely on additional lasers have been devised in the past
few years for the simulation of various types of gauge fields,
allowing the observation of spin-orbit coupling in a BEC
\cite{LJS11}.
\par
Currently, addressing individual atoms in optical lattices is
difficult because the separation between neighboring lattice sites
is comparable to the best achievable focusing widths of laser
beams (both typically being 0.5-0.8 $\mu$m), but recent progress
suggests that there may be methods for overcoming this difficulty
\cite{Nel07,Bak09,Wur09,She10,Gib10,Fuh10,Bak10,Weit11}.
\par
\begin{figure}
\includegraphics[width=0.5\textwidth ]{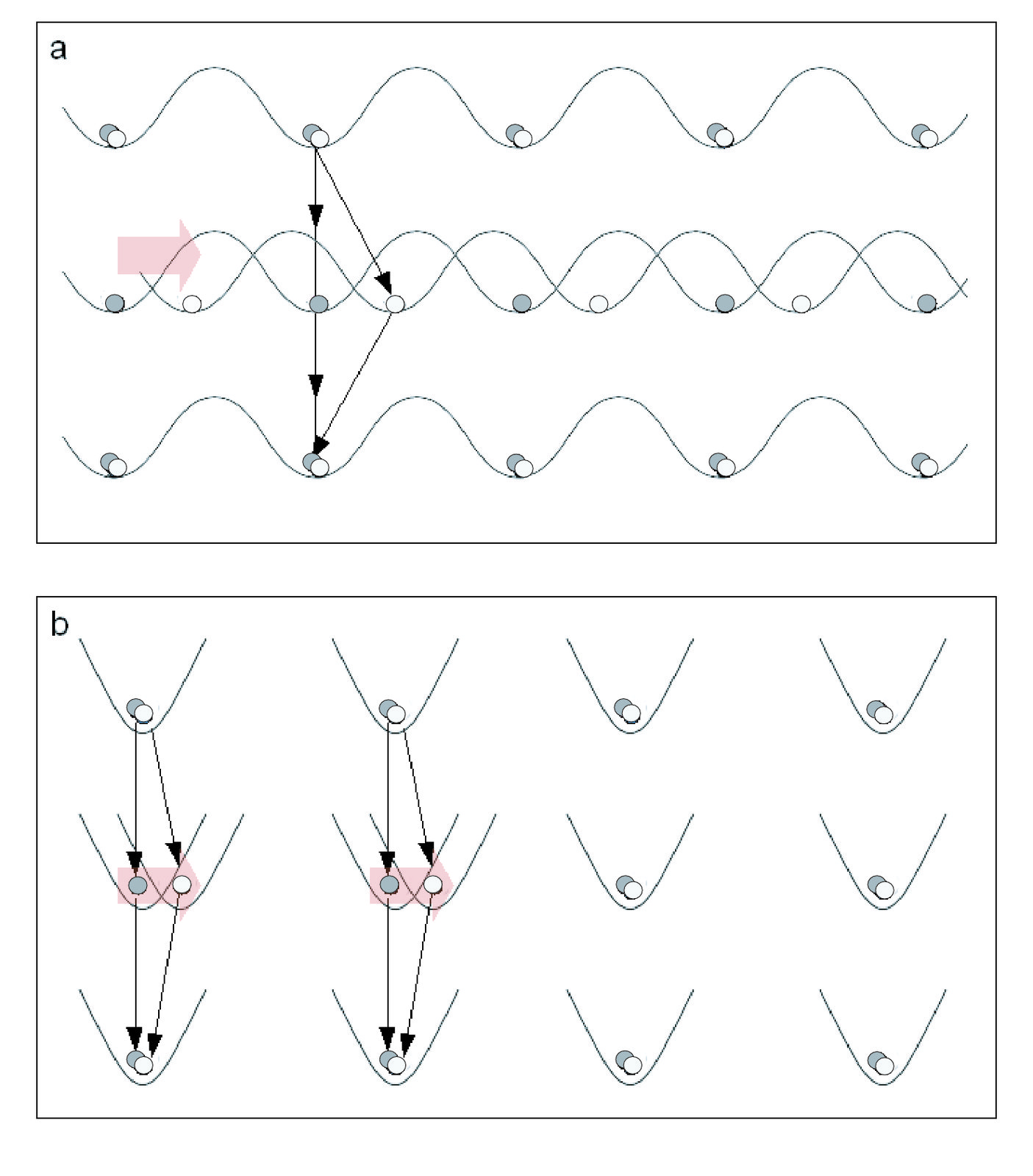}
\caption{\label{comp} (color online) Manipulation of atoms and
ions as proposed in \cite{Jan03}: (a) Neutral atoms in a double
optical potential: the interaction between neighboring atoms is
realized by displacing one of the lattices with respect to the
other. Gray circles denote the state $\ket{1}$ and white ones
denote the state $\ket{0}$. (b) Ions in independent trapping
potentials: the interaction between two neighboring ions is
achieved by conditionally displacing the corresponding ions
with a state-dependent force. Blue circles denote the state
$\ket{1}$ and white ones denote the state $\ket{0}$ (adapted
from \cite{Jan03}).}
\end{figure}

Atoms could also be used for DQS. One possible method for
implementing conditional quantum operations on atoms in optical
lattices \cite{Jan03} is schematically illustrated in Figure
\ref{comp}. Two optical lattice potentials are applied to the
atomic ensembles, one for each of the two internal atomic states
(which represent the qubit states). The interaction between
neighboring atoms is realized by displacing one of the lattices
with respect to the other [Figure \ref{comp} (a)]. With a
sufficiently large relative displacement of the two lattices,
interactions between more distant atoms can be achieved. Moreover,
thanks to their weak interaction with the environment, neutral
atoms have long decay times of the order of seconds.
\par
Alternative systems that can be used for quantum simulation
include \textbf{Rydberg atoms} \cite{Wei10} and
\textbf{polar molecules} \cite{Mic06,Lew06}. In the case of
Rydberg atoms in optical lattices or magnetic traps the lattice
spacing is $\sim\mu$m or higher, allowing single-site
addressability. Furthermore, dipole-dipole and van der Waals
interactions offer a means for implementing effective spin-spin
interactions.

\begin{figure}
\includegraphics[width=0.5\textwidth ]{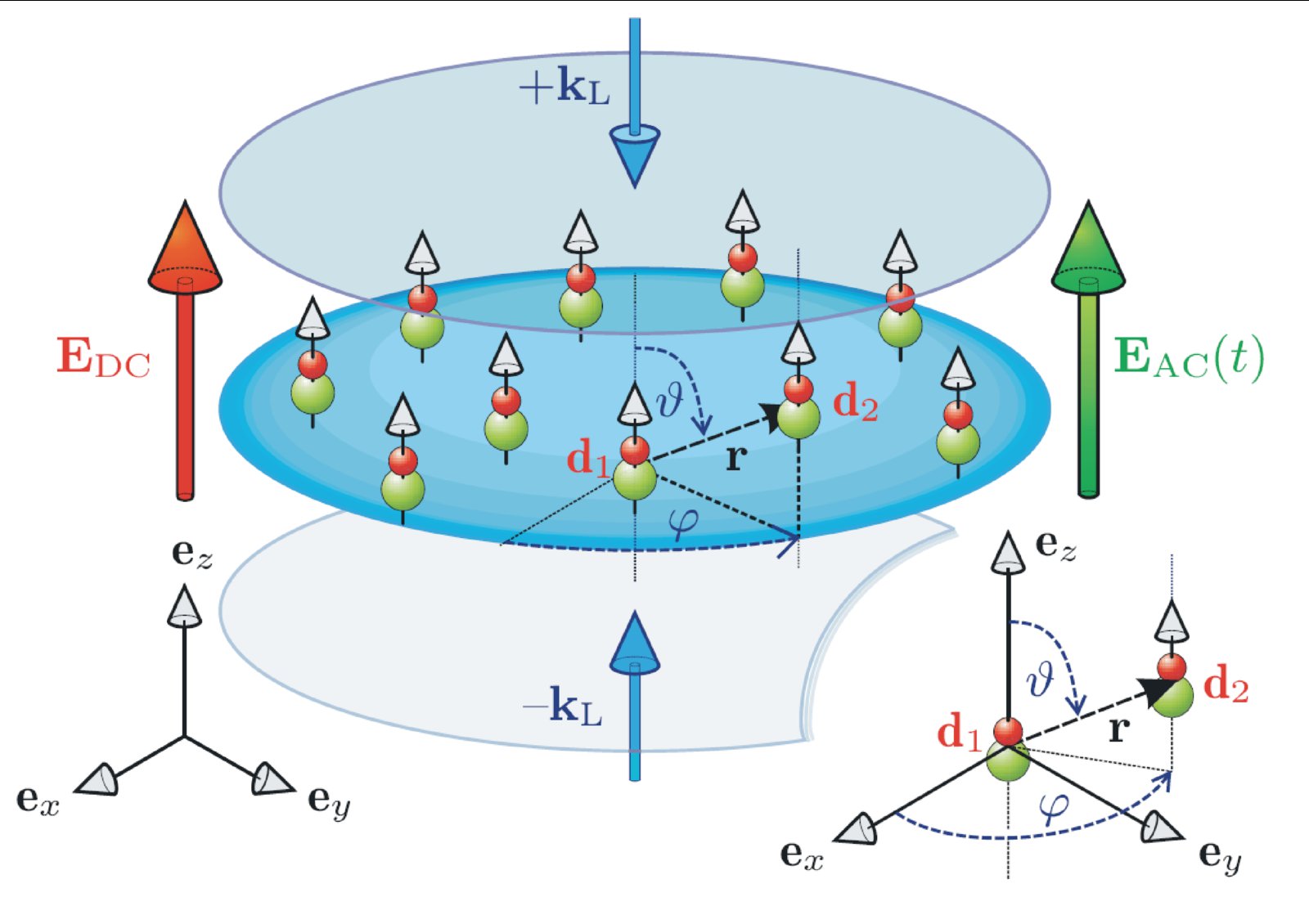}
\caption{\label{pm} (color online) Polar molecules trapped in a
plane by an optical potential created by two counter-propagating
laser beams with wavevectors indicated by blue arrows. The dipoles
are aligned perpendicular to the plane by a dc electric field,
indicated by the red arrow. The green arrow represents the ac
microwave field (reproduced from \cite{Pup08}).}
\end{figure}

With polar molecules \cite{Pup08} (see Figure \ref{pm}) microwave
excitations, dipole-dipole interactions, and spin-rotation
couplings, provide a universal toolbox for effective spin models.
The advantage of using polar molecules is that their large
electric dipole moments produce strong dipole-dipole interactions
that can be manipulated relatively easily via external dc and ac
microwave fields. This control may be used to study
strongly-correlated systems. Furthermore, extended Hubbard models
\cite{Ort09}, quantum phase transitions \cite{Cap10} and the
supersolid phase in a triangular lattice \cite{Pol10} could also
be simulated with these systems.

\textbf{Ions} can be trapped by electric (or magnetic) fields,
laser-cooled and manipulated with high precision for realizing
quantum simulation \cite{Sch04,SchFri08,Bla08,Jo09}. In fact,
one of the earliest theoretical studies on the physical
implementation of quantum simulation dealt with trapped ions
\cite{Win98}. Both the internal energy levels and the vibrational
modes of the trapped ions can be exploited for encoding quantum
information. In contrast with neutral atoms, which interact
weakly with each other, ions, being charged, interact rather
strongly via Coulomb repulsion. This facilitates the implementation
of two-qubit gates and the control of the qubit positions and motion.
Ion qubits also have long coherence times of the order of seconds,
and sequences of high-fidelity quantum gates have been demonstrated
in experiment \cite{Hann09,Lan11}.
\par
The quantum states of trapped ions are typically manipulated by
either resonantly driving transitions between different internal
states of the ions or resonantly driving sideband transitions
involving the internal states and the vibrational states of the
ions in the external trapping potential. For example, the Hamiltonian
describing the coupling between the internal and vibrational modes
due to the laser driving at the red-sideband frequency can be
written in the form:
\begin{equation}
H=i\hbar\eta\Omega[\exp(i\phi)\sigma_{+}a-\exp(-i\phi)\sigma_{-}a^{\dagger}],
\label{inthamilt}
\end{equation}
where $\Omega$ is the Rabi frequency of the transition between the
internal states, $\sigma_{+}$ and $\sigma_{-}$ are the two-level
atom transition operators, $\eta$ is the Lamb-Dicke parameter
(which is assumed to be small here), $a^\dagger$ and $a$ are the
creation and annihilation operators of the vibrational mode, and
$\phi$ is the laser phase. Using this Hamiltonian, as well as
those corresponding to blue-sideband driving and to resonant
driving of the ionic internal states, a variety of effective
Hamiltonians for AQS or quantum gates for DQS can be realized. The
high-fidelity one-, two-, and even three-qubit (Toffoli) gates
implemented with trapped ions have resulted in the most advanced
implementations of DQS to date \cite{Lan11,Bar11}. Analog quantum
simulations of frustrated spin systems \cite{Kim10} and of
relativistic single-particle motion \cite{Ger11} have also been
demonstrated recently.
\begin{figure}
\includegraphics[width=0.5\textwidth ]{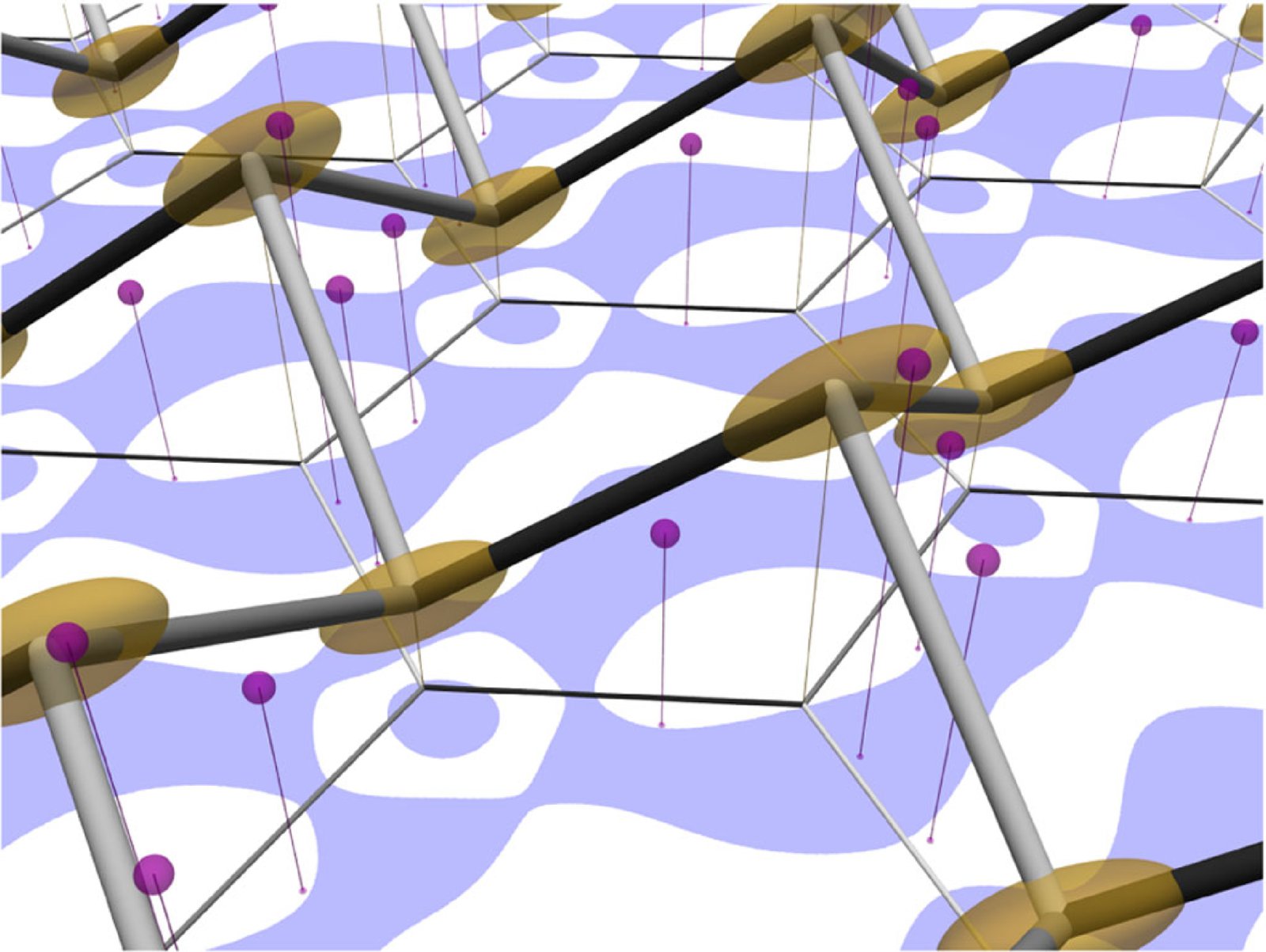}
\caption{\label{ions} (color online) Arrangement of electrodes for
the realization of a bilayer honeycomb lattice. The micro-trapping
regions are shown as ellipsoids, while the locations of unwanted
spurious microtraps are indicated with small spheres (reproduced
from \cite{Sch09}).}
\end{figure}
\par
Ions have generally been trapped using linear harmonic traps in
quantum-simulation experiments to date. It is possible, however,
to obtain different arrangements of many ions in anharmonic
one-dimensional traps (long ion strings \cite{Lin09}),
two-dimensional traps (planar crystals \cite{Por06,Bul08,Bie09} or
arrays of microtraps \cite{Chi08,Cla08,Sch09,Sch10,Lau12}) or
three-dimensional traps. It would also be possible to combine
trapped ions with optical lattices, as suggested in \cite{Sch08}.
With optimized electrode structures, various microtrap arrays
where ions are arranged in different lattice configurations (see
Figure \ref{ions}), can be constructed \cite{Sch09,Schm11}. In
fact a two-dimensional ion array was recently used to implement a
quantum simulation of a spin system with hundreds of ions
\cite{Bri12}.
\par
There are also alternative ways to generate two-qubit interactions
between ions. One such possibility is to use a state-dependent
displacement, which can be implemented by applying optical dipole
forces \cite{Bla08}. This method is particularly useful for ions
trapped in different harmonic potentials (e.g., in arrays of
microtraps), but in most of the experiments done to date it has
been realized with ions in the same potential. In Figure
\ref{comp} (b), the manipulation of ions in a one-dimensional
array of microtraps is depicted schematically. Two-qubit
interactions are usually realized with optical forces, but a
method for laserless simulation (avoiding the problem of
scattering) with ions in arrays of microtraps has been proposed in
\cite{Chi08} and very recently demonstrated in \cite{Osp11,Tim11}.

\begin{figure}
\includegraphics[width=0.4\textwidth ]{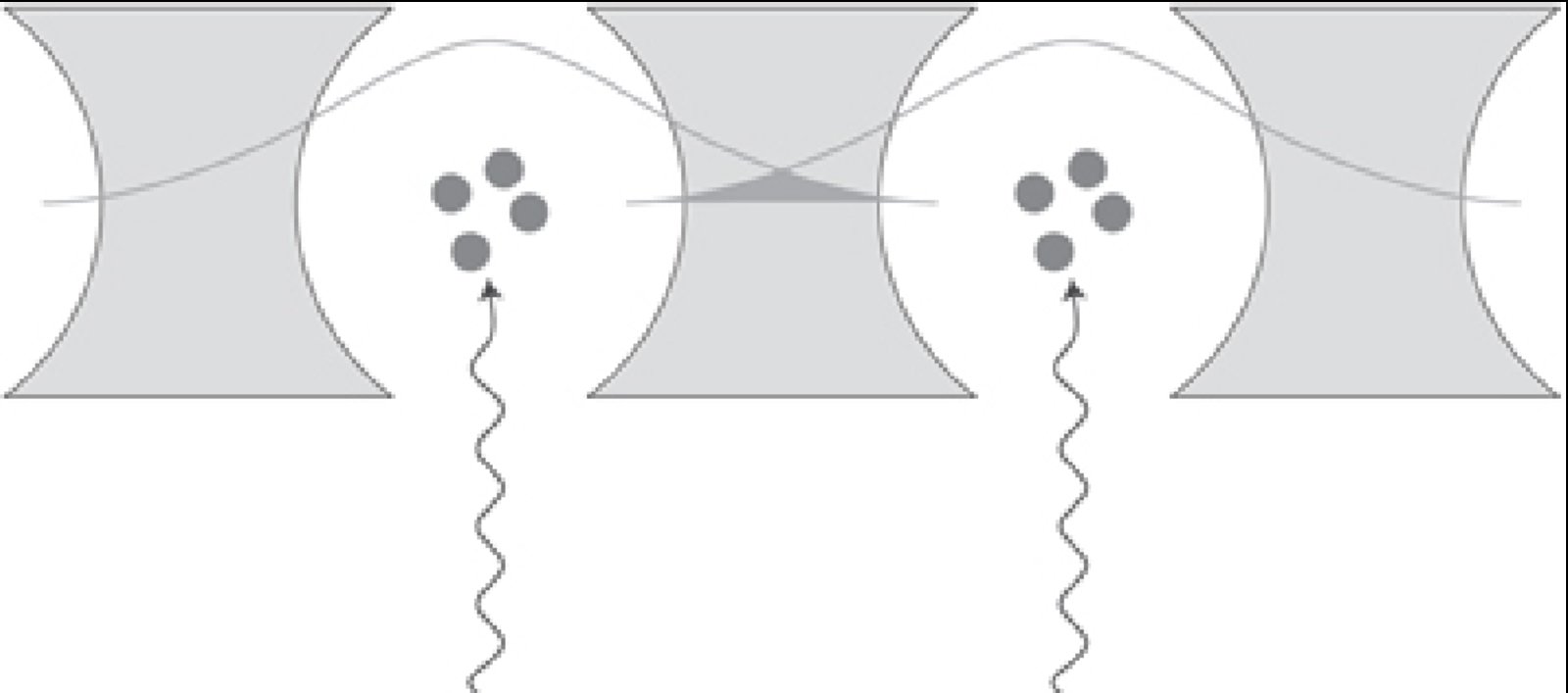}
\caption{\label{cavity} A one-dimensional array of cavities with
atoms trapped in the cavities. Photons can hop between the
different cavities because of the overlap between the light modes
in adjacent cavities. The atoms can be driven using externally
applied lasers. The atoms and cavities form hybrid excitations
(polaritons) that can hop between the different cavities and that
effectively interact with each other. The polariton-polariton
interaction strength can, for example, be tuned via the
atom-cavity detuning (reproduced from \cite{Har06}).}
\end{figure}
\textbf{Atoms in cavity arrays} (see Figure \ref{cavity})
\cite{Ang07,Har06,Gre06,Bra07} could also be used as quantum
simulators (see also \cite{Kim08}). This system provides an
alternative way of simulating the Bose-Hubbard model and quantum
phase transitions as well as spin models \cite{Kay08}.
\par
In the absence of atoms, the cavity array is described by the
Hamiltonian:
\begin{equation}
H=\omega_c\sum _{R} \left( a_R ^{\dagger}a_R +\frac{1}{2} \right)
+2\omega_c\alpha\sum _{R,R'} (a_R ^{\dagger}a_{R'}+ h.c.),
\end{equation}
where $a_R ^{\dagger}$ creates a photon in the cavity at site $R$,
$\omega_c$ is the frequency of the relevant cavity mode, and
$\alpha$ is the inter-cavity coupling coefficient \cite{Har06}.
One now introduces an atom or ensemble of atoms in each cavity,
and these atoms can be driven by external lasers. The atoms
trapped in a cavity together with the photons in the same cavity
form hybrid excitations called polaritons. The nonlinearity
introduced by the atoms results in effective polariton-polariton
interactions, which can be utilized for simulating, for example, the
Mott-insulator-superfluid phase transition.
\par
Arrays of cavities in an arbitrary geometry may be realized with
photonic bandgap cavities and toroidal or spherical microcavities
coupled via tapered optical fibers \cite{Gre06}. However, these
might be quite challenging to realize experimentally.
\par
A recent proposal suggested measurement and feedback control as
tools for realizing quantum simulation in atom-cavity systems
\cite{Voll08}, while others considered the possibility of using
atom-cavity systems for simulating the high-spin Heisenberg
model \cite{Cho08a} and the fractional quantum Hall effect
\cite{Cho08b}. Arrays of cavities could also be used to study
the quantum analogue of Fabry-Perot interferometers as suggested
in \cite{Lan08}.

\subsection{Nuclear and electronic spins}

\textbf{Nuclear spins} manipulated by means of NMR have been among
the first experimental systems to
implement small quantum algorithms and quantum simulation
\cite{Pen09,Pen10,Li11,Zha12}. Nuclear spin qubits have long
coherence times ($>1$ s), and high-fidelity quantum gates and the
coherent control of up to 12 qubits have been demonstrated.
\par
In the presence of a strong magnetic field pointing along the
z-axis, the general form of the NMR Hamiltonian is:
\begin{equation}
H=-\hbar \gamma B \sum_i I_i^z + \sum _{i>j} J_{ij} I_i^z I_j^z,
\end{equation}
where $\gamma$ is the gyromagnetic ratio, $B$ is the magnetic field,
$I$ is the angular momentum operator, and $J_{ij}$ are the
spin-spin coupling coefficients \cite{QCD}. The different transitions
between pairs of energy levels generally have distinct resonance
frequencies, allowing the addressing of the individual transitions
based on their frequencies. Using RF pulses various one-, two- and
possibly multi-qubit gates can be implemented. The field of NMR
benefits from very well developed control techniques. However, it
is not very flexible and its main disadvantage is the lack of
scalability, one of the main reasons being the spectral crowding
that occurs as the number of energy levels increases exponentially
with increasing number of spins. Although in solid-state NMR the
scalability drawback may be overcome to some extent, individual
addressing and measurement would still be impractical.
Nevertheless, nuclear spins provide a very good testbed for
various small simulation problems and allow the implementation of
both DQS and AQS. Furthermore, a recent proposal suggested that
nuclear spins attached to a diamond surface and addressed through
nitrogen-vacancy (NV) centers could offer an attractive route toward 
large-scale quantum simulator for strongly correlated systems
\cite{Cai13}.

\begin{figure}
\includegraphics[width=0.4\textwidth ]{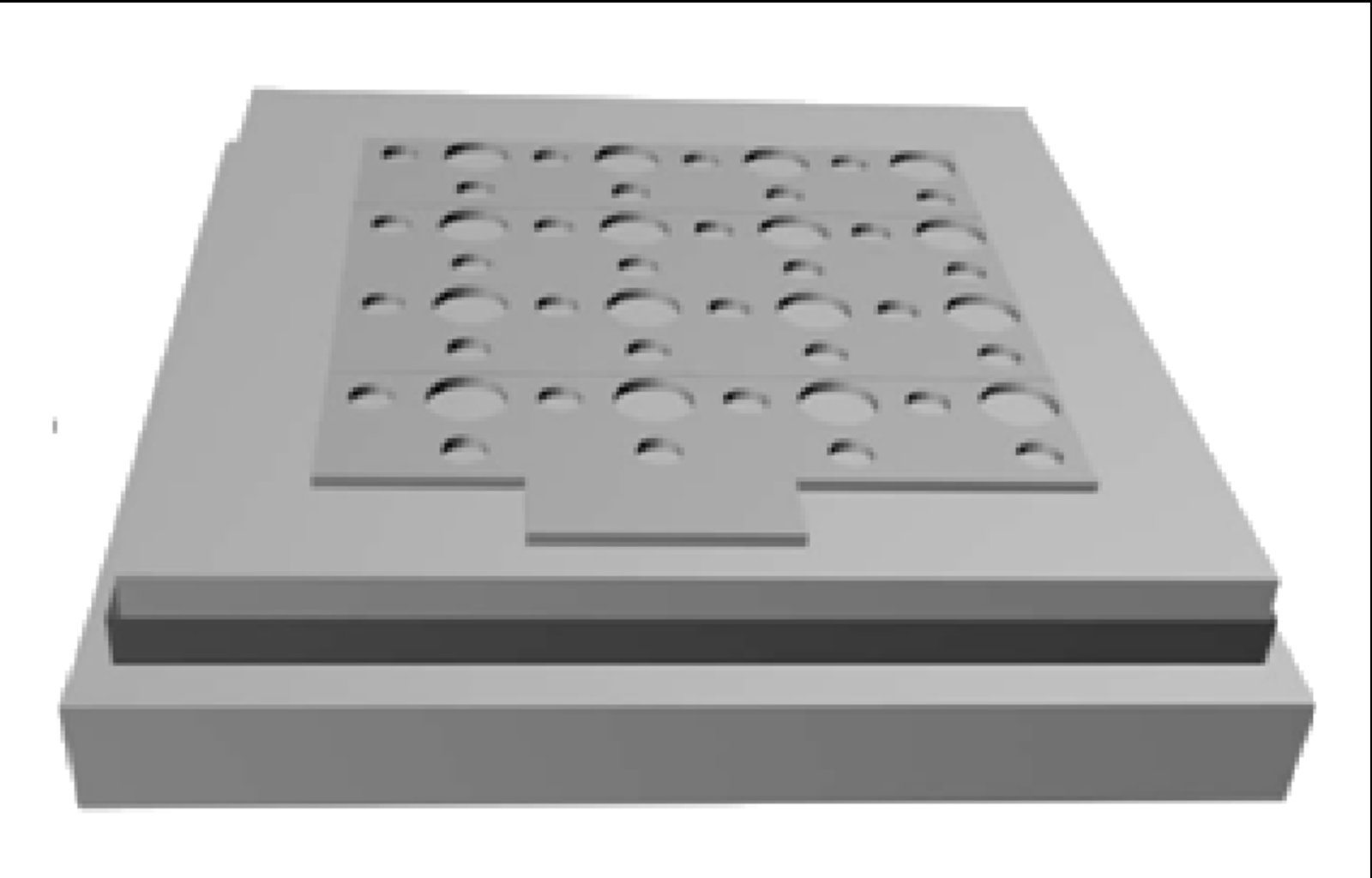}
\caption{\label{ad} An array of quantum dots realized by a
metallic gate with an array of two different size holes placed
on top of an Al$_x$Ga$_{1-x}$As/GaAs heterostructure. A negative
gate voltage is applied to this gate in order to create a potential
for the two-dimensional electron gas that is otherwise free to move
at the Al$_x$Ga$_{1-x}$As/GaAs interface. This was one of the early
theoretical proposals for AQS. (reproduced from \cite{Man02}).}
\end{figure}
Another system that could be used for quantum simulation is
\textbf{electron spins} in semiconductor quantum dots \cite{HA08}.
Quantum dots are semiconductor systems in which the excitations are
confined in a very small region in one or two dimensions. If the
region is roughly the same size as the wavelength of the charge
carrier, the energy levels are quantized and the quantum dot becomes
very similar to a real atom (and can therefore be referred to as an
``artificial atom''). Moreover, quantum dots allow flexible control
over the confinement potential and they can also be excited
optically. Furthermore, quantum dots with large tunnel coupling can
act as ``artificial molecules''. These features make electron spins
in quantum dots particularly attractive for quantum simulation.
\par
Quantum dots can be defined at fabrication or by applying bias
voltages using electrodes placed above a two-dimensional
electron gas. They can be designed to have certain characteristics
and assembled in large arrays. The manipulation and readout can be
done both electrically and optically. State-of-the-art quantum-dot
qubits now have long decay times of $>1$ s \cite{Ama08}.
\par
Arrays of quantum dots could be realized using two-dimensional
mesh gates \cite{Man02,Byr08} (see Figure \ref{ad}).
Alternatively, in \cite{Byr07} it was proposed that interfering
acoustic waves could be used to form an analogue of optical lattices
in a two-dimensional electron-gas, but this latter approach has
the disadvantage of heating and complicated engineering. By
adjusting the mesh-gate design and voltage, various lattice
geometries could be created. Electron spins in quantum dots may
provide an advantage over atoms in optical lattices due to the
very low temperatures (relative to the Fermi temperature) that can
be reached and the natural long-range Coulomb interaction. In a
recent experiment, the predictions of the two-dimensional
Mott-Hubbard model were tested for electrons in an artificial
honeycomb lattice \cite{Sin11}.
\par
The Hamiltonian for an array of quantum dots is given by
\begin{equation}
H=\sum _{j=1} ^{n} \mu _B g_j(t)B_j(t)\cdot S_j+\sum _{1\leq j
<k\leq n} J_{jk}(t)S_j\cdot S_k,
\end{equation}
where the first term is the energy due to an applied magnetic field
$B_j$, and the second term is the exchange interaction energy,
which is a result of virtual tunneling between the quantum dots.
Here, $S_j$ is the spin of the electric charge quanta of the $j$-th
dot \cite{QCD}. The interactions between the qubits can be
engineered by adjusting the gate voltages together with
a careful choice of the mesh hole sizes and doping \cite{Man02}.

\subsection{Superconducting circuits}
\begin{figure}
\includegraphics[width=0.5\textwidth ]{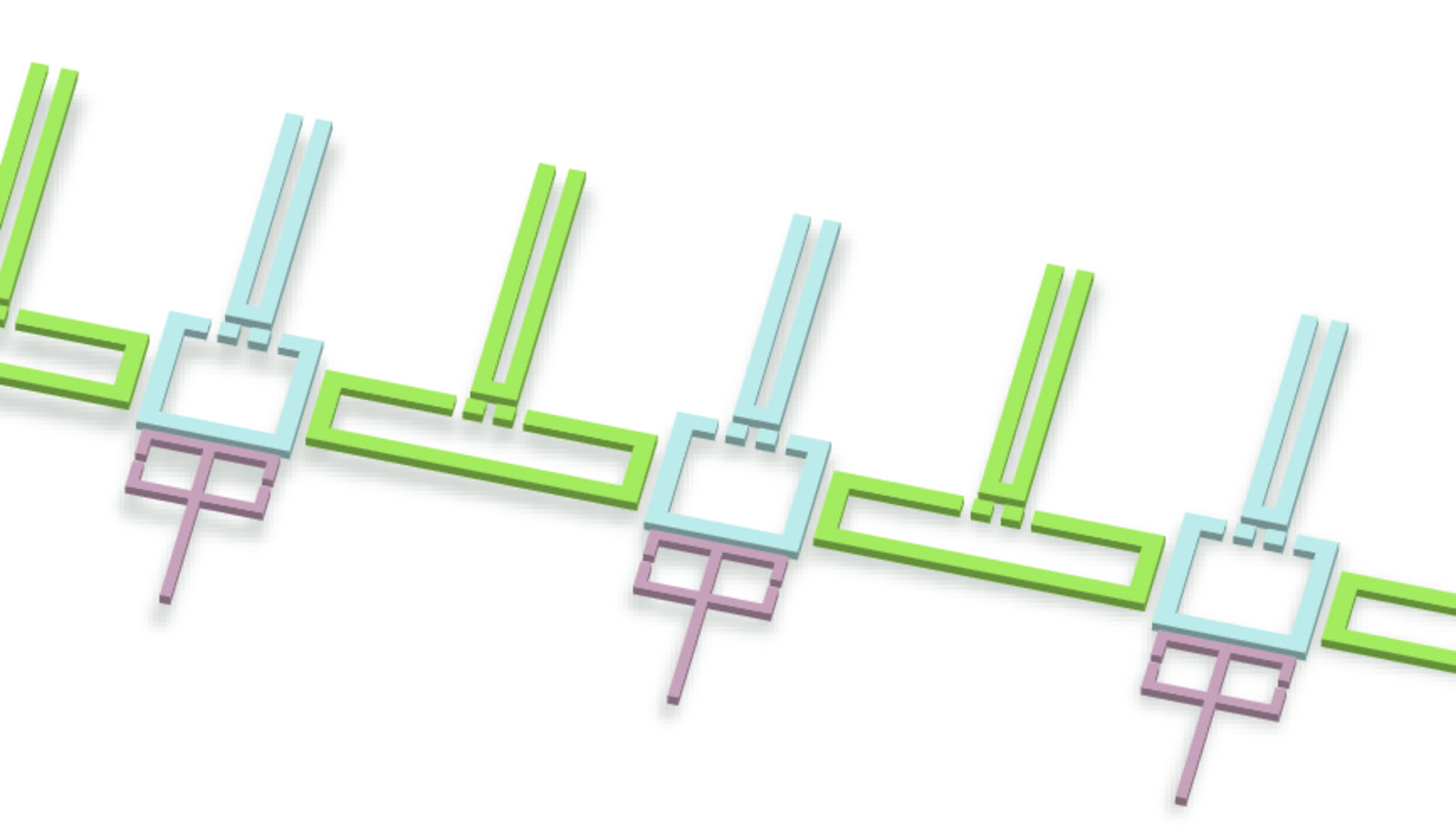}
\caption{\label{sc} (color online) Schematic diagram of an array
of superconducting flux qubits (see, e.g., \cite{Ash08}). The
green circuits are used to couple the flux qubits (blue), while
the brown circuits are used to read out the states of the qubits.
The gaps represent tunnel junctions.}
\end{figure}
Superconducting circuits \cite{You05,CW08,You11} (see Figure
\ref{sc}) can also be used as quantum simulators. Quantum
information can be encoded in different ways: in the number of
superconducting electrons on a small island, in the direction of a
current around a loop, or in oscillatory states of the circuit.
The circuit can be manipulated by applied voltages and currents
(including both dc and microwave-frequency ac signals) and
measured with high accuracy using integrated on-chip instruments.
Although macroscopic in size, these circuits can display quantum
behavior and can be seen  as ``artificial atoms''. The advantage
over real atoms is that superconducting circuits can be designed
and constructed to tailor their characteristic frequencies,
interaction strengths and so on. The frequencies can also be tuned
\textit{in situ} by adjusting an external parameter (typically an
external magnetic field), and the coupling energy between two
qubits can be turned on and off at will. Furthermore,
superconducting circuits can be coupled to ``cavities'', which are
actually electrical resonators (and the ``photons'' are, for the
most part, electron-density oscillations). This setup is very
useful for the study of electric-circuit analogues of cavity
quantum electrodynamics (circuit QED) \cite{You05,Sho08,You11}.
\par
State-of-the-art superconducting qubits have coherence times
exceeding $100\mu$s (i.e.~decoherence rates below 10 kHz), which
is quite high considering that other energy scales in the circuit
are typically in the range 10 MHz - 10 GHz. Individual control
and measurement have been demonstrated
\cite{Mar11}. Furthermore high-fidelity one-, two and three-qubit
quantum gates have been demonstrated. With this level of control,
DQS could be implemented in a superconducting circuit in the near
future.
\par
The Hamiltonian for $N$ charge (flux) qubits biased at their
symmetry points (which is optimal for quantum coherence) coupled
capacitively (inductively) is:
\begin{equation}
H=-\sum _{i=1} ^{N} \frac{\Delta _i}{2}\sigma_i ^z -\sum _{(i,j)}
J_{ij}\sigma_i ^x\sigma_j ^x,
\end{equation}
where $\Delta _i$ is the level splitting and $J_{ij}$ is the
strength of the coupling between qubits $i$ and $j$. It should be
noted, however, that superconducting circuits have more than two
energy levels, and these additional levels could also be utilized.
Indeed, a recent experiment demonstrated AQS of a spin larger than
1/2 using this approach \cite{Nee09}.
\par
\begin{figure}
\includegraphics[width=0.45\textwidth ]{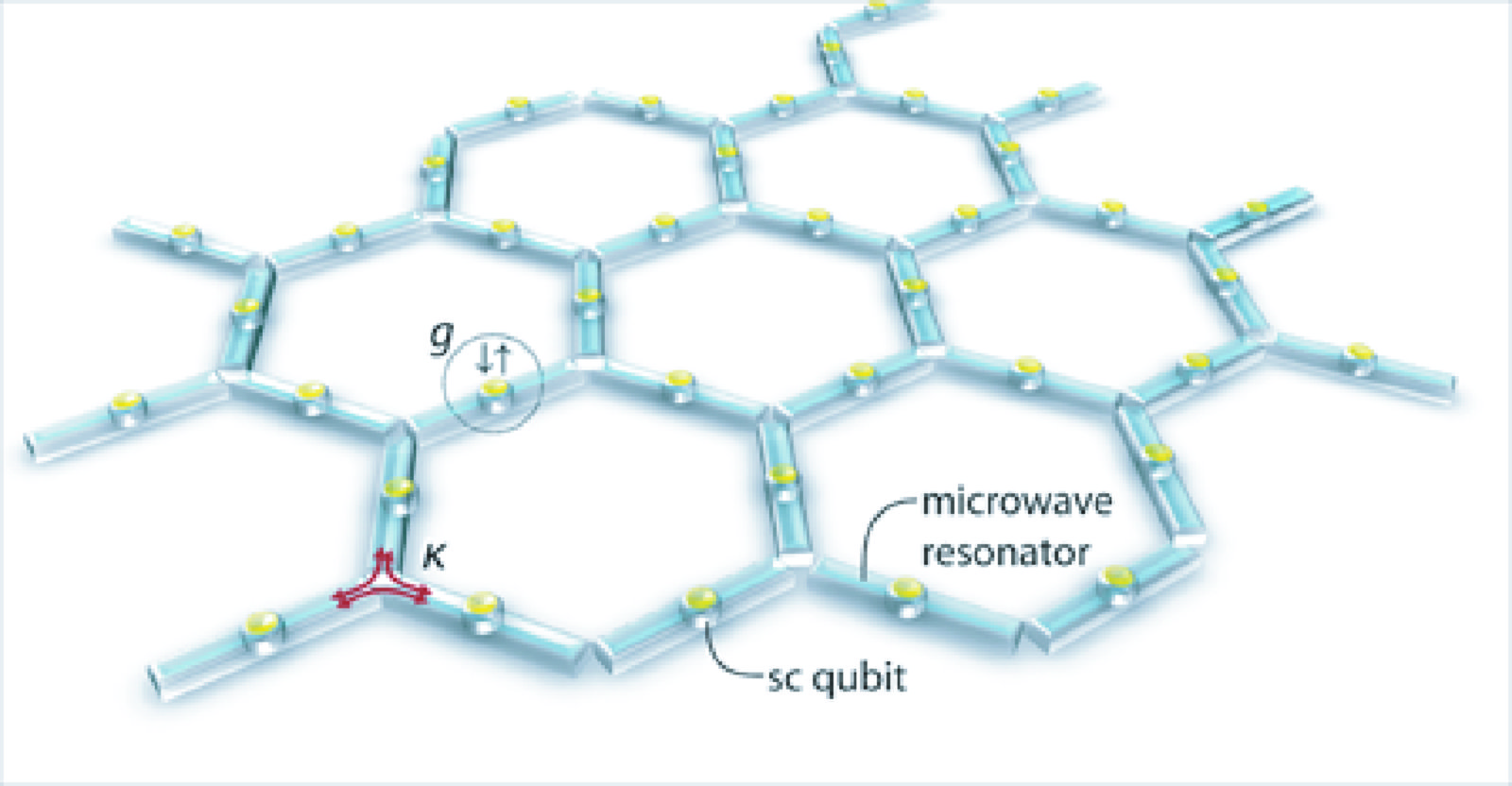}
\caption{\label{pho} (color online) A possible design for
implementing the Jaynes-Cummings lattice using superconducting
qubits. The blue strips are superconducting resonators and the
yellow dots are superconducting qubits. An effective
particle-particle interaction is created by the qubit-cavity
interactions. (reproduced from \cite{Koc10}).}
\end{figure}

As for scalability, circuits containing 512 qubits have been
fabricated \cite{Har10,Har12}, although quantum coherence was not
tested on these circuits in the same way that coherence is
commonly tested in other experiments using small numbers of
qubits. Furthermore, more than 200 superconducting resonators were
recently fabricated on a single chip \cite{Hou12}. If qubits are
integrated into such a circuit, it could realize the proposal of
atom-cavity arrays (or Jaynes-Cummings lattices), performed with
artificial atoms and cavities (see Figure \ref{pho}). It has also
been proposed that artificial gauge fields could be simulated with
such circuits \cite{Koc10}. A related study proposed an approach
to universal quantum computation and simulation using the
single-excitation subspace of an array of coupled superconducting
qubits \cite{Gel12}. Although unscalable, this approach may still 
enable a universal quantum simulation speedup relative to
present-day classical computer.
\par
The fact that superconducting circuits can be produced in large
numbers and ``wired'' together on a chip offers a rather
straightforward way of realizing various lattice geometries.
Examples include the Kitaev model on a honeycomb lattice
\cite{You08}, networks for simulating Anderson and Kondo models
\cite{JGR08}, highly-connected networks \cite{Tso08} and fractals
\cite{Tso10}.

\subsection{Photons}

Photons can carry quantum information over long distances, hardly
being affected by noise or decoherence. They naturally possess
the ability to encode qubit states, e.g.~in the polarization of
the photon, and one-qubit gates can be easily realized with linear
optical components. Although the difficulty in implementing
two-qubit gates is a serious drawback for photonic systems in the
context of quantum computation, there have been some notable
achievements for quantum simulation using photons.
\par
Photons have been used to calculate the possible fractional
statistics of anyons using a six-photon graph state \cite{Lu09},
to calculate the energy spectrum of the hydrogen molecule to 20
bits of precision \cite{Lan09} and to simulate frustrated spin
systems \cite{Ma11}. It has also been proposed that photons
propagating or trapped inside materials doped with atoms that
have suitable energy-level structures could be used for the
simulation of Luttinger liquids \cite{Ang11} and relativistic
field theories \cite{Ang13a,Ang13b}.
\par
Recently, it was shown that the propagation of photons in a network
of beam splitters is in general a computationally difficult task for
classical computers even for a few tens of photons \cite{Aar11},
and corresponding experiments with up to four photons were
performed \cite{Bro13,Spr13,Til13,Cre13}. However, with
limited flexibility and scalability, it remains to be seen how far
photon-based quantum simulation can go.

\subsection{Other systems}

One of the systems that are being considered as candidate platforms
for implementing quantum computation is NV centers in diamond
\cite{Bul10,Lad10}. Unlike other systems studied in the context
of quantum computation, however, NV centers in diamond have not
received much attention as potential quantum simulators, which might
be due to difficulties in coupling the NV centers to each other and
future scalability. Nevertheless, there have been recent studies
attempting to alleviate these difficulties (see, e.g., \cite{Wei13}).
\par
Another system that could be used for the quantum simulation of
condensed matter physics is electrons trapped on the surface of
liquid helium \cite{Mos08}. This setup could be used to simulate
the Ising model. In principle it could be scaled up, but the
control would be very difficult. A related system that was proposed
recently to implement DQS is a chain of molecular nanomagnets
controlled by external magnetic fields \cite{San11}.
\par
A rather unconventional example of a quantum simulator is a
two-component BEC in which the propagation of sound waves could
simulate some aspects of cosmic inflation \cite{Fis04}. Although
such a quantum simulator would be limited to a narrow class of
problems, it provides an alternative possibility for the
simulation of such systems.
\par
\begin{figure*}
\includegraphics[width=\textwidth ]{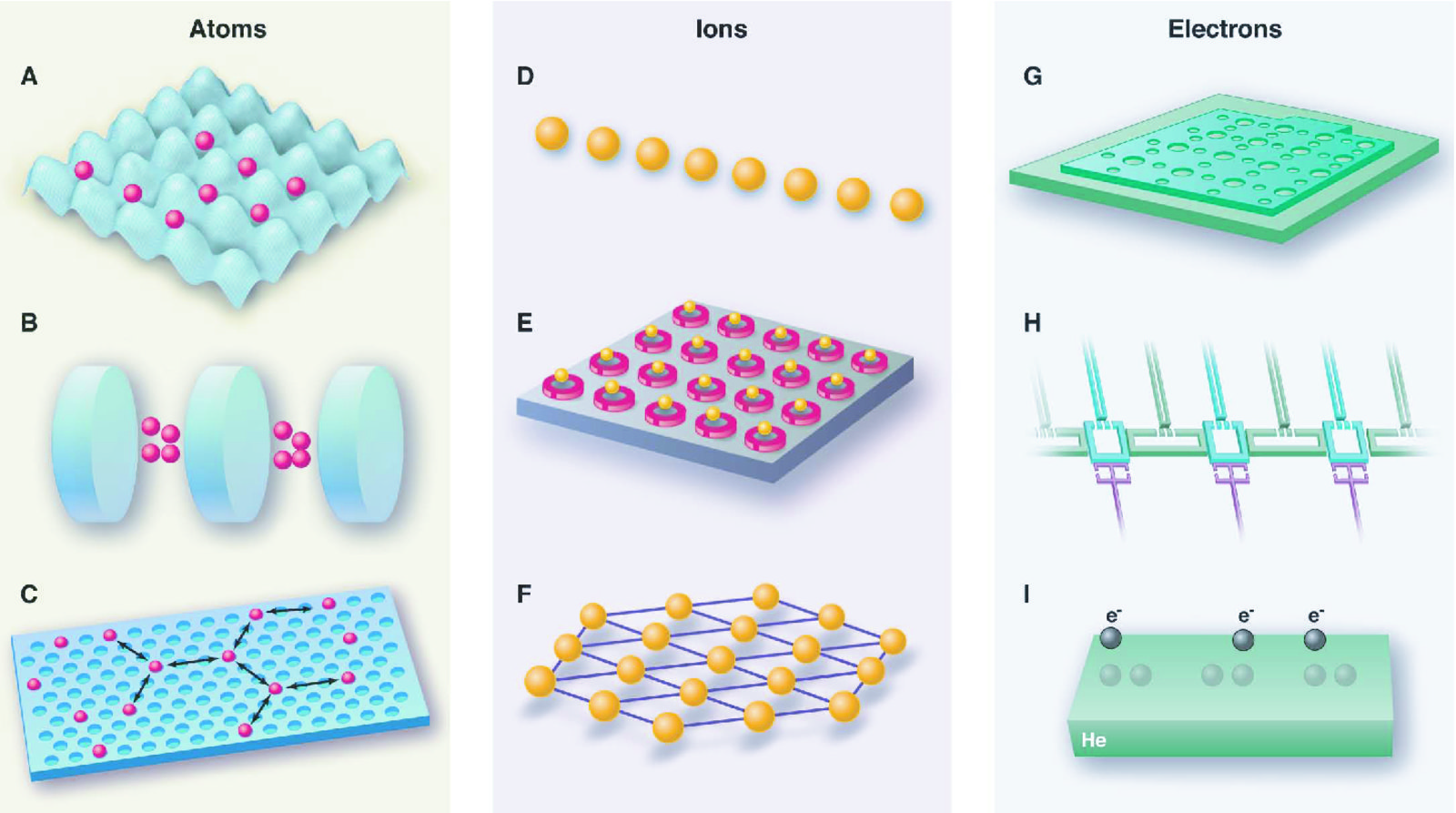}
\caption{\label{sims} (color online) One- or two-dimensional
arrays of qubits and controls can be used to simulate various
models in condensed-matter physics. Examples of such analog
quantum simulators include: \textbf{atoms} in
optical lattices \cite{Jak05} (A), one-dimensional (B) or
two-dimensional (C) arrays of cavities \cite{Har06,Gre06};
\textbf{ions} in linear chains (D), two-dimensional arrays of
planar traps \cite{Chi08} (E), or two-dimensional Coulomb crystals
\cite{Por06} (F); \textbf{electrons} in quantum dot arrays created
by a mesh gate \cite{Man02,Byr08} (G), in arrays of
superconducting circuits (H), or trapped on the surface of liquid
helium \cite{Mos08} (I). The average distance between the atoms in
optical lattices is less than 1 $\mu$m. Cavity arrays based on
photonic bandgap cavities would also have sub-micron inter-site
separations. As for the inter-ion distances in ion trap arrays,
they are about 10-50 $\mu$m and about the same for two-dimensional
Coulomb crystals. In arrays of quantum dots, the spacing between
dots is about $0.1$ $\mu$m. In superconducting circuits, the
distance between qubits is typically a few microns. In the case of
electrons on helium the distance between neighboring sites would
be about 1 $\mu$m. These inter-qubit distances (from 0.1 to 10
$\mu$m) are to be compared with the far smaller average
inter-atomic distances in solids, which are $\leq 1$ nm.  The
systems shown above realize a one- or two-dimensional array of
qubits which can be manipulated in different manners. The larger
distances between qubits make quantum simulators more controllable
and easier to measure. Therefore, they can be thought of as models
of the magnified lattice structure of a solid, with a
magnification factor of two to four orders of magnitude.
(reproduced from \cite{Bul09}).}
\end{figure*}
\begin{table*}[ht]
\caption{\label{tcomp} Strengths and weaknesses of some of the
proposed and demonstrated quantum simulators. An asterisk means
that the feature has been experimentally demonstrated. By
\textit{scaling} we mean controlling an array of at least a few
tens of qubits. By \textit{individual control} we refer to the
ability of controlling and measuring each individual qubit. The
weaknesses refer to the actual experimental implementations.
\vspace{8pt}}
\begin{tabular*}{\textwidth}{p{6cm} p{4.8cm} p{5.2cm}}
\hline
\textbf{Quantum simulator} &   \textbf{Strength} & \textbf{Weakness} \\[8pt]
\hline \textbf{Neutral atoms}  &  Scaling$^{*}$ & Individual control and readout\\
[8pt] \textbf{Trapped ions}  & Individual control and
readout$^{*}$ & Scaling\\ [8pt] \textbf{Cavity arrays} &
Individual control  and readout& Scaling \\ [8pt]
\textbf{Electronic spins (quantum dots)}
& Individual control and readout$^{*}$, tunability& Scaling\\
[8pt] \textbf{Superconducting circuits} & Individual control and
readout$^{*}$, tunability& Scaling (some recent progress) \\
[8pt] \textbf{Photons (linear
optics)}  & Flexibility$^{*}$& Scaling\\
[8pt]
\textbf{Nuclear spins (NMR)}  & Well-established, readily available technology$^{*}$& Scaling, no individual control\\
\hline \hline
\end{tabular*}
\end{table*}

\subsection{Current state of the art}

The above discussion of the physical systems that could implement
quantum simulation is summarized in Table \ref{tcomp} which lists
the strengths and weaknesses of each potential quantum simulator,
and in Figure \ref{sims}.
\par
Currently only with neutral atoms in an optical lattice is it
possible to perform quantum simulations with more than a few
particles, and these systems can at present be considered the most
advanced platform for AQS. Although individual control and readout
is not yet available, recent progress in this direction has been
made and alternative approaches using Rydberg atoms or polar
molecules are now being pursued. Meanwhile, recent experiments
with trapped ions have demonstrated exotic quantum simulations beyond
condensed-matter physics (e.g., Dirac particles \cite{Lam07,Ger10}
or the Klein paradox \cite{Cas10,Ger11}) and superconducting
circuits provide a way to study intriguing quantum phenomena
such as the dynamical Casimir effect \cite{Joh10,Lah13,Wil11}.
\par
With the current experimental state-of-the-art techniques, the most
advanced DQS has been realized with trapped ions \cite{Lan11}. In
these experiments, simulations of various spin models with up to
six spins have been demonstrated, with one experiment requiring
sequences of more than 100 quantum gates. Since no error correction
was used, the fidelity of the DQS implementation was somewhat
lower than for the AQS implementation in the same system (see
\cite{Fri08}). With error correction and improved control of
laser intensity fluctuations, quantum simulations with $>10$
qubits and hundreds of high-fidelity gates seem possible in the
coming years. Moreover, the DQS approach allows the simulation of
complex spin-spin interactions and could potentially be combined
with AQS techniques.

\section{Applications}

Quantum simulators will find numerous applications in diverse
areas of physics and chemistry \cite{Lan09}, and possibly even
biology \cite{Gho09,Gho10,Gho11}. Quantum simulation would lead
to new results that cannot be otherwise simulated, and would also
allow the testing of various theoretical models. In general, with
a quantum simulator one could address problems that are either
\textit{intractable on classical computers} or \textit{experimentally
challenging}. Moreover, being quantum systems themselves, quantum
simulators would be able to provide more insight into quantum phenomena
than classical simulators (e.g., the effects of decoherence and
the transition from quantum to classical). In this section we
discuss how different problems could be studied using the quantum
simulators described in the previous section. While some problems
are classically intractable (e.g., Hubbard models, spin
frustration and disorder, lattice gauge theories and quantum
chemistry calculations), others can be treated classically (e.g.,
James-Cummings Hamiltonian and interferometry) but are discussed
here as benchmarks for the progress of quantum simulation.

\subsection{Condensed-matter physics}

The difficulty in solving quantum many-body problems is reflected
in the open questions in condensed matter physics. Among the
best-known challenges in this field are understanding high-$T_c$
superconductivity and disordered and frustrated systems. In this
subsection we discuss the quantum simulation of these problems.

\subsubsection{Hubbard model}

The Hubbard model is the simplest model of interacting particles
on a lattice. However, for large numbers of particles in more than
one dimensions, the model is difficult to treat using classical
computers.

As discussed in Section \ref{AtomsAndIons}, an AQS of the
Bose-Hubbard model using atoms in optical lattices was proposed in
\cite{Jak98} and implemented in \cite{Gre02}. Recent experiments on
the subject include the realization of a Tonks-Girardeau gas in one
dimension \cite{Par04} and the investigation of atoms trapped in a
graphene-like lattice \cite{Ueh13}. There have also been
proposals to simulate this model with polar molecules
\cite{Ort09}, trapped ions \cite{PorCir04,PorCir06,Den08} or atoms
in arrays of coupled cavities \cite{Bra07,Lia10,Lan08,Zhou08}. The
simulation of the Fermi-Hubbard model in quantum-dot arrays was
proposed in \cite{Byr07,Byr08}. The simulation of the Hubbard model
with attractive interactions using atoms in optical lattices was
discussed in \cite{Ho09}. Proposals for the AQS of the related
Holstein model have also been put forward. These include
simulations using polar molecules \cite{Her11,Her13}, ions
\cite{Sto12} and superconducting circuits \cite{Mei13}.
\par
The simulation of the Hubbard model has also been considered in
the context of DQS. In \cite{Som02} it was shown how to obtain the
energy spectrum of the Fermi-Hubbard Hamiltonian:
\begin{align}
H_H=&-\sum_{(i,j);\sigma}[t_x (a^{\dagger}_{(i,j);\sigma}a_{(i+1,j);\sigma}+a^{\dagger}_{(i+1,j);\sigma}a_{(i,j);\sigma}) \nonumber\\
&+ t_y (a^{\dagger}_{(i,j);\sigma}a_{(i,j+1);\sigma}+a^{\dagger}_{(i,j+1);\sigma}a_{(i,j);\sigma})] \nonumber\\
&+ \tilde{U} \sum_{(i,j)}{n}_{(i,j);\uparrow }n_{(i,j);\downarrow },
\end{align}
where $t_x$ and $t_y$ are the hopping matrix elements allowing
fermions to move on the lattice. The first terms describe the
kinetic energy, and the last term ($\tilde{U}$) is the on-site
repulsion potential energy. The authors explicitly gave
the mapping between the operators of the Hubbard model and the
DQS; the initialization, evolution, and measurement steps were
described in detail. Note that the mapping between the fermionic
operators and the Pauli matrices employs the Jordan-Wigner
transformation. A possible implementation of DQS for the Holstein
model was analyzed in \cite{Mez12}.

\subsubsection{Spin models}

Spin systems are typically described by Hamiltonians of the form
\begin{equation}
\label{heis}
H_{\rm{XYZ}}=\sum_{i=1}^{N}[J_{x}\sigma_i ^x \sigma_{i+1} ^x + J_{y}\sigma_i ^y \sigma_{i+1} ^y +J_{z}\sigma_i ^z \sigma_{i+1} ^z].
\end{equation}
As mentioned in Section \ref{DQS}, along with the Hubbard
Hamiltonian spin Hamiltonians can be simulated on a DQS. It is also worth noting that
in certain limits the Hubbard model reduces to spin models. 
\par
The DQS implementation of various spin Hamiltonians with atoms in
optical lattices or trapped ions was discussed in \cite{Jan03}.
A recent trapped-ion experiment demonstrated the DQS of different
spin Hamiltonians using sequences of elementary quantum gates
\cite{Lan11}.
\par
\begin{figure}
\includegraphics[width=0.45\textwidth ]{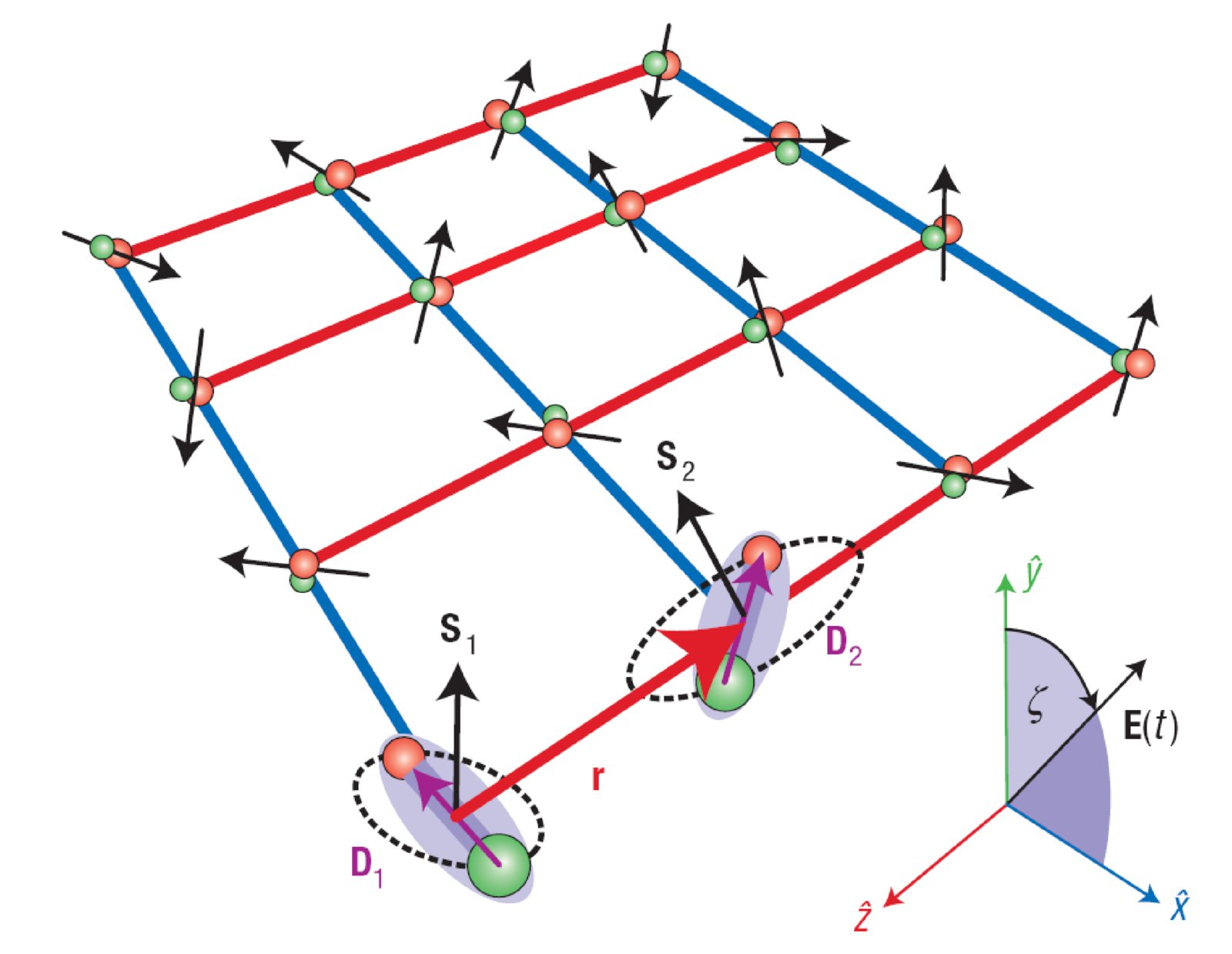}
\caption{\label{polar} (color online) Schematic diagram of a
proposal for the simulation of anisotropic spin models using polar
molecules trapped in an optical lattice: a two-dimensional square
lattice with nearest-neighbor orientation-dependent Ising
interactions. The effective interactions between the spins $S_1$
and $S_2$ of the molecules in their rovibrational ground states are
generated with a microwave field $E(t)$ inducing dipole-dipole
interactions between the molecules with dipole moments $D_1$ and
$D_2$, respectively. (reproduced from \cite{Mic06})}
\end{figure}

The AQS of spin models is also possible. For instance, with trapped
ions, an AQS for the Ising [Eq. (\ref{ising})], $XY$ [Eq. (\ref{heis})
with $J_z=0$], or $XYZ$ [Eq. (\ref{heis})] interactions can be realized
using the collective vibrational modes \cite{Por04}. These interactions
can be switched and tuned by lasers and by the choice of trapping
conditions. This direction has been pursued further in
\cite{Den05,Por05,PorCir05,Por08,Ber09,Lin11,Kor12} and put into
practice in the experiments of
\cite{Fri08,Edw10,Kim10,Kim11,Isl11,Bri12}. An earlier paper
\cite{Mil99} had shown that the conditional displacement of the
vibrational mode of trapped ions can be used to simulate nonlinear
collective and interacting spin systems. Furthermore, spin chains
and ladders can be investigated with atoms in optical lattices \cite{JGR04},
and a scheme to realize the anisotropic $XXZ$ [Eq. (\ref{heis})
with $J_x=J_y\neq J_z$] and isotropic $XXX$ [Eq. (\ref{heis}) with
$J_x=J_y=J_z$] Heisenberg spin Hamiltonians in an arbitrary array
of coupled cavities was proposed in \cite{Cho08a}. Reference
\cite{Mic06} discussed the possibility of engineering Hamiltonians
of spin lattice models with polar molecules stored in optical
lattices (see Figure \ref{polar}). The spins are represented by
single electrons of heteronuclear molecules. Using a
combination of microwave excitation, dipole-dipole
interactions and spin-rotation couplings enables the realization
of effective two-spin interactions with designable range, spatial
anisotropy and coupling strength.
\par
\begin{figure}
\includegraphics[width=0.45\textwidth ]{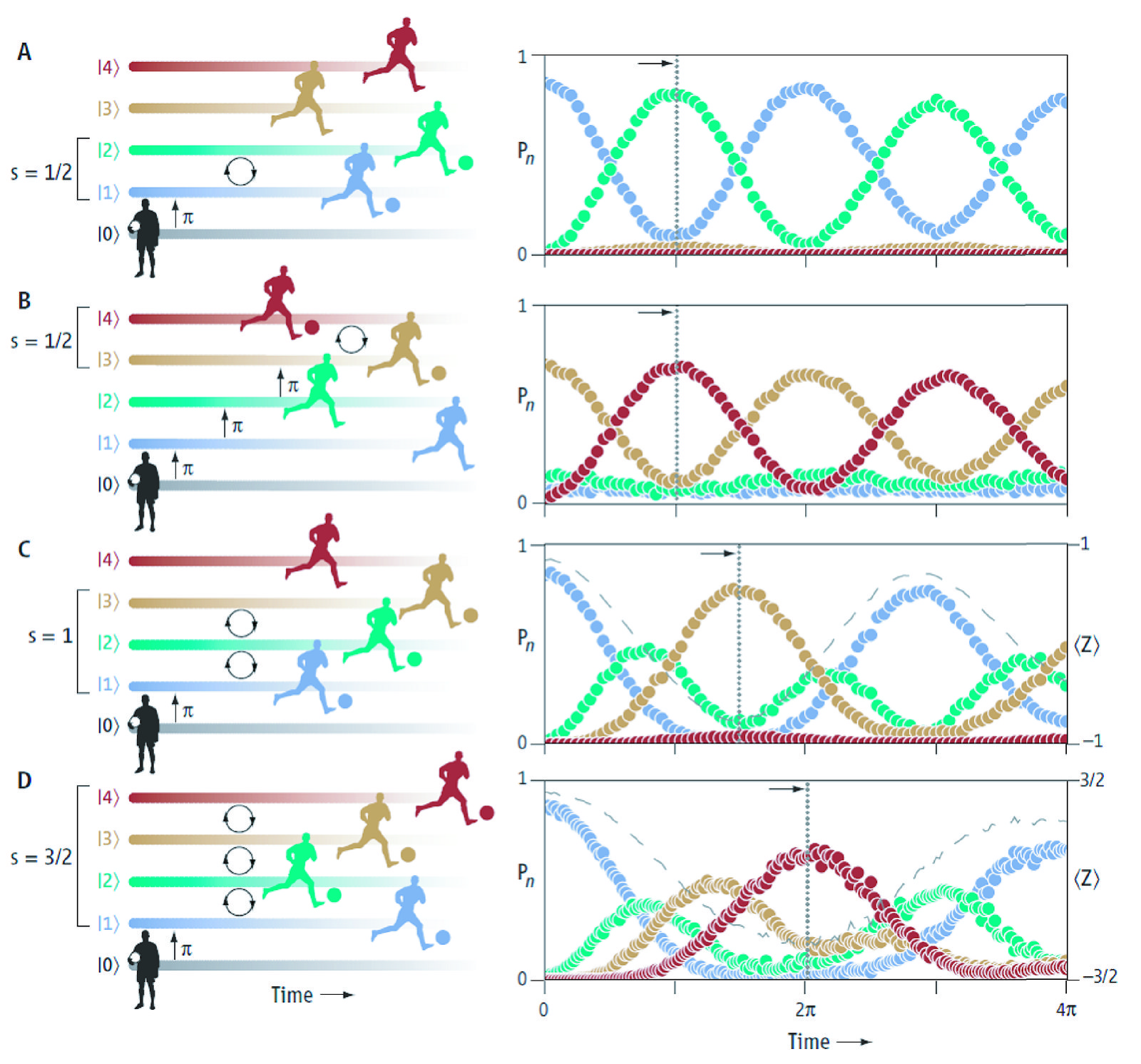}
\caption{\label{foot} (color online) Illustration of the
experiment of \cite{Nee09}, in which the states of a
superconducting circuit were used to emulate a particle with spin
$s$, up to $s=3/2$. For simulating a spin $s$, $2s+1$ quantum
states of the circuit are used. For example, in (D) four of the
circuit's quantum states are mapped onto the four quantum states
of a spin 3/2 particle, while a fifth state was used in the
experiment as a reference state for phase measurement. By applying
carefully tuned microwave-frequency drive signals, the circuit can
simulate the interaction between the spin and a magnetic field
with arbitrary strength and direction. An analogy with a classical
system (a football game), shown on the left side, can also be used
to give a simple explanation of the redistribution of the
population among the different energy levels in terms of football
players passing a ball among each other, with the goalkeeper
serving as the anchor. (reproduced from \cite{Nor09}).}
\end{figure}

Recently, in \cite{Tso10} the simulation of spin and Hubbard
models in higher or fractal dimensions with superconducting qubits
and resonators was proposed. The ability to access arbitrary dimensions
is made possible by the flexible connectivity, which derives from the
flexibility in designing tunable couplers between superconducting qubits
and resonators. Spin systems with $s>1/2$ can also be
naturally realized using superconducting circuits, because these
circuits generally have more than two quantum states that can be
employed in the simulation. In \cite{Nee09,Nor09}
the emulation of the dynamics of single spins with principal quantum
number $s = 1/2$, $1$ and $3/2$ was demonstrated (see Figure
\ref{foot}). The antiferromagnetic Heisenberg model with long-range
interactions could be realized with solid-state NMR
\cite{Rou07}. Itinerant ferromagnetism could be studied in a
two-component Fermi gas \cite{GB09}.

\subsubsection{Quantum phase transitions}

Having introduced the Hubbard model and spin models, we now
discuss quantum phase transitions, especially in these models.
Quantum phase transitions occur when one varies a physical parameter
at absolute zero temperature, when all thermal fluctuations have
ceased. The transition describes an abrupt change in the ground
state of the many-body system governed by its quantum
fluctuations. Quantum phase transitions are an interesting and
fundamental subject, but are difficult to investigate both by 
classical simulation and via experimental methods. Analog quantum
simulators can help in understanding this purely quantum
phenomenon, and the first steps in this direction have already
been explored.
\par
\begin{figure*}
\includegraphics{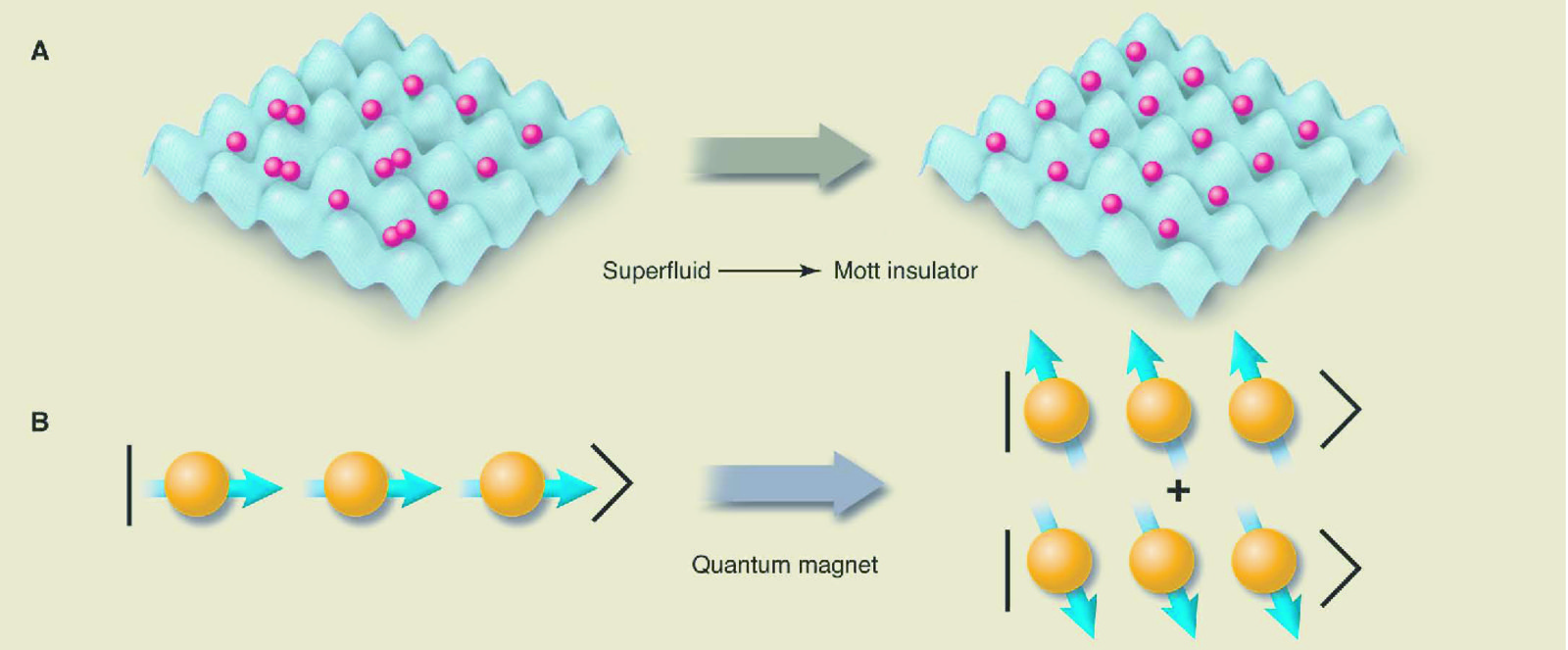}
\caption{\label{qpt} (color online) Examples of analog quantum
simulators of quantum phase transitions using ultracold neutral
atoms (a) and trapped ions (b). (a) The quantum phase transition
from a superfluid to a Mott insulator phase realized in
\cite{Gre02} using rubidium atoms trapped in an optical lattice.
The ratio between the tunneling energy and the on-site interaction
energy was controlled in \cite{Gre02} by adjusting the lattice
potential depth, such that the quantum phase transition was
observed. There are other alternative ways of simulating this
quantum phase transition with arrays of cavities \cite{Har06}, or
arrays of Josephson junctions \cite{Oud96}. (b) Magnetic quantum
phase transition simulated in \cite{Fri08} using trapped calcium
ions. The interactions between individual spins and an external
magnetic field were simulated by coupling the internal levels
(representing the spin-1/2 states) with a resonant RF field, while
the spin-spin interactions were simulated using a state-dependent
optical dipole force implemented by a walking wave (reproduced from
\cite{Bul09}).}
\end{figure*}

For example, the quantum phase transition from a superfluid to a
Mott insulator phase, predicted in \cite{Fis89}, was first
observed in 2002 \cite{Gre02} in an ensemble of atoms
trapped in an optical lattice. The physical model is described by
the Bose-Hubbard Hamiltonian given in Eq. (\ref{eqBH}). The phase
transition is perhaps easiest to understand in the case of unit
filling factor, i.e. when there are as many particles as lattice sites.
Deep in the superfluid phase, when $J \gg \tilde{U}$, delocalization
minimizes the kinetic energy and the system is said to be in the
weakly interacting regime (the ground state being approximately
given by $\left ( \sum _i \hat{b}_i^{\dagger} \right )^N
\ket{0}$, i.e. all the particles are in the single-particle ground
state). When $\tilde{U} \gg J$, the energy is minimized when the
particles are distributed evenly among the lattice sites and
the ground state is approximately given by $\prod _i
\hat{b}_i^{\dagger}\ket{0}$. This state is sometimes called strongly
correlated, and it corresponds to the Mott phase. By adjusting the
lattice potential depth, the ratio between the tunneling energy
$J$ and the on-site interaction energy $\tilde{U}$ can be controlled
and brought to the point where the transition between the Mott
insulator phase and the superfluid phase is induced. Then an
abrupt change in the ground state of the system is observed \cite{Gre02}.
A schematic diagram of this quantum phase transition is illustrated
in Figure \ref{qpt}a.
\par
In quantum magnets, the transition from a para- to an anti-ferromagnet
was emulated using two trapped calcium ions in \cite{Fri08}. The system
is described by the quantum Ising model. The Hamiltonian for a chain
(with two or more spins) is given by
\begin{equation}
\label{ising}
H_{I}=-B_x\sum_{i}\sigma_i^{x} +\sum_{i<j}J_{ij}\sigma_i ^z \sigma_j ^z,
\end{equation}
where $B_x$ is the magnetic field strength and $J_{ij}$ are the
spin-spin coupling coefficients. The first term, denoting the
interactions of individual spins with an external magnetic field,
is simulated by coupling the internal levels (representing the
two spin 1/2 states) with a resonant RF field. The spin-spin
interaction in the second term is simulated by a state-dependent
optical dipole force, implemented by a walking wave formed by two
perpendicular laser beams. When the strength of the spin-spin
interaction is increased adiabatically (increasing $J$ while keeping
$B_x$ constant), the system undergoes a transition from
paramagnetic ($\ket{\rightarrow} \ket{\rightarrow} \cdots \ket{\rightarrow}$)
to ferromagnetic ($\ket{\downarrow} \ket{\downarrow} \cdots \ket{\downarrow} +
\ket{\uparrow} \ket{\uparrow} \cdots \ket{\uparrow}$) or
anti-ferromagnetic order ($\ket{\downarrow} \ket{\uparrow} \cdots \ket{\downarrow} \ket{\uparrow} +
\ket{\uparrow} \ket{\downarrow} \cdots \ket{\uparrow} \ket{\downarrow}$). This is illustrated in Figure
\ref{qpt}b. The same technique for realizing the spin-spin
interaction, namely the state-dependent force, can be applied for
many ions  in a string or ion array. Further experimental
investigations in this direction have been performed in
\cite{Kim10,Isl11}. A recent experiment demonstrated the quantum
simulation of antiferromagnetic spin chains using neutral atoms in an optical lattice
\cite{Sim11}.

\subsubsection{Disordered and frustrated systems}

Disordered systems appear in many difficult problems in
condensed-matter physics such as transport, conductivity, spin
glasses and some models of high-$T_c$ superconductivity. The can
exhibit characteristic phenomena that are not present in perfectly
ordered systems. For example, the wave functions of particles can be
localized in disordered media in spite of the presence of hopping
terms in the Hamiltonian. This phenomenon can occur at the
single-particle level by coherent back scattering from random
impurities. Since the theoretical treatment of these problems is
particularly challenging, several proposals for quantum simulation
have been put forward.
\par
The Fano-Anderson Hamiltonian can be studied using DQS. This idea
was pursued theoretically and experimentally (with
liquid-state NMR) in \cite{Neg05}. The fermionic Fano-Anderson
model in one dimension consists of an $n$-site ring with an
impurity in the center. The fermions can hop between
nearest-neighbor sites on the ring or between a site on the ring
and the impurity (with matrix element $V/\sqrt{n}$ for the
latter). The Fano-Anderson Hamiltonian can be written as follows:
\begin{equation}
H_{\rm{FA}}=\sum_{l=0}^{n-1}\varepsilon_{k_l}c_{k_l}^{\dagger}c_{k_l} + \epsilon b^{\dagger}b + V(c_{k_0}^{\dagger}b+c_{k_0}b^{\dagger}),
\end{equation}
where the fermionic operators $c_{k_l}^{\dagger}$ ($c_{k_l}$) and
$b^{\dagger}$ ($b$) create (destroy) a spinless fermion in the
conduction mode $k_l$ and in the impurity, respectively. Note that in principle the Hamiltonian in Eq. (24) can be diagonalized exactly. Nevertheless, the experiment serves as a benchmark for the experimental progress of quantum simulators. Recently,
the decoherence-induced localization of the size of spin clusters
was investigated in an NMR quantum simulator \cite{Alv10}. Thermal states of frustrated spin systems were also simulated recently using NMR \cite{Zha12}.
\par
\begin{figure}
\includegraphics[width=0.45\textwidth ]{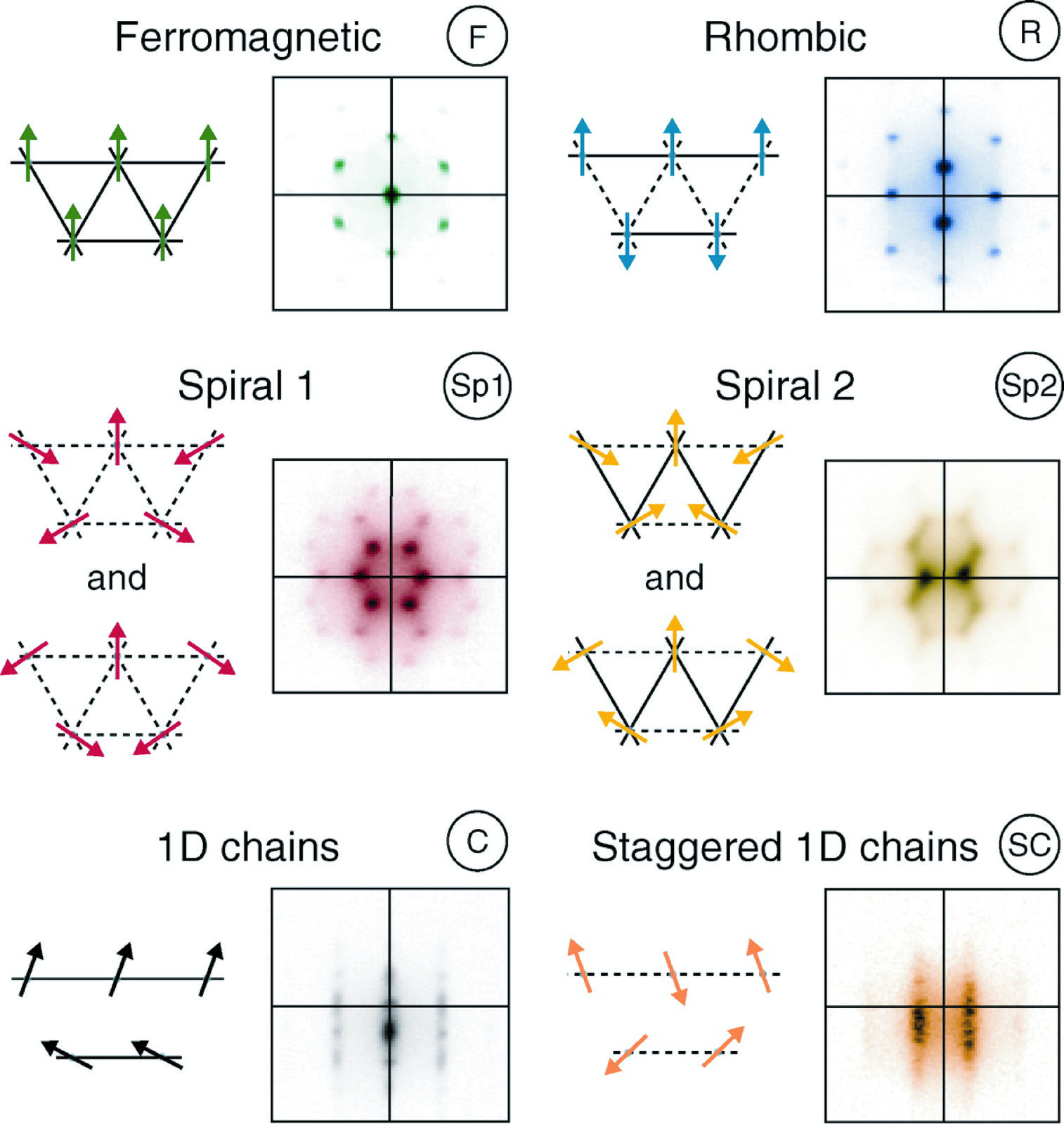}
\caption{\label{tri} (color online) Spin configurations in a
triangular lattice and their experimental signatures, simulated
using the motional degrees of freedom of atoms trapped in an
optical lattice, averaged over several experimental runs. The
coupling parameters can be tuned to ferro- (solid lines) or
antiferromagnetic values (dashed lines), and these determine the
resulting spin configuration (reproduced from \cite{Stu11}).}
\end{figure}

As for AQS, optical lattices have been used to experimentally realize a disordered system that may lead to the observation of a Bose glass phase \cite{Fal07}.
In a  recent experiment, the motional degrees of freedom of atoms
trapped in an optical lattice were used to simulate ferromagnetic,
antiferromagnetic and frustrated classical spin configurations \cite{Stu11}.
A rich phase diagram with different types of phase transitions was observed.
Figure \ref{tri} shows examples of spin configurations in a triangular lattice and their experimental signatures.
Furthermore, theoretical and experimental studies have
investigated possible routes towards Anderson-like localization of
atomic BECs in disordered potentials \cite{SchDre05,Bil08,Roa08}.
A review focused on the simulation of disordered quantum systems
with quantum gases can be found in \cite{Bou10}. A chain of
trapped ions could also be used to explore the physics of
disordered quantum systems \cite{Ber10}. In
\cite{JGR08} the authors proposed a mapping between
superconducting circuits and the Hamiltonians describing magnetic
impurities in conduction bands (Anderson and Kondo models). A
recent experiment on a driven superconducting qubit \cite{Gus13},
along with the appropriate mapping, exhibited an analogue of universal
conductance fluctuations, which are typically studied in the context
of particle propagation in two-dimensional disordered media.
\par
\begin{figure}
\includegraphics[width=0.45\textwidth ]{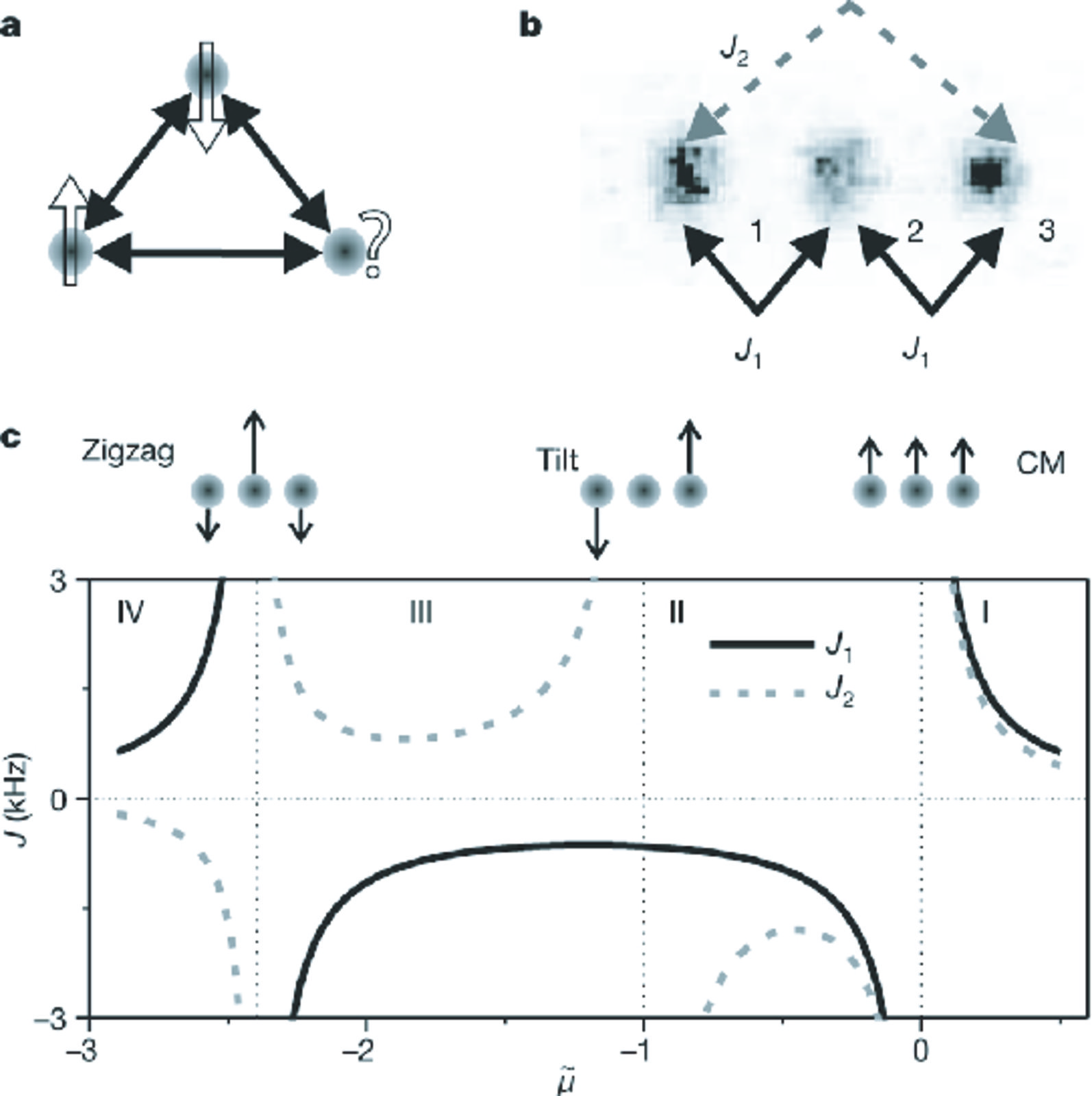}
\caption{\label{spi}An example of the simulation of the simplest
case of spin frustration: three antiferromagnetically coupled spins
(a). This simulation was realized with three ytterbium
ions in a one-dimensional trap (b). $J_1$ is the nearest-neighbor
coupling and $J_2$ the next-nearest-neighbor coupling. In (c) the
expected form of the Ising interactions $J_1$ and $J_2$,
controlled through the laser detuning, $\tilde{\mu}$, used to
generate an optical spin-dependent force, is depicted. The three
oscillation modes shown at the top of Panel C are used as markers,
as the parameter $\tilde\mu$ is defined relative to their
frequencies (reproduced from \cite{Kim10}).}
\end{figure}

Geometric frustration refers to the situation in which the
geometric properties of the crystal lattice forbid the simultaneous
minimization of all the interaction energies acting in a given region
(see Figure \ref{spi}). Well-known examples include the
antiferromagnetic Ising model on a triangular lattice or the
antiferromagnetic Heisenberg model on a Kagome lattice.
Frustrated antiferromagnets are materials whose quantum Monte
Carlo simulation suffers from a severe sign problem. There have been
studies on the AQS of frustrated spin models with trapped ions
\cite{Por06,Kim10,Iva11,Berm11}, NMR \cite{Pen09}, photons
\cite{Ma11} and atoms in optical lattices (see \cite{Lew07} and
references therein, \cite{Liu10,Bec10,Stu11}).

\subsubsection{Spin glasses}

Spin glasses typically occur when the interactions between spins
are ferromagnetic for some bonds and antiferromagnetic for others,
so the spin orientation cannot be uniform in space even at low
temperatures, and the spin orientation can become random and
almost frozen in time. Spin glasses can be efficiently simulated
using DQS. In \cite{Lid97,Lid04} an algorithm for the construction
of the Gibbs distribution of configurations in the Ising model was
developed. The algorithm can be applied to any dimension, to the
class of spin-glass Ising models with a finite portion of
frustrated plaquettes, to diluted Ising models and to models with
a magnetic field.
\par
An analog approach to the simulation of spin glasses using magnetic
impurities embedded in inert matrices, such as solid helium, was
proposed in \cite{Lem13}. A proposal for the analog simulation of
the Lipkin-Meshkov-Glick model and complex quantum systems, such as
Sherrington-Kirkpatrick (SK) spin glasses, using superconducting
qubits in circuit QED was given in \cite{Tso08}. Spin glasses could
also be studied using Fermi-Bose mixtures in inhomogeneous and
random optical lattices as suggested in \cite{San04,Ahu05}.

\subsubsection{Superconductivity}

The high-temperature superconductivity of compounds containing
copper-oxide planes is still a puzzle that might be solved using
large-scale simulations. The CuO$_2$ plane in a high-$T_c$
superconductor \cite{Ore00} could be simulated in an analog manner
by an array of electrostatically defined quantum dots, as suggested
in \cite{Man02} in one of the early proposals for AQS. Moreover, the
analog simulation of the $t$-$J$ model was proposed in \cite{Yam02}.
The $t$-$J$ model describes strongly correlated fermions on a lattice
and it is used in various attempts to understand high-$T_c$
superconductivity. Here, $t$ represents the size of the kinetic
energy of a hole disrupting an antiferromagnet with spin-spin
interaction energy $J$.
\par
The study of BCS pairing in superconductors could be done using
DQS. The general BCS pairing Hamiltonian has the form:
\begin{equation}
H_{\rm{BCS}}=\sum _{m=1} ^{N} \frac{\epsilon _m}{2}(n^F_m +n^F_{-m}) + \sum _{m,l=1} ^{N} V^{+}_{ml}c_m ^{\dagger}c_{-m} ^{\dagger}c_{-l}c_{l},
\end{equation}
where the fermionic $c_m ^{\dagger}$ ($c_m$) are creation (annihilation)
operators, $|m|=1,2,\dots ,N$, denotes all relevant quantum
numbers, the number operator $n^F _{\pm m}=c_{\pm m}
^{\dagger}c_{\pm m}$, and the matrix element $V^{+}_{ml} = \langle
m,-m|V|l, -l\rangle$. A polynomial-time algorithm to model this
BCS Hamiltonian has been proposed in \cite{Wu02,Wan06} and a
two-qubit version of the algorithm was experimentally realized
using NMR \cite{Yan06}.
\par
In another direction related to the quantum simulation of
superconductivity, several groups have observed signatures of the
crossover between a BCS and a BEC superfluid as the strength of
attractive interactions between fermionic particles is varied
(see, e.g., \cite{Reg04,Zwi05} and \cite{Lew07} and references
therein).

\subsubsection{Metamaterials}

The behavior of tunable metamaterials (regular structures obtained
from the periodic arrangement of mesoscopic building blocks) in
the quantum regime could be seen as a quantum simulation of
materials composed of regular atomic structures
\cite{Rak08,Bli08,Rak10}. One example of a quantum metamaterial is
an infinite chain of identical qubits inside a resonator. Such a
system offers new ways of controlling the propagation of
electromagnetic fields, which are not available to standard
materials.

\subsubsection{Topological order}

\begin{figure}
\includegraphics[width=0.45\textwidth ]{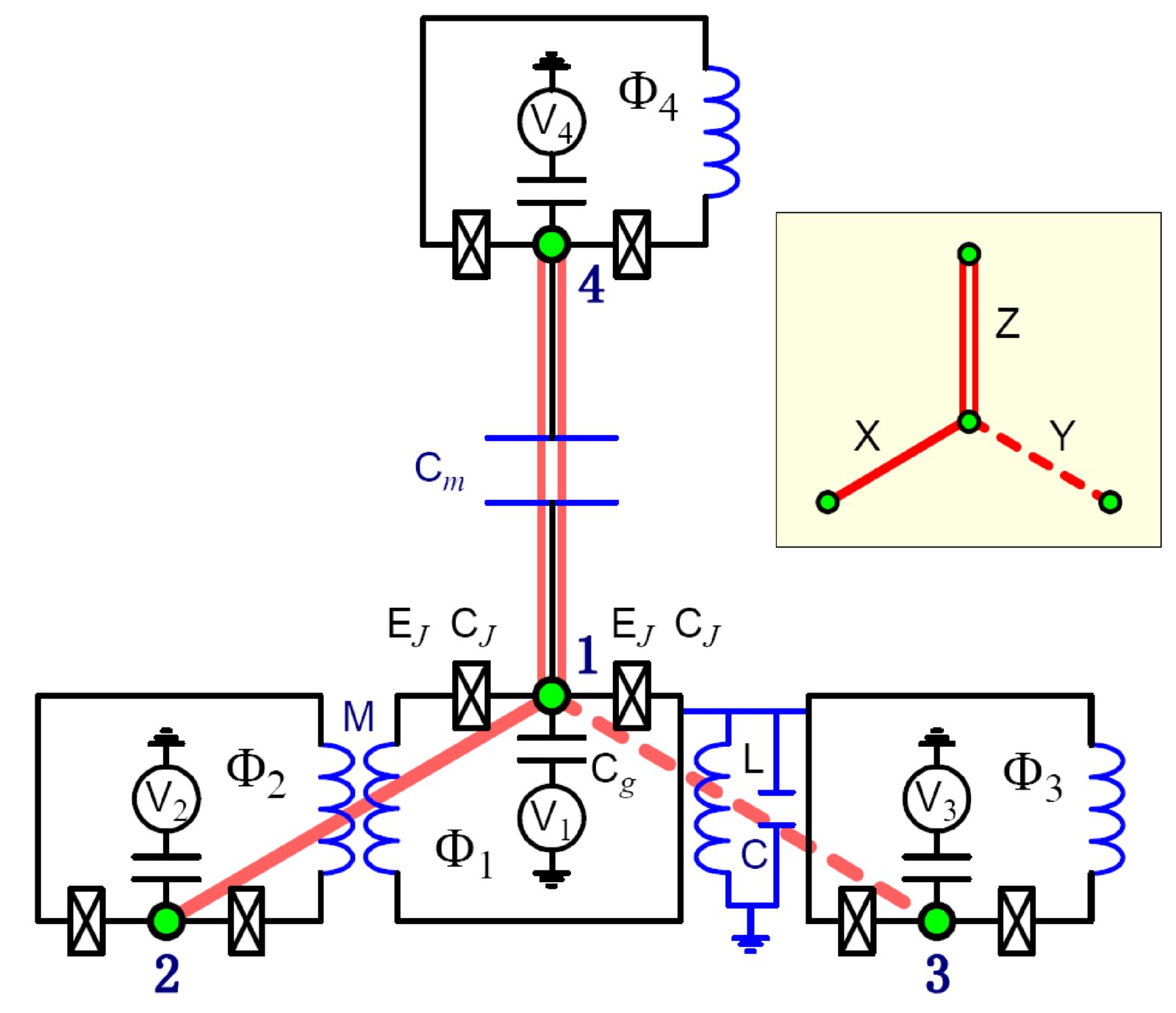}
\caption{\label{any} (color online) Schematic diagram of the basic
building block of a Kitaev lattice, consisting of four
superconducting charge qubits (labelled 1 to 4): (i) Qubits 1 and
2 are inductively coupled via a mutual inductance $M$; (ii) qubits
1 and 3 are coupled via an LC oscillator; and (iii) qubits 1 and 4
are capacitively coupled via a mutual capacitance $C_m$. Inset:
These three types of inter-qubit couplings are denoted as $X$,
$Y$ and $Z$ bonds (reproduced from \cite{You08}).}
\end{figure}
Anyons are two-dimensional particles whose quantum statistics are
neither bosonic nor fermionic. Anyons have been used to describe
two-dimensional systems such as sheets of graphene or the quantum
Hall effect. Moreover, they have been proposed as media for
implementing topological quantum computation \cite{Kit97}. The
fractional statistics of anyons in the Kitaev model \cite{Kit97}
involving four-body interactions could be studied using AQS with
cold atoms in optical lattices or alternatively using
DQS \cite{Han07}. The latter approach has been experimentally
realized with linear optics \cite{Lu09}. The Hamiltonian
\begin{equation}
H_K=-\sum _{v} A_v - \sum _{p} B_p,
\end{equation}
is the sum of mutually commuting stabilizers $A_v =\prod _{i\in v}
\sigma_i^x$ and $B_p=\prod _{i\in p} \sigma_i^z$, where $v$ runs
over the vertices and $p$ over the plaquettes (i.e. the small squares defined by four neighboring vertices) in a square lattice and
the products involve the qubits surrounding the vertices or
plaquettes (Note that in this setup the qubits are placed at the centers of lines connecting neighboring vertices). In optical lattices the creation and manipulation of
abelian and non-abelian anyons in topological lattice models could
be realized using ancilla particles \cite{Agu08}. Furthermore, it
could be possible to construct a Kitaev honeycomb lattice with
superconducting circuits \cite{You08} (see Figure \ref{any}).
Topological models have also been investigated in
\cite{Fre02,Mic06,Bre09,Sta09,Xue11,Kit11}. The implementation of
these proposals would also contribute to the development of
topological quantum computation.

\subsection{High-energy physics}

Another area for the application of quantum simulation that is
already showing promising developments is high-energy physics. The
study of relativistic quantum systems such as gauge fields or
Dirac fermions with quantum simulators was first suggested in
\cite{Bo98}.
\par
More recently, a mapping between the dynamics of the
$2+1$ Dirac oscillator and the Jaynes-Cummings model (in particular
in connection with trapped-ion experiments) was proposed in
\cite{Ber07}. Such mappings would allow the study of
relativistic quantum mechanics described by the Dirac equation in
a non-relativistic quantum system. A method for simulating the
Dirac equation in $3+1$ dimensions for a free spin-1/2 particle
using a single trapped ion was proposed in \cite{Lam07}. This
simulation offers the possibility of studying effects such as
Zitterbewegung and the Klein paradox. Zitterbewegung refers to the
rapid oscillatory motion of a free particle obeying the Dirac
equation. Zitterbewegung has never been observed in its
original form with free relativistic particles, but it has been
simulated with a trapped ion \cite{Ger10}. A recent paper extended
this work to the study of Zitterbewegung in a magnetic field
\cite{Rus10} (see also \cite{Wan10}). The Klein paradox refers to
the situation in relativistic quantum mechanics where a potential
barrier of the order of the electron mass becomes nearly
transparent for the electron. A quantum simulation of this
phenomenon was recently implemented with trapped ions \cite{Ger11}.
A proposal for a simulation using graphene was put forward in
\cite{Kat06}. Note that it is also possible to simulate
classically the Zitterbewegung of a free Dirac electron in an
optical superlattice \cite{Dre10}. Dirac particles could also be
investigated with neutral-atom quantum simulators
\cite{Gol09,Hou09,Cir10,Bra10,Wit10,Cas11}.
\par
\begin{figure}
\includegraphics[width=0.5\textwidth ]{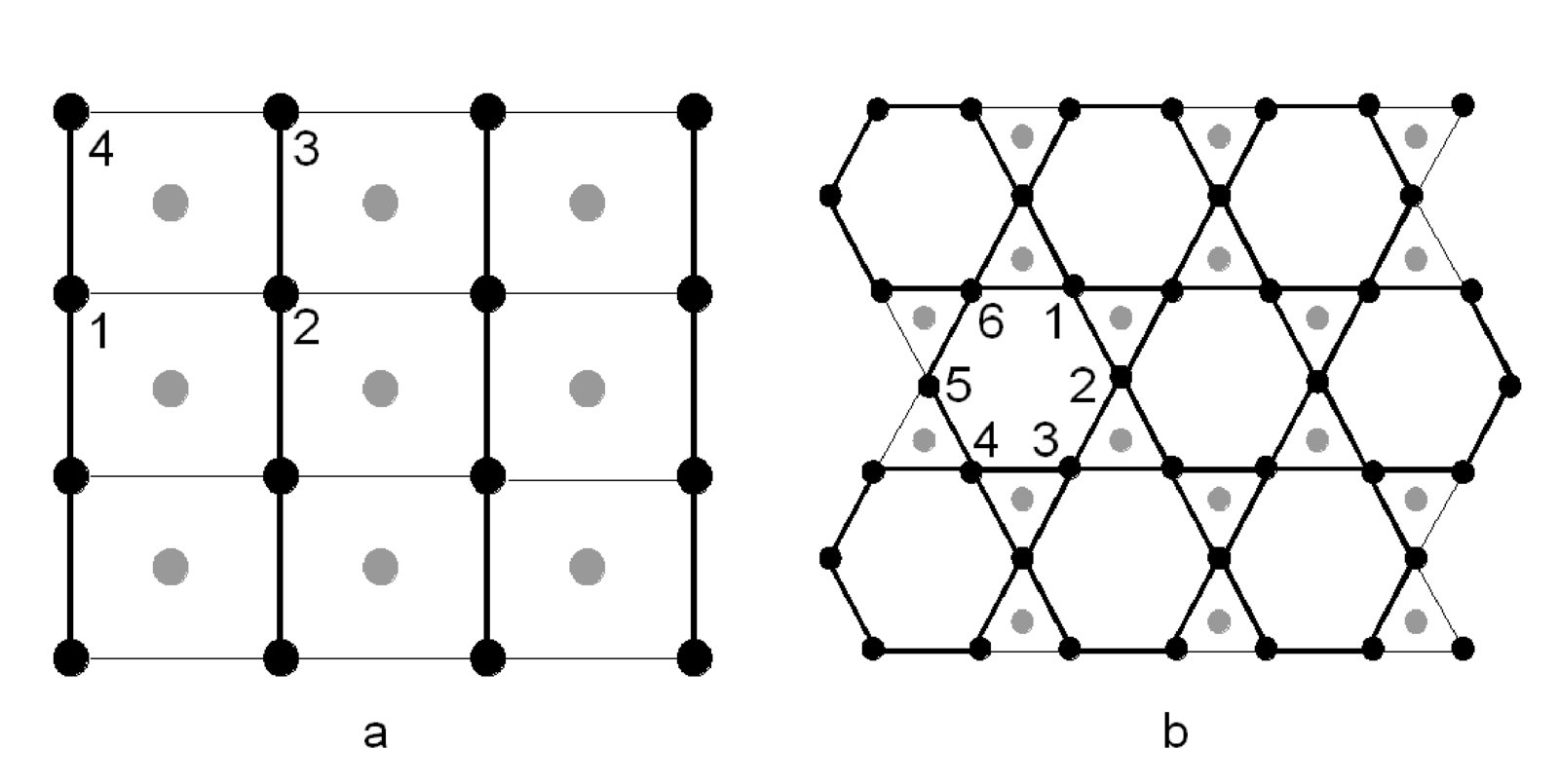}
\caption{\label{lgt} (a) Proposed setup for the simulation of
ring-exchange models \cite{Buc05}. The bosons (black dots) reside on
a square lattice with the molecules (gray dots) at the centers of
the plaquettes (adapted from \cite{Buc05}). (b) Simulation of
compact $U(1)$ lattice gauge theories \cite{Tew06}. Dipolar bosons
reside on the sites of the Kagome lattice (black dots); the
hexagonal dual lattice is the lattice formed by the centers of the
corner-sharing triangles (gray dots) (adapted from \cite{Tew06}).}
\end{figure}
The simulation of gauge theories can be a very computationally
intensive quantum many-body problem, and there have been several
recent proposals for the quantum simulation of such theories. For
example, several theoretical studies have considered abelian
\cite{Kap11,Zoh12,Ban12,Tag13} and non-abelian
\cite{Ban13,Zoh13,Tag12,Sta13} lattice gauge theories using
neutral atoms. A review on the subject was given in \cite{Wie13}.
There have also been related proposals using trapped ions
\cite{Cas11,Cas12}, cavity QED systems \cite{Bar13} and
superconducting circuits \cite{Mar13,Pad13}. A recent experiment
\cite{Lah13} used a superconducting circuit to simulate a massless
Klein-Gordon field with tunable speed of propagation, which was used
to simulate the generation of photons in the dynamical casimir effect.
In \cite{Hau13} the possible simulation of the Schwinger model using
a chain of trapped ions was considered. The simulation of lattice
gauge theories on a DQS was investigated in \cite{Byr06}. The
authors gave the mapping between the operators in the lattice
gauge Hamiltonian for the $U(1)$, $SU(2)$, and $SU(3)$ (which
is particularly important for quantum chromodynamics) lattice gauge theories and
spin operators, and they showed that the algorithm is efficient.
Another proposal for the DQS of field theories was given in
\cite{Jor12}.
\par
Atoms in optical lattices offer the possibility of realizing the
AQS of ring-exchange models \cite{Buc05,Tew06}.
Ring-exchange models describe elementary excitations in a solid,
where, e.g., a ring of hard-core bosons collectively rotate like
a ring around a central boson. The ring exchange Hamiltonian,
\begin{equation}
H_{\rm{RE}}=K\sum _{\rm{plaquettes}} \left( b_1^{\dagger}b_2b_3^{\dagger}b_4+b_4^{\dagger}b_3b_2^{\dagger}b_1 -n_1n_2 -n_3n_4 \right),
\end{equation}
can be realized using atoms with two internal states: one state
trapped in a square lattice (see Figure \ref{lgt}a) and
described by the simple Bose-Hubbard model, and a second one
trapped at the centers of the plaquettes. In trapped atomic gases
a microscopic Hamiltonian can be implemented and its phase diagram
can be studied experimentally via controlling the strength of the
interaction terms \cite{Buc05}. Furthermore, with this system it
is possible to simulate a certain class of strong coupling
Hamiltonians, and in doing so, study exotic phases with strong
correlations. By implementing a Hubbard model with an additional,
strong nearest-neighbor interaction on certain two- or
three-dimensional lattices (for example a two-dimensional Kagome
lattice) the simulation of a $U(1)$ lattice gauge theory would be
possible \cite{Tew06} (see Figure \ref{lgt}b). Non-interacting
relativistic fermionic theories or topological insulators could
also be investigated using these systems \cite{Maz12}.
\par
In \cite{Sem10} a quantum-circuit simulation of the state of a
nucleon was proposed. This simulation could be implemented using a
photonic network. Finally, we also note the possible simulation of
the $O(3)$ nonlinear sigma model using an array of superconducting
and insulating spheres with electrons trapped in the insulating
spheres, as discussed in \cite{Sch05}.

\subsection{Cosmology}

Quantum simulation could also find applications in analog
gravity/cosmology models. For example, acoustic waves in a
two-component BEC could be used to investigate scalar fields
within the curved space-time structure of an expanding universe
\cite{Fis04}. The simulation would be performed by varying the
interparticle coupling and/or expanding the condensate in a
temporal ramp. This idea might be experimentally challenging but
it opens up a new possible way to study cosmology. The study of the
analogue of cosmological particle creation with trapped ions was
proposed in \cite{Sch07}, and more recently, the analogue of
quantum field effects in cosmological spacetimes was
investigated in \cite{Men10}. There are also numerous similarities
between the behavior of superfluid helium and cosmological
phenomena, such as processes in the early universe, as discussed
in detail in \cite{Vol09}. These similarities could be exploited,
and a system of liquid helium could be used for the quantum simulation
of problems in cosmology.
\par
With analog models it would be possible to test predicted
phenomena that have not yet been observed in experiment. Examples
include the possible observation of an Unruh-like effect (i.e.,
the observation by an accelerating observer of a thermal flux of
particles in vacuum) using the phonon excitation of trapped ions
\cite{Als05} and the simulation of the Schwinger effect (i.e., the
production of electron-positron pairs from the vacuum under the
action of a strong electric field) with atoms in an optical
lattice \cite{Szp11,SS11}. Furthermore, analogues of Hawking
radiation can be investigated with atoms \cite{Gio05}, ions
\cite{Hor09} and superconducting circuits \cite{Nat09},
exciton-polariton superfluids in semiconductors \cite{Ger12} or
even with ultrashort pulses of light in microstructured optical
fibers \cite{Phi08} (see also \cite{Nat11}). Recently it has been
demonstrated that Hawking radiation can be measured in a classical
analog system, namely water surface waves \cite{Wei11}.

\subsection{Atomic physics}

\begin{figure}
\includegraphics[width=0.5\textwidth ]{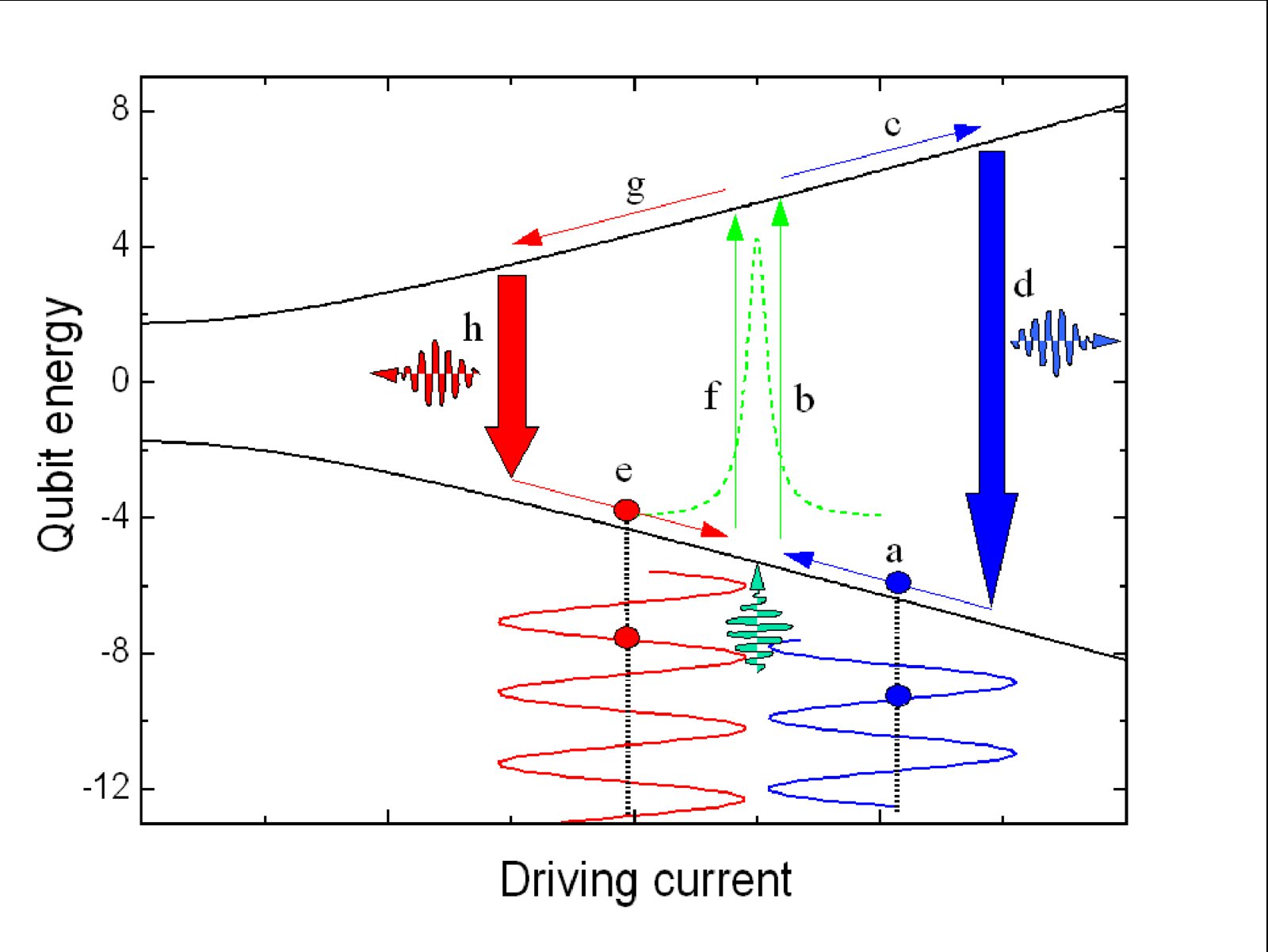}
\caption{\label{cooling} (color online) Sisyphus cooling and
amplification in a superconducting circuit \cite{Gra08,Nor08}. The
two energy levels of a superconducting qubit are shown as a
function of the current applied to the circuit. The current in a
nearby superconducting resonator drives the qubit, and a microwave
drive signal (green) is applied to the qubit. The cycle $a
\rightarrow b \rightarrow c \rightarrow d$ on the right side
mimics Sisyphus cooling in atomic physics: the energy of emitted
photons (blue) is higher than the energy of photons absorbed from
the drive signal, causing a constant removal of energy from the
resonator. In the analogy with Sisyphus in Greek mythology he
current in the resonator plays the role of Sisyphus and the qubit
plays the role of the rock. Sisyphus is constantly pushing the
rock uphill (as can be seen in steps a and c), and the rock keeps
going back to the bottom of the hill. The cycle $e \rightarrow f
\rightarrow g \rightarrow h$ on the left side represents the
opposite scenario, Sisyphus amplification, where energy is
constantly added to the resonator (adapted from \cite{Nor08}).}
\end{figure}
There is a deep analogy between natural atoms and the
artificial atoms formed, for example, by electrons in small
superconducting circuits \cite{You05,You11}. Both have discrete
energy levels and exhibit coherent quantum oscillations between
those levels. But whereas natural atoms are
driven using visible or microwave photons that excite electrons
from one state to another, the artificial atoms in the circuits
are driven by currents, voltages, and microwave photons. The
resulting electric and magnetic fields control the tunneling of
electrons across Josephson junctions. The effects of those fields
on the circuits are the analogues of the Stark and Zeeman effects
in atoms. Differences between quantum circuits and natural atoms
include how strongly each system couples to its environment (the
coupling is weak for atoms and strong for circuits) and the energy
scales of the two systems. In contrast with naturally occurring
atoms, artificial atoms can be lithographically designed to have
specific characteristics, such as a large dipole moment or
particular transition frequencies. This tunability is an important
advantage over natural atoms. Superconducting circuits can also be
used for new types of masers, lasers and single-photon generators
\cite{You07}. These circuits can provide analogues of sideband
cooling and Sisyphus cooling \cite{Gra08,Nor08} (see Figure
\ref{cooling}). Moreover, they can be used to test Bell
inequalities \cite{Ans09}, produce Schr\"odinger-cat states, study
Landau-Zener-St\"uckelberg interferometry \cite{Shev10} and
simulate the Einstein-Podolsky-Rosen experiment \cite{You05}.
Superconducting circuits have also been used to engineer selection
rules and thus create combinations of selection rules that are not
possible with natural atoms \cite{Liu05,Dep08,Har09,deg10}.
\par
One of the most important models in atomic physics and quantum
optics is the Jaynes-Cummings Hamiltonian, which describes the
interaction of a single quantized mode of the electromagnetic
field with a two-level atom:
\begin{equation}
H_{JC}=i\hbar g_0\left( a^{\dagger}\sigma_{-} -a\sigma_{+} \right),
\end{equation}
where $g_0$ is the atom-field coupling strength, the operators
$a^{\dagger}$ and $a$ are the bosonic creation and annihilation
operators, and $\sigma_{+}$ and $\sigma_{-}$ are the atomic
raising and lowering operators.
\par
The Jaynes-Cummings Hamiltonian can be realized with
superconducting circuits (see,
e.g.~\cite{You03,Wal04,You05,Sho08,Zue09,You11}).

\subsection{Quantum chemistry}

Quantum simulators could also have an important contribution in
quantum chemistry (see \cite{Kas10}). For example, an efficient
algorithm for calculating the thermal rate constant was given  in
\cite{Lid99}. The algorithm involves the initialization of the DQS
to an equal superposition of position eigenstates, followed by a
unitary evolution that makes use of the quantum Fourier transform
and finally, a sequence of measurements yielding the energy
spectrum and eigenstates. The algorithm would run in polynomial
time while any exact classical simulation is bound to exhibit
exponential scaling. Another example is the multiconfiguration
self-consistent field wave function-based quantum algorithm
proposed in \cite{Wan08} for obtaining the energy spectrum of
a molecular system.
\par
Moreover, a quantum computer could be used to simulate the
static and dynamical chemical properties
of molecules. In \cite{Guz05} it was shown how to compute
molecular energies using DQS and the required number of qubits
was estimated (see also Figure \ref{molecenerg}). The number of
qubits scales linearly with the number of basis functions, and the
number of gates grows polynomially with the number of
qubits, so simulations with a few tens or a hundred qubits would
already outperform classical computers. The simulation of chemical
dynamics is also achievable in polynomial time \cite{Kas08}. The
molecular energies of the hydrogen molecule have been calculated
using a small photonic DQS \cite{Lan09}. More recently, chemical
reaction dynamics have been investigated using NMR DQS
\cite{Lu11}.
\par
Using AQS it would be possible to simulate chemical reactions.
In \cite{Smi07} a proposal for using the redistribution of
electrons between semiconductor quantum dots to simulate the
redistribution of electrons in a chemical reaction was put
forward. The authors showed how a quantum dot with one electron
can be considered an artificial hydrogen atom and a quantum dot
containing four electrons can be viewed as an artificial oxygen
atom. Depending on the tunnel coupling strengths, electron
distribution, and shell structure, the dots can form both ionic-
and covalent-like bonds. Various reaction regimes and different
reaction products can be obtained by varying the speed of voltage
changes applied to the gates forming the quantum dots. This
promises the modeling of various chemical reactions \cite{Smi07}.
Furthermore, a recent proposal suggests that chemical
reactions can also be simulated with ultracold atoms in a
waveguide \cite{Tor11}.

\subsection{Open quantum systems}

Simulating the dynamics of open quantum systems is even more
costly than that of closed quantum systems because solving the
Lindblad equation requires quadratically more resources than
solving the Schr\"odinger equation for the same physical system.
Simulating open quantum systems with quantum simulators can be
done in two ways. First, one could exploit the natural decoherence
of the simulator, as first suggested in \cite{lloyd96}
and investigated experimentally in \cite{Tse00} (see also the
discussion in Section \ref{deco}). Second, it would also be
possible to simulate open quantum systems with closed quantum
systems. A simulation of decoherence caused by classical noise
that was artificially added to the control signal of a superconducting
qubit was reported in \cite{Li13}. In \cite{Pii06} it was
theoretically shown that a driven harmonic oscillator can
act as a quantum simulator for the non-Markovian damped harmonic
oscillator. Other studies considered the Dicke
model \cite{Sch01,Che07}, the open-system dynamics of a harmonic
oscillator coupled to an artificially engineered reservoir
\cite{Lut98,Man04} or open quantum systems in thermal equilibrium
\cite{Ter02}. Reference \cite{Weim11} discussed an open-system quantum
simulator with Rydberg states of neutral atoms in an optical
lattice. General methods for simulating the markovian dynamics of open
quantum systems have been investigated in \cite{Bac01,Wang11,Ter00}.
A recent experiment with trapped ions demonstrated the possibility of
engineering open-system dynamics using the dissipative preparation of
entangled states \cite{Bar11}. Another recent experiment simulated
non-markovian Dynamics using a linear-optics setup \cite{Chi12}.

\subsection{Quantum chaos}

An application of DQS with a few qubits is the study of the
dynamics of simple quantum maps \cite{Sch98,Ter02,Lev04,Geo04}.
For instance, the quantum baker's map, a prototypical example in
quantum chaos, has an efficient realization in terms of quantum
gates \cite{Sch98}. The quantum baker's map has been
experimentally realized in NMR \cite{Wei02} and with linear optics
\cite{How00}. Another example is the kicked Harper model
\cite{Lev04}. It has been shown that in some cases the quantum
approach to the kicked Harper model can provide a polynomial gain
as compared to classical algorithms. It should exhibit observable
interesting behavior even with only eight qubits.

\subsection{Nuclear physics}

In nuclear physics one must solve an $N$-body quantum problem.
Even though $N$ is not as large as in condensed matter physics (in
this case $N$, the atomic mass number, is smaller than 300), the
calculation of the nuclear force is difficult and therefore
nuclear physics simulations require significant computing power.
Several phenomenological models have been developed, one of which
is the superfluid model of the atomic nucleus. As suggested in
\cite{Nuclear}, this model could be simulated using an analog
controllable system: a superfluid gas of fermionic atoms. Given the
ease of rotating atomic clouds, this approach offers the possibility
of studying the response of nuclei to increasing angular momentum
with unprecedented control over the relevant parameters such as the
interaction strength, particle number, external trapping potential
and rotation frequency. The reviews \cite{Zin08,Zin13} detail
similarities between atomic nuclei and cold atomic gases, which
could be utilized for designing quantum simulators of nuclei using
atomic ensembles.

\subsection{Interferometry}

Nonlinear interferometers were first investigated in trapped-ion
experiments \cite{Lei02}. This was among the first experimental
realizations of quantum simulation. More recently, the study of
Mach-Zehnder interferometry with an array of trapped ions was
explored in \cite{Hu12}.
\par
Boson sampling \cite{Aar11} is another interferometry problem
that is closely related to quantum simulation. Several recent
experiments have demonstrated boson sampling using photons
\cite{Bro13,Spr13,Til13,Cre13}, and Ref. \cite{Lau12} proposed an
implementation using trapped ions.
\par
Superconducting qubits can be used, for example, to perform
Landau-Zener-St\"uckelberg interferometry \cite{Shev10} and
Fano and Fabry-Perot interferometry
\cite{Lan08,Lia10,Hou10}. These latter phenomena can be realized
using quasi-one-dimensional open systems where photons are
injected from one side and move towards the opposite side of the
device. Along the way, the photons interact with either one or two
qubits acting as tunable mirrors, controlled by changing the
applied electric and/or magnetic fields on the qubits. These
qubits, operating as tunable mirrors, can change the reflection
and transmission coefficients of photons confined in waveguides.

\subsection{Other applications}

Some recent topics in physics research, such as Majorana fermions
\cite{YWZ11,Cas11}, graphene \cite{Gib09} or neutrino oscillations
\cite{Noh12} are now discussed in the context of quantum
simulation.
\par
The possibility of simulating the Schr\"odinger equation on a DQS
was first discussed in 1996 \cite{lloyd96}, with concrete
algorithms given in 1998 \cite{Bog98}, where a quantum lattice-gas
model for the many-particle Schr\"odinger equation in $d$
dimensions was proposed. More recently, the single-particle
Schr\"odinger equation was considered in more detail in
\cite{Ben08}, where it was shown that six to ten qubits would be
sufficient for a simulation of the single-particle
Schr\"odinger equation.
\par
Another direction of interest is quantum heat engines, quantum
Brownian motion \cite{Han09,Han05} and quantum thermodynamics
\cite{Mar08}. The quantum versions of the Carnot and other heat
engines were discussed in \cite{Qua06}. A proposal for the
implementation of a quantum heat engine assisted by a Maxwell
demon with superconducting circuits was given in \cite{Quan06}.
\par
There are many potential applications for quantum simulators in
various fields of physics and chemistry. As mentioned at the
beginning of this section, with quantum simulation one could
address problems that are either classically intractable (such as
the numerous examples from condensed-matter physics discussed
above) or experimentally difficult or inaccessible (such as the
examples from high-energy physics and cosmology).
As practical quantum simulators become available, it is to be
expected that more disciplines will wish to add quantum simulation
to their toolbox of research methods and many new applications
will be uncovered.

\begin{table*}
\caption{\label{tapp} Potential applications of quantum simulators
and the physical systems in which they could be implemented, along
with relevant references. Asterisks denote experimental realizations.
We note that this is not an exhaustive list.\vspace{8pt}} \small
\begin{tabular*}{\textwidth}{l*{3}{l}r}
\hline
\textbf{Application} &  &\textbf{Proposed implementation} \\[8pt] \hline
\textbf{Condensed matter physics}:  &   & \\[8pt]
  &  Hubbard models  &  Atoms \cite{Jak98},\cite{Gre02}* \\ 
  &    &  Ions \cite{Den08}\\
  &    &  Polar molecules \cite{Ort09} \\
  &    &  Quantum dots \cite{Byr08} \\
  &    &  Cavities \cite{Har06,Gre06}, \\
  &    &  \cite{Ang07} \\[8pt]
  &  Spin models  &  Atoms \cite{Jan03,JGR04},\\
  &    &  \cite{Sim11,Stu11}* \\
  &    &  Ions \cite{Jan03,Por04},\\
  &    &  \cite{Den05,Ber09},\\
  &    &  \cite{Edw10,Lan11}*,\\
  &    &  \cite{Kim11,Bri12}*\\
  &    &  Cavities \cite{Cho08a,Che10}\\
  &    &  Nuclear spins on diamond surface \cite{Cai13}\\ 
  &    &  Superconducting circuits \cite{Tso10}\\
  &    &  Electrons on helium \cite{Mos08} \\[8pt]
  &  Quantum phase transitions  &  Atoms \cite{Gre02}* \\
  &    &  Polar molecules \cite{Cap10},\\
  &    &  \cite{Pol10} \\
  &    &  Ions \cite{Ret08,GPG10},\\
  &    &  \cite{Iva09}, \cite{Fri08}*,\\
  &    &  NMR, \cite{Rou07,Pen05},\\
  &    &  \cite{Zha08} \\
  &    &  Superconducting circuits \cite{Oud96}* \\[8pt]
  &  Spin glasses  &  DQS \cite{Lid97} \\
  &    &  Superconducting circuits \cite{Tso08} \\[8pt]
  &  Disordered systems & Atoms \cite{SchDre05,Fal07}*, \\
  &    &  \cite{Bil08,Roa08}*\\
  &    &  Ions \cite{Ber10}\\
  &    &  Superconducting circuits \cite{JGR08}\\
  &    &  NMR \cite{Neg05,Alv10}*\\[8pt]
  &  Frustrated systems  &  Ions \cite{Por06}, \cite{Kim10}* \\
  &    &  Photons \cite{Ma11}* \\ [8pt]
  &  High-$T_c$ superconductivity  &  DQS \cite{Yam02}\\
  &    &  Quantum dots \cite{Man02} \\[8pt]
  &  BCS pairing & NMR \cite{Yan06}* \\[8pt]
  &  BCS-BEC crossover  &  Atoms \cite{Reg04,Zwi05}* \\[8pt]
  &  Metamaterials  &  Superconducting circuits \cite{Rak08} \\[8pt]
  &  Time-symmetry breaking  &  Superconducting circuits \cite{Koc10} \\[8pt]
  &  Topological order  &  Atoms \cite{Agu08} \\
  &    &  Polar molecules \cite{Mic06}\\
  &    &  Linear optics \cite{Lu09}*\\
  &    &  Superconducting circuits \cite{You08}  \\[8pt]
\hline
\hline
\end{tabular*}
\end{table*}
\begin{table*}
\caption{\label{tapp1} Continuation of Table II, but
focused on applications other than condensed-matter physics. As in
Table \ref{tapp}, this is not an exhaustive list.\vspace{8pt}}
\begin{tabular*}{\textwidth}{l*{3}{l}r}
\hline
\textbf{Application} &  &\textbf{Proposed implementation} \\[8pt] \hline
\textbf{High-energy physics}: &   & \\[8pt]
  &  Lattice gauge theories  &  DQS \cite{Byr06}\\
  &    &  Atoms \cite{Buc05} \\[8pt]
  &  Dirac particles  &  Ions \cite{Lam07,Rus10},\\
  &    &  \cite{Ger10}*, \cite{Cas10,Cas11} \\
  &    &  Atoms \cite{Gol09,Hou09,Cir10} \\[8pt]
  &  Nucleons &  Photons \cite{Sem10}\\ [8pt]\hline
\textbf{Cosmology}:   &  &  \\[8pt]
  &  Unruh effect  &  Ions \cite{Als05}\\[8pt]
  &  Hawking radiation  &  Atoms \cite{Gio05} \\
  &    &  Ions \cite{Hor09} \\
  &    &  Superconducting circuits \cite{Nat09}\\[8pt]
  &  Universe expansion  &  BEC \cite{Fis04} \\
  &    &  Ions \cite{Sch05,Men10}\\[8pt] \hline
\textbf{Atomic physics}:   &  &  \\[8pt]
  &  Cavity QED & Superconducting circuits \cite{You03}, \cite{Wal04}*\\
  &  Cooling & Superconducting circuits \cite{Gra08}*, \cite{You11}\\[8pt] \hline
\textbf{Open systems}: &  & \\[8pt]
  &    &  NMR \cite{Tse00}*\\
  &    &  Ions \cite{Pii06},\cite{Bar11}*\\
  &    &  Superconducting circuits \cite{Li13}*\\[8pt]\hline
\textbf{Chemistry}: &  &  \\[8pt]
  &  Thermal rate calculations & DQS \cite{Lid99}\\[8pt]
  &  Molecular energies  &  DQS \cite{Guz05} \\
  &    &  Linear optics \cite{Lan09}*, NMR \cite{Du09}*\\[8pt]
  &  Chemical reactions  &  DQS \cite{Kas08}\\
  &    &  Quantum dots \cite{Smi07}\\[8pt] \hline
\textbf{Quantum chaos}: &   & \\[8pt]
  &    &  NMR \cite{Wei02}* \\
  &    &  Linear optics \cite{How00} \\[8pt] \hline
\textbf{Interferometry}: &   & \\[8pt]
  &    &  Ions \cite{Lei02}*,\cite{Hu12,Lau12}\\
  &    &  Photons \cite{Aar11},\cite{Bro13}*, \\
  &    &  \cite{Spr13,Til13,Cre13}* \\
  &    &  Superconducting circuits \cite{Lan08,Lia10} \\[8pt] \hline
\textbf{Other applications}:& & \\[8pt]
  &  Schr\"odinger equation  &  DQS \cite{Bog98}\\[8pt]
  &  Quantum thermodynamics  &  Superconducting circuits \cite{Qua06,Quan06}\\[8pt]
\hline
\hline
\end{tabular*}
\end{table*}

\section{Concluding remarks}

Recent theoretical and experimental results on quantum simulation
lead us to believe that practical quantum simulators will be built
in the near future. The demonstration of quantum simulations using
more than a few tens of qubits would mark the point where
quantum computers (whether digital or analog) surpass their
classical counterparts, at least for certain applications. This
would be a milestone for physics, computer science, and science in
general.
\par
However, there are still issues that must be addressed. From the
experimental point of view, in all proposed quantum simulators
improved controllability and scalability are required. With the
exception of atoms in optical lattices, quantum simulators cannot
yet handle large arrays of qubits. On the other hand, individual
control and readout are difficult to realize for atoms in optical
lattices, while for other systems that is not a problem.
We note here that for some problems where bulk properties are of
interest, individual control and measurement may not be required.
It is also important to note that, as some recent experiments with
trapped ions and superconducting qubits have demonstrated, even
with a small-scale quantum simulator exciting physical phenomena
could be explored.
\par
From the theoretical point of view, further studies of decoherence
and control would be very useful, especially the estimation of the
experimental requirements for each quantum simulator. It is also
of both theoretical and practical importance to investigate when
and to what extent one can make use of the simulator's
decoherence. 
\par
Quantum simulators would not only provide insights into new
physical phenomena, but also help solve difficult quantum
many-body problems. Moreover, theoretical and experimental
progress in quantum simulation will also have a positive impact on
the development of other fields such as adiabatic quantum
computation \cite{Far01}, measurement-based quantum computation
\cite{Rau03}, topological quantum computation \cite{Kit97} and the
theory of quantum computation \cite{Vol08}. For example, adiabatic
quantum computation can be viewed as a special case of quantum
simulation where the ground state of the simulated Hamiltonian
encodes the solution to a computational problem. The ability to
simulate various Hamiltonians could then be useful for realizing
practical adiabatic quantum computation \cite{Ash06}. Progress on
the experimental implementation of quantum simulation would also
be relevant for measurement-based quantum computation. For instance,
ions in planar Coulomb crystals \cite{Tay08,Wun09} and atoms in
optical lattices \cite{Kay06} have been proposed for implementing
measurement-based quantum computation. The study of entanglement
in many-body systems and its relation with quantum phase transitions
\cite{Ami08} should also be mentioned as an exciting direction
closely related to quantum simulation. Finally, the ability to
incorporate non-physical operations into quantum simulations of
physical systems \cite{Mez12} could lead to new possibilities
for studying quantum simulation and quantum systems in general.
\par
Quantum simulation will profoundly impact physics research. It
will provide a new tool for testing physical theories or
predicting the behavior of physical systems in various possible
conditions, in addition to allowing access to new physical
regimes that are currently beyond experimental reach. Even in the
absence of a theory to be tested or realistically motivated
conditions under which the behavior of a physical system is to be
predicted, quantum simulators will also offer scientists a new
realm of exploration. It might very well happen that unpredicted
discoveries are made through the curiosity-driven experimentation
with future quantum simulators.

\begin{acknowledgments}
We thank N. Lambert and P. Nation for useful comments on the
manuscript. This work was partially supported by the ARO, RIKEN
iTHES Project, MURI Center for Dynamic Magneto-Optics, JSPS-RFBR
contract no. 12-02-92100, Grant-in-Aid for Scientific Research
(S), MEXT Kakenhi on Quantum Cybernetics and the JSPS via its
FIRST program.

\end{acknowledgments}
\bibliographystyle{apsrmplong}
\bibliography{qsim}

\begin{thebibliography}{377}
\expandafter\ifx\csname natexlab\endcsname\relax\def\natexlab#1{#1}\fi
\expandafter\ifx\csname bibnamefont\endcsname\relax
  \def\bibnamefont#1{#1}\fi
\expandafter\ifx\csname bibfnamefont\endcsname\relax
  \def\bibfnamefont#1{#1}\fi
\expandafter\ifx\csname citenamefont\endcsname\relax
  \def\citenamefont#1{#1}\fi
\expandafter\ifx\csname url\endcsname\relax
  \def\url#1{\texttt{#1}}\fi
\expandafter\ifx\csname urlprefix\endcsname\relax\def\urlprefix{URL }\fi
\providecommand{\bibinfo}[2]{#2}
\providecommand{\eprint}[2][]{\url{#2}}

\bibitem[{\citenamefont{Aaronson and Arkhipov}(2011)}]{Aar11}
\bibinfo{author}{\bibnamefont{Aaronson}, \bibfnamefont{S.}}, and
  \bibinfo{author}{\bibfnamefont{A.}~\bibnamefont{Arkhipov}},
  \bibinfo{year}{2011}, {``}\bibinfo{title}{The computational complexity of
  linear optics},{''} \bibinfo{journal}{Proceedings of the 43rd Annual ACM
  Symposium on Theory of Computing}  \bibinfo{pages}{333 -- 342}.

\bibitem[{\citenamefont{Abrams and Lloyd}(1997)}]{Abr97}
\bibinfo{author}{\bibnamefont{Abrams}, \bibfnamefont{D.}}, and
  \bibinfo{author}{\bibfnamefont{S.}~\bibnamefont{Lloyd}},
  \bibinfo{year}{1997}, {``}\bibinfo{title}{Simulation of many-body Fermi
  systems on a universal quantum computer},{''} \bibinfo{journal}{Phys. Rev.
  Lett.} \textbf{\bibinfo{volume}{79}},  \bibinfo{pages}{2586--2589}.

\bibitem[{\citenamefont{Abrams and Lloyd}(1999)}]{Abr99}
\bibinfo{author}{\bibnamefont{Abrams}, \bibfnamefont{D.~S.}}, and
  \bibinfo{author}{\bibfnamefont{S.}~\bibnamefont{Lloyd}},
  \bibinfo{year}{1999}, {``}\bibinfo{title}{Quantum algorithm providing
  exponential speed increase for finding eigenvalues and eigenvectors},{''}
  \bibinfo{journal}{Phys. Rev. Lett.} \textbf{\bibinfo{volume}{83}},
  \bibinfo{pages}{5162--5165}.

\bibitem[{\citenamefont{Aguado} \emph{et~al.}(2008)\citenamefont{Aguado,
  Brennen, Verstraete, and Cirac}}]{Agu08}
\bibinfo{author}{\bibnamefont{Aguado}, \bibfnamefont{M.}},
  \bibinfo{author}{\bibfnamefont{G.~K.} \bibnamefont{Brennen}},
  \bibinfo{author}{\bibfnamefont{F.}~\bibnamefont{Verstraete}}, and
  \bibinfo{author}{\bibfnamefont{J.~I.} \bibnamefont{Cirac}},
  \bibinfo{year}{2008}, {``}\bibinfo{title}{Creation, manipulation, and
  detection of abelian and non-abelian anyons in optical lattices},{''}
  \bibinfo{journal}{Phys. Rev. Lett.} \textbf{\bibinfo{volume}{101}},
  \bibinfo{pages}{260501}.

\bibitem[{\citenamefont{Ahufinger} \emph{et~al.}(2005)\citenamefont{Ahufinger,
  Sanchez-Palencia, Kantian, Sanpera, and Lewenstein}}]{Ahu05}
\bibinfo{author}{\bibnamefont{Ahufinger}, \bibfnamefont{V.}},
  \bibinfo{author}{\bibfnamefont{L.}~\bibnamefont{Sanchez-Palencia}},
  \bibinfo{author}{\bibfnamefont{A.}~\bibnamefont{Kantian}},
  \bibinfo{author}{\bibfnamefont{A.}~\bibnamefont{Sanpera}}, and
  \bibinfo{author}{\bibfnamefont{M.}~\bibnamefont{Lewenstein}},
  \bibinfo{year}{2005}, {``}\bibinfo{title}{Disordered ultracold atomic gases
  in optical lattices: A case study of Fermi-Bose mixtures},{''}
  \bibinfo{journal}{Phys. Rev. A} \textbf{\bibinfo{volume}{72}},
  \bibinfo{pages}{063616}.

\bibitem[{\citenamefont{Alsing} \emph{et~al.}(2005)\citenamefont{Alsing,
  Dowling, and Milburn}}]{Als05}
\bibinfo{author}{\bibnamefont{Alsing}, \bibfnamefont{P.}},
  \bibinfo{author}{\bibfnamefont{J.}~\bibnamefont{Dowling}}, and
  \bibinfo{author}{\bibfnamefont{G.}~\bibnamefont{Milburn}},
  \bibinfo{year}{2005}, {``}\bibinfo{title}{Ion trap simulations of quantum
  fields in an expanding universe},{''} \bibinfo{journal}{Phys. Rev. Lett.}
  \textbf{\bibinfo{volume}{94}},  \bibinfo{pages}{220401}.

\bibitem[{\citenamefont{\'Alvarez and Suter}(2010)}]{Alv10}
\bibinfo{author}{\bibnamefont{\'Alvarez}, \bibfnamefont{G.~A.}}, and
  \bibinfo{author}{\bibfnamefont{D.}~\bibnamefont{Suter}},
  \bibinfo{year}{2010}, {``}\bibinfo{title}{NMR quantum simulation of
  localization effects induced by decoherence},{''} \bibinfo{journal}{Phys.
  Rev. Lett.} \textbf{\bibinfo{volume}{104}},  \bibinfo{pages}{230403}.

\bibitem[{\citenamefont{Amasha} \emph{et~al.}(2008)\citenamefont{Amasha,
  MacLean, Radu, Zumbuhl, Kastner, Hanson, and Gossard}}]{Ama08}
\bibinfo{author}{\bibnamefont{Amasha}, \bibfnamefont{S.}},
  \bibinfo{author}{\bibfnamefont{K.}~\bibnamefont{MacLean}},
  \bibinfo{author}{\bibfnamefont{I.~P.} \bibnamefont{Radu}},
  \bibinfo{author}{\bibfnamefont{D.~M.} \bibnamefont{Zumbuhl}},
  \bibinfo{author}{\bibfnamefont{M.~A.} \bibnamefont{Kastner}},
  \bibinfo{author}{\bibfnamefont{M.~P.} \bibnamefont{Hanson}}, and
  \bibinfo{author}{\bibfnamefont{A.~C.} \bibnamefont{Gossard}},
  \bibinfo{year}{2008}, {``}\bibinfo{title}{Electrical control of spin
  relaxation in a quantum dot},{''} \bibinfo{journal}{Phys. Rev. Lett.}
  \textbf{\bibinfo{volume}{100}},  \bibinfo{pages}{046803}.

\bibitem[{\citenamefont{Amico} \emph{et~al.}(2008)\citenamefont{Amico, Fazio,
  Osterloh, and Vedral}}]{Ami08}
\bibinfo{author}{\bibnamefont{Amico}, \bibfnamefont{L.}},
  \bibinfo{author}{\bibfnamefont{R.}~\bibnamefont{Fazio}},
  \bibinfo{author}{\bibfnamefont{A.}~\bibnamefont{Osterloh}}, and
  \bibinfo{author}{\bibfnamefont{V.}~\bibnamefont{Vedral}},
  \bibinfo{year}{2008}, {``}\bibinfo{title}{Entanglement in many-body
  systems},{''} \bibinfo{journal}{Rev. Mod. Phys.}
  \textbf{\bibinfo{volume}{80}},  \bibinfo{pages}{517}.

\bibitem[{\citenamefont{Angelakis} \emph{et~al.}(2013)\citenamefont{Angelakis,
  Huo, Chang, Kwek, and Korepin}}]{Ang13a}
\bibinfo{author}{\bibnamefont{Angelakis}, \bibfnamefont{D.~G.}},
  \bibinfo{author}{\bibfnamefont{M.}~\bibnamefont{Huo}},
  \bibinfo{author}{\bibfnamefont{D.}~\bibnamefont{Chang}},
  \bibinfo{author}{\bibfnamefont{L.~C.} \bibnamefont{Kwek}}, and
  \bibinfo{author}{\bibfnamefont{V.}~\bibnamefont{Korepin}},
  \bibinfo{year}{2013}, {``}\bibinfo{title}{Mimicking interacting relativistic
  theories with stationary light,},{''} \bibinfo{journal}{Phys. Rev. Lett.}
  \textbf{\bibinfo{volume}{110}},  \bibinfo{pages}{100502}.

\bibitem[{\citenamefont{Angelakis} \emph{et~al.}(2011)\citenamefont{Angelakis,
  Huo, Kyoseva, and Kwek}}]{Ang11}
\bibinfo{author}{\bibnamefont{Angelakis}, \bibfnamefont{D.~G.}},
  \bibinfo{author}{\bibfnamefont{M.}~\bibnamefont{Huo}},
  \bibinfo{author}{\bibfnamefont{E.}~\bibnamefont{Kyoseva}}, and
  \bibinfo{author}{\bibfnamefont{L.~C.} \bibnamefont{Kwek}},
  \bibinfo{year}{2011}, {``}\bibinfo{title}{A photonic Luttinger liquid and
  spin-charge separation in a quantum optical system},{''}
  \bibinfo{journal}{Phys. Rev. Lett.} \textbf{\bibinfo{volume}{106}},
  \bibinfo{pages}{153601}.

\bibitem[{\citenamefont{Angelakis and Noh}(2013)}]{Ang13b}
\bibinfo{author}{\bibnamefont{Angelakis}, \bibfnamefont{D.~G.}}, and
  \bibinfo{author}{\bibfnamefont{C.}~\bibnamefont{Noh}}, \bibinfo{year}{2013},
  {``}\bibinfo{title}{Quantum simulation of the Jackiw-Rebbi model with
  photons},{''} \eprint{arXiv:1306.2179}.

\bibitem[{\citenamefont{Angelakis} \emph{et~al.}(2007)\citenamefont{Angelakis,
  Santos, and Bose}}]{Ang07}
\bibinfo{author}{\bibnamefont{Angelakis}, \bibfnamefont{D.~G.}},
  \bibinfo{author}{\bibfnamefont{M.~F.} \bibnamefont{Santos}}, and
  \bibinfo{author}{\bibfnamefont{S.}~\bibnamefont{Bose}}, \bibinfo{year}{2007},
  {``}\bibinfo{title}{Photon-blockade-induced Mott transitions and $XY$ spin
  models in coupled cavity arrays},{''} \bibinfo{journal}{Phys. Rev. A}
  \textbf{\bibinfo{volume}{76}},  \bibinfo{pages}{031805}.

\bibitem[{\citenamefont{Ansmann} \emph{et~al.}(2009)\citenamefont{Ansmann,
  Wang, Bialczak, Hofheinz, Lucero, Neeley, O'Connell, Sank, Weides, Wenner,
  Cleland, and Martinis}}]{Ans09}
\bibinfo{author}{\bibnamefont{Ansmann}, \bibfnamefont{M.}},
  \bibinfo{author}{\bibfnamefont{H.}~\bibnamefont{Wang}},
  \bibinfo{author}{\bibfnamefont{R.~C.} \bibnamefont{Bialczak}},
  \bibinfo{author}{\bibfnamefont{M.}~\bibnamefont{Hofheinz}},
  \bibinfo{author}{\bibfnamefont{E.}~\bibnamefont{Lucero}},
  \bibinfo{author}{\bibfnamefont{M.}~\bibnamefont{Neeley}},
  \bibinfo{author}{\bibfnamefont{A.~D.} \bibnamefont{O'Connell}},
  \bibinfo{author}{\bibfnamefont{D.}~\bibnamefont{Sank}},
  \bibinfo{author}{\bibfnamefont{M.}~\bibnamefont{Weides}},
  \bibinfo{author}{\bibfnamefont{J.}~\bibnamefont{Wenner}},
  \bibinfo{author}{\bibfnamefont{A.~N.} \bibnamefont{Cleland}}, and
  \bibinfo{author}{\bibfnamefont{J.~M.} \bibnamefont{Martinis}},
  \bibinfo{year}{2009}, {``}\bibinfo{title}{Violation of Bell's inequality in
  Josephson phase qubits},{''} \bibinfo{journal}{Nature}
  \textbf{\bibinfo{volume}{461}},  \bibinfo{pages}{504--506}.

\bibitem[{\citenamefont{Ashhab} \emph{et~al.}(2006)\citenamefont{Ashhab,
  Johansson, and Nori}}]{Ash06}
\bibinfo{author}{\bibnamefont{Ashhab}, \bibfnamefont{S.}},
  \bibinfo{author}{\bibfnamefont{J.}~\bibnamefont{Johansson}}, and
  \bibinfo{author}{\bibfnamefont{F.}~\bibnamefont{Nori}}, \bibinfo{year}{2006},
  {``}\bibinfo{title}{Decoherence in a scalable adiabatic quantum
  computer},{''} \bibinfo{journal}{Phys. Rev. A} \textbf{\bibinfo{volume}{74}},
   \bibinfo{pages}{052330}.

\bibitem[{\citenamefont{Ashhab} \emph{et~al.}(2008)\citenamefont{Ashhab,
  Niskanen, Harrabi, Nakamura, Picot, de~Groot, Harmans, Mooij, and
  Nori}}]{Ash08}
\bibinfo{author}{\bibnamefont{Ashhab}, \bibfnamefont{S.}},
  \bibinfo{author}{\bibfnamefont{A.~O.} \bibnamefont{Niskanen}},
  \bibinfo{author}{\bibfnamefont{K.}~\bibnamefont{Harrabi}},
  \bibinfo{author}{\bibfnamefont{Y.}~\bibnamefont{Nakamura}},
  \bibinfo{author}{\bibfnamefont{T.}~\bibnamefont{Picot}},
  \bibinfo{author}{\bibfnamefont{P.~C.} \bibnamefont{de~Groot}},
  \bibinfo{author}{\bibfnamefont{C.~J. P.~M.} \bibnamefont{Harmans}},
  \bibinfo{author}{\bibfnamefont{J.~E.} \bibnamefont{Mooij}}, and
  \bibinfo{author}{\bibfnamefont{F.}~\bibnamefont{Nori}}, \bibinfo{year}{2008},
  {``}\bibinfo{title}{Interqubit coupling mediated by a high-excitation-energy
  quantum object},{''} \bibinfo{journal}{Phys. Rev. B}
  \textbf{\bibinfo{volume}{77}},  \bibinfo{pages}{014510}.

\bibitem[{\citenamefont{Aspuru-Guzik}
  \emph{et~al.}(2005)\citenamefont{Aspuru-Guzik, Dutoi, Love, and
  Head-Gordon}}]{Guz05}
\bibinfo{author}{\bibnamefont{Aspuru-Guzik}, \bibfnamefont{A.}},
  \bibinfo{author}{\bibfnamefont{A.~D.} \bibnamefont{Dutoi}},
  \bibinfo{author}{\bibfnamefont{P.~J.} \bibnamefont{Love}}, and
  \bibinfo{author}{\bibfnamefont{M.}~\bibnamefont{Head-Gordon}},
  \bibinfo{year}{2005}, {``}\bibinfo{title}{Simulated quantum computation of
  molecular energies},{''} \bibinfo{journal}{Science}
  \textbf{\bibinfo{volume}{309}},  \bibinfo{pages}{1704--1707}.

\bibitem[{\citenamefont{Aspuru-Guzik and Walther}(2012)}]{Asp12}
\bibinfo{author}{\bibnamefont{Aspuru-Guzik}, \bibfnamefont{A.}}, and
  \bibinfo{author}{\bibfnamefont{P.}~\bibnamefont{Walther}},
  \bibinfo{year}{2012}, {``}\bibinfo{title}{Photonic quantum simulators},{''}
  \bibinfo{journal}{Nature Physics} \textbf{\bibinfo{volume}{8}},
  \bibinfo{pages}{285--291}.

\bibitem[{\citenamefont{Bacon} \emph{et~al.}(2001)\citenamefont{Bacon, Childs,
  Chuang, Kempe, Leung, and Zhou}}]{Bac01}
\bibinfo{author}{\bibnamefont{Bacon}, \bibfnamefont{D.}},
  \bibinfo{author}{\bibfnamefont{A.~M.} \bibnamefont{Childs}},
  \bibinfo{author}{\bibfnamefont{I.~L.} \bibnamefont{Chuang}},
  \bibinfo{author}{\bibfnamefont{J.}~\bibnamefont{Kempe}},
  \bibinfo{author}{\bibfnamefont{D.~W.} \bibnamefont{Leung}}, and
  \bibinfo{author}{\bibfnamefont{X.}~\bibnamefont{Zhou}}, \bibinfo{year}{2001},
  {``}\bibinfo{title}{Universal simulation of Markovian quantum dynamics},{''}
  \bibinfo{journal}{Phys. Rev. A} \textbf{\bibinfo{volume}{64}},
  \bibinfo{pages}{062302}.

\bibitem[{\citenamefont{Bakr} \emph{et~al.}(2009)\citenamefont{Bakr, Gillen,
  Peng, F\"olling, and Greiner}}]{Bak09}
\bibinfo{author}{\bibnamefont{Bakr}, \bibfnamefont{W.}},
  \bibinfo{author}{\bibfnamefont{J.}~\bibnamefont{Gillen}},
  \bibinfo{author}{\bibfnamefont{A.}~\bibnamefont{Peng}},
  \bibinfo{author}{\bibfnamefont{S.}~\bibnamefont{F\"olling}}, and
  \bibinfo{author}{\bibfnamefont{M.}~\bibnamefont{Greiner}},
  \bibinfo{year}{2009}, {``}\bibinfo{title}{A quantum gas microscope for
  detecting single atoms in a Hubbard-regime optical lattice},{''}
  \bibinfo{journal}{Nature} \textbf{\bibinfo{volume}{462}},
  \bibinfo{pages}{74--77}.

\bibitem[{\citenamefont{Bakr} \emph{et~al.}(2010)\citenamefont{Bakr, Peng, Tai,
  Ma, Simon, Gillen, F\"olling, Pollet, and Greiner}}]{Bak10}
\bibinfo{author}{\bibnamefont{Bakr}, \bibfnamefont{W.~S.}},
  \bibinfo{author}{\bibfnamefont{A.}~\bibnamefont{Peng}},
  \bibinfo{author}{\bibfnamefont{M.~E.} \bibnamefont{Tai}},
  \bibinfo{author}{\bibfnamefont{R.}~\bibnamefont{Ma}},
  \bibinfo{author}{\bibfnamefont{J.}~\bibnamefont{Simon}},
  \bibinfo{author}{\bibfnamefont{J.~I.} \bibnamefont{Gillen}},
  \bibinfo{author}{\bibfnamefont{S.}~\bibnamefont{F\"olling}},
  \bibinfo{author}{\bibfnamefont{L.}~\bibnamefont{Pollet}}, and
  \bibinfo{author}{\bibfnamefont{M.}~\bibnamefont{Greiner}},
  \bibinfo{year}{2010}, {``}\bibinfo{title}{Probing the Superfluid to Mott
  Insulator Transition at the Single--Atom Level},{''}
  \bibinfo{journal}{Science} \textbf{\bibinfo{volume}{329}},
  \bibinfo{pages}{547--550}.

\bibitem[{\citenamefont{Banerjee} \emph{et~al.}(2013)\citenamefont{Banerjee,
  B\"ogli, Dalmonte, Rico, Stebler, Wiese, and Zoller}}]{Ban13}
\bibinfo{author}{\bibnamefont{Banerjee}, \bibfnamefont{D.}},
  \bibinfo{author}{\bibfnamefont{M.}~\bibnamefont{B\"ogli}},
  \bibinfo{author}{\bibfnamefont{M.}~\bibnamefont{Dalmonte}},
  \bibinfo{author}{\bibfnamefont{E.}~\bibnamefont{Rico}},
  \bibinfo{author}{\bibfnamefont{P.}~\bibnamefont{Stebler}},
  \bibinfo{author}{\bibfnamefont{U.-J.} \bibnamefont{Wiese}}, and
  \bibinfo{author}{\bibfnamefont{P.}~\bibnamefont{Zoller}},
  \bibinfo{year}{2013}, {``}\bibinfo{title}{Atomic Quantum Simulation of
  $\mathbf{U}(N)$ and $\mathrm{SU}(N)$ Non-Abelian Lattice Gauge Theories},{''}
  \bibinfo{journal}{Phys. Rev. Lett.} \textbf{\bibinfo{volume}{110}},
  \bibinfo{pages}{125303}.

\bibitem[{\citenamefont{Banerjee} \emph{et~al.}(2012)\citenamefont{Banerjee,
  Dalmonte, M\"uller, Rico, Stebler, Wiese, and Zoller}}]{Ban12}
\bibinfo{author}{\bibnamefont{Banerjee}, \bibfnamefont{D.}},
  \bibinfo{author}{\bibfnamefont{M.}~\bibnamefont{Dalmonte}},
  \bibinfo{author}{\bibfnamefont{M.}~\bibnamefont{M\"uller}},
  \bibinfo{author}{\bibfnamefont{E.}~\bibnamefont{Rico}},
  \bibinfo{author}{\bibfnamefont{P.}~\bibnamefont{Stebler}},
  \bibinfo{author}{\bibfnamefont{U.-J.} \bibnamefont{Wiese}}, and
  \bibinfo{author}{\bibfnamefont{P.}~\bibnamefont{Zoller}},
  \bibinfo{year}{2012}, {``}\bibinfo{title}{Atomic Quantum Simulation of
  Dynamical Gauge Fields Coupled to Fermionic Matter: From String Breaking to
  Evolution after a Quench},{''} \bibinfo{journal}{Phys. Rev. Lett.}
  \textbf{\bibinfo{volume}{109}},  \bibinfo{pages}{175302}.

\bibitem[{\citenamefont{Barreiro} \emph{et~al.}(2011)\citenamefont{Barreiro,
  Mueller, Schindler, Nigg, Monz, Chwalla, Hennrich, Roos, Zoller, and
  Blatt}}]{Bar11}
\bibinfo{author}{\bibnamefont{Barreiro}, \bibfnamefont{J.~T.}},
  \bibinfo{author}{\bibfnamefont{M.}~\bibnamefont{Mueller}},
  \bibinfo{author}{\bibfnamefont{P.}~\bibnamefont{Schindler}},
  \bibinfo{author}{\bibfnamefont{D.}~\bibnamefont{Nigg}},
  \bibinfo{author}{\bibfnamefont{T.}~\bibnamefont{Monz}},
  \bibinfo{author}{\bibfnamefont{M.}~\bibnamefont{Chwalla}},
  \bibinfo{author}{\bibfnamefont{M.}~\bibnamefont{Hennrich}},
  \bibinfo{author}{\bibfnamefont{C.~F.} \bibnamefont{Roos}},
  \bibinfo{author}{\bibfnamefont{P.}~\bibnamefont{Zoller}}, and
  \bibinfo{author}{\bibfnamefont{R.}~\bibnamefont{Blatt}},
  \bibinfo{year}{2011}, {``}\bibinfo{title}{An open-system quantum simulator
  with trapped ions},{''} \bibinfo{journal}{Nature}
  \textbf{\bibinfo{volume}{470}},  \bibinfo{pages}{486}.

\bibitem[{\citenamefont{Barrett} \emph{et~al.}(2013)\citenamefont{Barrett,
  Hammerer, Harrison, Northup, and Osborne}}]{Bar13}
\bibinfo{author}{\bibnamefont{Barrett}, \bibfnamefont{S.}},
  \bibinfo{author}{\bibfnamefont{K.}~\bibnamefont{Hammerer}},
  \bibinfo{author}{\bibfnamefont{S.}~\bibnamefont{Harrison}},
  \bibinfo{author}{\bibfnamefont{T.~E.} \bibnamefont{Northup}}, and
  \bibinfo{author}{\bibfnamefont{T.~J.} \bibnamefont{Osborne}},
  \bibinfo{year}{2013}, {``}\bibinfo{title}{Simulating Quantum Fields with
  Cavity QED},{''} \bibinfo{journal}{Phys. Rev. Lett.}
  \textbf{\bibinfo{volume}{110}},  \bibinfo{pages}{090501}.

\bibitem[{\citenamefont{Becker} \emph{et~al.}(2010)\citenamefont{Becker,
  Soltan-Panahi, Kronjager, Doscher, Bongs, and Sengstock}}]{Bec10}
\bibinfo{author}{\bibnamefont{Becker}, \bibfnamefont{C.}},
  \bibinfo{author}{\bibfnamefont{P.}~\bibnamefont{Soltan-Panahi}},
  \bibinfo{author}{\bibfnamefont{J.}~\bibnamefont{Kronjager}},
  \bibinfo{author}{\bibfnamefont{S.}~\bibnamefont{Doscher}},
  \bibinfo{author}{\bibfnamefont{K.}~\bibnamefont{Bongs}}, and
  \bibinfo{author}{\bibfnamefont{K.}~\bibnamefont{Sengstock}},
  \bibinfo{year}{2010}, {``}\bibinfo{title}{Ultracold quantum gases in
  triangular optical lattices},{''} \bibinfo{journal}{New J. Phys.}
  \textbf{\bibinfo{volume}{12}},  \bibinfo{pages}{065025}.

\bibitem[{\citenamefont{Benenti and Strini}(2008)}]{Ben08}
\bibinfo{author}{\bibnamefont{Benenti}, \bibfnamefont{G.}}, and
  \bibinfo{author}{\bibfnamefont{G.}~\bibnamefont{Strini}},
  \bibinfo{year}{2008}, {``}\bibinfo{title}{Quantum simulation of the
  single-particle Schr\"odinger equation},{''} \bibinfo{journal}{Am. J. Phys.}
  \textbf{\bibinfo{volume}{76}},  \bibinfo{pages}{657--662}.

\bibitem[{\citenamefont{Bennett} \emph{et~al.}(2002)\citenamefont{Bennett,
  Cirac, Leifer, Leung, Linden, Popescu, and Vidal}}]{Ben02}
\bibinfo{author}{\bibnamefont{Bennett}, \bibfnamefont{C.~H.}},
  \bibinfo{author}{\bibfnamefont{J.~I.} \bibnamefont{Cirac}},
  \bibinfo{author}{\bibfnamefont{M.~S.} \bibnamefont{Leifer}},
  \bibinfo{author}{\bibfnamefont{D.~W.} \bibnamefont{Leung}},
  \bibinfo{author}{\bibfnamefont{N.}~\bibnamefont{Linden}},
  \bibinfo{author}{\bibfnamefont{S.}~\bibnamefont{Popescu}}, and
  \bibinfo{author}{\bibfnamefont{G.}~\bibnamefont{Vidal}},
  \bibinfo{year}{2002}, {``}\bibinfo{title}{Optimal simulation of two-qubit
  Hamiltonians using general local operations},{''} \bibinfo{journal}{Phys.
  Rev. A} \textbf{\bibinfo{volume}{66}},  \bibinfo{pages}{012305}.

\bibitem[{\citenamefont{Bermudez} \emph{et~al.}(2011)\citenamefont{Bermudez,
  Almeida, Schmidt-Kaler, Retzker, and Plenio}}]{Berm11}
\bibinfo{author}{\bibnamefont{Bermudez}, \bibfnamefont{A.}},
  \bibinfo{author}{\bibfnamefont{J.}~\bibnamefont{Almeida}},
  \bibinfo{author}{\bibfnamefont{F.}~\bibnamefont{Schmidt-Kaler}},
  \bibinfo{author}{\bibfnamefont{A.}~\bibnamefont{Retzker}}, and
  \bibinfo{author}{\bibfnamefont{M.~B.} \bibnamefont{Plenio}},
  \bibinfo{year}{2011}, {``}\bibinfo{title}{Frustrated Quantum Spin Models with
  Cold Coulomb Crystals},{''} \bibinfo{journal}{Phys. Rev. Lett.}
  \textbf{\bibinfo{volume}{107}},  \bibinfo{pages}{207209}.

\bibitem[{\citenamefont{Bermudez} \emph{et~al.}(2010)\citenamefont{Bermudez,
  Martin-Delgado, and Porras}}]{Ber10}
\bibinfo{author}{\bibnamefont{Bermudez}, \bibfnamefont{A.}},
  \bibinfo{author}{\bibfnamefont{M.}~\bibnamefont{Martin-Delgado}}, and
  \bibinfo{author}{\bibfnamefont{D.}~\bibnamefont{Porras}},
  \bibinfo{year}{2010}, {``}\bibinfo{title}{Localization of phonons in ion
  traps with controlled quantum disorder},{''} \bibinfo{journal}{New J. Phys.}
  \textbf{\bibinfo{volume}{12}},  \bibinfo{pages}{123016}.

\bibitem[{\citenamefont{Bermudez} \emph{et~al.}(2007)\citenamefont{Bermudez,
  Martin-Delgado, and Solano}}]{Ber07}
\bibinfo{author}{\bibnamefont{Bermudez}, \bibfnamefont{A.}},
  \bibinfo{author}{\bibfnamefont{M.~A.} \bibnamefont{Martin-Delgado}}, and
  \bibinfo{author}{\bibfnamefont{E.}~\bibnamefont{Solano}},
  \bibinfo{year}{2007}, {``}\bibinfo{title}{Exact mapping of the 2+1 Dirac
  oscillator onto the Jaynes-Cummings model: ion-trap experimental
  proposal},{''} \bibinfo{journal}{Phys. Rev. A} \textbf{\bibinfo{volume}{76}},
   \bibinfo{pages}{041801}.

\bibitem[{\citenamefont{Bermudez} \emph{et~al.}(2009)\citenamefont{Bermudez,
  Porras, and Martin-Delgado}}]{Ber09}
\bibinfo{author}{\bibnamefont{Bermudez}, \bibfnamefont{A.}},
  \bibinfo{author}{\bibfnamefont{D.}~\bibnamefont{Porras}}, and
  \bibinfo{author}{\bibfnamefont{M.~A.} \bibnamefont{Martin-Delgado}},
  \bibinfo{year}{2009}, {``}\bibinfo{title}{Competing many-body interactions in
  systems of trapped ions},{''} \bibinfo{journal}{Phys. Rev. A}
  \textbf{\bibinfo{volume}{79}},  \bibinfo{pages}{060303}.

\bibitem[{\citenamefont{Berry} \emph{et~al.}(2007)\citenamefont{Berry, Ahokas,
  Cleve, and Sanders}}]{Berr07}
\bibinfo{author}{\bibnamefont{Berry}, \bibfnamefont{D.~W.}},
  \bibinfo{author}{\bibfnamefont{G.}~\bibnamefont{Ahokas}},
  \bibinfo{author}{\bibfnamefont{R.}~\bibnamefont{Cleve}}, and
  \bibinfo{author}{\bibfnamefont{B.~C.} \bibnamefont{Sanders}},
  \bibinfo{year}{2007}, {``}\bibinfo{title}{Efficient quantum algorithms for
  simulating sparse Hamiltonians},{''} \bibinfo{journal}{Comm. Math. Phys.}
  \textbf{\bibinfo{volume}{270}},  \bibinfo{pages}{359--371}.

\bibitem[{\citenamefont{Biamonte} \emph{et~al.}(2011)\citenamefont{Biamonte,
  Bergholm, Whitfield, Fitzsimons, and Aspuru-Guzik}}]{Bia11}
\bibinfo{author}{\bibnamefont{Biamonte}, \bibfnamefont{J.}},
  \bibinfo{author}{\bibfnamefont{V.}~\bibnamefont{Bergholm}},
  \bibinfo{author}{\bibfnamefont{J.}~\bibnamefont{Whitfield}},
  \bibinfo{author}{\bibfnamefont{J.}~\bibnamefont{Fitzsimons}}, and
  \bibinfo{author}{\bibfnamefont{A.}~\bibnamefont{Aspuru-Guzik}},
  \bibinfo{year}{2011}, {``}\bibinfo{title}{Adiabatic Quantum Simulators},{''}
  \bibinfo{journal}{AIP Advances} \textbf{\bibinfo{volume}{1}},
  \bibinfo{pages}{022126}.

\bibitem[{\citenamefont{Biercuk} \emph{et~al.}(2009)\citenamefont{Biercuk, Uys,
  Vandevender, Shiga, Itano, and Bollinger}}]{Bie09}
\bibinfo{author}{\bibnamefont{Biercuk}, \bibfnamefont{M.~J.}},
  \bibinfo{author}{\bibfnamefont{H.}~\bibnamefont{Uys}},
  \bibinfo{author}{\bibfnamefont{A.}~\bibnamefont{Vandevender}},
  \bibinfo{author}{\bibfnamefont{N.}~\bibnamefont{Shiga}},
  \bibinfo{author}{\bibfnamefont{W.~M.} \bibnamefont{Itano}}, and
  \bibinfo{author}{\bibfnamefont{J.~J.} \bibnamefont{Bollinger}},
  \bibinfo{year}{2009}, {``}\bibinfo{title}{High-fidelity quantum control using
  ion crystals in a Penning trap},{''} \bibinfo{journal}{Quantum Inform.
  Comput.} \textbf{\bibinfo{volume}{9}},  \bibinfo{pages}{920 -- 949}.

\bibitem[{\citenamefont{Billy} \emph{et~al.}(2008)\citenamefont{Billy, Josse,
  Zuo, Bernard, Hambrecht, Lugan, Clement, Sanchez-Palencia, Bouyer, and
  Aspect}}]{Bil08}
\bibinfo{author}{\bibnamefont{Billy}, \bibfnamefont{J.}},
  \bibinfo{author}{\bibfnamefont{V.}~\bibnamefont{Josse}},
  \bibinfo{author}{\bibfnamefont{Z.}~\bibnamefont{Zuo}},
  \bibinfo{author}{\bibfnamefont{A.}~\bibnamefont{Bernard}},
  \bibinfo{author}{\bibfnamefont{B.}~\bibnamefont{Hambrecht}},
  \bibinfo{author}{\bibfnamefont{P.}~\bibnamefont{Lugan}},
  \bibinfo{author}{\bibfnamefont{D.}~\bibnamefont{Clement}},
  \bibinfo{author}{\bibfnamefont{L.}~\bibnamefont{Sanchez-Palencia}},
  \bibinfo{author}{\bibfnamefont{P.}~\bibnamefont{Bouyer}}, and
  \bibinfo{author}{\bibfnamefont{A.}~\bibnamefont{Aspect}},
  \bibinfo{year}{2008}, {``}\bibinfo{title}{Direct observation of Anderson
  localization of matter waves in a controlled disorder},{''}
  \bibinfo{journal}{Nature} \textbf{\bibinfo{volume}{453}},
  \bibinfo{pages}{891--894}.

\bibitem[{\citenamefont{Blatt and Roos}(2012)}]{Bla12}
\bibinfo{author}{\bibnamefont{Blatt}, \bibfnamefont{R.}}, and
  \bibinfo{author}{\bibfnamefont{C.~F.} \bibnamefont{Roos}},
  \bibinfo{year}{2012}, {``}\bibinfo{title}{Quantum simulations with trapped
  ions},{''} \bibinfo{journal}{Nature Physics} \textbf{\bibinfo{volume}{8}},
  \bibinfo{pages}{277--284}.

\bibitem[{\citenamefont{Blatt and Wineland}(2008)}]{Bla08}
\bibinfo{author}{\bibnamefont{Blatt}, \bibfnamefont{R.}}, and
  \bibinfo{author}{\bibfnamefont{D.}~\bibnamefont{Wineland}},
  \bibinfo{year}{2008}, {``}\bibinfo{title}{Entangled states of trapped atomic
  ions},{''} \bibinfo{journal}{Nature} \textbf{\bibinfo{volume}{453}},
  \bibinfo{pages}{1008--1015}.

\bibitem[{\citenamefont{Bliokh} \emph{et~al.}(2008)\citenamefont{Bliokh,
  Bliokh, Freilikher, Savel'ev, and Nori}}]{Bli08}
\bibinfo{author}{\bibnamefont{Bliokh}, \bibfnamefont{K.~Y.}},
  \bibinfo{author}{\bibfnamefont{Y.~P.} \bibnamefont{Bliokh}},
  \bibinfo{author}{\bibfnamefont{V.}~\bibnamefont{Freilikher}},
  \bibinfo{author}{\bibfnamefont{S.}~\bibnamefont{Savel'ev}}, and
  \bibinfo{author}{\bibfnamefont{F.}~\bibnamefont{Nori}}, \bibinfo{year}{2008},
  {``}\bibinfo{title}{Unusual resonators: plasmonics, metamaterials, and random
  media},{''} \bibinfo{journal}{Rev. Mod. Phys.} \textbf{\bibinfo{volume}{80}},
   \bibinfo{pages}{1201--1213}.

\bibitem[{\citenamefont{Bloch} \emph{et~al.}(2012)\citenamefont{Bloch,
  Dalibard, and Nascimbene}}]{Blo12}
\bibinfo{author}{\bibnamefont{Bloch}, \bibfnamefont{I.}},
  \bibinfo{author}{\bibfnamefont{J.}~\bibnamefont{Dalibard}}, and
  \bibinfo{author}{\bibfnamefont{S.}~\bibnamefont{Nascimbene}},
  \bibinfo{year}{2012}, {``}\bibinfo{title}{Quantum simulations with ultracold
  quantum gases},{''} \bibinfo{journal}{Nature Physics}
  \textbf{\bibinfo{volume}{8}},  \bibinfo{pages}{267--276}.

\bibitem[{\citenamefont{Bloch} \emph{et~al.}(2008)\citenamefont{Bloch,
  Dalibard, and Zwerger}}]{Blo08}
\bibinfo{author}{\bibnamefont{Bloch}, \bibfnamefont{I.}},
  \bibinfo{author}{\bibfnamefont{J.}~\bibnamefont{Dalibard}}, and
  \bibinfo{author}{\bibfnamefont{W.}~\bibnamefont{Zwerger}},
  \bibinfo{year}{2008}, {``}\bibinfo{title}{Many-body physics with ultracold
  gases},{''} \bibinfo{journal}{Rev. Mod. Phys.} \textbf{\bibinfo{volume}{80}},
   \bibinfo{pages}{885}.

\bibitem[{\citenamefont{Boghosian and Taylor}(1998{\natexlab{a}})}]{Bog98}
\bibinfo{author}{\bibnamefont{Boghosian}, \bibfnamefont{B.~M.}}, and
  \bibinfo{author}{\bibfnamefont{W.}~\bibnamefont{Taylor}},
  \bibinfo{year}{1998}{\natexlab{a}}, {``}\bibinfo{title}{Quantum lattice-gas
  model for the many-particle Schr\"odinger equation in $d$ dimensions},{''}
  \bibinfo{journal}{Phys. Rev. E} \textbf{\bibinfo{volume}{57}},
  \bibinfo{pages}{54--66}.

\bibitem[{\citenamefont{Boghosian and Taylor}(1998{\natexlab{b}})}]{Bo98}
\bibinfo{author}{\bibnamefont{Boghosian}, \bibfnamefont{B.~M.}}, and
  \bibinfo{author}{\bibfnamefont{W.}~\bibnamefont{Taylor}},
  \bibinfo{year}{1998}{\natexlab{b}}, {``}\bibinfo{title}{Simulating quantum
  mechanics on a quantum computer},{''} \bibinfo{journal}{Physica D}
  \textbf{\bibinfo{volume}{120}},  \bibinfo{pages}{30--42}.

\bibitem[{\citenamefont{Bouyer}(2010)}]{Bou10}
\bibinfo{author}{\bibnamefont{Bouyer}, \bibfnamefont{P.}},
  \bibinfo{year}{2010}, {``}\bibinfo{title}{Quantum gases and optical speckle:
  a new tool to simulate disordered quantum systems},{''}
  \bibinfo{journal}{Rep. Progr. Phys.} \textbf{\bibinfo{volume}{73}},
  \bibinfo{pages}{062401}.

\bibitem[{\citenamefont{Brandao} \emph{et~al.}(2007)\citenamefont{Brandao,
  Hartmann, and Plenio}}]{Bra07}
\bibinfo{author}{\bibnamefont{Brandao}, \bibfnamefont{F.~G. S.~L.}},
  \bibinfo{author}{\bibfnamefont{M.~J.} \bibnamefont{Hartmann}}, and
  \bibinfo{author}{\bibfnamefont{M.~B.} \bibnamefont{Plenio}},
  \bibinfo{year}{2007}, {``}\bibinfo{title}{Quantum phase transitions with
  photons and polaritons},{''} \eprint{arXiv:quant-ph/0702003}.

\bibitem[{\citenamefont{Braun}(2010)}]{Bra10}
\bibinfo{author}{\bibnamefont{Braun}, \bibfnamefont{D.}}, \bibinfo{year}{2010},
  {``}\bibinfo{title}{Versatile cold-atom simulator of non-Abelian gauge
  potentials},{''} \bibinfo{journal}{Phys. Rev. A}
  \textbf{\bibinfo{volume}{82}},  \bibinfo{pages}{013617}.

\bibitem[{\citenamefont{Bravyi} \emph{et~al.}(2008)\citenamefont{Bravyi,
  DiVincenzo, Loss, and Terhal}}]{Bra08}
\bibinfo{author}{\bibnamefont{Bravyi}, \bibfnamefont{S.}},
  \bibinfo{author}{\bibfnamefont{D.~P.} \bibnamefont{DiVincenzo}},
  \bibinfo{author}{\bibfnamefont{D.}~\bibnamefont{Loss}}, and
  \bibinfo{author}{\bibfnamefont{B.~M.} \bibnamefont{Terhal}},
  \bibinfo{year}{2008}, {``}\bibinfo{title}{Quantum simulation of many-body
  Hamiltonians using perturbation theory with bounded-strength
  interactions},{''} \bibinfo{journal}{Phys. Rev. Lett.}
  \textbf{\bibinfo{volume}{101}},  \bibinfo{pages}{070503}.

\bibitem[{\citenamefont{Bremner} \emph{et~al.}(2005)\citenamefont{Bremner,
  Bacon, and Nielsen}}]{Bre05}
\bibinfo{author}{\bibnamefont{Bremner}, \bibfnamefont{M.~J.}},
  \bibinfo{author}{\bibfnamefont{D.}~\bibnamefont{Bacon}}, and
  \bibinfo{author}{\bibfnamefont{M.~A.} \bibnamefont{Nielsen}},
  \bibinfo{year}{2005}, {``}\bibinfo{title}{Simulating Hamiltonian dynamics
  using many-qudit Hamiltonians and local unitary control},{''}
  \bibinfo{journal}{Phys. Rev. A} \textbf{\bibinfo{volume}{71}},
  \bibinfo{pages}{052312}.

\bibitem[{\citenamefont{Brennen} \emph{et~al.}(2009)\citenamefont{Brennen,
  Aguado, and Cirac}}]{Bre09}
\bibinfo{author}{\bibnamefont{Brennen}, \bibfnamefont{G.}},
  \bibinfo{author}{\bibfnamefont{M.}~\bibnamefont{Aguado}}, and
  \bibinfo{author}{\bibfnamefont{J.}~\bibnamefont{Cirac}},
  \bibinfo{year}{2009}, {``}\bibinfo{title}{Simulations of quantum double
  models},{''} \bibinfo{journal}{New J. Phys.} \textbf{\bibinfo{volume}{11}},
  \bibinfo{pages}{053009}.

\bibitem[{\citenamefont{Britton} \emph{et~al.}(2012)\citenamefont{Britton,
  Sawyer, Keith, Wang, Freericks, Uys, Biercuk, and Bollinger}}]{Bri12}
\bibinfo{author}{\bibnamefont{Britton}, \bibfnamefont{J.~W.}},
  \bibinfo{author}{\bibfnamefont{B.~C.} \bibnamefont{Sawyer}},
  \bibinfo{author}{\bibfnamefont{A.~C.} \bibnamefont{Keith}},
  \bibinfo{author}{\bibfnamefont{C.~C.~J.} \bibnamefont{Wang}},
  \bibinfo{author}{\bibfnamefont{J.~K.} \bibnamefont{Freericks}},
  \bibinfo{author}{\bibfnamefont{H.}~\bibnamefont{Uys}},
  \bibinfo{author}{\bibfnamefont{M.~J.} \bibnamefont{Biercuk}}, and
  \bibinfo{author}{\bibfnamefont{J.~J.} \bibnamefont{Bollinger}},
  \bibinfo{year}{2012}, {``}\bibinfo{title}{Engineered two-dimensional Ising
  interactions in a trapped-ion quantum simulator with hundreds of spins},{''}
  \bibinfo{journal}{Nature} \textbf{\bibinfo{volume}{484}},
  \bibinfo{pages}{489 -- 492}.

\bibitem[{\citenamefont{Broome} \emph{et~al.}(2013)\citenamefont{Broome,
  Fedrizzi, Rahimi-Keshari, Dove, Aaronson, Ralph, and White}}]{Bro13}
\bibinfo{author}{\bibnamefont{Broome}, \bibfnamefont{M.~A.}},
  \bibinfo{author}{\bibfnamefont{A.}~\bibnamefont{Fedrizzi}},
  \bibinfo{author}{\bibfnamefont{S.}~\bibnamefont{Rahimi-Keshari}},
  \bibinfo{author}{\bibfnamefont{J.}~\bibnamefont{Dove}},
  \bibinfo{author}{\bibfnamefont{S.}~\bibnamefont{Aaronson}},
  \bibinfo{author}{\bibfnamefont{T.~C.} \bibnamefont{Ralph}}, and
  \bibinfo{author}{\bibfnamefont{A.~G.} \bibnamefont{White}},
  \bibinfo{year}{2013}, {``}\bibinfo{title}{Photonic Boson Sampling in a
  Tunable Circuit},{''} \bibinfo{journal}{Science}
  \textbf{\bibinfo{volume}{339}},  \bibinfo{pages}{794 -- 798}.

\bibitem[{\citenamefont{Brown} \emph{et~al.}(2011)\citenamefont{Brown, De,
  Kendon, and Munro}}]{Bro11}
\bibinfo{author}{\bibnamefont{Brown}, \bibfnamefont{K.~L.}},
  \bibinfo{author}{\bibfnamefont{S.}~\bibnamefont{De}},
  \bibinfo{author}{\bibfnamefont{V.~M.} \bibnamefont{Kendon}}, and
  \bibinfo{author}{\bibfnamefont{W.~J.} \bibnamefont{Munro}},
  \bibinfo{year}{2011}, {``}\bibinfo{title}{Ancilla-based quantum
  simulation},{''} \bibinfo{journal}{New J. Phys.}
  \textbf{\bibinfo{volume}{13}},  \bibinfo{pages}{095007}.

\bibitem[{\citenamefont{Brown} \emph{et~al.}(2010)\citenamefont{Brown, Munro,
  and Kendon}}]{Bro10}
\bibinfo{author}{\bibnamefont{Brown}, \bibfnamefont{K.~L.}},
  \bibinfo{author}{\bibfnamefont{W.~J.} \bibnamefont{Munro}}, and
  \bibinfo{author}{\bibfnamefont{V.~M.} \bibnamefont{Kendon}},
  \bibinfo{year}{2010}, {``}\bibinfo{title}{Using Quantum Computers for Quantum
  Simulation},{''} \bibinfo{journal}{Entropy} \textbf{\bibinfo{volume}{12}},
  \bibinfo{pages}{2268}.

\bibitem[{\citenamefont{Brown}(2007)}]{Bro07}
\bibinfo{author}{\bibnamefont{Brown}, \bibfnamefont{K.~R.}},
  \bibinfo{year}{2007}, {``}\bibinfo{title}{Energy protection arguments fail in
  the interaction picture},{''} \bibinfo{journal}{Phys. Rev. A}
  \textbf{\bibinfo{volume}{76}},  \bibinfo{pages}{022327}.

\bibitem[{\citenamefont{Brown} \emph{et~al.}(2006)\citenamefont{Brown, Clark,
  and Chuang}}]{Bro06}
\bibinfo{author}{\bibnamefont{Brown}, \bibfnamefont{K.~R.}},
  \bibinfo{author}{\bibfnamefont{R.~J.} \bibnamefont{Clark}}, and
  \bibinfo{author}{\bibfnamefont{I.~L.} \bibnamefont{Chuang}},
  \bibinfo{year}{2006}, {``}\bibinfo{title}{Limitations of quantum simulation
  examined by simulating a pairing Hamiltonian using Nuclear Magnetic
  Resonance},{''} \bibinfo{journal}{Phys. Rev. Lett.}
  \textbf{\bibinfo{volume}{97}},  \bibinfo{pages}{050504}.

\bibitem[{\citenamefont{B\"uchler} \emph{et~al.}(2005)\citenamefont{B\"uchler,
  Hermele, Huber, Fisher, and Zoller}}]{Buc05}
\bibinfo{author}{\bibnamefont{B\"uchler}, \bibfnamefont{H.~P.}},
  \bibinfo{author}{\bibfnamefont{M.}~\bibnamefont{Hermele}},
  \bibinfo{author}{\bibfnamefont{S.~D.} \bibnamefont{Huber}},
  \bibinfo{author}{\bibfnamefont{M.~P.~A.} \bibnamefont{Fisher}}, and
  \bibinfo{author}{\bibfnamefont{P.}~\bibnamefont{Zoller}},
  \bibinfo{year}{2005}, {``}\bibinfo{title}{Atomic quantum simulator for
  lattice gauge theories and ring exchange models},{''} \bibinfo{journal}{Phys.
  Rev. Lett.} \textbf{\bibinfo{volume}{95}},  \bibinfo{pages}{040402}.

\bibitem[{\citenamefont{Buluta} \emph{et~al.}(2011)\citenamefont{Buluta,
  Ashhab, and Nori}}]{Bul10}
\bibinfo{author}{\bibnamefont{Buluta}, \bibfnamefont{I.}},
  \bibinfo{author}{\bibfnamefont{S.}~\bibnamefont{Ashhab}}, and
  \bibinfo{author}{\bibfnamefont{F.}~\bibnamefont{Nori}}, \bibinfo{year}{2011},
  {``}\bibinfo{title}{Natural and artificial atoms for quantum
  computation},{''} \bibinfo{journal}{Rep. Progr. Phys.}
  \textbf{\bibinfo{volume}{74}},  \bibinfo{pages}{104401}.

\bibitem[{\citenamefont{Buluta} \emph{et~al.}(2008)\citenamefont{Buluta,
  Kitaoka, Georgescu, and Hasegawa}}]{Bul08}
\bibinfo{author}{\bibnamefont{Buluta}, \bibfnamefont{I.~M.}},
  \bibinfo{author}{\bibfnamefont{M.}~\bibnamefont{Kitaoka}},
  \bibinfo{author}{\bibfnamefont{S.}~\bibnamefont{Georgescu}}, and
  \bibinfo{author}{\bibfnamefont{S.}~\bibnamefont{Hasegawa}},
  \bibinfo{year}{2008}, {``}\bibinfo{title}{Investigation of planar Coulomb
  crystals for quantum simulation and computation},{''} \bibinfo{journal}{Phys.
  Rev. A} \textbf{\bibinfo{volume}{77}},  \bibinfo{pages}{062320}.

\bibitem[{\citenamefont{Buluta and Nori}(2009)}]{Bul09}
\bibinfo{author}{\bibnamefont{Buluta}, \bibfnamefont{I.~M.}}, and
  \bibinfo{author}{\bibfnamefont{F.}~\bibnamefont{Nori}}, \bibinfo{year}{2009},
  {``}\bibinfo{title}{Quantum simulators},{''} \bibinfo{journal}{Science}
  \textbf{\bibinfo{volume}{326}},  \bibinfo{pages}{108 -- 111}.

\bibitem[{\citenamefont{Byrnes} \emph{et~al.}(2008)\citenamefont{Byrnes, Kim,
  Kusudo, and Yamamoto}}]{Byr08}
\bibinfo{author}{\bibnamefont{Byrnes}, \bibfnamefont{T.}},
  \bibinfo{author}{\bibfnamefont{N.~Y.} \bibnamefont{Kim}},
  \bibinfo{author}{\bibfnamefont{K.}~\bibnamefont{Kusudo}}, and
  \bibinfo{author}{\bibfnamefont{Y.}~\bibnamefont{Yamamoto}},
  \bibinfo{year}{2008}, {``}\bibinfo{title}{Quantum simulation of Fermi-Hubbard
  models in semiconductor quantum-dot arrays},{''} \bibinfo{journal}{Phys. Rev.
  B} \textbf{\bibinfo{volume}{78}},  \bibinfo{pages}{075320}.

\bibitem[{\citenamefont{Byrnes} \emph{et~al.}(2007)\citenamefont{Byrnes,
  Recher, Kim, Utsunomiya, and Yamamoto}}]{Byr07}
\bibinfo{author}{\bibnamefont{Byrnes}, \bibfnamefont{T.}},
  \bibinfo{author}{\bibfnamefont{P.}~\bibnamefont{Recher}},
  \bibinfo{author}{\bibfnamefont{N.~Y.} \bibnamefont{Kim}},
  \bibinfo{author}{\bibfnamefont{S.}~\bibnamefont{Utsunomiya}}, and
  \bibinfo{author}{\bibfnamefont{Y.}~\bibnamefont{Yamamoto}},
  \bibinfo{year}{2007}, {``}\bibinfo{title}{Quantum simulator for the Hubbard
  model with long-range Coulomb interactions using Surface Acoustic Waves},{''}
  \bibinfo{journal}{Phys. Rev. Lett.} \textbf{\bibinfo{volume}{99}},
  \bibinfo{pages}{016405}.

\bibitem[{\citenamefont{Byrnes and Yamamoto}(2006)}]{Byr06}
\bibinfo{author}{\bibnamefont{Byrnes}, \bibfnamefont{T.}}, and
  \bibinfo{author}{\bibfnamefont{Y.}~\bibnamefont{Yamamoto}},
  \bibinfo{year}{2006}, {``}\bibinfo{title}{Simulating lattice gauge theories
  on a quantum computer},{''} \bibinfo{journal}{Phys. Rev. A}
  \textbf{\bibinfo{volume}{73}},  \bibinfo{pages}{022328}.

\bibitem[{\citenamefont{Cai} \emph{et~al.}(2013)\citenamefont{Cai, Retzker,
  Jelezko, and Plenio}}]{Cai13}
\bibinfo{author}{\bibnamefont{Cai}, \bibfnamefont{J.~M.}},
  \bibinfo{author}{\bibfnamefont{A.}~\bibnamefont{Retzker}},
  \bibinfo{author}{\bibfnamefont{F.}~\bibnamefont{Jelezko}}, and
  \bibinfo{author}{\bibfnamefont{M.~B.} \bibnamefont{Plenio}},
  \bibinfo{year}{2013}, {``}\bibinfo{title}{A large-scale quantum simulator on
  a diamond surface at room temperature},{''} \bibinfo{journal}{Nature Physics}
  \textbf{\bibinfo{volume}{9}},  \bibinfo{pages}{168 -- 173}.

\bibitem[{\citenamefont{Capogrosso-Sansone}
  \emph{et~al.}(2010)\citenamefont{Capogrosso-Sansone, Trefzger, Lewenstein,
  Zoller, and Pupillo}}]{Cap10}
\bibinfo{author}{\bibnamefont{Capogrosso-Sansone}, \bibfnamefont{B.}},
  \bibinfo{author}{\bibfnamefont{C.}~\bibnamefont{Trefzger}},
  \bibinfo{author}{\bibfnamefont{M.}~\bibnamefont{Lewenstein}},
  \bibinfo{author}{\bibfnamefont{P.}~\bibnamefont{Zoller}}, and
  \bibinfo{author}{\bibfnamefont{G.}~\bibnamefont{Pupillo}},
  \bibinfo{year}{2010}, {``}\bibinfo{title}{Quantum phases of cold polar
  molecules in 2D optical lattices},{''} \bibinfo{journal}{Phys. Rev. Lett.}
  \textbf{\bibinfo{volume}{104}},  \bibinfo{pages}{125301}.

\bibitem[{\citenamefont{Casanova} \emph{et~al.}(2010)\citenamefont{Casanova,
  Garcia-Ripoll, Gerritsma, Roos, and Solano}}]{Cas10}
\bibinfo{author}{\bibnamefont{Casanova}, \bibfnamefont{J.}},
  \bibinfo{author}{\bibfnamefont{J.}~\bibnamefont{Garcia-Ripoll}},
  \bibinfo{author}{\bibfnamefont{R.}~\bibnamefont{Gerritsma}},
  \bibinfo{author}{\bibfnamefont{C.}~\bibnamefont{Roos}}, and
  \bibinfo{author}{\bibfnamefont{E.}~\bibnamefont{Solano}},
  \bibinfo{year}{2010}, {``}\bibinfo{title}{Klein tunneling and Dirac
  potentials in trapped ions},{''} \bibinfo{journal}{Phys. Rev. A}
  \textbf{\bibinfo{volume}{82}},  \bibinfo{pages}{020101}.

\bibitem[{\citenamefont{Casanova} \emph{et~al.}(2012)\citenamefont{Casanova,
  Mezzacapo, Lamata, and Solano}}]{Cas12}
\bibinfo{author}{\bibnamefont{Casanova}, \bibfnamefont{J.}},
  \bibinfo{author}{\bibfnamefont{A.}~\bibnamefont{Mezzacapo}},
  \bibinfo{author}{\bibfnamefont{L.}~\bibnamefont{Lamata}}, and
  \bibinfo{author}{\bibfnamefont{E.}~\bibnamefont{Solano}},
  \bibinfo{year}{2012}, {``}\bibinfo{title}{Quantum Simulation of Interacting
  Fermion Lattice Models in Trapped Ions},{''} \bibinfo{journal}{Phys. Rev.
  Lett.} \textbf{\bibinfo{volume}{108}},  \bibinfo{pages}{190502}.

\bibitem[{\citenamefont{Casanova} \emph{et~al.}(2011)\citenamefont{Casanova,
  Sabin, Leon, Egusquiza, Gerritsma, Roos, Garcia-Ripoll, and Solano}}]{Cas11}
\bibinfo{author}{\bibnamefont{Casanova}, \bibfnamefont{J.}},
  \bibinfo{author}{\bibfnamefont{C.}~\bibnamefont{Sabin}},
  \bibinfo{author}{\bibfnamefont{J.}~\bibnamefont{Leon}},
  \bibinfo{author}{\bibfnamefont{I.~L.} \bibnamefont{Egusquiza}},
  \bibinfo{author}{\bibfnamefont{R.}~\bibnamefont{Gerritsma}},
  \bibinfo{author}{\bibfnamefont{C.~F.} \bibnamefont{Roos}},
  \bibinfo{author}{\bibfnamefont{J.~J.} \bibnamefont{Garcia-Ripoll}}, and
  \bibinfo{author}{\bibfnamefont{E.}~\bibnamefont{Solano}},
  \bibinfo{year}{2011}, {``}\bibinfo{title}{Quantum simulation of the Majorana
  equation and unphysical operations},{''} \bibinfo{journal}{Phys. Rev. X}
  \textbf{\bibinfo{volume}{1}},  \bibinfo{pages}{021018}.

\bibitem[{\citenamefont{Stojanovi\ifmmode~\acute{c}\else \'{c}\fi{}}
  \emph{et~al.}(2012)\citenamefont{Stojanovi\ifmmode~\acute{c}\else \'{c}\fi{},
  Shi, Bruder, and Cirac}}]{Sto12}
\bibinfo{author}{\bibnamefont{Stojanovi\ifmmode~\acute{c}\else \'{c}\fi{}},
  \bibfnamefont{V.~M.}}, \bibinfo{author}{\bibfnamefont{T.}~\bibnamefont{Shi}},
  \bibinfo{author}{\bibfnamefont{C.}~\bibnamefont{Bruder}}, and
  \bibinfo{author}{\bibfnamefont{J.~I.} \bibnamefont{Cirac}},
  \bibinfo{year}{2012}, {``}\bibinfo{title}{Quantum Simulation of Small-Polaron
  Formation with Trapped Ions},{''} \bibinfo{journal}{Phys. Rev. Lett.}
  \textbf{\bibinfo{volume}{109}},  \bibinfo{pages}{250501}.

\bibitem[{\citenamefont{Chen}
  \emph{et~al.}(2007{\natexlab{a}})\citenamefont{Chen, Chen, and
  Liang}}]{Che07}
\bibinfo{author}{\bibnamefont{Chen}, \bibfnamefont{G.}},
  \bibinfo{author}{\bibfnamefont{Z.}~\bibnamefont{Chen}}, and
  \bibinfo{author}{\bibfnamefont{J.}~\bibnamefont{Liang}},
  \bibinfo{year}{2007}{\natexlab{a}}, {``}\bibinfo{title}{Simulation of the
  superradiant quantum phase transition in the superconducting charge qubits
  inside a cavity},{''} \bibinfo{journal}{Phys. Rev. A}
  \textbf{\bibinfo{volume}{76}},  \bibinfo{pages}{055803}.

\bibitem[{\citenamefont{Chen}
  \emph{et~al.}(2007{\natexlab{b}})\citenamefont{Chen, Church, Englert, Henkel,
  Rohwedder, Scully, and Zubairy}}]{QCD}
\bibinfo{author}{\bibnamefont{Chen}, \bibfnamefont{G.}},
  \bibinfo{author}{\bibfnamefont{D.}~\bibnamefont{Church}},
  \bibinfo{author}{\bibfnamefont{B.-G.} \bibnamefont{Englert}},
  \bibinfo{author}{\bibfnamefont{C.}~\bibnamefont{Henkel}},
  \bibinfo{author}{\bibfnamefont{B.}~\bibnamefont{Rohwedder}},
  \bibinfo{author}{\bibfnamefont{M.}~\bibnamefont{Scully}}, and
  \bibinfo{author}{\bibfnamefont{M.~S.} \bibnamefont{Zubairy}},
  \bibinfo{year}{2007}{\natexlab{b}}, \emph{\bibinfo{title}{Quantum computing
  devices: principles, designs and analysis}} (\bibinfo{publisher}{Chapman and
  Hall}).

\bibitem[{\citenamefont{Chen} \emph{et~al.}(2010)\citenamefont{Chen, Zhou,
  Shou, Zhou, and Guo}}]{Che10}
\bibinfo{author}{\bibnamefont{Chen}, \bibfnamefont{Z.}},
  \bibinfo{author}{\bibfnamefont{Z.}~\bibnamefont{Zhou}},
  \bibinfo{author}{\bibfnamefont{X.}~\bibnamefont{Shou}},
  \bibinfo{author}{\bibfnamefont{X.}~\bibnamefont{Zhou}}, and
  \bibinfo{author}{\bibfnamefont{G.}~\bibnamefont{Guo}}, \bibinfo{year}{2010},
  {``}\bibinfo{title}{Quantum simulation of Heinsenberg spin chains with
  next-nearest-neighbor interations in coupled cavities},{''}
  \bibinfo{journal}{Phys. Rev. A} \textbf{\bibinfo{volume}{81}},
  \bibinfo{pages}{022303}.

\bibitem[{\citenamefont{Chiaverini and Lybarger}(2008)}]{Chi08}
\bibinfo{author}{\bibnamefont{Chiaverini}, \bibfnamefont{J.}}, and
  \bibinfo{author}{\bibfnamefont{W.~E.} \bibnamefont{Lybarger}},
  \bibinfo{year}{2008}, {``}\bibinfo{title}{Laserless trapped-ion quantum
  simulations without spontaneous scattering using microtrap arrays},{''}
  \bibinfo{journal}{Phys. Rev. A} \textbf{\bibinfo{volume}{77}},
  \bibinfo{pages}{022324}.

\bibitem[{\citenamefont{Chiuri} \emph{et~al.}(2012)\citenamefont{Chiuri,
  Greganti, Mazzola, Paternostro, and Mataloni}}]{Chi12}
\bibinfo{author}{\bibnamefont{Chiuri}, \bibfnamefont{A.}},
  \bibinfo{author}{\bibfnamefont{C.}~\bibnamefont{Greganti}},
  \bibinfo{author}{\bibfnamefont{L.}~\bibnamefont{Mazzola}},
  \bibinfo{author}{\bibfnamefont{M.}~\bibnamefont{Paternostro}}, and
  \bibinfo{author}{\bibfnamefont{P.}~\bibnamefont{Mataloni}},
  \bibinfo{year}{2012}, {``}\bibinfo{title}{Linear Optics Simulation of
  Non-Markovian Quantum Dynamics},{''} \bibinfo{journal}{Sci. Rep.}
  \textbf{\bibinfo{volume}{2}},  \bibinfo{pages}{968}.

\bibitem[{\citenamefont{Cho}
  \emph{et~al.}(2008{\natexlab{a}})\citenamefont{Cho, Angelakis, and
  Bose}}]{Cho08b}
\bibinfo{author}{\bibnamefont{Cho}, \bibfnamefont{J.}},
  \bibinfo{author}{\bibfnamefont{D.~G.} \bibnamefont{Angelakis}}, and
  \bibinfo{author}{\bibfnamefont{S.}~\bibnamefont{Bose}},
  \bibinfo{year}{2008}{\natexlab{a}}, {``}\bibinfo{title}{Fractional Quantum
  Hall State in Coupled Cavities},{''} \bibinfo{journal}{Phys. Rev. Lett.}
  \textbf{\bibinfo{volume}{101}},  \bibinfo{pages}{246809}.

\bibitem[{\citenamefont{Cho}
  \emph{et~al.}(2008{\natexlab{b}})\citenamefont{Cho, Angelakis, and
  Bose}}]{Cho08a}
\bibinfo{author}{\bibnamefont{Cho}, \bibfnamefont{J.}},
  \bibinfo{author}{\bibfnamefont{D.~G.} \bibnamefont{Angelakis}}, and
  \bibinfo{author}{\bibfnamefont{S.}~\bibnamefont{Bose}},
  \bibinfo{year}{2008}{\natexlab{b}}, {``}\bibinfo{title}{Simulation of
  high-spin Heisenberg models in coupled cavities},{''} \bibinfo{journal}{Phys.
  Rev. A} \textbf{\bibinfo{volume}{78}},  \bibinfo{pages}{062338}.

\bibitem[{\citenamefont{Cirac} \emph{et~al.}(2010)\citenamefont{Cirac, Maraner,
  and Pachos}}]{Cir10}
\bibinfo{author}{\bibnamefont{Cirac}, \bibfnamefont{J.}},
  \bibinfo{author}{\bibfnamefont{P.}~\bibnamefont{Maraner}}, and
  \bibinfo{author}{\bibfnamefont{J.}~\bibnamefont{Pachos}},
  \bibinfo{year}{2010}, {``}\bibinfo{title}{Cold atom simulation of interacting
  quantum field theories},{''} \bibinfo{journal}{Phys. Rev. Lett.}
  \textbf{\bibinfo{volume}{105}},  \bibinfo{pages}{190403}.

\bibitem[{\citenamefont{Cirac and Verstraete}(2009)}]{Cir09}
\bibinfo{author}{\bibnamefont{Cirac}, \bibfnamefont{J.~I.}}, and
  \bibinfo{author}{\bibfnamefont{F.}~\bibnamefont{Verstraete}},
  \bibinfo{year}{2009}, {``}\bibinfo{title}{Renormalization and tensor product
  states in spin chains and lattices},{''} \bibinfo{journal}{Journal of Physics
  A} \textbf{\bibinfo{volume}{42}}, ISSN \bibinfo{issn}{{1751-8113}}.

\bibitem[{\citenamefont{Cirac and Zoller}(2003)}]{Cir03}
\bibinfo{author}{\bibnamefont{Cirac}, \bibfnamefont{J.~I.}}, and
  \bibinfo{author}{\bibfnamefont{P.}~\bibnamefont{Zoller}},
  \bibinfo{year}{2003}, {``}\bibinfo{title}{How to manipulate cold atoms},{''}
  \bibinfo{journal}{Science} \textbf{\bibinfo{volume}{301}},
  \bibinfo{pages}{176--177}.

\bibitem[{\citenamefont{Clark}
  \emph{et~al.}(2009{\natexlab{a}})\citenamefont{Clark, Metodi, Gasster, and
  Brown}}]{Cla09}
\bibinfo{author}{\bibnamefont{Clark}, \bibfnamefont{C.~R.}},
  \bibinfo{author}{\bibfnamefont{T.~S.} \bibnamefont{Metodi}},
  \bibinfo{author}{\bibfnamefont{S.~D.} \bibnamefont{Gasster}}, and
  \bibinfo{author}{\bibfnamefont{K.~R.} \bibnamefont{Brown}},
  \bibinfo{year}{2009}{\natexlab{a}}, {``}\bibinfo{title}{Resource requirements
  for fault-tolerant quantum simulation: the transverse Ising model ground
  state},{''} \bibinfo{journal}{Phys. Rev. A} \textbf{\bibinfo{volume}{79}},
  \bibinfo{pages}{062314}.

\bibitem[{\citenamefont{Clark}
  \emph{et~al.}(2009{\natexlab{b}})\citenamefont{Clark, Lin, Brown, and
  Chuang}}]{Cla08}
\bibinfo{author}{\bibnamefont{Clark}, \bibfnamefont{R.~J.}},
  \bibinfo{author}{\bibfnamefont{T.}~\bibnamefont{Lin}},
  \bibinfo{author}{\bibfnamefont{K.~R.} \bibnamefont{Brown}}, and
  \bibinfo{author}{\bibfnamefont{I.~L.} \bibnamefont{Chuang}},
  \bibinfo{year}{2009}{\natexlab{b}}, {``}\bibinfo{title}{A two-dimensional
  lattice ion trap for quantum simulation},{''} \bibinfo{journal}{J. Appl.
  Phys.} \textbf{\bibinfo{volume}{105}},  \bibinfo{pages}{013114}.

\bibitem[{\citenamefont{Clarke and Wilhelm}(2008)}]{CW08}
\bibinfo{author}{\bibnamefont{Clarke}, \bibfnamefont{J.}}, and
  \bibinfo{author}{\bibfnamefont{F.~K.} \bibnamefont{Wilhelm}},
  \bibinfo{year}{2008}, {``}\bibinfo{title}{Superconducting quantum bits},{''}
  \bibinfo{journal}{Nature} \textbf{\bibinfo{volume}{453}},
  \bibinfo{pages}{1031--1042}.

\bibitem[{\citenamefont{Crespi} \emph{et~al.}(2013)\citenamefont{Crespi,
  Osellame, Ramponi, Brod, Galvao, Spagnolo, Vitelli, Maiorino, Mataloni, and
  Sciarrino}}]{Cre13}
\bibinfo{author}{\bibnamefont{Crespi}, \bibfnamefont{A.}},
  \bibinfo{author}{\bibfnamefont{R.}~\bibnamefont{Osellame}},
  \bibinfo{author}{\bibfnamefont{R.}~\bibnamefont{Ramponi}},
  \bibinfo{author}{\bibfnamefont{D.~J.} \bibnamefont{Brod}},
  \bibinfo{author}{\bibfnamefont{E.~F.} \bibnamefont{Galvao}},
  \bibinfo{author}{\bibfnamefont{N.}~\bibnamefont{Spagnolo}},
  \bibinfo{author}{\bibfnamefont{C.}~\bibnamefont{Vitelli}},
  \bibinfo{author}{\bibfnamefont{E.}~\bibnamefont{Maiorino}},
  \bibinfo{author}{\bibfnamefont{P.}~\bibnamefont{Mataloni}}, and
  \bibinfo{author}{\bibfnamefont{F.}~\bibnamefont{Sciarrino}},
  \bibinfo{year}{2013}, {``}\bibinfo{title}{Integrated multimode
  interferometers with arbitrary designs for photonic boson sampling},{''}
  \bibinfo{journal}{Nature Photonics} \textbf{\bibinfo{volume}{7}},
  \bibinfo{pages}{545 -- 549}.

\bibitem[{\citenamefont{Cucchietti}
  \emph{et~al.}(2007)\citenamefont{Cucchietti, Fernandez-Vidal, and
  Paz}}]{Cuc07}
\bibinfo{author}{\bibnamefont{Cucchietti}, \bibfnamefont{F.~M.}},
  \bibinfo{author}{\bibfnamefont{S.}~\bibnamefont{Fernandez-Vidal}}, and
  \bibinfo{author}{\bibfnamefont{J.~P.} \bibnamefont{Paz}},
  \bibinfo{year}{2007}, {``}\bibinfo{title}{Universal decoherence induced by an
  environmental quantum phase transition},{''} \bibinfo{journal}{Phys. Rev. A}
  \textbf{\bibinfo{volume}{75}},  \bibinfo{pages}{032337}.

\bibitem[{\citenamefont{Dalibard} \emph{et~al.}(2011)\citenamefont{Dalibard,
  Gerbier, Juzeli\ifmmode~\bar{u}\else \={u}\fi{}nas, and \"Ohberg}}]{dal11}
\bibinfo{author}{\bibnamefont{Dalibard}, \bibfnamefont{J.}},
  \bibinfo{author}{\bibfnamefont{F.}~\bibnamefont{Gerbier}},
  \bibinfo{author}{\bibfnamefont{G.}~\bibnamefont{Juzeli\ifmmode~\bar{u}\else
  \={u}\fi{}nas}}, and
  \bibinfo{author}{\bibfnamefont{P.}~\bibnamefont{\"Ohberg}},
  \bibinfo{year}{2011}, {``}\bibinfo{title}{Artificial gauge potentials for
  neutral atoms},{''} \bibinfo{journal}{Rev. Mod. Phys.}
  \textbf{\bibinfo{volume}{83}},  \bibinfo{pages}{1523--1543}.

\bibitem[{\citenamefont{D'Ariano} \emph{et~al.}(2003)\citenamefont{D'Ariano,
  Paris, and Sacchi}}]{Dar03}
\bibinfo{author}{\bibnamefont{D'Ariano}, \bibfnamefont{G.~M.}},
  \bibinfo{author}{\bibfnamefont{M.~G.~A.} \bibnamefont{Paris}}, and
  \bibinfo{author}{\bibfnamefont{M.~F.} \bibnamefont{Sacchi}},
  \bibinfo{year}{2003}, {``}\bibinfo{title}{Quantum tomography},{''}
  \bibinfo{journal}{Advances in Imaging and Electron Physics}
  \textbf{\bibinfo{volume}{128}},  \bibinfo{pages}{205--308}.

\bibitem[{\citenamefont{Daskin and Kais}(2011)}]{Das11}
\bibinfo{author}{\bibnamefont{Daskin}, \bibfnamefont{A.}}, and
  \bibinfo{author}{\bibfnamefont{S.}~\bibnamefont{Kais}}, \bibinfo{year}{2011},
  {``}\bibinfo{title}{Decomposition of unitary matrices for finding quantum
  circuits: Application to molecular Hamiltonians},{''} \bibinfo{journal}{J.
  Chem. Phys.} \textbf{\bibinfo{volume}{134}},  \bibinfo{pages}{144112}.

\bibitem[{\citenamefont{Deng} \emph{et~al.}(2005)\citenamefont{Deng, Porras,
  and Cirac}}]{Den05}
\bibinfo{author}{\bibnamefont{Deng}, \bibfnamefont{X.-L.}},
  \bibinfo{author}{\bibfnamefont{D.}~\bibnamefont{Porras}}, and
  \bibinfo{author}{\bibfnamefont{J.~I.} \bibnamefont{Cirac}},
  \bibinfo{year}{2005}, {``}\bibinfo{title}{Effective spin quantum phases in
  systems of trapped ions},{''} \bibinfo{journal}{Phys. Rev. A}
  \textbf{\bibinfo{volume}{72}},  \bibinfo{pages}{063407}.

\bibitem[{\citenamefont{Deng} \emph{et~al.}(2008)\citenamefont{Deng, Porras,
  and Cirac}}]{Den08}
\bibinfo{author}{\bibnamefont{Deng}, \bibfnamefont{X.-L.}},
  \bibinfo{author}{\bibfnamefont{D.}~\bibnamefont{Porras}}, and
  \bibinfo{author}{\bibfnamefont{J.~I.} \bibnamefont{Cirac}},
  \bibinfo{year}{2008}, {``}\bibinfo{title}{Quantum phases of interacting
  phonons in ion traps},{''} \bibinfo{journal}{Phys. Rev. A}
  \textbf{\bibinfo{volume}{77}},  \bibinfo{pages}{033403}.

\bibitem[{\citenamefont{Deppe} \emph{et~al.}(2008)\citenamefont{Deppe,
  Mariantoni, Menzel, Marx, Kakuyanagi, Tanaka, Meno, Semba, Solano, and
  Gross}}]{Dep08}
\bibinfo{author}{\bibnamefont{Deppe}, \bibfnamefont{F.}},
  \bibinfo{author}{\bibfnamefont{M.}~\bibnamefont{Mariantoni}},
  \bibinfo{author}{\bibfnamefont{E.~P.} \bibnamefont{Menzel}},
  \bibinfo{author}{\bibfnamefont{A.}~\bibnamefont{Marx}},
  \bibinfo{author}{\bibfnamefont{S.~S.~K.} \bibnamefont{Kakuyanagi}},
  \bibinfo{author}{\bibfnamefont{H.}~\bibnamefont{Tanaka}},
  \bibinfo{author}{\bibfnamefont{T.}~\bibnamefont{Meno}},
  \bibinfo{author}{\bibfnamefont{K.}~\bibnamefont{Semba}},
  \bibinfo{author}{\bibfnamefont{H.~T.~E.} \bibnamefont{Solano}}, and
  \bibinfo{author}{\bibfnamefont{R.}~\bibnamefont{Gross}},
  \bibinfo{year}{2008}, {``}\bibinfo{title}{Two-photon probe of the
  Jaynes-Cummings model and symmetry breaking in circuit QED},{''}
  \bibinfo{journal}{Nature Phys.} \textbf{\bibinfo{volume}{4}},
  \bibinfo{pages}{686--691}.

\bibitem[{\citenamefont{Dodd} \emph{et~al.}(2002)\citenamefont{Dodd, Nielsen,
  Bremner, and Thew}}]{Dod02}
\bibinfo{author}{\bibnamefont{Dodd}, \bibfnamefont{J.~L.}},
  \bibinfo{author}{\bibfnamefont{M.~A.} \bibnamefont{Nielsen}},
  \bibinfo{author}{\bibfnamefont{M.~J.} \bibnamefont{Bremner}}, and
  \bibinfo{author}{\bibfnamefont{R.~T.} \bibnamefont{Thew}},
  \bibinfo{year}{2002}, {``}\bibinfo{title}{Universal quantum computation and
  simulation using any entangling Hamiltonian and local unitaries},{''}
  \bibinfo{journal}{Phys. Rev. A} \textbf{\bibinfo{volume}{65}},
  \bibinfo{pages}{040301}.

\bibitem[{\citenamefont{Dreisow} \emph{et~al.}(2010)\citenamefont{Dreisow,
  Heinrich, Keil, T\"unnermann, Nolte, Longhi, and Szameit}}]{Dre10}
\bibinfo{author}{\bibnamefont{Dreisow}, \bibfnamefont{F.}},
  \bibinfo{author}{\bibfnamefont{M.}~\bibnamefont{Heinrich}},
  \bibinfo{author}{\bibfnamefont{R.}~\bibnamefont{Keil}},
  \bibinfo{author}{\bibfnamefont{A.}~\bibnamefont{T\"unnermann}},
  \bibinfo{author}{\bibfnamefont{S.}~\bibnamefont{Nolte}},
  \bibinfo{author}{\bibfnamefont{S.}~\bibnamefont{Longhi}}, and
  \bibinfo{author}{\bibfnamefont{A.}~\bibnamefont{Szameit}},
  \bibinfo{year}{2010}, {``}\bibinfo{title}{Classical simulation of
  relativistic Zitterbewegung in photonic lattices},{''}
  \bibinfo{journal}{Phys. Rev. Lett.} \textbf{\bibinfo{volume}{105}},
  \bibinfo{pages}{143902}.

\bibitem[{\citenamefont{Du} \emph{et~al.}(2010)\citenamefont{Du, Xu, Peng,
  Wang, Wu, and Lu}}]{Du09}
\bibinfo{author}{\bibnamefont{Du}, \bibfnamefont{J.}},
  \bibinfo{author}{\bibfnamefont{N.}~\bibnamefont{Xu}},
  \bibinfo{author}{\bibfnamefont{X.}~\bibnamefont{Peng}},
  \bibinfo{author}{\bibfnamefont{P.}~\bibnamefont{Wang}},
  \bibinfo{author}{\bibfnamefont{S.}~\bibnamefont{Wu}}, and
  \bibinfo{author}{\bibfnamefont{D.}~\bibnamefont{Lu}}, \bibinfo{year}{2010},
  {``}\bibinfo{title}{NMR implementation of a molecualr hydrogen quantum
  simulation with adiabatic state preparation},{''} \bibinfo{journal}{Phys.
  Rev. Lett.} \textbf{\bibinfo{volume}{104}},  \bibinfo{pages}{030502}.

\bibitem[{\citenamefont{D\"ur} \emph{et~al.}(2008)\citenamefont{D\"ur, Bremne,
  and Briegel}}]{Dur08}
\bibinfo{author}{\bibnamefont{D\"ur}, \bibfnamefont{W.}},
  \bibinfo{author}{\bibfnamefont{M.~J.} \bibnamefont{Bremne}}, and
  \bibinfo{author}{\bibfnamefont{H.~J.} \bibnamefont{Briegel}},
  \bibinfo{year}{2008}, {``}\bibinfo{title}{Quantum simulation of interacting
  high-dimensional systems: the influence of noise},{''}
  \bibinfo{journal}{Phys. Rev. A} \textbf{\bibinfo{volume}{78}},
  \bibinfo{pages}{052325}.

\bibitem[{\citenamefont{Edwards} \emph{et~al.}(2010)\citenamefont{Edwards,
  Korenblit, Kim, Islam, Chang, Freericks, Lin, Duan, and Monroe}}]{Edw10}
\bibinfo{author}{\bibnamefont{Edwards}, \bibfnamefont{E.}},
  \bibinfo{author}{\bibfnamefont{S.}~\bibnamefont{Korenblit}},
  \bibinfo{author}{\bibfnamefont{K.}~\bibnamefont{Kim}},
  \bibinfo{author}{\bibfnamefont{R.}~\bibnamefont{Islam}},
  \bibinfo{author}{\bibfnamefont{M.}~\bibnamefont{Chang}},
  \bibinfo{author}{\bibfnamefont{J.}~\bibnamefont{Freericks}},
  \bibinfo{author}{\bibfnamefont{G.}~\bibnamefont{Lin}},
  \bibinfo{author}{\bibfnamefont{L.}~\bibnamefont{Duan}}, and
  \bibinfo{author}{\bibfnamefont{C.}~\bibnamefont{Monroe}},
  \bibinfo{year}{2010}, {``}\bibinfo{title}{Quantum simulation and phase
  diagram of the transverse-field Ising-model with three atomic spins},{''}
  \bibinfo{journal}{Phys. Rev. B} \textbf{\bibinfo{volume}{82}},
  \bibinfo{pages}{060412}.

\bibitem[{\citenamefont{Fallani} \emph{et~al.}(2007)\citenamefont{Fallani, Lye,
  Guarrera, Fort, and Inguscio}}]{Fal07}
\bibinfo{author}{\bibnamefont{Fallani}, \bibfnamefont{L.}},
  \bibinfo{author}{\bibfnamefont{J.~E.} \bibnamefont{Lye}},
  \bibinfo{author}{\bibfnamefont{V.}~\bibnamefont{Guarrera}},
  \bibinfo{author}{\bibfnamefont{C.}~\bibnamefont{Fort}}, and
  \bibinfo{author}{\bibfnamefont{M.}~\bibnamefont{Inguscio}},
  \bibinfo{year}{2007}, {``}\bibinfo{title}{Ultracold atoms in a disordered
  crystal of light: towards a Bose glass},{''} \bibinfo{journal}{Phys. Rev.
  Lett.} \textbf{\bibinfo{volume}{98}},  \bibinfo{pages}{130404}.

\bibitem[{\citenamefont{Farhi} \emph{et~al.}(2001)\citenamefont{Farhi,
  Goldstone, Gutmann, Lapan, Lundgren, and Preda}}]{Far01}
\bibinfo{author}{\bibnamefont{Farhi}, \bibfnamefont{E.}},
  \bibinfo{author}{\bibfnamefont{J.}~\bibnamefont{Goldstone}},
  \bibinfo{author}{\bibfnamefont{S.}~\bibnamefont{Gutmann}},
  \bibinfo{author}{\bibfnamefont{J.}~\bibnamefont{Lapan}},
  \bibinfo{author}{\bibfnamefont{A.}~\bibnamefont{Lundgren}}, and
  \bibinfo{author}{\bibfnamefont{D.}~\bibnamefont{Preda}},
  \bibinfo{year}{2001}, {``}\bibinfo{title}{A quantum adiabatic evolution
  algorithm applied to random instances of an NP-complete problem},{''}
  \bibinfo{journal}{Science} \textbf{\bibinfo{volume}{292}},
  \bibinfo{pages}{472--475}.

\bibitem[{\citenamefont{Fetter and Walecka}(2003)}]{FetWal}
\bibinfo{author}{\bibnamefont{Fetter}, \bibfnamefont{A.~L.}}, and
  \bibinfo{author}{\bibfnamefont{J.~D.} \bibnamefont{Walecka}},
  \bibinfo{year}{2003}, \emph{\bibinfo{title}{Quantum theory of many-particle
  systems}} (\bibinfo{publisher}{Dover Publications}).

\bibitem[{\citenamefont{Feynman}(1982)}]{feynman82}
\bibinfo{author}{\bibnamefont{Feynman}, \bibfnamefont{R.}},
  \bibinfo{year}{1982}, {``}\bibinfo{title}{Simulating physics with
  computers},{''} \bibinfo{journal}{Int. J. Theor. Phys.}
  \textbf{\bibinfo{volume}{21}},  \bibinfo{pages}{467--488}.

\bibitem[{\citenamefont{Fischer and Sch\"utzhold}(2004)}]{Fis04}
\bibinfo{author}{\bibnamefont{Fischer}, \bibfnamefont{U.~R.}}, and
  \bibinfo{author}{\bibfnamefont{R.}~\bibnamefont{Sch\"utzhold}},
  \bibinfo{year}{2004}, {``}\bibinfo{title}{Quantum simulation of cosmic
  inflation in two-component Bose-Einstein condensates},{''}
  \bibinfo{journal}{Phys. Rev. A} \textbf{\bibinfo{volume}{70}},
  \bibinfo{pages}{063615}.

\bibitem[{\citenamefont{Fisher} \emph{et~al.}(1989)\citenamefont{Fisher,
  Weichman, Grinstein, and Fisher}}]{Fis89}
\bibinfo{author}{\bibnamefont{Fisher}, \bibfnamefont{M.~P.~A.}},
  \bibinfo{author}{\bibfnamefont{P.~B.} \bibnamefont{Weichman}},
  \bibinfo{author}{\bibfnamefont{G.}~\bibnamefont{Grinstein}}, and
  \bibinfo{author}{\bibfnamefont{D.~S.} \bibnamefont{Fisher}},
  \bibinfo{year}{1989}, {``}\bibinfo{title}{Boson localization and the
  superfluid-insulator transition},{''} \bibinfo{journal}{Phys. Rev. B}
  \textbf{\bibinfo{volume}{40}},  \bibinfo{pages}{546--570}.

\bibitem[{\citenamefont{Freedman} \emph{et~al.}(2002)\citenamefont{Freedman,
  Kitaev, and Wang}}]{Fre02}
\bibinfo{author}{\bibnamefont{Freedman}, \bibfnamefont{M.}},
  \bibinfo{author}{\bibfnamefont{A.}~\bibnamefont{Kitaev}}, and
  \bibinfo{author}{\bibfnamefont{Z.}~\bibnamefont{Wang}}, \bibinfo{year}{2002},
  {``}\bibinfo{title}{Simulation of topological field theories by quantum
  computers},{''} \bibinfo{journal}{Comm. Math. Phys.}
  \textbf{\bibinfo{volume}{227}},  \bibinfo{pages}{587--603}.

\bibitem[{\citenamefont{Friedenauer}
  \emph{et~al.}(2008)\citenamefont{Friedenauer, Schmitz, Gl\"uckert, Porras,
  and Sch\"atz}}]{Fri08}
\bibinfo{author}{\bibnamefont{Friedenauer}, \bibfnamefont{A.}},
  \bibinfo{author}{\bibfnamefont{H.}~\bibnamefont{Schmitz}},
  \bibinfo{author}{\bibfnamefont{J.~T.} \bibnamefont{Gl\"uckert}},
  \bibinfo{author}{\bibfnamefont{D.}~\bibnamefont{Porras}}, and
  \bibinfo{author}{\bibfnamefont{T.}~\bibnamefont{Sch\"atz}},
  \bibinfo{year}{2008}, {``}\bibinfo{title}{Simulating a quantum magnet with
  trapped ions},{''} \bibinfo{journal}{Nature Physics}
  \textbf{\bibinfo{volume}{4}},  \bibinfo{pages}{757--761}.

\bibitem[{\citenamefont{Fuhrmanek} \emph{et~al.}(2011)\citenamefont{Fuhrmanek,
  Bourgain, Sortais, and Browaeys}}]{Fuh10}
\bibinfo{author}{\bibnamefont{Fuhrmanek}, \bibfnamefont{A.}},
  \bibinfo{author}{\bibfnamefont{R.}~\bibnamefont{Bourgain}},
  \bibinfo{author}{\bibfnamefont{Y.~R.~P.} \bibnamefont{Sortais}}, and
  \bibinfo{author}{\bibfnamefont{A.}~\bibnamefont{Browaeys}},
  \bibinfo{year}{2011}, {``}\bibinfo{title}{Free-space lossless state-detection
  of a single trapped atom},{''} \bibinfo{journal}{Phys. Rev. Lett.}
  \textbf{\bibinfo{volume}{106}},  \bibinfo{pages}{133003}.

\bibitem[{\citenamefont{Garcia-Ripoll}
  \emph{et~al.}(2004)\citenamefont{Garcia-Ripoll, Martin-Delgado, and
  Cirac}}]{JGR04}
\bibinfo{author}{\bibnamefont{Garcia-Ripoll}, \bibfnamefont{J.~J.}},
  \bibinfo{author}{\bibfnamefont{M.~A.} \bibnamefont{Martin-Delgado}}, and
  \bibinfo{author}{\bibfnamefont{J.~I.} \bibnamefont{Cirac}},
  \bibinfo{year}{2004}, {``}\bibinfo{title}{Implementation of spin Hamiltonians
  in optical lattices},{''} \bibinfo{journal}{Phys. Rev. Lett.}
  \textbf{\bibinfo{volume}{93}},  \bibinfo{pages}{250405}.

\bibitem[{\citenamefont{Garcia-Ripoll}
  \emph{et~al.}(2008)\citenamefont{Garcia-Ripoll, Solano, and
  Martin-Delgado}}]{JGR08}
\bibinfo{author}{\bibnamefont{Garcia-Ripoll}, \bibfnamefont{J.~J.}},
  \bibinfo{author}{\bibfnamefont{E.}~\bibnamefont{Solano}}, and
  \bibinfo{author}{\bibfnamefont{M.~A.} \bibnamefont{Martin-Delgado}},
  \bibinfo{year}{2008}, {``}\bibinfo{title}{Quantum simulation of Anderson and
  Kondo lattices with superconducting qubits},{''} \bibinfo{journal}{Phys. Rev.
  B} \textbf{\bibinfo{volume}{77}},  \bibinfo{pages}{024522}.

\bibitem[{\citenamefont{Geller} \emph{et~al.}(2012)\citenamefont{Geller,
  Martinis, Sornborger, Stancil, Pritchett, and Galiautdinov}}]{Gel12}
\bibinfo{author}{\bibnamefont{Geller}, \bibfnamefont{M.~R.}},
  \bibinfo{author}{\bibfnamefont{J.~M.} \bibnamefont{Martinis}},
  \bibinfo{author}{\bibfnamefont{A.~T.} \bibnamefont{Sornborger}},
  \bibinfo{author}{\bibfnamefont{P.~C.} \bibnamefont{Stancil}},
  \bibinfo{author}{\bibfnamefont{E.~J.} \bibnamefont{Pritchett}}, and
  \bibinfo{author}{\bibfnamefont{A.}~\bibnamefont{Galiautdinov}},
  \bibinfo{year}{2012}, {``}\bibinfo{title}{Universal quantum simulation with
  pre-threshold superconducting qubits: Single-excitation subspace method},{''}
  \eprint{arXiv:1210.5260}.

\bibitem[{\citenamefont{Georgeot}(2004)}]{Geo04}
\bibinfo{author}{\bibnamefont{Georgeot}, \bibfnamefont{B.}},
  \bibinfo{year}{2004}, {``}\bibinfo{title}{Quantum computing of Poincar\'e
  recurrences and periodic orbits},{''} \bibinfo{journal}{Phys. Rev. A}
  \textbf{\bibinfo{volume}{69}},  \bibinfo{pages}{032301}.

\bibitem[{\citenamefont{Georgescu} \emph{et~al.}(2011)\citenamefont{Georgescu,
  Ashhab, Nakatsukasa, and Nori}}]{Nuclear}
\bibinfo{author}{\bibnamefont{Georgescu}, \bibfnamefont{I.~M.}},
  \bibinfo{author}{\bibfnamefont{S.}~\bibnamefont{Ashhab}},
  \bibinfo{author}{\bibfnamefont{T.}~\bibnamefont{Nakatsukasa}}, and
  \bibinfo{author}{\bibfnamefont{F.}~\bibnamefont{Nori}}, \bibinfo{year}{2011},
  {``}\bibinfo{title}{Analog quantum simulation of the atomic nucleus with a
  fermionic condensate},{''} \eprint{in preparation}.

\bibitem[{\citenamefont{Gerace and Carusotto}(2012)}]{Ger12}
\bibinfo{author}{\bibnamefont{Gerace}, \bibfnamefont{D.}}, and
  \bibinfo{author}{\bibfnamefont{I.}~\bibnamefont{Carusotto}},
  \bibinfo{year}{2012}, {``}\bibinfo{title}{Analog Hawking radiation from an
  acoustic black hole in a flowing polariton superfluid},{''}
  \bibinfo{journal}{Phys. Rev. B} \textbf{\bibinfo{volume}{86}},
  \bibinfo{pages}{144505}.

\bibitem[{\citenamefont{Gerritsma} \emph{et~al.}(2010)\citenamefont{Gerritsma,
  Kirchmair, Zahringer, Solano, Blatt, and Roos}}]{Ger10}
\bibinfo{author}{\bibnamefont{Gerritsma}, \bibfnamefont{R.}},
  \bibinfo{author}{\bibfnamefont{G.}~\bibnamefont{Kirchmair}},
  \bibinfo{author}{\bibfnamefont{F.}~\bibnamefont{Zahringer}},
  \bibinfo{author}{\bibfnamefont{E.}~\bibnamefont{Solano}},
  \bibinfo{author}{\bibfnamefont{R.}~\bibnamefont{Blatt}}, and
  \bibinfo{author}{\bibfnamefont{C.}~\bibnamefont{Roos}}, \bibinfo{year}{2010},
  {``}\bibinfo{title}{Quantum simulation of the Dirac equation},{''}
  \bibinfo{journal}{Nature} \textbf{\bibinfo{volume}{463}},
  \bibinfo{pages}{68--71}.

\bibitem[{\citenamefont{Gerritsma} \emph{et~al.}(2011)\citenamefont{Gerritsma,
  Lanyon, Kirchmair, Z\"ahringer, Hempel, Casanova, Garc\'\i{}a-Ripoll, Solano,
  Blatt, and Roos}}]{Ger11}
\bibinfo{author}{\bibnamefont{Gerritsma}, \bibfnamefont{R.}},
  \bibinfo{author}{\bibfnamefont{B.~P.} \bibnamefont{Lanyon}},
  \bibinfo{author}{\bibfnamefont{G.}~\bibnamefont{Kirchmair}},
  \bibinfo{author}{\bibfnamefont{F.}~\bibnamefont{Z\"ahringer}},
  \bibinfo{author}{\bibfnamefont{C.}~\bibnamefont{Hempel}},
  \bibinfo{author}{\bibfnamefont{J.}~\bibnamefont{Casanova}},
  \bibinfo{author}{\bibfnamefont{J.~J.} \bibnamefont{Garc\'\i{}a-Ripoll}},
  \bibinfo{author}{\bibfnamefont{E.}~\bibnamefont{Solano}},
  \bibinfo{author}{\bibfnamefont{R.}~\bibnamefont{Blatt}}, and
  \bibinfo{author}{\bibfnamefont{C.~F.} \bibnamefont{Roos}},
  \bibinfo{year}{2011}, {``}\bibinfo{title}{Quantum simulation of the Klein
  paradox with trapped ions},{''} \bibinfo{journal}{Phys. Rev. Lett.}
  \textbf{\bibinfo{volume}{106}},  \bibinfo{pages}{060503}.

\bibitem[{\citenamefont{Ghosh} \emph{et~al.}(2009)\citenamefont{Ghosh, Smirnov,
  and Nori}}]{Gho09}
\bibinfo{author}{\bibnamefont{Ghosh}, \bibfnamefont{P.}},
  \bibinfo{author}{\bibfnamefont{A.}~\bibnamefont{Smirnov}}, and
  \bibinfo{author}{\bibfnamefont{F.}~\bibnamefont{Nori}}, \bibinfo{year}{2009},
  {``}\bibinfo{title}{Modelling light-driven proton pumps in artificial
  photosynthetic reaction centers},{''} \bibinfo{journal}{J. Chem. Phys.}
  \textbf{\bibinfo{volume}{131}},  \bibinfo{pages}{035102}.

\bibitem[{\citenamefont{Ghosh}
  \emph{et~al.}(2011{\natexlab{a}})\citenamefont{Ghosh, Smirnov, and
  Nori}}]{Gho10}
\bibinfo{author}{\bibnamefont{Ghosh}, \bibfnamefont{P.}},
  \bibinfo{author}{\bibfnamefont{A.}~\bibnamefont{Smirnov}}, and
  \bibinfo{author}{\bibfnamefont{F.}~\bibnamefont{Nori}},
  \bibinfo{year}{2011}{\natexlab{a}}, {``}\bibinfo{title}{Artificial
  photosynthetic reaction centers coupled to light-harvesting antennas},{''}
  \bibinfo{journal}{Phys. Rev. E} \textbf{\bibinfo{volume}{84}},
  \bibinfo{pages}{061138}.

\bibitem[{\citenamefont{Ghosh}
  \emph{et~al.}(2011{\natexlab{b}})\citenamefont{Ghosh, Smirnov, and
  Nori}}]{Gho11}
\bibinfo{author}{\bibnamefont{Ghosh}, \bibfnamefont{P.}},
  \bibinfo{author}{\bibfnamefont{A.}~\bibnamefont{Smirnov}}, and
  \bibinfo{author}{\bibfnamefont{F.}~\bibnamefont{Nori}},
  \bibinfo{year}{2011}{\natexlab{b}}, {``}\bibinfo{title}{Quantum effects in
  energy and charge transfer in an artificial photosynthetic complex},{''}
  \bibinfo{journal}{J. Chem. Phys.} \textbf{\bibinfo{volume}{134}},
  \bibinfo{pages}{244103}.

\bibitem[{\citenamefont{Gibbons} \emph{et~al.}(2011)\citenamefont{Gibbons,
  Hamley, Shih, and Chapman}}]{Gib10}
\bibinfo{author}{\bibnamefont{Gibbons}, \bibfnamefont{M.~J.}},
  \bibinfo{author}{\bibfnamefont{C.~D.} \bibnamefont{Hamley}},
  \bibinfo{author}{\bibfnamefont{C.-Y.} \bibnamefont{Shih}}, and
  \bibinfo{author}{\bibfnamefont{M.~S.} \bibnamefont{Chapman}},
  \bibinfo{year}{2011}, {``}\bibinfo{title}{Nondestructive fluorescent state
  detection of single neutral atom qubits},{''} \bibinfo{journal}{Phys. Rev.
  Lett.} \textbf{\bibinfo{volume}{106}},  \bibinfo{pages}{133002}.

\bibitem[{\citenamefont{Gibertini} \emph{et~al.}(2009)\citenamefont{Gibertini,
  Singha, Pellegrini, Polini, Vignale, Pinczuk, Pfeiffer, and West}}]{Gib09}
\bibinfo{author}{\bibnamefont{Gibertini}, \bibfnamefont{M.}},
  \bibinfo{author}{\bibfnamefont{A.}~\bibnamefont{Singha}},
  \bibinfo{author}{\bibfnamefont{V.}~\bibnamefont{Pellegrini}},
  \bibinfo{author}{\bibfnamefont{M.}~\bibnamefont{Polini}},
  \bibinfo{author}{\bibfnamefont{G.}~\bibnamefont{Vignale}},
  \bibinfo{author}{\bibfnamefont{A.}~\bibnamefont{Pinczuk}},
  \bibinfo{author}{\bibfnamefont{L.~N.} \bibnamefont{Pfeiffer}}, and
  \bibinfo{author}{\bibfnamefont{K.~W.} \bibnamefont{West}},
  \bibinfo{year}{2009}, {``}\bibinfo{title}{Engineering artificial graphene in
  a two-dimensional electron gas},{''} \bibinfo{journal}{Phys. Rev. B}
  \textbf{\bibinfo{volume}{79}},  \bibinfo{pages}{241406}.

\bibitem[{\citenamefont{Giorgi} \emph{et~al.}(2010)\citenamefont{Giorgi,
  Paganelli, and Galve}}]{GPG10}
\bibinfo{author}{\bibnamefont{Giorgi}, \bibfnamefont{G.}},
  \bibinfo{author}{\bibfnamefont{S.}~\bibnamefont{Paganelli}}, and
  \bibinfo{author}{\bibfnamefont{F.}~\bibnamefont{Galve}},
  \bibinfo{year}{2010}, {``}\bibinfo{title}{Ion-trap simulation of the quantum
  phase transition in an exactly solvable model of spins couple to bosons},{''}
  \bibinfo{journal}{Phys. Rev. A} \textbf{\bibinfo{volume}{81}},
  \bibinfo{pages}{052118}.

\bibitem[{\citenamefont{Giovanazzi}(2005)}]{Gio05}
\bibinfo{author}{\bibnamefont{Giovanazzi}, \bibfnamefont{S.}},
  \bibinfo{year}{2005}, {``}\bibinfo{title}{Hawking radiation in sonic black
  holes},{''} \bibinfo{journal}{Phys. Rev. Lett.}
  \textbf{\bibinfo{volume}{94}}(\bibinfo{number}{6}),  \bibinfo{pages}{061302}.

\bibitem[{\citenamefont{Goldman} \emph{et~al.}(2009)\citenamefont{Goldman,
  Kubasiak, Bermudez, Gaspard, Lewenstein, and Martin-Delgado}}]{Gol09}
\bibinfo{author}{\bibnamefont{Goldman}, \bibfnamefont{N.}},
  \bibinfo{author}{\bibfnamefont{A.}~\bibnamefont{Kubasiak}},
  \bibinfo{author}{\bibfnamefont{A.}~\bibnamefont{Bermudez}},
  \bibinfo{author}{\bibfnamefont{P.}~\bibnamefont{Gaspard}},
  \bibinfo{author}{\bibfnamefont{M.}~\bibnamefont{Lewenstein}}, and
  \bibinfo{author}{\bibfnamefont{M.}~\bibnamefont{Martin-Delgado}},
  \bibinfo{year}{2009}, {``}\bibinfo{title}{Non-Abelian optical lattices:
  anomalous quantum Hall effect and Dirac fermions},{''}
  \bibinfo{journal}{Phys. Rev. Lett.} \textbf{\bibinfo{volume}{103}},
  \bibinfo{pages}{035301}.

\bibitem[{\citenamefont{Grajcar} \emph{et~al.}(2008)\citenamefont{Grajcar,
  van~der Ploeg, Izmalkov, Il'ichev, Meyer, Fedorov, Shnirman, and
  Schon}}]{Gra08}
\bibinfo{author}{\bibnamefont{Grajcar}, \bibfnamefont{M.}},
  \bibinfo{author}{\bibfnamefont{S.~H.~W.} \bibnamefont{van~der Ploeg}},
  \bibinfo{author}{\bibfnamefont{A.}~\bibnamefont{Izmalkov}},
  \bibinfo{author}{\bibfnamefont{E.}~\bibnamefont{Il'ichev}},
  \bibinfo{author}{\bibfnamefont{H.-G.} \bibnamefont{Meyer}},
  \bibinfo{author}{\bibfnamefont{A.}~\bibnamefont{Fedorov}},
  \bibinfo{author}{\bibfnamefont{A.}~\bibnamefont{Shnirman}}, and
  \bibinfo{author}{\bibfnamefont{G.}~\bibnamefont{Schon}},
  \bibinfo{year}{2008}, {``}\bibinfo{title}{Sisyphus cooling and amplification
  by a superconducting qubit},{''} \bibinfo{journal}{Nature Physics}
  \textbf{\bibinfo{volume}{4}},  \bibinfo{pages}{612--616}.

\bibitem[{\citenamefont{Greentree} \emph{et~al.}(2006)\citenamefont{Greentree,
  Tahan, Cole, and Hollenberg}}]{Gre06}
\bibinfo{author}{\bibnamefont{Greentree}, \bibfnamefont{A.~D.}},
  \bibinfo{author}{\bibfnamefont{C.}~\bibnamefont{Tahan}},
  \bibinfo{author}{\bibfnamefont{J.~H.} \bibnamefont{Cole}}, and
  \bibinfo{author}{\bibfnamefont{L.~C.~L.} \bibnamefont{Hollenberg}},
  \bibinfo{year}{2006}, {``}\bibinfo{title}{Quantum phase transitions of
  light},{''} \bibinfo{journal}{Nature Physics} \textbf{\bibinfo{volume}{2}},
  \bibinfo{pages}{856--861}.

\bibitem[{\citenamefont{Greiner and F\"olling}(2008)}]{Gre08}
\bibinfo{author}{\bibnamefont{Greiner}, \bibfnamefont{M.}}, and
  \bibinfo{author}{\bibfnamefont{S.}~\bibnamefont{F\"olling}},
  \bibinfo{year}{2008}, {``}\bibinfo{title}{Optical lattices},{''}
  \bibinfo{journal}{Nature} \textbf{\bibinfo{volume}{453}},
  \bibinfo{pages}{736--738}.

\bibitem[{\citenamefont{Greiner} \emph{et~al.}(2002)\citenamefont{Greiner,
  Mandel, Esslinger, H\"ansch, and Bloch}}]{Gre02}
\bibinfo{author}{\bibnamefont{Greiner}, \bibfnamefont{M.}},
  \bibinfo{author}{\bibfnamefont{O.}~\bibnamefont{Mandel}},
  \bibinfo{author}{\bibfnamefont{T.}~\bibnamefont{Esslinger}},
  \bibinfo{author}{\bibfnamefont{T.~W.} \bibnamefont{H\"ansch}}, and
  \bibinfo{author}{\bibfnamefont{I.}~\bibnamefont{Bloch}},
  \bibinfo{year}{2002}, {``}\bibinfo{title}{Quantum phase transition from a
  superfluid to a Mott insulator in a gas of ultracold atoms},{''}
  \bibinfo{journal}{Nature} \textbf{\bibinfo{volume}{415}},
  \bibinfo{pages}{39--44}.

\bibitem[{\citenamefont{de~Groot} \emph{et~al.}(2010)\citenamefont{de~Groot,
  Lisenfeld, Schouten, Ashhab, Lupascu, Harmans, and Mooij}}]{deg10}
\bibinfo{author}{\bibnamefont{de~Groot}, \bibfnamefont{P.~C.}},
  \bibinfo{author}{\bibfnamefont{J.}~\bibnamefont{Lisenfeld}},
  \bibinfo{author}{\bibfnamefont{R.~N.} \bibnamefont{Schouten}},
  \bibinfo{author}{\bibfnamefont{S.}~\bibnamefont{Ashhab}},
  \bibinfo{author}{\bibfnamefont{A.}~\bibnamefont{Lupascu}},
  \bibinfo{author}{\bibfnamefont{C.~J. P.~M.} \bibnamefont{Harmans}}, and
  \bibinfo{author}{\bibfnamefont{J.~E.} \bibnamefont{Mooij}},
  \bibinfo{year}{2010}, {``}\bibinfo{title}{Selective darkening of degenerate
  transitions demonstrated with two superconducting quantum bits},{''}
  \bibinfo{journal}{Nature Phys.} \textbf{\bibinfo{volume}{6}},
  \bibinfo{pages}{763--766}.

\bibitem[{\citenamefont{Gustavsson}
  \emph{et~al.}(2013)\citenamefont{Gustavsson, Bylander, and Oliver}}]{Gus13}
\bibinfo{author}{\bibnamefont{Gustavsson}, \bibfnamefont{S.}},
  \bibinfo{author}{\bibfnamefont{J.}~\bibnamefont{Bylander}}, and
  \bibinfo{author}{\bibfnamefont{W.~D.} \bibnamefont{Oliver}},
  \bibinfo{year}{2013}, {``}\bibinfo{title}{Time-Reversal Symmetry and
  Universal Conductance Fluctuations in a Driven Two-Level System,},{''}
  \bibinfo{journal}{Phys. Rev. Lett.} \textbf{\bibinfo{volume}{110}},
  \bibinfo{pages}{016603}.

\bibitem[{\citenamefont{Han} \emph{et~al.}(2007)\citenamefont{Han, Raussendorf,
  and Duan}}]{Han07}
\bibinfo{author}{\bibnamefont{Han}, \bibfnamefont{Y.-J.}},
  \bibinfo{author}{\bibfnamefont{R.}~\bibnamefont{Raussendorf}}, and
  \bibinfo{author}{\bibfnamefont{L.-M.} \bibnamefont{Duan}},
  \bibinfo{year}{2007}, {``}\bibinfo{title}{Scheme for demonstration of
  fractional statistics of anyons in an exactly solvable model},{''}
  \bibinfo{journal}{Phys. Rev. Lett.} \textbf{\bibinfo{volume}{98}},
  \bibinfo{pages}{150404}.

\bibitem[{\citenamefont{H\"anggi and Marchesoni}(2009)}]{Han09}
\bibinfo{author}{\bibnamefont{H\"anggi}, \bibfnamefont{P.}}, and
  \bibinfo{author}{\bibfnamefont{F.}~\bibnamefont{Marchesoni}},
  \bibinfo{year}{2009}, {``}\bibinfo{title}{Artificial Brownian motors:
  Controlling transport on the nanoscale},{''} \bibinfo{journal}{Rev. Mod.
  Phys.} \textbf{\bibinfo{volume}{81}},  \bibinfo{pages}{387}.

\bibitem[{\citenamefont{H\"anggi} \emph{et~al.}(2005)\citenamefont{H\"anggi,
  Marchesoni, and Nori}}]{Han05}
\bibinfo{author}{\bibnamefont{H\"anggi}, \bibfnamefont{P.}},
  \bibinfo{author}{\bibfnamefont{F.}~\bibnamefont{Marchesoni}}, and
  \bibinfo{author}{\bibfnamefont{F.}~\bibnamefont{Nori}}, \bibinfo{year}{2005},
  {``}\bibinfo{title}{Brownian motors},{''} \bibinfo{journal}{Annalen der
  Physik} \textbf{\bibinfo{volume}{14}}, ~\bibinfo{pages}{51}.

\bibitem[{\citenamefont{Hanneke} \emph{et~al.}(2009)\citenamefont{Hanneke,
  Home, Jost, Amini, Leibfried, and Wineland}}]{Hann09}
\bibinfo{author}{\bibnamefont{Hanneke}, \bibfnamefont{D.}},
  \bibinfo{author}{\bibfnamefont{J.~P.} \bibnamefont{Home}},
  \bibinfo{author}{\bibfnamefont{J.~D.} \bibnamefont{Jost}},
  \bibinfo{author}{\bibfnamefont{J.~M.} \bibnamefont{Amini}},
  \bibinfo{author}{\bibfnamefont{D.}~\bibnamefont{Leibfried}}, and
  \bibinfo{author}{\bibfnamefont{D.~J.} \bibnamefont{Wineland}},
  \bibinfo{year}{2009}, {``}\bibinfo{title}{Realisation of a programmable
  two-qubit quantum processor},{''} \bibinfo{journal}{Nature Physics}
  \textbf{\bibinfo{volume}{6}},  \bibinfo{pages}{13--16}.

\bibitem[{\citenamefont{Hanson and Awschalom}(2008)}]{HA08}
\bibinfo{author}{\bibnamefont{Hanson}, \bibfnamefont{R.}}, and
  \bibinfo{author}{\bibfnamefont{D.~D.} \bibnamefont{Awschalom}},
  \bibinfo{year}{2008}, {``}\bibinfo{title}{Coherent manipulation of single
  spins in semiconductors},{''} \bibinfo{journal}{Nature}
  \textbf{\bibinfo{volume}{453}},  \bibinfo{pages}{1043--1049}.

\bibitem[{\citenamefont{Harrabi} \emph{et~al.}(2009)\citenamefont{Harrabi,
  Yoshihara, Niskanen, Nakamura, and Tsai}}]{Har09}
\bibinfo{author}{\bibnamefont{Harrabi}, \bibfnamefont{K.}},
  \bibinfo{author}{\bibfnamefont{F.}~\bibnamefont{Yoshihara}},
  \bibinfo{author}{\bibfnamefont{A.~O.} \bibnamefont{Niskanen}},
  \bibinfo{author}{\bibfnamefont{Y.}~\bibnamefont{Nakamura}}, and
  \bibinfo{author}{\bibfnamefont{J.~S.} \bibnamefont{Tsai}},
  \bibinfo{year}{2009}, {``}\bibinfo{title}{Engineered selection rules for
  tunable coupling in a superconducting quantum circuit},{''}
  \bibinfo{journal}{Phys. Rev. B} \textbf{\bibinfo{volume}{79}},
  \bibinfo{pages}{020507}.

\bibitem[{\citenamefont{Harris}(2012)}]{Har12}
\bibinfo{author}{\bibnamefont{Harris}, \bibfnamefont{R.}},
  \bibinfo{year}{2012}, {``}\bibinfo{title}{private communication},{''}.

\bibitem[{\citenamefont{Harris} \emph{et~al.}(2010)\citenamefont{Harris,
  Johnson, Lanting, Berkley, Johansson, Bunyk, Tolkacheva, Ladizinsky,
  Ladizinsky, Oh, Cioata, Perminov} \emph{et~al.}}]{Har10}
\bibinfo{author}{\bibnamefont{Harris}, \bibfnamefont{R.}},
  \bibinfo{author}{\bibfnamefont{M.~W.} \bibnamefont{Johnson}},
  \bibinfo{author}{\bibfnamefont{T.}~\bibnamefont{Lanting}},
  \bibinfo{author}{\bibfnamefont{A.~J.} \bibnamefont{Berkley}},
  \bibinfo{author}{\bibfnamefont{J.}~\bibnamefont{Johansson}},
  \bibinfo{author}{\bibfnamefont{P.}~\bibnamefont{Bunyk}},
  \bibinfo{author}{\bibfnamefont{E.}~\bibnamefont{Tolkacheva}},
  \bibinfo{author}{\bibfnamefont{E.}~\bibnamefont{Ladizinsky}},
  \bibinfo{author}{\bibfnamefont{N.}~\bibnamefont{Ladizinsky}},
  \bibinfo{author}{\bibfnamefont{T.}~\bibnamefont{Oh}},
  \bibinfo{author}{\bibfnamefont{F.}~\bibnamefont{Cioata}},
  \bibinfo{author}{\bibfnamefont{I.}~\bibnamefont{Perminov}}, \emph{et~al.},
  \bibinfo{year}{2010}, {``}\bibinfo{title}{Experimental investigation of an
  eight-qubit unit cell in a superconducting optimization processor},{''}
  \bibinfo{journal}{Phys. Rev. B} \textbf{\bibinfo{volume}{82}},
  \bibinfo{pages}{024511}.

\bibitem[{\citenamefont{Hartmann} \emph{et~al.}(2006)\citenamefont{Hartmann,
  Brandao, and Plenio}}]{Har06}
\bibinfo{author}{\bibnamefont{Hartmann}, \bibfnamefont{M.~J.}},
  \bibinfo{author}{\bibfnamefont{F.~G. S.~L.} \bibnamefont{Brandao}}, and
  \bibinfo{author}{\bibfnamefont{M.~B.} \bibnamefont{Plenio}},
  \bibinfo{year}{2006}, {``}\bibinfo{title}{Strongly interacting polaritons in
  coupled arrays of cavities},{''} \bibinfo{journal}{Nature Physics}
  \textbf{\bibinfo{volume}{2}},  \bibinfo{pages}{849--855}.

\bibitem[{\citenamefont{Hastings}(2006)}]{Has06}
\bibinfo{author}{\bibnamefont{Hastings}, \bibfnamefont{M.~B.}},
  \bibinfo{year}{2006}, {``}\bibinfo{title}{Solving gapped Hamiltonians
  locally},{''} \bibinfo{journal}{Phys. Rev. B} \textbf{\bibinfo{volume}{73}},
  \bibinfo{pages}{085115}.

\bibitem[{\citenamefont{Hauke} \emph{et~al.}(2012)\citenamefont{Hauke,
  Cucchietti, Tagliacozzo, Deutsch, and Lewenstein}}]{Hau12}
\bibinfo{author}{\bibnamefont{Hauke}, \bibfnamefont{P.}},
  \bibinfo{author}{\bibfnamefont{F.~M.} \bibnamefont{Cucchietti}},
  \bibinfo{author}{\bibfnamefont{L.}~\bibnamefont{Tagliacozzo}},
  \bibinfo{author}{\bibfnamefont{I.}~\bibnamefont{Deutsch}}, and
  \bibinfo{author}{\bibfnamefont{M.}~\bibnamefont{Lewenstein}},
  \bibinfo{year}{2012}, {``}\bibinfo{title}{Can one trust quantum
  simulators?},{''} \bibinfo{journal}{Rep. Prog. Phys.}
  \textbf{\bibinfo{volume}{75}},  \bibinfo{pages}{082401}.

\bibitem[{\citenamefont{Hauke} \emph{et~al.}(2013)\citenamefont{Hauke, Marcos,
  Dalmonte, and Zoller}}]{Hau13}
\bibinfo{author}{\bibnamefont{Hauke}, \bibfnamefont{P.}},
  \bibinfo{author}{\bibfnamefont{D.}~\bibnamefont{Marcos}},
  \bibinfo{author}{\bibfnamefont{M.}~\bibnamefont{Dalmonte}}, and
  \bibinfo{author}{\bibfnamefont{P.}~\bibnamefont{Zoller}},
  \bibinfo{year}{2013}, {``}\bibinfo{title}{Quantum simulation of a lattice
  Schwinger model in a chain of trapped ions},{''} \bibinfo{journal}{Phys. Rev.
  X} \textbf{\bibinfo{volume}{3}},  \bibinfo{pages}{041018}.

\bibitem[{\citenamefont{Herrera and Krems}(2011)}]{Her11}
\bibinfo{author}{\bibnamefont{Herrera}, \bibfnamefont{F.}}, and
  \bibinfo{author}{\bibfnamefont{R.~V.} \bibnamefont{Krems}},
  \bibinfo{year}{2011}, {``}\bibinfo{title}{Tunable Holstein model with cold
  polar molecules},{''} \bibinfo{journal}{Phys. Rev. A}
  \textbf{\bibinfo{volume}{84}},  \bibinfo{pages}{051401}.

\bibitem[{\citenamefont{Herrera} \emph{et~al.}(2013)\citenamefont{Herrera,
  Madison, Krems, and Berciu}}]{Her13}
\bibinfo{author}{\bibnamefont{Herrera}, \bibfnamefont{F.}},
  \bibinfo{author}{\bibfnamefont{K.~W.} \bibnamefont{Madison}},
  \bibinfo{author}{\bibfnamefont{R.~V.} \bibnamefont{Krems}}, and
  \bibinfo{author}{\bibfnamefont{M.}~\bibnamefont{Berciu}},
  \bibinfo{year}{2013}, {``}\bibinfo{title}{Investigating Polaron Transitions
  with Polar Molecules},{''} \bibinfo{journal}{Phys. Rev. Lett.}
  \textbf{\bibinfo{volume}{110}},  \bibinfo{pages}{223002}.

\bibitem[{\citenamefont{Hilbert and Lopez}(2011)}]{Hil11}
\bibinfo{author}{\bibnamefont{Hilbert}, \bibfnamefont{M.}}, and
  \bibinfo{author}{\bibfnamefont{P.}~\bibnamefont{Lopez}},
  \bibinfo{year}{2011}, {``}\bibinfo{title}{The world's technological capacity
  to store, communicate and compute information},{''}
  \bibinfo{journal}{Science} \textbf{\bibinfo{volume}{332}},
  ~\bibinfo{pages}{60}.

\bibitem[{\citenamefont{Ho} \emph{et~al.}(2009)\citenamefont{Ho, Cazalilla, and
  Giamarchi}}]{Ho09}
\bibinfo{author}{\bibnamefont{Ho}, \bibfnamefont{A.~F.}},
  \bibinfo{author}{\bibfnamefont{M.~A.} \bibnamefont{Cazalilla}}, and
  \bibinfo{author}{\bibfnamefont{T.}~\bibnamefont{Giamarchi}},
  \bibinfo{year}{2009}, {``}\bibinfo{title}{Quantum simulation of the Hubbard
  model: The attractive route},{''} \bibinfo{journal}{Phys. Rev. A}
  \textbf{\bibinfo{volume}{79}},  \bibinfo{pages}{033620}.

\bibitem[{\citenamefont{Horstmann} \emph{et~al.}(2010)\citenamefont{Horstmann,
  Reznik, Fagnocchi, and Cirac}}]{Hor09}
\bibinfo{author}{\bibnamefont{Horstmann}, \bibfnamefont{B.}},
  \bibinfo{author}{\bibfnamefont{B.}~\bibnamefont{Reznik}},
  \bibinfo{author}{\bibfnamefont{S.}~\bibnamefont{Fagnocchi}}, and
  \bibinfo{author}{\bibfnamefont{J.~I.} \bibnamefont{Cirac}},
  \bibinfo{year}{2010}, {``}\bibinfo{title}{Hawking radiation from an acoustic
  black hole on an ion ring},{''} \bibinfo{journal}{Phys. Rev. Lett.}
  \textbf{\bibinfo{volume}{104}},  \bibinfo{pages}{250403}.

\bibitem[{\citenamefont{Hou} \emph{et~al.}(2009)\citenamefont{Hou, Yang, and
  Liu}}]{Hou09}
\bibinfo{author}{\bibnamefont{Hou}, \bibfnamefont{J.-M.}},
  \bibinfo{author}{\bibfnamefont{W.-X.} \bibnamefont{Yang}}, and
  \bibinfo{author}{\bibfnamefont{X.-J.} \bibnamefont{Liu}},
  \bibinfo{year}{2009}, {``}\bibinfo{title}{Massless Dirac fermions in a square
  optical lattice},{''} \bibinfo{journal}{Phys. Rev. A}
  \textbf{\bibinfo{volume}{79}},  \bibinfo{pages}{043621}.

\bibitem[{\citenamefont{Houck} \emph{et~al.}(2012)\citenamefont{Houck,
  T\"ureci, and Koch}}]{Hou12}
\bibinfo{author}{\bibnamefont{Houck}, \bibfnamefont{A.~A.}},
  \bibinfo{author}{\bibfnamefont{H.~E.} \bibnamefont{T\"ureci}}, and
  \bibinfo{author}{\bibfnamefont{J.}~\bibnamefont{Koch}}, \bibinfo{year}{2012},
  {``}\bibinfo{title}{On-chip quantum simulation with superconducting
  circuits},{''} \bibinfo{journal}{Nature Physics}
  \textbf{\bibinfo{volume}{8}},  \bibinfo{pages}{292--299}.

\bibitem[{\citenamefont{Howell and Yeaze}(2000)}]{How00}
\bibinfo{author}{\bibnamefont{Howell}, \bibfnamefont{J.~C.}}, and
  \bibinfo{author}{\bibfnamefont{J.~A.} \bibnamefont{Yeaze}},
  \bibinfo{year}{2000}, {``}\bibinfo{title}{Linear optics simulations of the
  quantum baker's map},{''} \bibinfo{journal}{Phys. Rev. A}
  \textbf{\bibinfo{volume}{61}},  \bibinfo{pages}{012304}.

\bibitem[{\citenamefont{Hu} \emph{et~al.}(2012)\citenamefont{Hu, Feng, and
  Lee}}]{Hu12}
\bibinfo{author}{\bibnamefont{Hu}, \bibfnamefont{Y.-M.}},
  \bibinfo{author}{\bibfnamefont{M.}~\bibnamefont{Feng}}, and
  \bibinfo{author}{\bibfnamefont{C.}~\bibnamefont{Lee}}, \bibinfo{year}{2012},
  {``}\bibinfo{title}{Adiabatic Mach-Zehnder interferometer via an array of
  trapped ions},{''} \bibinfo{journal}{Phys. Rev. A}
  \textbf{\bibinfo{volume}{85}},  \bibinfo{pages}{043604}.

\bibitem[{\citenamefont{Ian} \emph{et~al.}(2009)\citenamefont{Ian, Liu, and
  Nori}}]{Hou10}
\bibinfo{author}{\bibnamefont{Ian}, \bibfnamefont{H.}},
  \bibinfo{author}{\bibfnamefont{Y.-X.} \bibnamefont{Liu}}, and
  \bibinfo{author}{\bibfnamefont{F.}~\bibnamefont{Nori}}, \bibinfo{year}{2009},
  {``}\bibinfo{title}{Tunable electromagnetically induced transparency and
  absorption with dressed superconducting qubits},{''} \bibinfo{journal}{Phys.
  Rev. A} \textbf{\bibinfo{volume}{81}},  \bibinfo{pages}{063823}.

\bibitem[{\citenamefont{Islam} \emph{et~al.}(2011)\citenamefont{Islam, Edward,
  Kim, Korenblit, Noh, Carmichael, G.-D.Lin, Duan, Wang, Freericks, and
  Monroe}}]{Isl11}
\bibinfo{author}{\bibnamefont{Islam}, \bibfnamefont{R.}},
  \bibinfo{author}{\bibfnamefont{E.~E.} \bibnamefont{Edward}},
  \bibinfo{author}{\bibfnamefont{K.}~\bibnamefont{Kim}},
  \bibinfo{author}{\bibfnamefont{S.}~\bibnamefont{Korenblit}},
  \bibinfo{author}{\bibfnamefont{C.}~\bibnamefont{Noh}},
  \bibinfo{author}{\bibfnamefont{H.}~\bibnamefont{Carmichael}},
  \bibinfo{author}{\bibnamefont{G.-D.Lin}},
  \bibinfo{author}{\bibfnamefont{L.-M.} \bibnamefont{Duan}},
  \bibinfo{author}{\bibfnamefont{C.-C.~J.} \bibnamefont{Wang}},
  \bibinfo{author}{\bibfnamefont{J.~K.} \bibnamefont{Freericks}}, and
  \bibinfo{author}{\bibfnamefont{C.}~\bibnamefont{Monroe}},
  \bibinfo{year}{2011}, {``}\bibinfo{title}{Onset of a quantum phase transition
  with a trapped ion quantum simulator},{''} \bibinfo{journal}{Nature
  Communications} \textbf{\bibinfo{volume}{2}},  \bibinfo{pages}{377}.

\bibitem[{\citenamefont{Ivanov} \emph{et~al.}(2009)\citenamefont{Ivanov,
  Ivanov, Vitanov, Mering, Fleischhauer, and Singer}}]{Iva09}
\bibinfo{author}{\bibnamefont{Ivanov}, \bibfnamefont{P.~A.}},
  \bibinfo{author}{\bibfnamefont{S.~S.} \bibnamefont{Ivanov}},
  \bibinfo{author}{\bibfnamefont{N.~V.} \bibnamefont{Vitanov}},
  \bibinfo{author}{\bibfnamefont{A.}~\bibnamefont{Mering}},
  \bibinfo{author}{\bibfnamefont{M.}~\bibnamefont{Fleischhauer}}, and
  \bibinfo{author}{\bibfnamefont{K.}~\bibnamefont{Singer}},
  \bibinfo{year}{2009}, {``}\bibinfo{title}{Simulation of a quantum phase
  transition of polaritons with trapped ions},{''} \bibinfo{journal}{Phys. Rev.
  A} \textbf{\bibinfo{volume}{80}},  \bibinfo{pages}{060301}.

\bibitem[{\citenamefont{Ivanov and Schmidt-Kaler}(2011)}]{Iva11}
\bibinfo{author}{\bibnamefont{Ivanov}, \bibfnamefont{P.~A.}}, and
  \bibinfo{author}{\bibfnamefont{F.}~\bibnamefont{Schmidt-Kaler}},
  \bibinfo{year}{2011}, {``}\bibinfo{title}{Simulation of quantum magnetism in
  mixed-spin systems with impurity-doped ion crystals},{''}
  \bibinfo{journal}{New Journal of Physics} \textbf{\bibinfo{volume}{13}},
  \bibinfo{pages}{125008}.

\bibitem[{\citenamefont{Jaksch} \emph{et~al.}(1998)\citenamefont{Jaksch,
  Bruder, Cirac, Gardiner, and Zoller}}]{Jak98}
\bibinfo{author}{\bibnamefont{Jaksch}, \bibfnamefont{D.}},
  \bibinfo{author}{\bibfnamefont{C.}~\bibnamefont{Bruder}},
  \bibinfo{author}{\bibfnamefont{J.~I.} \bibnamefont{Cirac}},
  \bibinfo{author}{\bibfnamefont{C.~W.} \bibnamefont{Gardiner}}, and
  \bibinfo{author}{\bibfnamefont{P.}~\bibnamefont{Zoller}},
  \bibinfo{year}{1998}, {``}\bibinfo{title}{Cold bosonic atoms in optical
  lattices},{''} \bibinfo{journal}{Phys. Rev. Lett.}
  \textbf{\bibinfo{volume}{81}},  \bibinfo{pages}{3108--3111}.

\bibitem[{\citenamefont{Jaksch and Zoller}(2005)}]{Jak05}
\bibinfo{author}{\bibnamefont{Jaksch}, \bibfnamefont{D.}}, and
  \bibinfo{author}{\bibfnamefont{P.}~\bibnamefont{Zoller}},
  \bibinfo{year}{2005}, {``}\bibinfo{title}{The cold atom Hubbard toolbox},{''}
  \bibinfo{journal}{Annals of Physics} \textbf{\bibinfo{volume}{315}},
  \bibinfo{pages}{52--79}.

\bibitem[{\citenamefont{Jan\'{e}} \emph{et~al.}(2003)\citenamefont{Jan\'{e},
  Vidal, D\"ur, Zoller, and Cirac}}]{Jan03}
\bibinfo{author}{\bibnamefont{Jan\'{e}}, \bibfnamefont{E.}},
  \bibinfo{author}{\bibfnamefont{G.}~\bibnamefont{Vidal}},
  \bibinfo{author}{\bibfnamefont{W.}~\bibnamefont{D\"ur}},
  \bibinfo{author}{\bibfnamefont{P.}~\bibnamefont{Zoller}}, and
  \bibinfo{author}{\bibfnamefont{J.}~\bibnamefont{Cirac}},
  \bibinfo{year}{2003}, {``}\bibinfo{title}{Simulation of quantum dynamics with
  quantum optical systems},{''} \bibinfo{journal}{Quant. Inf. Comput.}
  \textbf{\bibinfo{volume}{3}},  \bibinfo{pages}{15--29}.

\bibitem[{\citenamefont{Jo} \emph{et~al.}(2009)\citenamefont{Jo, Lee, Choi,
  Christensen, Kim, Thywissen, Pritchard, and Ketterle}}]{GB09}
\bibinfo{author}{\bibnamefont{Jo}, \bibfnamefont{G.-B.}},
  \bibinfo{author}{\bibfnamefont{Y.-R.} \bibnamefont{Lee}},
  \bibinfo{author}{\bibfnamefont{J.-H.} \bibnamefont{Choi}},
  \bibinfo{author}{\bibfnamefont{C.~A.} \bibnamefont{Christensen}},
  \bibinfo{author}{\bibfnamefont{T.~H.} \bibnamefont{Kim}},
  \bibinfo{author}{\bibfnamefont{J.~H.} \bibnamefont{Thywissen}},
  \bibinfo{author}{\bibfnamefont{D.~E.} \bibnamefont{Pritchard}}, and
  \bibinfo{author}{\bibfnamefont{W.}~\bibnamefont{Ketterle}},
  \bibinfo{year}{2009}, {``}\bibinfo{title}{Itinerant ferromagnetism in a Fermi
  gas of ultracold atoms},{''} \bibinfo{journal}{Science}
  \textbf{\bibinfo{volume}{325}},  \bibinfo{pages}{1521--1524}.

\bibitem[{\citenamefont{Johanning} \emph{et~al.}(2009)\citenamefont{Johanning,
  Varon, and Wunderlich}}]{Jo09}
\bibinfo{author}{\bibnamefont{Johanning}, \bibfnamefont{M.}},
  \bibinfo{author}{\bibfnamefont{A.}~\bibnamefont{Varon}}, and
  \bibinfo{author}{\bibfnamefont{C.}~\bibnamefont{Wunderlich}},
  \bibinfo{year}{2009}, {``}\bibinfo{title}{Quantum simulations with cold
  trapped ions},{''} \bibinfo{journal}{J. Phys. B: At. Mol. Opt. Phys.}
  \textbf{\bibinfo{volume}{42}},  \bibinfo{pages}{154009}.

\bibitem[{\citenamefont{Johansson} \emph{et~al.}(2010)\citenamefont{Johansson,
  Johansson, Wilson, and Nori}}]{Joh10}
\bibinfo{author}{\bibnamefont{Johansson}, \bibfnamefont{J.~R.}},
  \bibinfo{author}{\bibfnamefont{G.}~\bibnamefont{Johansson}},
  \bibinfo{author}{\bibfnamefont{C.~M.} \bibnamefont{Wilson}}, and
  \bibinfo{author}{\bibfnamefont{F.}~\bibnamefont{Nori}}, \bibinfo{year}{2010},
  {``}\bibinfo{title}{Dynamical Casimir effect in superconducting microwave
  circuits},{''} \bibinfo{journal}{Phys. Rev. A} \textbf{\bibinfo{volume}{82}},
   \bibinfo{pages}{052509}.

\bibitem[{\citenamefont{Jordan} \emph{et~al.}(2012)\citenamefont{Jordan, Lee,
  and Preskill}}]{Jor12}
\bibinfo{author}{\bibnamefont{Jordan}, \bibfnamefont{S.~P.}},
  \bibinfo{author}{\bibfnamefont{K.~S.~M.} \bibnamefont{Lee}}, and
  \bibinfo{author}{\bibfnamefont{J.}~\bibnamefont{Preskill}},
  \bibinfo{year}{2012}, {``}\bibinfo{title}{Quantum Algorithms for Quantum
  Field Theories},{''} \bibinfo{journal}{Science}
  \textbf{\bibinfo{volume}{336}},  \bibinfo{pages}{1130--1133}.

\bibitem[{\citenamefont{Kapit and Mueller}(2011)}]{Kap11}
\bibinfo{author}{\bibnamefont{Kapit}, \bibfnamefont{E.}}, and
  \bibinfo{author}{\bibfnamefont{E.}~\bibnamefont{Mueller}},
  \bibinfo{year}{2011}, {``}\bibinfo{title}{Optical-lattice Hamiltonians for
  relativistic quantum electrodynamics},{''} \bibinfo{journal}{Phys. Rev. A}
  \textbf{\bibinfo{volume}{83}},  \bibinfo{pages}{033625}.

\bibitem[{\citenamefont{Kassal} \emph{et~al.}(2008)\citenamefont{Kassal,
  Jordan, Love, Mohseni, and Aspuru-Guzik}}]{Kas08}
\bibinfo{author}{\bibnamefont{Kassal}, \bibfnamefont{I.}},
  \bibinfo{author}{\bibfnamefont{S.~P.} \bibnamefont{Jordan}},
  \bibinfo{author}{\bibfnamefont{P.~J.} \bibnamefont{Love}},
  \bibinfo{author}{\bibfnamefont{M.}~\bibnamefont{Mohseni}}, and
  \bibinfo{author}{\bibfnamefont{A.}~\bibnamefont{Aspuru-Guzik}},
  \bibinfo{year}{2008}, {``}\bibinfo{title}{Quantum algorithms for the
  simulation of chemical dynamics},{''} \bibinfo{journal}{PNAS}
  \textbf{\bibinfo{volume}{105}},  \bibinfo{pages}{18681--18686}.

\bibitem[{\citenamefont{Kassal} \emph{et~al.}(2011)\citenamefont{Kassal,
  Whitfield, Perdomo-Ortiz, Yung, and Aspuru-Guzik}}]{Kas10}
\bibinfo{author}{\bibnamefont{Kassal}, \bibfnamefont{I.}},
  \bibinfo{author}{\bibfnamefont{J.~D.} \bibnamefont{Whitfield}},
  \bibinfo{author}{\bibfnamefont{A.}~\bibnamefont{Perdomo-Ortiz}},
  \bibinfo{author}{\bibfnamefont{M.-H.} \bibnamefont{Yung}}, and
  \bibinfo{author}{\bibfnamefont{A.}~\bibnamefont{Aspuru-Guzik}},
  \bibinfo{year}{2011}, {``}\bibinfo{title}{Simulating Chemistry Using Quantum
  Computers},{''} \bibinfo{journal}{Annu. Rev. Phys. Chem.}
  \textbf{\bibinfo{volume}{62}},  \bibinfo{pages}{185}.

\bibitem[{\citenamefont{Katsnelson}
  \emph{et~al.}(2006)\citenamefont{Katsnelson, Novoselov, and Geim}}]{Kat06}
\bibinfo{author}{\bibnamefont{Katsnelson}, \bibfnamefont{M.~I.}},
  \bibinfo{author}{\bibfnamefont{K.~S.} \bibnamefont{Novoselov}}, and
  \bibinfo{author}{\bibfnamefont{A.~K.} \bibnamefont{Geim}},
  \bibinfo{year}{2006}, {``}\bibinfo{title}{Chiral tunnelling and the Klein
  paradox in graphene},{''} \bibinfo{journal}{Nature Physics}
  \textbf{\bibinfo{volume}{2}},  \bibinfo{pages}{620 -- 625}.

\bibitem[{\citenamefont{Kay and Angelakis}(2008)}]{Kay08}
\bibinfo{author}{\bibnamefont{Kay}, \bibfnamefont{A.}}, and
  \bibinfo{author}{\bibfnamefont{D.}~\bibnamefont{Angelakis}},
  \bibinfo{year}{2008}, {``}\bibinfo{title}{Reproducing spin lattice models in
  strongly coupled atom-cavity systems},{''} \bibinfo{journal}{EPL}
  \textbf{\bibinfo{volume}{84}},  \bibinfo{pages}{20001}.

\bibitem[{\citenamefont{Kay} \emph{et~al.}(2006)\citenamefont{Kay, Pachos, and
  Adams}}]{Kay06}
\bibinfo{author}{\bibnamefont{Kay}, \bibfnamefont{A.}},
  \bibinfo{author}{\bibfnamefont{J.~K.} \bibnamefont{Pachos}}, and
  \bibinfo{author}{\bibfnamefont{C.~S.} \bibnamefont{Adams}},
  \bibinfo{year}{2006}, {``}\bibinfo{title}{Graph-state preparation and quantum
  computation with global addressing of optical lattices},{''}
  \bibinfo{journal}{Phys. Rev. A} \textbf{\bibinfo{volume}{73}},
  \bibinfo{pages}{022310}.

\bibitem[{\citenamefont{Kim} \emph{et~al.}(2010)\citenamefont{Kim, Chang,
  Korenblit, Islam, Edwards, Freericks, Lin, Duan, and Monroe}}]{Kim10}
\bibinfo{author}{\bibnamefont{Kim}, \bibfnamefont{K.}},
  \bibinfo{author}{\bibfnamefont{M.-S.} \bibnamefont{Chang}},
  \bibinfo{author}{\bibfnamefont{S.}~\bibnamefont{Korenblit}},
  \bibinfo{author}{\bibfnamefont{R.}~\bibnamefont{Islam}},
  \bibinfo{author}{\bibfnamefont{E.~E.} \bibnamefont{Edwards}},
  \bibinfo{author}{\bibfnamefont{J.~K.} \bibnamefont{Freericks}},
  \bibinfo{author}{\bibfnamefont{G.-D.} \bibnamefont{Lin}},
  \bibinfo{author}{\bibfnamefont{L.-M.} \bibnamefont{Duan}}, and
  \bibinfo{author}{\bibfnamefont{C.}~\bibnamefont{Monroe}},
  \bibinfo{year}{2010}, {``}\bibinfo{title}{Quantum simulation of frustrated
  Ising spins with trapped ions},{''} \bibinfo{journal}{Nature}
  \textbf{\bibinfo{volume}{465}},  \bibinfo{pages}{590}.

\bibitem[{\citenamefont{Kim} \emph{et~al.}(2011)\citenamefont{Kim, Korenblit,
  Islam, Edwards, Chang, Noh, Carmichael, Lin, Duan, Wang, Freericks, and
  Monroe}}]{Kim11}
\bibinfo{author}{\bibnamefont{Kim}, \bibfnamefont{K.}},
  \bibinfo{author}{\bibfnamefont{S.}~\bibnamefont{Korenblit}},
  \bibinfo{author}{\bibfnamefont{R.}~\bibnamefont{Islam}},
  \bibinfo{author}{\bibfnamefont{E.~E.} \bibnamefont{Edwards}},
  \bibinfo{author}{\bibfnamefont{M.-S.} \bibnamefont{Chang}},
  \bibinfo{author}{\bibfnamefont{C.}~\bibnamefont{Noh}},
  \bibinfo{author}{\bibfnamefont{H.}~\bibnamefont{Carmichael}},
  \bibinfo{author}{\bibfnamefont{G.-D.} \bibnamefont{Lin}},
  \bibinfo{author}{\bibfnamefont{L.-M.} \bibnamefont{Duan}},
  \bibinfo{author}{\bibfnamefont{C.~C.~J.} \bibnamefont{Wang}},
  \bibinfo{author}{\bibfnamefont{J.~K.} \bibnamefont{Freericks}}, and
  \bibinfo{author}{\bibfnamefont{C.}~\bibnamefont{Monroe}},
  \bibinfo{year}{2011}, {``}\bibinfo{title}{Quantum simulation of the
  transverse Ising model with trapped ions},{''} \bibinfo{journal}{New Journal
  of Physics} \textbf{\bibinfo{volume}{13}},  \bibinfo{pages}{105003}.

\bibitem[{\citenamefont{Kimble}(2008)}]{Kim08}
\bibinfo{author}{\bibnamefont{Kimble}, \bibfnamefont{H.~J.}},
  \bibinfo{year}{2008}, {``}\bibinfo{title}{The quantum internet},{''}
  \bibinfo{journal}{Nature} \textbf{\bibinfo{volume}{453}},
  \bibinfo{pages}{1023--1030}.

\bibitem[{\citenamefont{Kitaev}(2003)}]{Kit97}
\bibinfo{author}{\bibnamefont{Kitaev}, \bibfnamefont{A.}},
  \bibinfo{year}{2003}, {``}\bibinfo{title}{Fault-tolerant quantum computation
  by anyons},{''} \bibinfo{journal}{Ann. Phys.} \textbf{\bibinfo{volume}{303}},
   \bibinfo{pages}{2--30}.

\bibitem[{\citenamefont{Kitagawa}(2012)}]{Kit11}
\bibinfo{author}{\bibnamefont{Kitagawa}, \bibfnamefont{T.}},
  \bibinfo{year}{2012}, {``}\bibinfo{title}{Topological phenomena in quantum
  walks; elementary introduction to the physics of topological phases},{''}
  \bibinfo{journal}{Quantum Information Processing}
  \textbf{\bibinfo{volume}{11}},  \bibinfo{pages}{1107--1148}.

\bibitem[{\citenamefont{Koch} \emph{et~al.}(2010)\citenamefont{Koch, Houck,
  Hur, and Girvin}}]{Koc10}
\bibinfo{author}{\bibnamefont{Koch}, \bibfnamefont{J.}},
  \bibinfo{author}{\bibfnamefont{A.~A.} \bibnamefont{Houck}},
  \bibinfo{author}{\bibfnamefont{K.~L.} \bibnamefont{Hur}}, and
  \bibinfo{author}{\bibfnamefont{S.~M.} \bibnamefont{Girvin}},
  \bibinfo{year}{2010}, {``}\bibinfo{title}{Time-reversal-symmetry breaking in
  circuit-QED-based photon lattices},{''} \bibinfo{journal}{Phys. Rev. A}
  \textbf{\bibinfo{volume}{82}},  \bibinfo{pages}{043811}.

\bibitem[{\citenamefont{Korenblit} \emph{et~al.}(2012)\citenamefont{Korenblit,
  Kafri, Campbell, Islam, Edwards, Gong, Lin, Duan, Kim, Kim, and
  Monroe}}]{Kor12}
\bibinfo{author}{\bibnamefont{Korenblit}, \bibfnamefont{S.}},
  \bibinfo{author}{\bibfnamefont{D.}~\bibnamefont{Kafri}},
  \bibinfo{author}{\bibfnamefont{W.~C.} \bibnamefont{Campbell}},
  \bibinfo{author}{\bibfnamefont{R.}~\bibnamefont{Islam}},
  \bibinfo{author}{\bibfnamefont{E.~E.} \bibnamefont{Edwards}},
  \bibinfo{author}{\bibfnamefont{Z.-X.} \bibnamefont{Gong}},
  \bibinfo{author}{\bibfnamefont{G.-D.} \bibnamefont{Lin}},
  \bibinfo{author}{\bibfnamefont{L.-M.} \bibnamefont{Duan}},
  \bibinfo{author}{\bibfnamefont{J.}~\bibnamefont{Kim}},
  \bibinfo{author}{\bibfnamefont{K.}~\bibnamefont{Kim}}, and
  \bibinfo{author}{\bibfnamefont{C.}~\bibnamefont{Monroe}},
  \bibinfo{year}{2012}, {``}\bibinfo{title}{Quantum Simulation of Spin Models
  on an Arbitrary Lattice with Trapped Ions},{''} \bibinfo{journal}{New J.
  Phys.} \textbf{\bibinfo{volume}{14}},  \bibinfo{pages}{095024}.

\bibitem[{\citenamefont{Kraus} \emph{et~al.}(2007)\citenamefont{Kraus, Wolf,
  and Cirac}}]{Kra07}
\bibinfo{author}{\bibnamefont{Kraus}, \bibfnamefont{C.~V.}},
  \bibinfo{author}{\bibfnamefont{M.~M.} \bibnamefont{Wolf}}, and
  \bibinfo{author}{\bibfnamefont{J.~I.} \bibnamefont{Cirac}},
  \bibinfo{year}{2007}, {``}\bibinfo{title}{Quantum simulations under
  translational symmetry},{''} \bibinfo{journal}{Phys. Rev. A}
  \textbf{\bibinfo{volume}{75}},  \bibinfo{pages}{022303}.

\bibitem[{\citenamefont{Ladd} \emph{et~al.}(2010)\citenamefont{Ladd, Jelezko,
  Laflamme, Nakamura, Monroe, and O'Brien}}]{Lad10}
\bibinfo{author}{\bibnamefont{Ladd}, \bibfnamefont{T.~D.}},
  \bibinfo{author}{\bibfnamefont{F.}~\bibnamefont{Jelezko}},
  \bibinfo{author}{\bibfnamefont{R.}~\bibnamefont{Laflamme}},
  \bibinfo{author}{\bibfnamefont{Y.}~\bibnamefont{Nakamura}},
  \bibinfo{author}{\bibfnamefont{C.}~\bibnamefont{Monroe}}, and
  \bibinfo{author}{\bibfnamefont{J.~L.} \bibnamefont{O'Brien}},
  \bibinfo{year}{2010}, {``}\bibinfo{title}{Quantum computers},{''}
  \bibinfo{journal}{Nature} \textbf{\bibinfo{volume}{464}},
  ~\bibinfo{pages}{45}.

\bibitem[{\citenamefont{L\"ahteenm\"aki}
  \emph{et~al.}(2013)\citenamefont{L\"ahteenm\"aki, Paraoanu, Hassel, and
  Hakonen}}]{Lah13}
\bibinfo{author}{\bibnamefont{L\"ahteenm\"aki}, \bibfnamefont{P.}},
  \bibinfo{author}{\bibfnamefont{G.~S.} \bibnamefont{Paraoanu}},
  \bibinfo{author}{\bibfnamefont{J.}~\bibnamefont{Hassel}}, and
  \bibinfo{author}{\bibfnamefont{P.~J.} \bibnamefont{Hakonen}},
  \bibinfo{year}{2013}, {``}\bibinfo{title}{Dynamical Casimir effect in a
  Josephson metamaterial},{''} \bibinfo{journal}{PNAS}
  \textbf{\bibinfo{volume}{110}},  \bibinfo{pages}{4234--4238}.

\bibitem[{\citenamefont{Lamata} \emph{et~al.}(2007)\citenamefont{Lamata,
  Le\'on, Sch\"atz, and Solano}}]{Lam07}
\bibinfo{author}{\bibnamefont{Lamata}, \bibfnamefont{L.}},
  \bibinfo{author}{\bibfnamefont{J.}~\bibnamefont{Le\'on}},
  \bibinfo{author}{\bibfnamefont{T.}~\bibnamefont{Sch\"atz}}, and
  \bibinfo{author}{\bibfnamefont{E.}~\bibnamefont{Solano}},
  \bibinfo{year}{2007}, {``}\bibinfo{title}{Dirac equation and quantum
  relativistic effects in a single trapped ion},{''} \bibinfo{journal}{Phys.
  Rev. Lett.} \textbf{\bibinfo{volume}{98}},  \bibinfo{pages}{253005}.

\bibitem[{\citenamefont{Lanyon} \emph{et~al.}(2011)\citenamefont{Lanyon,
  Hempel, Nigg, M\"uller, Gerritsma, Z\"ahringer, Schindler, Barreiro, Rambach,
  Kirchmair, Hennrich, Zoller} \emph{et~al.}}]{Lan11}
\bibinfo{author}{\bibnamefont{Lanyon}, \bibfnamefont{B.~P.}},
  \bibinfo{author}{\bibfnamefont{C.}~\bibnamefont{Hempel}},
  \bibinfo{author}{\bibfnamefont{D.}~\bibnamefont{Nigg}},
  \bibinfo{author}{\bibfnamefont{M.}~\bibnamefont{M\"uller}},
  \bibinfo{author}{\bibfnamefont{R.}~\bibnamefont{Gerritsma}},
  \bibinfo{author}{\bibfnamefont{F.}~\bibnamefont{Z\"ahringer}},
  \bibinfo{author}{\bibfnamefont{P.}~\bibnamefont{Schindler}},
  \bibinfo{author}{\bibfnamefont{J.~T.} \bibnamefont{Barreiro}},
  \bibinfo{author}{\bibfnamefont{M.}~\bibnamefont{Rambach}},
  \bibinfo{author}{\bibfnamefont{G.}~\bibnamefont{Kirchmair}},
  \bibinfo{author}{\bibfnamefont{M.}~\bibnamefont{Hennrich}},
  \bibinfo{author}{\bibfnamefont{P.}~\bibnamefont{Zoller}}, \emph{et~al.},
  \bibinfo{year}{2011}, {``}\bibinfo{title}{Universal digital quantum
  simulation with trapped ions},{''} \bibinfo{journal}{Science}
  \textbf{\bibinfo{volume}{334}},  \bibinfo{pages}{57--61}.

\bibitem[{\citenamefont{Lanyon} \emph{et~al.}(2010)\citenamefont{Lanyon,
  Whitfield, Gillett, Goggin, Almeida, Kassal, Biamonte, Mohseni, Powell,
  Barbieri, Aspuru-Guzik, and White}}]{Lan09}
\bibinfo{author}{\bibnamefont{Lanyon}, \bibfnamefont{B.~P.}},
  \bibinfo{author}{\bibfnamefont{J.~D.} \bibnamefont{Whitfield}},
  \bibinfo{author}{\bibfnamefont{G.~G.} \bibnamefont{Gillett}},
  \bibinfo{author}{\bibfnamefont{M.~E.} \bibnamefont{Goggin}},
  \bibinfo{author}{\bibfnamefont{M.~P.} \bibnamefont{Almeida}},
  \bibinfo{author}{\bibfnamefont{I.}~\bibnamefont{Kassal}},
  \bibinfo{author}{\bibfnamefont{J.~D.} \bibnamefont{Biamonte}},
  \bibinfo{author}{\bibfnamefont{M.}~\bibnamefont{Mohseni}},
  \bibinfo{author}{\bibfnamefont{B.~J.} \bibnamefont{Powell}},
  \bibinfo{author}{\bibfnamefont{M.}~\bibnamefont{Barbieri}},
  \bibinfo{author}{\bibfnamefont{A.}~\bibnamefont{Aspuru-Guzik}}, and
  \bibinfo{author}{\bibfnamefont{A.~G.} \bibnamefont{White}},
  \bibinfo{year}{2010}, {``}\bibinfo{title}{Towards quantum chemistry on a
  quantum computer},{''} \bibinfo{journal}{Nature Chemistry}
  \textbf{\bibinfo{volume}{2}},  \bibinfo{pages}{106 -- 111}.

\bibitem[{\citenamefont{Lau and James}(2012)}]{Lau12}
\bibinfo{author}{\bibnamefont{Lau}, \bibfnamefont{H.-K.}}, and
  \bibinfo{author}{\bibfnamefont{D.~F.~V.} \bibnamefont{James}},
  \bibinfo{year}{2012}, {``}\bibinfo{title}{Proposal for a scalable universal
  bosonic simulator using individually trapped ions},{''}
  \bibinfo{journal}{Phys. Rev. A} \textbf{\bibinfo{volume}{85}},
  \bibinfo{pages}{062329}.

\bibitem[{\citenamefont{Leibfried} \emph{et~al.}(2002)\citenamefont{Leibfried,
  DeMarco, Meyer, Rowe, Ben-Kish, Britton, Itano, Jelenkovic, Langer,
  Rosenband, and Wineland}}]{Lei02}
\bibinfo{author}{\bibnamefont{Leibfried}, \bibfnamefont{D.}},
  \bibinfo{author}{\bibfnamefont{B.}~\bibnamefont{DeMarco}},
  \bibinfo{author}{\bibfnamefont{V.}~\bibnamefont{Meyer}},
  \bibinfo{author}{\bibfnamefont{M.}~\bibnamefont{Rowe}},
  \bibinfo{author}{\bibfnamefont{A.}~\bibnamefont{Ben-Kish}},
  \bibinfo{author}{\bibfnamefont{J.}~\bibnamefont{Britton}},
  \bibinfo{author}{\bibfnamefont{W.~M.} \bibnamefont{Itano}},
  \bibinfo{author}{\bibfnamefont{B.}~\bibnamefont{Jelenkovic}},
  \bibinfo{author}{\bibfnamefont{C.}~\bibnamefont{Langer}},
  \bibinfo{author}{\bibfnamefont{T.}~\bibnamefont{Rosenband}}, and
  \bibinfo{author}{\bibfnamefont{D.~J.} \bibnamefont{Wineland}},
  \bibinfo{year}{2002}, {``}\bibinfo{title}{Trapped-ion quantum simulator :
  experimental application to nonlinear interferometers},{''}
  \bibinfo{journal}{Phys. Rev. Lett.} \textbf{\bibinfo{volume}{89}},
  \bibinfo{pages}{247901}.

\bibitem[{\citenamefont{Lemeshko} \emph{et~al.}(2013)\citenamefont{Lemeshko,
  Yao, Gorshkov, Weimer, Bennett, Momose, and Gopalakrishnan}}]{Lem13}
\bibinfo{author}{\bibnamefont{Lemeshko}, \bibfnamefont{M.}},
  \bibinfo{author}{\bibfnamefont{N.~Y.} \bibnamefont{Yao}},
  \bibinfo{author}{\bibfnamefont{A.~V.} \bibnamefont{Gorshkov}},
  \bibinfo{author}{\bibfnamefont{H.}~\bibnamefont{Weimer}},
  \bibinfo{author}{\bibfnamefont{S.~D.} \bibnamefont{Bennett}},
  \bibinfo{author}{\bibfnamefont{T.}~\bibnamefont{Momose}}, and
  \bibinfo{author}{\bibfnamefont{S.}~\bibnamefont{Gopalakrishnan}},
  \bibinfo{year}{2013}, {``}\bibinfo{title}{Controllable quantum spin glasses
  with magnetic impurities embedded in quantum solids},{''}
  \bibinfo{journal}{Phys. Rev. B} \textbf{\bibinfo{volume}{88}},
  \bibinfo{pages}{014426}.

\bibitem[{\citenamefont{Levi and Georgeot}(2004)}]{Lev04}
\bibinfo{author}{\bibnamefont{Levi}, \bibfnamefont{B.}}, and
  \bibinfo{author}{\bibfnamefont{B.}~\bibnamefont{Georgeot}},
  \bibinfo{year}{2004}, {``}\bibinfo{title}{Quantum computation of a complex
  system: the kicked Harper model},{''} \bibinfo{journal}{Phys. Rev. E}
  \textbf{\bibinfo{volume}{70}},  \bibinfo{pages}{056218}.

\bibitem[{\citenamefont{Lewenstein}(2006)}]{Lew06}
\bibinfo{author}{\bibnamefont{Lewenstein}, \bibfnamefont{M.}},
  \bibinfo{year}{2006}, {``}\bibinfo{title}{Atomic and molecular physics: Polar
  molecules in topological order},{''} \bibinfo{journal}{Nature Physics}
  \textbf{\bibinfo{volume}{2}},  \bibinfo{pages}{309--310}.

\bibitem[{\citenamefont{Lewenstein}
  \emph{et~al.}(2007)\citenamefont{Lewenstein, Sanpera, Ahufinger, Damski,
  Sen(De), and Sen}}]{Lew07}
\bibinfo{author}{\bibnamefont{Lewenstein}, \bibfnamefont{M.}},
  \bibinfo{author}{\bibfnamefont{A.}~\bibnamefont{Sanpera}},
  \bibinfo{author}{\bibfnamefont{V.}~\bibnamefont{Ahufinger}},
  \bibinfo{author}{\bibfnamefont{B.}~\bibnamefont{Damski}},
  \bibinfo{author}{\bibfnamefont{A.}~\bibnamefont{Sen(De)}}, and
  \bibinfo{author}{\bibfnamefont{U.}~\bibnamefont{Sen}}, \bibinfo{year}{2007},
  {``}\bibinfo{title}{Ultracold atomic gases in optical lattices: mimicking
  condensed matter physics and beyond},{''} \bibinfo{journal}{Advances in
  Physics} \textbf{\bibinfo{volume}{56}},  \bibinfo{pages}{243--379}.

\bibitem[{\citenamefont{Li} \emph{et~al.}(2013)\citenamefont{Li, Silveri,
  Kumar, Pirkkalainen, Veps\"al\"ainen, Chien, andM. A.~Sillanp\"a\"a, Hakonen,
  Thuneberg, and Paraoanu}}]{Li13}
\bibinfo{author}{\bibnamefont{Li}, \bibfnamefont{J.}},
  \bibinfo{author}{\bibfnamefont{M.~P.} \bibnamefont{Silveri}},
  \bibinfo{author}{\bibfnamefont{K.~S.} \bibnamefont{Kumar}},
  \bibinfo{author}{\bibfnamefont{J.-M.} \bibnamefont{Pirkkalainen}},
  \bibinfo{author}{\bibfnamefont{A.}~\bibnamefont{Veps\"al\"ainen}},
  \bibinfo{author}{\bibfnamefont{W.~C.} \bibnamefont{Chien}},
  \bibinfo{author}{\bibfnamefont{J.~T.} \bibnamefont{andM. A.~Sillanp\"a\"a}},
  \bibinfo{author}{\bibfnamefont{P.~J.} \bibnamefont{Hakonen}},
  \bibinfo{author}{\bibfnamefont{E.~V.} \bibnamefont{Thuneberg}}, and
  \bibinfo{author}{\bibfnamefont{G.~S.} \bibnamefont{Paraoanu}},
  \bibinfo{year}{2013}, {``}\bibinfo{title}{Motional Averaging in a
  Superconducting Qubit},{''} \bibinfo{journal}{Nature Commun.}
  \textbf{\bibinfo{volume}{4}},  \bibinfo{pages}{1420}.

\bibitem[{\citenamefont{Li} \emph{et~al.}(2011)\citenamefont{Li, Yung, Chen,
  Lu, Whitfield, Peng, Aspuru-Guzik, and Du}}]{Li11}
\bibinfo{author}{\bibnamefont{Li}, \bibfnamefont{Z.}},
  \bibinfo{author}{\bibfnamefont{M.-H.} \bibnamefont{Yung}},
  \bibinfo{author}{\bibfnamefont{H.}~\bibnamefont{Chen}},
  \bibinfo{author}{\bibfnamefont{D.}~\bibnamefont{Lu}},
  \bibinfo{author}{\bibfnamefont{J.~D.} \bibnamefont{Whitfield}},
  \bibinfo{author}{\bibfnamefont{X.}~\bibnamefont{Peng}},
  \bibinfo{author}{\bibfnamefont{A.}~\bibnamefont{Aspuru-Guzik}}, and
  \bibinfo{author}{\bibfnamefont{J.}~\bibnamefont{Du}}, \bibinfo{year}{2011},
  {``}\bibinfo{title}{Solving Quantum Ground-State Problems with Nuclear
  Magnetic Resonance},{''} \bibinfo{journal}{Scientific Reports}
  \textbf{\bibinfo{volume}{1}}, ~\bibinfo{pages}{88}.

\bibitem[{\citenamefont{Liao} \emph{et~al.}(2010)\citenamefont{Liao, Gong,
  Zhou, Liu, Sun, and Nori}}]{Lia10}
\bibinfo{author}{\bibnamefont{Liao}, \bibfnamefont{J.~Q.}},
  \bibinfo{author}{\bibfnamefont{Z.~R.} \bibnamefont{Gong}},
  \bibinfo{author}{\bibfnamefont{L.}~\bibnamefont{Zhou}},
  \bibinfo{author}{\bibfnamefont{Y.~X.} \bibnamefont{Liu}},
  \bibinfo{author}{\bibfnamefont{C.~P.} \bibnamefont{Sun}}, and
  \bibinfo{author}{\bibfnamefont{F.}~\bibnamefont{Nori}}, \bibinfo{year}{2010},
  {``}\bibinfo{title}{Controlling the transport of single photons by tuning the
  frequency of either one or two cavities in an array of coupled cavities},{''}
  \bibinfo{journal}{Phys. Rev. A} \textbf{\bibinfo{volume}{81}},
  \bibinfo{pages}{042304}.

\bibitem[{\citenamefont{Lidar and Wang}(1999)}]{Lid99}
\bibinfo{author}{\bibnamefont{Lidar}, \bibfnamefont{D.}}, and
  \bibinfo{author}{\bibfnamefont{H.}~\bibnamefont{Wang}}, \bibinfo{year}{1999},
  {``}\bibinfo{title}{Calculating the thermal rate constant with exponential
  speedup on a quantum computer},{''} \bibinfo{journal}{Phys. Rev. E}
  \textbf{\bibinfo{volume}{59}},  \bibinfo{pages}{2429--2438}.

\bibitem[{\citenamefont{Lidar}(2004)}]{Lid04}
\bibinfo{author}{\bibnamefont{Lidar}, \bibfnamefont{D.~A.}},
  \bibinfo{year}{2004}, {``}\bibinfo{title}{On the quantum computational
  complexity of the Ising spin glass partition function and of knot
  invariants},{''} \bibinfo{journal}{New J. Phys.}
  \textbf{\bibinfo{volume}{6}},  \bibinfo{pages}{167}.

\bibitem[{\citenamefont{Lidar and Biham}(1997)}]{Lid97}
\bibinfo{author}{\bibnamefont{Lidar}, \bibfnamefont{D.~A.}}, and
  \bibinfo{author}{\bibfnamefont{O.}~\bibnamefont{Biham}},
  \bibinfo{year}{1997}, {``}\bibinfo{title}{Simulating Ising spin glasses on a
  quantum computer},{''} \bibinfo{journal}{Phys. Rev. E}
  \textbf{\bibinfo{volume}{56}},  \bibinfo{pages}{3661--3681}.

\bibitem[{\citenamefont{Lin}
  \emph{et~al.}(2011{\natexlab{a}})\citenamefont{Lin, Monroe, and
  Duan}}]{Lin11}
\bibinfo{author}{\bibnamefont{Lin}, \bibfnamefont{G.-D.}},
  \bibinfo{author}{\bibfnamefont{C.}~\bibnamefont{Monroe}}, and
  \bibinfo{author}{\bibfnamefont{L.-M.} \bibnamefont{Duan}},
  \bibinfo{year}{2011}{\natexlab{a}}, {``}\bibinfo{title}{Sharp Phase
  Transitions in a Small Frustrated Network of Trapped Ion Spins},{''}
  \bibinfo{journal}{Phys. Rev. Lett.} \textbf{\bibinfo{volume}{106}},
  \bibinfo{pages}{230402}.

\bibitem[{\citenamefont{Lin} \emph{et~al.}(2009)\citenamefont{Lin, Zhu, Islam,
  Monroe, and Duan}}]{Lin09}
\bibinfo{author}{\bibnamefont{Lin}, \bibfnamefont{G.-D.}},
  \bibinfo{author}{\bibfnamefont{S.-L.} \bibnamefont{Zhu}},
  \bibinfo{author}{\bibfnamefont{R.}~\bibnamefont{Islam}},
  \bibinfo{author}{\bibfnamefont{C.}~\bibnamefont{Monroe}}, and
  \bibinfo{author}{\bibfnamefont{L.-M.} \bibnamefont{Duan}},
  \bibinfo{year}{2009}, {``}\bibinfo{title}{Large scale quantum computation in
  an anharmonic linear ion trap},{''} \bibinfo{journal}{Europhys. Lett.}
  \textbf{\bibinfo{volume}{86}},  \bibinfo{pages}{60004}.

\bibitem[{\citenamefont{Lin}
  \emph{et~al.}(2011{\natexlab{b}})\citenamefont{Lin, Jim\'enez-Garcia, and
  Spielman}}]{LJS11}
\bibinfo{author}{\bibnamefont{Lin}, \bibfnamefont{Y.}},
  \bibinfo{author}{\bibfnamefont{K.}~\bibnamefont{Jim\'enez-Garcia}}, and
  \bibinfo{author}{\bibfnamefont{I.~B.} \bibnamefont{Spielman}},
  \bibinfo{year}{2011}{\natexlab{b}}, {``}\bibinfo{title}{Spin-orbit-coupled
  Bose-Einstein condensates},{''} \bibinfo{journal}{Nature}
  \textbf{\bibinfo{volume}{471}},  \bibinfo{pages}{83 -- 86}.

\bibitem[{\citenamefont{Liu} \emph{et~al.}(2010)\citenamefont{Liu, Zhu, Jiang,
  and W.M}}]{Liu10}
\bibinfo{author}{\bibnamefont{Liu}, \bibfnamefont{G.}},
  \bibinfo{author}{\bibfnamefont{S.}~\bibnamefont{Zhu}},
  \bibinfo{author}{\bibfnamefont{S.~F.} \bibnamefont{Jiang},
  \bibfnamefont{S.J.}}, and
  \bibinfo{author}{\bibfnamefont{L.}~\bibnamefont{W.M}}, \bibinfo{year}{2010},
  {``}\bibinfo{title}{Simulating and detecting the quantum spin Hall effect in
  the kagome optical lattice},{''} \bibinfo{journal}{Phys. Rev. A}
  \textbf{\bibinfo{volume}{82}},  \bibinfo{pages}{053605}.

\bibitem[{\citenamefont{Liu} \emph{et~al.}(2005)\citenamefont{Liu, You, Wei,
  Sun, and Nori}}]{Liu05}
\bibinfo{author}{\bibnamefont{Liu}, \bibfnamefont{Y.-X.}},
  \bibinfo{author}{\bibfnamefont{J.~Q.} \bibnamefont{You}},
  \bibinfo{author}{\bibfnamefont{L.~F.} \bibnamefont{Wei}},
  \bibinfo{author}{\bibfnamefont{C.~P.} \bibnamefont{Sun}}, and
  \bibinfo{author}{\bibfnamefont{F.}~\bibnamefont{Nori}}, \bibinfo{year}{2005},
  {``}\bibinfo{title}{Optical selection rules and phase-dependent adiabatic
  state control in a superconducting quantum circuit},{''}
  \bibinfo{journal}{Phys. Rev. Lett} \textbf{\bibinfo{volume}{95}},
  \bibinfo{pages}{087001}.

\bibitem[{\citenamefont{Lloyd}(1996)}]{lloyd96}
\bibinfo{author}{\bibnamefont{Lloyd}, \bibfnamefont{S.}}, \bibinfo{year}{1996},
  {``}\bibinfo{title}{Universal quantum simulators},{''}
  \bibinfo{journal}{Science} \textbf{\bibinfo{volume}{273}},
  \bibinfo{pages}{1073--1078}.

\bibitem[{\citenamefont{Lu} \emph{et~al.}(2009)\citenamefont{Lu, Gao, G\"uhne,
  Zhou, Chen, and Pan}}]{Lu09}
\bibinfo{author}{\bibnamefont{Lu}, \bibfnamefont{C.-Y.}},
  \bibinfo{author}{\bibfnamefont{W.-B.} \bibnamefont{Gao}},
  \bibinfo{author}{\bibfnamefont{O.}~\bibnamefont{G\"uhne}},
  \bibinfo{author}{\bibfnamefont{X.-Q.} \bibnamefont{Zhou}},
  \bibinfo{author}{\bibfnamefont{Z.-B.} \bibnamefont{Chen}}, and
  \bibinfo{author}{\bibfnamefont{J.-W.} \bibnamefont{Pan}},
  \bibinfo{year}{2009}, {``}\bibinfo{title}{Demonstrating anyonic fractional
  statistics with a six-qubit quantum simulator},{''} \bibinfo{journal}{Phys.
  Rev. Lett.} \textbf{\bibinfo{volume}{102}},  \bibinfo{pages}{030502}.

\bibitem[{\citenamefont{Lu} \emph{et~al.}(2012)\citenamefont{Lu, Xu, Xu, Li,
  Chen, Peng, Xu, and Du}}]{Lu12}
\bibinfo{author}{\bibnamefont{Lu}, \bibfnamefont{D.}},
  \bibinfo{author}{\bibfnamefont{B.}~\bibnamefont{Xu}},
  \bibinfo{author}{\bibfnamefont{N.}~\bibnamefont{Xu}},
  \bibinfo{author}{\bibfnamefont{Z.}~\bibnamefont{Li}},
  \bibinfo{author}{\bibfnamefont{H.}~\bibnamefont{Chen}},
  \bibinfo{author}{\bibfnamefont{X.}~\bibnamefont{Peng}},
  \bibinfo{author}{\bibfnamefont{R.}~\bibnamefont{Xu}}, and
  \bibinfo{author}{\bibfnamefont{J.}~\bibnamefont{Du}}, \bibinfo{year}{2012},
  {``}\bibinfo{title}{Quantum Chemistry Simulation on Quantum Computers:
  Theories and Experiments},{''} \bibinfo{journal}{Phys. Chem. Chem. Phys.}
  \textbf{\bibinfo{volume}{14}},  \bibinfo{pages}{9411--9420}.

\bibitem[{\citenamefont{Lu} \emph{et~al.}(2011)\citenamefont{Lu, Xu, Xu, Chen,
  Gong, Peng, and Du}}]{Lu11}
\bibinfo{author}{\bibnamefont{Lu}, \bibfnamefont{D.}},
  \bibinfo{author}{\bibfnamefont{N.}~\bibnamefont{Xu}},
  \bibinfo{author}{\bibfnamefont{R.}~\bibnamefont{Xu}},
  \bibinfo{author}{\bibfnamefont{H.}~\bibnamefont{Chen}},
  \bibinfo{author}{\bibfnamefont{J.}~\bibnamefont{Gong}},
  \bibinfo{author}{\bibfnamefont{X.}~\bibnamefont{Peng}}, and
  \bibinfo{author}{\bibfnamefont{J.}~\bibnamefont{Du}}, \bibinfo{year}{2011},
  {``}\bibinfo{title}{Simulation of chemical isomerization reaction dynamics on
  a NMR quantum simulator},{''} \bibinfo{journal}{Phys. Rev. Lett.}
  \textbf{\bibinfo{volume}{107}},  \bibinfo{pages}{020501}.

\bibitem[{\citenamefont{L\"utkenhaus}
  \emph{et~al.}(1998)\citenamefont{L\"utkenhaus, Cirac, and Zoller}}]{Lut98}
\bibinfo{author}{\bibnamefont{L\"utkenhaus}, \bibfnamefont{N.}},
  \bibinfo{author}{\bibfnamefont{J.~I.} \bibnamefont{Cirac}}, and
  \bibinfo{author}{\bibfnamefont{P.}~\bibnamefont{Zoller}},
  \bibinfo{year}{1998}, {``}\bibinfo{title}{Mimicking a squeezed-bath
  interaction: quantum-reservoir engineering with atoms},{''}
  \bibinfo{journal}{Phys. Rev. A}
  \textbf{\bibinfo{volume}{57}}(\bibinfo{number}{1}),
  \bibinfo{pages}{548--558}.

\bibitem[{\citenamefont{Ma} \emph{et~al.}(2011)\citenamefont{Ma, Dakic, Naylor,
  Zeilinger, and Walther}}]{Ma11}
\bibinfo{author}{\bibnamefont{Ma}, \bibfnamefont{X.}},
  \bibinfo{author}{\bibfnamefont{B.}~\bibnamefont{Dakic}},
  \bibinfo{author}{\bibfnamefont{W.}~\bibnamefont{Naylor}},
  \bibinfo{author}{\bibfnamefont{A.}~\bibnamefont{Zeilinger}}, and
  \bibinfo{author}{\bibfnamefont{P.}~\bibnamefont{Walther}},
  \bibinfo{year}{2011}, {``}\bibinfo{title}{Quantum simulation of the
  wavefunction to probe frustrated Heisenberg spin systems},{''}
  \bibinfo{journal}{Nature Physics} \textbf{\bibinfo{volume}{7}},
  \bibinfo{pages}{399--405}.

\bibitem[{\citenamefont{Ma} \emph{et~al.}(2012)\citenamefont{Ma, Dakic, Naylor,
  Chan, Gong, Duan, Zeilinger, and Walther}}]{Ma12}
\bibinfo{author}{\bibnamefont{Ma}, \bibfnamefont{X.}},
  \bibinfo{author}{\bibfnamefont{S.}~\bibnamefont{Dakic},
  \bibfnamefont{B.and~Kropatsche}},
  \bibinfo{author}{\bibfnamefont{W.}~\bibnamefont{Naylor}},
  \bibinfo{author}{\bibfnamefont{Y.}~\bibnamefont{Chan}},
  \bibinfo{author}{\bibfnamefont{Z.}~\bibnamefont{Gong}},
  \bibinfo{author}{\bibfnamefont{L.}~\bibnamefont{Duan}},
  \bibinfo{author}{\bibfnamefont{A.}~\bibnamefont{Zeilinger}}, and
  \bibinfo{author}{\bibfnamefont{P.}~\bibnamefont{Walther}},
  \bibinfo{year}{2012}, {``}\bibinfo{title}{Photonic quantum simulation of
  ground state configurations of Heisenberg square and checkerboard lattice
  spin systems},{''} \eprint{arXiv:1205.2801}.

\bibitem[{\citenamefont{Manin}(1980)}]{Man80}
\bibinfo{author}{\bibnamefont{Manin}, \bibfnamefont{Y.}}, \bibinfo{year}{1980},
  \emph{\bibinfo{title}{Computable and uncomputable}}
  (\bibinfo{publisher}{Sovetskoye Radio Press (in Russian)}).

\bibitem[{\citenamefont{Maniscalco}
  \emph{et~al.}(2004)\citenamefont{Maniscalco, Piilo, Intravaia, Petruccione,
  and Messina}}]{Man04}
\bibinfo{author}{\bibnamefont{Maniscalco}, \bibfnamefont{S.}},
  \bibinfo{author}{\bibfnamefont{J.}~\bibnamefont{Piilo}},
  \bibinfo{author}{\bibfnamefont{F.}~\bibnamefont{Intravaia}},
  \bibinfo{author}{\bibfnamefont{F.}~\bibnamefont{Petruccione}}, and
  \bibinfo{author}{\bibfnamefont{A.}~\bibnamefont{Messina}},
  \bibinfo{year}{2004}, {``}\bibinfo{title}{Simulating quantum Brownian motion
  with single trapped ions},{''} \bibinfo{journal}{Phys. Rev. A}
  \textbf{\bibinfo{volume}{69}},  \bibinfo{pages}{052101}.

\bibitem[{\citenamefont{Manousakis}(2002)}]{Man02}
\bibinfo{author}{\bibnamefont{Manousakis}, \bibfnamefont{E.}},
  \bibinfo{year}{2002}, {``}\bibinfo{title}{A quantum-dot array as model for
  copper-oxide superconductors: a dedicated quantum simulator for the
  many-fermion problem},{''} \bibinfo{journal}{J. Low Temp. Phys.}
  \textbf{\bibinfo{volume}{126}},  \bibinfo{pages}{1501--1513}.

\bibitem[{\citenamefont{Marcos} \emph{et~al.}(2013)\citenamefont{Marcos, Rabl,
  Rico, and Zoller}}]{Mar13}
\bibinfo{author}{\bibnamefont{Marcos}, \bibfnamefont{D.}},
  \bibinfo{author}{\bibfnamefont{P.}~\bibnamefont{Rabl}},
  \bibinfo{author}{\bibfnamefont{E.}~\bibnamefont{Rico}}, and
  \bibinfo{author}{\bibfnamefont{P.}~\bibnamefont{Zoller}},
  \bibinfo{year}{2013}, {``}\bibinfo{title}{Superconducting Circuits for
  Quantum Simulation of Dynamical Gauge Fields},{''} \bibinfo{journal}{Phys.
  Rev. Lett} \textbf{\bibinfo{volume}{111}},  \bibinfo{pages}{110504}.

\bibitem[{\citenamefont{Mariantoni}
  \emph{et~al.}(2011)\citenamefont{Mariantoni, Wang, Yamamoto, Neeley,
  Bialczak, Chen, Lenander, Lucero, O'Connell, Sank, Weides, Wenner}
  \emph{et~al.}}]{Mar11}
\bibinfo{author}{\bibnamefont{Mariantoni}, \bibfnamefont{M.}},
  \bibinfo{author}{\bibfnamefont{H.}~\bibnamefont{Wang}},
  \bibinfo{author}{\bibfnamefont{T.}~\bibnamefont{Yamamoto}},
  \bibinfo{author}{\bibfnamefont{M.}~\bibnamefont{Neeley}},
  \bibinfo{author}{\bibfnamefont{R.~C.} \bibnamefont{Bialczak}},
  \bibinfo{author}{\bibfnamefont{Y.}~\bibnamefont{Chen}},
  \bibinfo{author}{\bibfnamefont{M.}~\bibnamefont{Lenander}},
  \bibinfo{author}{\bibfnamefont{E.}~\bibnamefont{Lucero}},
  \bibinfo{author}{\bibfnamefont{A.~D.} \bibnamefont{O'Connell}},
  \bibinfo{author}{\bibfnamefont{D.}~\bibnamefont{Sank}},
  \bibinfo{author}{\bibfnamefont{M.}~\bibnamefont{Weides}},
  \bibinfo{author}{\bibfnamefont{J.}~\bibnamefont{Wenner}}, \emph{et~al.},
  \bibinfo{year}{2011}, {``}\bibinfo{title}{Implementing the Quantum von
  Neumann Architecture with Superconducting Circuits},{''}
  \bibinfo{journal}{Science} \textbf{\bibinfo{volume}{334}},
  \bibinfo{pages}{61--65}.

\bibitem[{\citenamefont{Maruyama} \emph{et~al.}(2009)\citenamefont{Maruyama,
  Nori, and Vedral}}]{Mar08}
\bibinfo{author}{\bibnamefont{Maruyama}, \bibfnamefont{K.}},
  \bibinfo{author}{\bibfnamefont{F.}~\bibnamefont{Nori}}, and
  \bibinfo{author}{\bibfnamefont{V.}~\bibnamefont{Vedral}},
  \bibinfo{year}{2009}, {``}\bibinfo{title}{Physics of Maxwell's demon and
  information},{''} \bibinfo{journal}{Rev. Mod. Phys.}
  \textbf{\bibinfo{volume}{81}}, ~\bibinfo{pages}{1}.

\bibitem[{\citenamefont{Marzuoli and Rasetti}(2002)}]{Mar02}
\bibinfo{author}{\bibnamefont{Marzuoli}, \bibfnamefont{A.}}, and
  \bibinfo{author}{\bibfnamefont{M.}~\bibnamefont{Rasetti}},
  \bibinfo{year}{2002}, {``}\bibinfo{title}{Spin network quantum
  simulator},{''} \bibinfo{journal}{Phys. Lett. A}
  \textbf{\bibinfo{volume}{306}},  \bibinfo{pages}{79--87}.

\bibitem[{\citenamefont{Mazza} \emph{et~al.}(2012)\citenamefont{Mazza,
  Bermudez, Goldman, Rizzi, Martin-Delgado, and Lewenstein}}]{Maz12}
\bibinfo{author}{\bibnamefont{Mazza}, \bibfnamefont{L.}},
  \bibinfo{author}{\bibfnamefont{A.}~\bibnamefont{Bermudez}},
  \bibinfo{author}{\bibfnamefont{N.}~\bibnamefont{Goldman}},
  \bibinfo{author}{\bibfnamefont{M.}~\bibnamefont{Rizzi}},
  \bibinfo{author}{\bibfnamefont{M.~A.} \bibnamefont{Martin-Delgado}}, and
  \bibinfo{author}{\bibfnamefont{M.}~\bibnamefont{Lewenstein}},
  \bibinfo{year}{2012}, {``}\bibinfo{title}{An optical-lattice-based quantum
  simulator for relativistic field theories and topological insulators},{''}
  \bibinfo{journal}{New J. Phys.} \textbf{\bibinfo{volume}{14}},
  \bibinfo{pages}{015007}.

\bibitem[{\citenamefont{McKague} \emph{et~al.}(2009)\citenamefont{McKague,
  Mosca, and Gisin}}]{Mck09}
\bibinfo{author}{\bibnamefont{McKague}, \bibfnamefont{M.}},
  \bibinfo{author}{\bibfnamefont{M.}~\bibnamefont{Mosca}}, and
  \bibinfo{author}{\bibfnamefont{N.}~\bibnamefont{Gisin}},
  \bibinfo{year}{2009}, {``}\bibinfo{title}{Simulating quantum systems using
  real Hilbert spaces},{''} \bibinfo{journal}{Phys. Rev. Lett.}
  \textbf{\bibinfo{volume}{102}},  \bibinfo{pages}{020505}.

\bibitem[{\citenamefont{Mei} \emph{et~al.}(2013)\citenamefont{Mei, Stojanovic,
  Siddiqi, and Tian}}]{Mei13}
\bibinfo{author}{\bibnamefont{Mei}, \bibfnamefont{F.}},
  \bibinfo{author}{\bibfnamefont{V.~M.} \bibnamefont{Stojanovic}},
  \bibinfo{author}{\bibfnamefont{I.}~\bibnamefont{Siddiqi}}, and
  \bibinfo{author}{\bibfnamefont{L.}~\bibnamefont{Tian}}, \bibinfo{year}{2013},
  {``}\bibinfo{title}{Analog Superconducting Quantum Simulator for Holstein
  Polarons},{''} \bibinfo{journal}{Phys. Rev. B} \textbf{\bibinfo{volume}{88}},
   \bibinfo{pages}{224502}.

\bibitem[{\citenamefont{Menicucci} \emph{et~al.}(2010)\citenamefont{Menicucci,
  Olson, and Milburn}}]{Men10}
\bibinfo{author}{\bibnamefont{Menicucci}, \bibfnamefont{N.}},
  \bibinfo{author}{\bibfnamefont{S.}~\bibnamefont{Olson}}, and
  \bibinfo{author}{\bibfnamefont{G.}~\bibnamefont{Milburn}},
  \bibinfo{year}{2010}, {``}\bibinfo{title}{Simulating quantum effects of
  cosmological expansion using a static ion trap},{''} \bibinfo{journal}{New J.
  Phys.} \textbf{\bibinfo{volume}{12}},  \bibinfo{pages}{095019}.

\bibitem[{\citenamefont{Meyer}(2002)}]{Mey01}
\bibinfo{author}{\bibnamefont{Meyer}, \bibfnamefont{D.}}, \bibinfo{year}{2002},
  {``}\bibinfo{title}{Quantum computing classical physics},{''}
  \bibinfo{journal}{Phil. Trans. Roy. Soc. A} \textbf{\bibinfo{volume}{360}},
  \bibinfo{pages}{395--405}.

\bibitem[{\citenamefont{Mezzacapo} \emph{et~al.}(2012)\citenamefont{Mezzacapo,
  Casanova, Lamata, and Solano}}]{Mez12}
\bibinfo{author}{\bibnamefont{Mezzacapo}, \bibfnamefont{A.}},
  \bibinfo{author}{\bibfnamefont{J.}~\bibnamefont{Casanova}},
  \bibinfo{author}{\bibfnamefont{L.}~\bibnamefont{Lamata}}, and
  \bibinfo{author}{\bibfnamefont{E.}~\bibnamefont{Solano}},
  \bibinfo{year}{2012}, {``}\bibinfo{title}{Digital Quantum Simulation of the
  Holstein Model in Trapped Ions},{''} \bibinfo{journal}{Phys. Rev. Lett.}
  \textbf{\bibinfo{volume}{109}},  \bibinfo{pages}{200501}.

\bibitem[{\citenamefont{Micheli} \emph{et~al.}(2006)\citenamefont{Micheli,
  Brennen, and Zoller}}]{Mic06}
\bibinfo{author}{\bibnamefont{Micheli}, \bibfnamefont{A.}},
  \bibinfo{author}{\bibfnamefont{G.~K.} \bibnamefont{Brennen}}, and
  \bibinfo{author}{\bibfnamefont{P.}~\bibnamefont{Zoller}},
  \bibinfo{year}{2006}, {``}\bibinfo{title}{A toolbox for lattice-spin models
  with polar molecules},{''} \bibinfo{journal}{Nature Physics}
  \textbf{\bibinfo{volume}{2}},  \bibinfo{pages}{341--347}.

\bibitem[{\citenamefont{Milburn}(1999)}]{Mil99}
\bibinfo{author}{\bibnamefont{Milburn}, \bibfnamefont{G.}},
  \bibinfo{year}{1999}, {``}\bibinfo{title}{Simulating nonlinear spin models in
  an ion trap},{''} \eprint{arXiv:quant-ph/9908037}.

\bibitem[{\citenamefont{Montangero}(2004)}]{Mon04}
\bibinfo{author}{\bibnamefont{Montangero}, \bibfnamefont{S.}},
  \bibinfo{year}{2004}, {``}\bibinfo{title}{Dynamically localized systems:
  exponential sensitivity of entanglement and efficient quantum
  simulations},{''} \bibinfo{journal}{Phys. Rev. A}
  \textbf{\bibinfo{volume}{70}},  \bibinfo{pages}{032311}.

\bibitem[{\citenamefont{Mostame and Sch\"utzhold}(2008)}]{Mos08}
\bibinfo{author}{\bibnamefont{Mostame}, \bibfnamefont{S.}}, and
  \bibinfo{author}{\bibfnamefont{R.}~\bibnamefont{Sch\"utzhold}},
  \bibinfo{year}{2008}, {``}\bibinfo{title}{Quantum simulator for the Ising
  model with electrons floating on a helium film},{''} \bibinfo{journal}{Phys.
  Rev. Lett.} \textbf{\bibinfo{volume}{101}},  \bibinfo{pages}{220501}.

\bibitem[{\citenamefont{Nation} \emph{et~al.}(2009)\citenamefont{Nation,
  Blencowe, Rimberg, and Buks}}]{Nat09}
\bibinfo{author}{\bibnamefont{Nation}, \bibfnamefont{P.~D.}},
  \bibinfo{author}{\bibfnamefont{M.~P.} \bibnamefont{Blencowe}},
  \bibinfo{author}{\bibfnamefont{A.~J.} \bibnamefont{Rimberg}}, and
  \bibinfo{author}{\bibfnamefont{E.}~\bibnamefont{Buks}}, \bibinfo{year}{2009},
  {``}\bibinfo{title}{Analogue Hawking radiation in a dc-SQUID array coplanar
  waveguide},{''} \bibinfo{journal}{Phys. Rev. Lett.}
  \textbf{\bibinfo{volume}{103}},  \bibinfo{pages}{087004}.

\bibitem[{\citenamefont{Nation} \emph{et~al.}(2012)\citenamefont{Nation,
  Johansson, Blencowe, and Nori}}]{Nat11}
\bibinfo{author}{\bibnamefont{Nation}, \bibfnamefont{P.~D.}},
  \bibinfo{author}{\bibfnamefont{J.~R.} \bibnamefont{Johansson}},
  \bibinfo{author}{\bibfnamefont{M.~P.} \bibnamefont{Blencowe}}, and
  \bibinfo{author}{\bibfnamefont{F.}~\bibnamefont{Nori}}, \bibinfo{year}{2012},
  {``}\bibinfo{title}{Stimulating uncertainty: amplifying the quantum vacuum
  with superconducting circuits},{''} \bibinfo{journal}{Rev. Mod. Phys.}
  \textbf{\bibinfo{volume}{84}},  \bibinfo{pages}{1--24}.

\bibitem[{\citenamefont{Neeley} \emph{et~al.}(2009)\citenamefont{Neeley,
  Ansmann, Bialczak, Hofheinz, Lucero, O'Connell, Sank, Wang, J.Wenner,
  Cleland, Geller, and Martinis}}]{Nee09}
\bibinfo{author}{\bibnamefont{Neeley}, \bibfnamefont{M.}},
  \bibinfo{author}{\bibfnamefont{M.}~\bibnamefont{Ansmann}},
  \bibinfo{author}{\bibfnamefont{R.~C.} \bibnamefont{Bialczak}},
  \bibinfo{author}{\bibfnamefont{M.}~\bibnamefont{Hofheinz}},
  \bibinfo{author}{\bibfnamefont{E.}~\bibnamefont{Lucero}},
  \bibinfo{author}{\bibfnamefont{A.}~\bibnamefont{O'Connell}},
  \bibinfo{author}{\bibfnamefont{D.}~\bibnamefont{Sank}},
  \bibinfo{author}{\bibfnamefont{H.}~\bibnamefont{Wang}},
  \bibinfo{author}{\bibnamefont{J.Wenner}},
  \bibinfo{author}{\bibfnamefont{A.~N.} \bibnamefont{Cleland}},
  \bibinfo{author}{\bibfnamefont{M.~R.} \bibnamefont{Geller}}, and
  \bibinfo{author}{\bibfnamefont{J.~M.} \bibnamefont{Martinis}},
  \bibinfo{year}{2009}, {``}\bibinfo{title}{Emulation of a quantum spin with a
  superconducting phase qudit},{''} \bibinfo{journal}{Science}
  \textbf{\bibinfo{volume}{325}},  \bibinfo{pages}{722 -- 725}.

\bibitem[{\citenamefont{Negrevergne}
  \emph{et~al.}(2005)\citenamefont{Negrevergne, Somma, Ortiz, Knill, and
  Laflamme}}]{Neg05}
\bibinfo{author}{\bibnamefont{Negrevergne}, \bibfnamefont{C.}},
  \bibinfo{author}{\bibfnamefont{R.}~\bibnamefont{Somma}},
  \bibinfo{author}{\bibfnamefont{G.}~\bibnamefont{Ortiz}},
  \bibinfo{author}{\bibfnamefont{E.}~\bibnamefont{Knill}}, and
  \bibinfo{author}{\bibfnamefont{R.}~\bibnamefont{Laflamme}},
  \bibinfo{year}{2005}, {``}\bibinfo{title}{Liquid-state NMR simulations of
  quantum many-body problems},{''} \bibinfo{journal}{Phys. Rev. A}
  \textbf{\bibinfo{volume}{71}},  \bibinfo{pages}{032344}.

\bibitem[{\citenamefont{Nelson} \emph{et~al.}(2007)\citenamefont{Nelson, Li,
  and Weiss}}]{Nel07}
\bibinfo{author}{\bibnamefont{Nelson}, \bibfnamefont{K.~D.}},
  \bibinfo{author}{\bibfnamefont{X.}~\bibnamefont{Li}}, and
  \bibinfo{author}{\bibfnamefont{D.~S.} \bibnamefont{Weiss}},
  \bibinfo{year}{2007}, {``}\bibinfo{title}{Imaging single atoms in a
  three--dimensional array},{''} \bibinfo{journal}{Nature Physics}
  \textbf{\bibinfo{volume}{3}},  \bibinfo{pages}{556--560}.

\bibitem[{\citenamefont{Nielsen} \emph{et~al.}(2002)\citenamefont{Nielsen,
  Bremner, Dodd, Childs, and Dawson}}]{Nie02}
\bibinfo{author}{\bibnamefont{Nielsen}, \bibfnamefont{M.~A.}},
  \bibinfo{author}{\bibfnamefont{M.~J.} \bibnamefont{Bremner}},
  \bibinfo{author}{\bibfnamefont{J.~L.} \bibnamefont{Dodd}},
  \bibinfo{author}{\bibfnamefont{A.~M.} \bibnamefont{Childs}}, and
  \bibinfo{author}{\bibfnamefont{C.~M.} \bibnamefont{Dawson}},
  \bibinfo{year}{2002}, {``}\bibinfo{title}{Universal simulation of Hamiltonian
  dynamics for quantum systems with finite-dimensional state spaces},{''}
  \bibinfo{journal}{Phys. Rev. A} \textbf{\bibinfo{volume}{66}},
  \bibinfo{pages}{022317}.

\bibitem[{\citenamefont{Nielsen and Chuang}(2000)}]{NC}
\bibinfo{author}{\bibnamefont{Nielsen}, \bibfnamefont{M.~A.}}, and
  \bibinfo{author}{\bibfnamefont{I.~L.} \bibnamefont{Chuang}},
  \bibinfo{year}{2000}, \emph{\bibinfo{title}{Quantum computation and quantum
  information}} (\bibinfo{publisher}{Cambridge University press}).

\bibitem[{\citenamefont{Noh} \emph{et~al.}(2012)\citenamefont{Noh,
  Rodriguez-Lara, and Angelakis}}]{Noh12}
\bibinfo{author}{\bibnamefont{Noh}, \bibfnamefont{C.}},
  \bibinfo{author}{\bibfnamefont{B.~M.} \bibnamefont{Rodriguez-Lara}}, and
  \bibinfo{author}{\bibfnamefont{D.}~\bibnamefont{Angelakis}},
  \bibinfo{year}{2012}, {``}\bibinfo{title}{Quantum simulation of neutrino
  oscillations with trapped ions},{''} \bibinfo{journal}{New J. Phys.}
  \textbf{\bibinfo{volume}{14}},  \bibinfo{pages}{033028}.

\bibitem[{\citenamefont{Nori}(2008)}]{Nor08}
\bibinfo{author}{\bibnamefont{Nori}, \bibfnamefont{F.}}, \bibinfo{year}{2008},
  {``}\bibinfo{title}{Atomic physics with a circuit},{''}
  \bibinfo{journal}{Nature Physics} \textbf{\bibinfo{volume}{4}},
  \bibinfo{pages}{589}.

\bibitem[{\citenamefont{Nori}(2009)}]{Nor09}
\bibinfo{author}{\bibnamefont{Nori}, \bibfnamefont{F.}}, \bibinfo{year}{2009},
  {``}\bibinfo{title}{Quantum football},{''} \bibinfo{journal}{Science}
  \textbf{\bibinfo{volume}{325}},  \bibinfo{pages}{689}.

\bibitem[{\citenamefont{O'Hara} \emph{et~al.}(2002)\citenamefont{O'Hara,
  Hemmer, Gehm, Granade, and Thomas}}]{Oha02}
\bibinfo{author}{\bibnamefont{O'Hara}, \bibfnamefont{K.}},
  \bibinfo{author}{\bibfnamefont{S.}~\bibnamefont{Hemmer}},
  \bibinfo{author}{\bibfnamefont{M.}~\bibnamefont{Gehm}},
  \bibinfo{author}{\bibfnamefont{S.}~\bibnamefont{Granade}}, and
  \bibinfo{author}{\bibfnamefont{J.}~\bibnamefont{Thomas}},
  \bibinfo{year}{2002}, {``}\bibinfo{title}{Observation of a strongly
  interacting degenerate Fermi gas of atoms},{''} \bibinfo{journal}{Science}
  \textbf{\bibinfo{volume}{298}},  \bibinfo{pages}{2179 -- 2182}.

\bibitem[{\citenamefont{Orenstein and Millis}(2000)}]{Ore00}
\bibinfo{author}{\bibnamefont{Orenstein}, \bibfnamefont{J.}}, and
  \bibinfo{author}{\bibfnamefont{A.~J.} \bibnamefont{Millis}},
  \bibinfo{year}{2000}, {``}\bibinfo{title}{Advances in the physics of
  high-temperature superconductivity},{''} \bibinfo{journal}{Science}
  \textbf{\bibinfo{volume}{288}},  \bibinfo{pages}{468--474}.

\bibitem[{\citenamefont{Ortiz} \emph{et~al.}(2001)\citenamefont{Ortiz,
  Gubernatis, Knill, and Laflamme}}]{Ort01}
\bibinfo{author}{\bibnamefont{Ortiz}, \bibfnamefont{G.}},
  \bibinfo{author}{\bibfnamefont{J.~E.} \bibnamefont{Gubernatis}},
  \bibinfo{author}{\bibfnamefont{E.}~\bibnamefont{Knill}}, and
  \bibinfo{author}{\bibfnamefont{R.}~\bibnamefont{Laflamme}},
  \bibinfo{year}{2001}, {``}\bibinfo{title}{Quantum algorithms for fermionic
  simulations},{''} \bibinfo{journal}{Phys. Rev. A}
  \textbf{\bibinfo{volume}{64}},  \bibinfo{pages}{022319}.

\bibitem[{\citenamefont{Ortiz} \emph{et~al.}(2002)\citenamefont{Ortiz,
  Gubernatis, Knill, and Laflamme}}]{Ort01a}
\bibinfo{author}{\bibnamefont{Ortiz}, \bibfnamefont{G.}},
  \bibinfo{author}{\bibfnamefont{J.~E.} \bibnamefont{Gubernatis}},
  \bibinfo{author}{\bibfnamefont{E.}~\bibnamefont{Knill}}, and
  \bibinfo{author}{\bibfnamefont{R.}~\bibnamefont{Laflamme}},
  \bibinfo{year}{2002}, {``}\bibinfo{title}{Simulating fermions on a quantum
  computer},{''} \bibinfo{journal}{Comp. Phys. Comm.}
  \textbf{\bibinfo{volume}{146}},  \bibinfo{pages}{302--316}.

\bibitem[{\citenamefont{Ortner} \emph{et~al.}(2009)\citenamefont{Ortner,
  Micheli, Pupillo, and Zoller}}]{Ort09}
\bibinfo{author}{\bibnamefont{Ortner}, \bibfnamefont{M.}},
  \bibinfo{author}{\bibfnamefont{A.}~\bibnamefont{Micheli}},
  \bibinfo{author}{\bibfnamefont{G.}~\bibnamefont{Pupillo}}, and
  \bibinfo{author}{\bibfnamefont{P.}~\bibnamefont{Zoller}},
  \bibinfo{year}{2009}, {``}\bibinfo{title}{Quantum simulations of extended
  Hubbard models with dipolar crystals},{''} \bibinfo{journal}{New J. Phys.}
  \textbf{\bibinfo{volume}{11}},  \bibinfo{pages}{055045}.

\bibitem[{\citenamefont{Ospelkaus} \emph{et~al.}(2011)\citenamefont{Ospelkaus,
  Warring, Colombe, Brown, Amini, Leibfried, and Wineland}}]{Osp11}
\bibinfo{author}{\bibnamefont{Ospelkaus}, \bibfnamefont{C.}},
  \bibinfo{author}{\bibfnamefont{U.}~\bibnamefont{Warring}},
  \bibinfo{author}{\bibfnamefont{Y.}~\bibnamefont{Colombe}},
  \bibinfo{author}{\bibfnamefont{K.~R.} \bibnamefont{Brown}},
  \bibinfo{author}{\bibfnamefont{J.~M.} \bibnamefont{Amini}},
  \bibinfo{author}{\bibfnamefont{D.}~\bibnamefont{Leibfried}}, and
  \bibinfo{author}{\bibfnamefont{D.~J.} \bibnamefont{Wineland}},
  \bibinfo{year}{2011}, {``}\bibinfo{title}{Microwave quantum logic gates for
  trapped ions},{''} \bibinfo{journal}{Nature} \textbf{\bibinfo{volume}{476}},
  \bibinfo{pages}{181--184}.

\bibitem[{\citenamefont{van Oudenaarden and Mooij}(1996)}]{Oud96}
\bibinfo{author}{\bibnamefont{van Oudenaarden}, \bibfnamefont{A.}}, and
  \bibinfo{author}{\bibfnamefont{J.~E.} \bibnamefont{Mooij}},
  \bibinfo{year}{1996}, {``}\bibinfo{title}{One-dimensional Mott insulator
  formed by quantum vortices in Josephson Junction arrays},{''}
  \bibinfo{journal}{Phys. Rev. Lett.} \textbf{\bibinfo{volume}{76}},
  \bibinfo{pages}{4947--4950}.

\bibitem[{\citenamefont{Paredes} \emph{et~al.}(2004)\citenamefont{Paredes,
  Widera, Murg, Mandel, F\"olling, Cirac, Shlyapnikov, H\"ansch, and
  Bloch}}]{Par04}
\bibinfo{author}{\bibnamefont{Paredes}, \bibfnamefont{B.}},
  \bibinfo{author}{\bibfnamefont{A.}~\bibnamefont{Widera}},
  \bibinfo{author}{\bibfnamefont{V.}~\bibnamefont{Murg}},
  \bibinfo{author}{\bibfnamefont{O.}~\bibnamefont{Mandel}},
  \bibinfo{author}{\bibfnamefont{S.}~\bibnamefont{F\"olling}},
  \bibinfo{author}{\bibfnamefont{I.}~\bibnamefont{Cirac}},
  \bibinfo{author}{\bibfnamefont{G.}~\bibnamefont{Shlyapnikov}},
  \bibinfo{author}{\bibfnamefont{T.}~\bibnamefont{H\"ansch}}, and
  \bibinfo{author}{\bibfnamefont{I.}~\bibnamefont{Bloch}},
  \bibinfo{year}{2004}, {``}\bibinfo{title}{Tonks-Girardeau gas of ultracold
  atoms in an optical lattice},{''} \bibinfo{journal}{Nature}
  \textbf{\bibinfo{volume}{429}},  \bibinfo{pages}{277--281}.

\bibitem[{\citenamefont{Pedernales}
  \emph{et~al.}(2013)\citenamefont{Pedernales, Candia, Ballester, and
  Solano}}]{Pad13}
\bibinfo{author}{\bibnamefont{Pedernales}, \bibfnamefont{J.~S.}},
  \bibinfo{author}{\bibfnamefont{R.~D.} \bibnamefont{Candia}},
  \bibinfo{author}{\bibfnamefont{D.}~\bibnamefont{Ballester}}, and
  \bibinfo{author}{\bibfnamefont{E.}~\bibnamefont{Solano}},
  \bibinfo{year}{2013}, {``}\bibinfo{title}{Quantum Simulations of Relativistic
  Quantum Physics in Circuit QED,},{''} \bibinfo{journal}{New J. Phys}
  \textbf{\bibinfo{volume}{15}},  \bibinfo{pages}{055008}.

\bibitem[{\citenamefont{Peng} \emph{et~al.}(2005)\citenamefont{Peng, Du, and
  Suter}}]{Pen05}
\bibinfo{author}{\bibnamefont{Peng}, \bibfnamefont{X.}},
  \bibinfo{author}{\bibfnamefont{J.}~\bibnamefont{Du}}, and
  \bibinfo{author}{\bibfnamefont{D.}~\bibnamefont{Suter}},
  \bibinfo{year}{2005}, {``}\bibinfo{title}{Quantum phase transition of
  ground-state entanglement in a Heisenberg spin chain simulated in an NMR
  quantum computer},{''} \bibinfo{journal}{Phys. Rev. A}
  \textbf{\bibinfo{volume}{71}},  \bibinfo{pages}{012307}.

\bibitem[{\citenamefont{Peng and Suter}(2010)}]{Pen10}
\bibinfo{author}{\bibnamefont{Peng}, \bibfnamefont{X.}}, and
  \bibinfo{author}{\bibfnamefont{D.}~\bibnamefont{Suter}},
  \bibinfo{year}{2010}, {``}\bibinfo{title}{Spin qubits for quantum
  simulations},{''} \bibinfo{journal}{Front. Phys. China}
  \textbf{\bibinfo{volume}{5}},  \bibinfo{pages}{1--25}.

\bibitem[{\citenamefont{Peng} \emph{et~al.}(2009)\citenamefont{Peng, Zhang, Du,
  and Suter}}]{Pen09}
\bibinfo{author}{\bibnamefont{Peng}, \bibfnamefont{X.}},
  \bibinfo{author}{\bibfnamefont{J.}~\bibnamefont{Zhang}},
  \bibinfo{author}{\bibfnamefont{J.}~\bibnamefont{Du}}, and
  \bibinfo{author}{\bibfnamefont{D.}~\bibnamefont{Suter}},
  \bibinfo{year}{2009}, {``}\bibinfo{title}{Quantum simulation of a system with
  competing two- and three-body interactions},{''} \bibinfo{journal}{Phys. Rev.
  Lett.} \textbf{\bibinfo{volume}{103}},  \bibinfo{pages}{140501}.

\bibitem[{\citenamefont{Philbin} \emph{et~al.}(2008)\citenamefont{Philbin,
  Kuklewicz, Robertson, Hill, Konig, and Leonhardt}}]{Phi08}
\bibinfo{author}{\bibnamefont{Philbin}, \bibfnamefont{T.~G.}},
  \bibinfo{author}{\bibfnamefont{C.}~\bibnamefont{Kuklewicz}},
  \bibinfo{author}{\bibfnamefont{S.}~\bibnamefont{Robertson}},
  \bibinfo{author}{\bibfnamefont{S.}~\bibnamefont{Hill}},
  \bibinfo{author}{\bibfnamefont{F.}~\bibnamefont{Konig}}, and
  \bibinfo{author}{\bibfnamefont{U.}~\bibnamefont{Leonhardt}},
  \bibinfo{year}{2008}, {``}\bibinfo{title}{Fiber-optical analog of the event
  horizon},{''} \bibinfo{journal}{Science} \textbf{\bibinfo{volume}{319}},
  \bibinfo{pages}{1367--1370}.

\bibitem[{\citenamefont{Piilo and Maniscalco}(2006)}]{Pii06}
\bibinfo{author}{\bibnamefont{Piilo}, \bibfnamefont{J.}}, and
  \bibinfo{author}{\bibfnamefont{S.}~\bibnamefont{Maniscalco}},
  \bibinfo{year}{2006}, {``}\bibinfo{title}{Driven harmonic oscillator as a
  quantum simulator for open systems},{''} \bibinfo{journal}{Phys. Rev. A}
  \textbf{\bibinfo{volume}{74}},  \bibinfo{pages}{032303}.

\bibitem[{\citenamefont{Pollet} \emph{et~al.}(2010)\citenamefont{Pollet, Picon,
  B\"uchler, and Troyer}}]{Pol10}
\bibinfo{author}{\bibnamefont{Pollet}, \bibfnamefont{L.}},
  \bibinfo{author}{\bibfnamefont{J.~D.} \bibnamefont{Picon}},
  \bibinfo{author}{\bibfnamefont{H.~P.} \bibnamefont{B\"uchler}}, and
  \bibinfo{author}{\bibfnamefont{M.}~\bibnamefont{Troyer}},
  \bibinfo{year}{2010}, {``}\bibinfo{title}{Supersolid phase with cold polar
  molecules on a triangular lattice},{''} \bibinfo{journal}{Phys. Rev. Lett.}
  \textbf{\bibinfo{volume}{104}},  \bibinfo{pages}{125302}.

\bibitem[{\citenamefont{Porras and Cirac}(2004{\natexlab{a}})}]{PorCir04}
\bibinfo{author}{\bibnamefont{Porras}, \bibfnamefont{D.}}, and
  \bibinfo{author}{\bibfnamefont{J.}~\bibnamefont{Cirac}},
  \bibinfo{year}{2004}{\natexlab{a}}, {``}\bibinfo{title}{Bose-Einstein
  condensation and strong-correlation behavior of phonons in ion traps},{''}
  \bibinfo{journal}{Phys. Rev. Lett.} \textbf{\bibinfo{volume}{93}},
  \bibinfo{pages}{263602}.

\bibitem[{\citenamefont{Porras and Cirac}(2005{\natexlab{a}})}]{Por05}
\bibinfo{author}{\bibnamefont{Porras}, \bibfnamefont{D.}}, and
  \bibinfo{author}{\bibfnamefont{J.}~\bibnamefont{Cirac}},
  \bibinfo{year}{2005}{\natexlab{a}}, {``}\bibinfo{title}{Simulation of quantum
  spin models and phase transitions with trapped ions},{''}
  \bibinfo{journal}{Laser Physics} \textbf{\bibinfo{volume}{15}},
  \bibinfo{pages}{88--94}.

\bibitem[{\citenamefont{Porras and Cirac}(2004{\natexlab{b}})}]{Por04}
\bibinfo{author}{\bibnamefont{Porras}, \bibfnamefont{D.}}, and
  \bibinfo{author}{\bibfnamefont{J.~I.} \bibnamefont{Cirac}},
  \bibinfo{year}{2004}{\natexlab{b}}, {``}\bibinfo{title}{Effective quantum
  spin systems with trapped ions},{''} \bibinfo{journal}{Phys. Rev. Lett.}
  \textbf{\bibinfo{volume}{92}},  \bibinfo{pages}{207901}.

\bibitem[{\citenamefont{Porras and Cirac}(2005{\natexlab{b}})}]{PorCir05}
\bibinfo{author}{\bibnamefont{Porras}, \bibfnamefont{D.}}, and
  \bibinfo{author}{\bibfnamefont{J.~I.} \bibnamefont{Cirac}},
  \bibinfo{year}{2005}{\natexlab{b}}, {``}\bibinfo{title}{Simulation of quantum
  magnetism with trapped ions},{''} \bibinfo{journal}{Proc. SPIE}
  \textbf{\bibinfo{volume}{5833}},  \bibinfo{pages}{127}.

\bibitem[{\citenamefont{Porras and Cirac}(2006{\natexlab{a}})}]{PorCir06}
\bibinfo{author}{\bibnamefont{Porras}, \bibfnamefont{D.}}, and
  \bibinfo{author}{\bibfnamefont{J.~I.} \bibnamefont{Cirac}},
  \bibinfo{year}{2006}{\natexlab{a}}, {``}\bibinfo{title}{Phonon superfluids in
  sets of trapped ions},{''} \bibinfo{journal}{Foundations of Physics}
  \textbf{\bibinfo{volume}{36}},  \bibinfo{pages}{465}.

\bibitem[{\citenamefont{Porras and Cirac}(2006{\natexlab{b}})}]{Por06}
\bibinfo{author}{\bibnamefont{Porras}, \bibfnamefont{D.}}, and
  \bibinfo{author}{\bibfnamefont{J.~I.} \bibnamefont{Cirac}},
  \bibinfo{year}{2006}{\natexlab{b}}, {``}\bibinfo{title}{Quantum manipulation
  of trapped ions in two dimensional Coulomb crystals},{''}
  \bibinfo{journal}{Phys. Rev. Lett.} \textbf{\bibinfo{volume}{96}},
  \bibinfo{pages}{250501}.

\bibitem[{\citenamefont{Porras} \emph{et~al.}(2008)\citenamefont{Porras,
  Marquardt, von Delft, and Cirac}}]{Por08}
\bibinfo{author}{\bibnamefont{Porras}, \bibfnamefont{D.}},
  \bibinfo{author}{\bibfnamefont{F.}~\bibnamefont{Marquardt}},
  \bibinfo{author}{\bibfnamefont{J.}~\bibnamefont{von Delft}}, and
  \bibinfo{author}{\bibfnamefont{J.~I.} \bibnamefont{Cirac}},
  \bibinfo{year}{2008}, {``}\bibinfo{title}{Mesoscopic spin-boson models of
  trapped ions},{''} \bibinfo{journal}{Phys. Rev. A}
  \textbf{\bibinfo{volume}{78}},  \bibinfo{pages}{010101}.

\bibitem[{\citenamefont{Poulin} \emph{et~al.}(2011)\citenamefont{Poulin, Qarry,
  Somma, and Verstraete}}]{Pou11}
\bibinfo{author}{\bibnamefont{Poulin}, \bibfnamefont{D.}},
  \bibinfo{author}{\bibfnamefont{A.}~\bibnamefont{Qarry}},
  \bibinfo{author}{\bibfnamefont{R.}~\bibnamefont{Somma}}, and
  \bibinfo{author}{\bibfnamefont{F.}~\bibnamefont{Verstraete}},
  \bibinfo{year}{2011}, {``}\bibinfo{title}{Quantum simulation of
  time-dependent Hamiltonians and the convenient Illusion of Hilbert
  space},{''} \bibinfo{journal}{Phys. Rev. Lett.}
  \textbf{\bibinfo{volume}{106}},  \bibinfo{pages}{170501}.

\bibitem[{\citenamefont{Pupillo} \emph{et~al.}(2009)\citenamefont{Pupillo,
  Micheli, B\"uchler, and Zoller}}]{Pup08}
\bibinfo{author}{\bibnamefont{Pupillo}, \bibfnamefont{G.}},
  \bibinfo{author}{\bibfnamefont{A.}~\bibnamefont{Micheli}},
  \bibinfo{author}{\bibfnamefont{H.}~\bibnamefont{B\"uchler}}, and
  \bibinfo{author}{\bibfnamefont{P.}~\bibnamefont{Zoller}},
  \bibinfo{year}{2009}, \emph{\bibinfo{title}{Condensed matter physics with
  cold polar molecules}}, volume \bibinfo{volume}{Cold Molecules: Theory,
  Experiment, Applications} (\bibinfo{publisher}{CRC Press}),
  \eprint{arXiv:0805.1896}.

\bibitem[{\citenamefont{Quan} \emph{et~al.}(2007)\citenamefont{Quan, Liu, Sun,
  and Nori}}]{Qua06}
\bibinfo{author}{\bibnamefont{Quan}, \bibfnamefont{H.~T.}},
  \bibinfo{author}{\bibfnamefont{Y.~X.} \bibnamefont{Liu}},
  \bibinfo{author}{\bibfnamefont{C.~P.} \bibnamefont{Sun}}, and
  \bibinfo{author}{\bibfnamefont{F.}~\bibnamefont{Nori}}, \bibinfo{year}{2007},
  {``}\bibinfo{title}{Quantum thermodynamic cycles and quantum heat
  engines},{''} \bibinfo{journal}{Phys. Rev. E} \textbf{\bibinfo{volume}{76}},
  \bibinfo{pages}{031105}.

\bibitem[{\citenamefont{Quan} \emph{et~al.}(2006)\citenamefont{Quan, Wang, Liu,
  Sun, and Nori}}]{Quan06}
\bibinfo{author}{\bibnamefont{Quan}, \bibfnamefont{H.~T.}},
  \bibinfo{author}{\bibfnamefont{Y.~D.} \bibnamefont{Wang}},
  \bibinfo{author}{\bibfnamefont{Y.~X.} \bibnamefont{Liu}},
  \bibinfo{author}{\bibfnamefont{C.~P.} \bibnamefont{Sun}}, and
  \bibinfo{author}{\bibfnamefont{F.}~\bibnamefont{Nori}}, \bibinfo{year}{2006},
  {``}\bibinfo{title}{Maxwell's demon assisted thermodynamic cycle in
  superconducting quantum circuits},{''} \bibinfo{journal}{Phys. Rev. Lett.}
  \textbf{\bibinfo{volume}{97}},  \bibinfo{pages}{180402}.

\bibitem[{\citenamefont{de~Raedt} \emph{et~al.}(2007)\citenamefont{de~Raedt,
  Michielsen, de~Raedt, Trieu, Arnold, Lippert, Watanabe, and Ito}}]{Rae07}
\bibinfo{author}{\bibnamefont{de~Raedt}, \bibfnamefont{K.}},
  \bibinfo{author}{\bibfnamefont{K.}~\bibnamefont{Michielsen}},
  \bibinfo{author}{\bibfnamefont{H.}~\bibnamefont{de~Raedt}},
  \bibinfo{author}{\bibfnamefont{B.}~\bibnamefont{Trieu}},
  \bibinfo{author}{\bibfnamefont{G.}~\bibnamefont{Arnold}},
  \bibinfo{author}{\bibfnamefont{M.~R.~T.} \bibnamefont{Lippert}},
  \bibinfo{author}{\bibfnamefont{H.}~\bibnamefont{Watanabe}}, and
  \bibinfo{author}{\bibfnamefont{N.}~\bibnamefont{Ito}}, \bibinfo{year}{2007},
  {``}\bibinfo{title}{Massive parallel quantum computer simulator},{''}
  \bibinfo{journal}{Comp. Phys. Comm.} \textbf{\bibinfo{volume}{176}},
  \bibinfo{pages}{121--136}.

\bibitem[{\citenamefont{Raeisi} \emph{et~al.}(2012)\citenamefont{Raeisi, Wiebe,
  and Sanders}}]{Rae12}
\bibinfo{author}{\bibnamefont{Raeisi}, \bibfnamefont{S.}},
  \bibinfo{author}{\bibfnamefont{N.}~\bibnamefont{Wiebe}}, and
  \bibinfo{author}{\bibfnamefont{B.~C.} \bibnamefont{Sanders}},
  \bibinfo{year}{2012}, {``}\bibinfo{title}{Quantum-circuit design for
  efficient simulations of many-body quantum dynamics},{''}
  \bibinfo{journal}{New J. Phys.} \textbf{\bibinfo{volume}{14}},
  \bibinfo{pages}{103017}.

\bibitem[{\citenamefont{Rakhmanov} \emph{et~al.}(2010)\citenamefont{Rakhmanov,
  Yampol'skii, Fan, Capasso, and Nori}}]{Rak10}
\bibinfo{author}{\bibnamefont{Rakhmanov}, \bibfnamefont{A.~L.}},
  \bibinfo{author}{\bibfnamefont{V.~A.} \bibnamefont{Yampol'skii}},
  \bibinfo{author}{\bibfnamefont{J.~A.} \bibnamefont{Fan}},
  \bibinfo{author}{\bibfnamefont{F.}~\bibnamefont{Capasso}}, and
  \bibinfo{author}{\bibfnamefont{F.}~\bibnamefont{Nori}}, \bibinfo{year}{2010},
  {``}\bibinfo{title}{Layered superconductors as negative-refractive-index
  metamaterials},{''} \bibinfo{journal}{Phys. Rev. B}
  \textbf{\bibinfo{volume}{81}},  \bibinfo{pages}{075101}.

\bibitem[{\citenamefont{Rakhmanov} \emph{et~al.}(2008)\citenamefont{Rakhmanov,
  Zagoskin, Savel'ev, and Nori}}]{Rak08}
\bibinfo{author}{\bibnamefont{Rakhmanov}, \bibfnamefont{A.~L.}},
  \bibinfo{author}{\bibfnamefont{A.~M.} \bibnamefont{Zagoskin}},
  \bibinfo{author}{\bibfnamefont{S.}~\bibnamefont{Savel'ev}}, and
  \bibinfo{author}{\bibfnamefont{F.}~\bibnamefont{Nori}}, \bibinfo{year}{2008},
  {``}\bibinfo{title}{Quantum metamaterials: electromagnetic waves in a
  Josephson qubit line},{''} \bibinfo{journal}{Phys. Rev. B}
  \textbf{\bibinfo{volume}{77}},  \bibinfo{pages}{144507}.

\bibitem[{\citenamefont{Raussendorf}
  \emph{et~al.}(2003)\citenamefont{Raussendorf, Browne, and Briegel}}]{Rau03}
\bibinfo{author}{\bibnamefont{Raussendorf}, \bibfnamefont{R.}},
  \bibinfo{author}{\bibfnamefont{D.~E.} \bibnamefont{Browne}}, and
  \bibinfo{author}{\bibfnamefont{H.~J.} \bibnamefont{Briegel}},
  \bibinfo{year}{2003}, {``}\bibinfo{title}{Measurement-based quantum
  computation on cluster states},{''} \bibinfo{journal}{Phys. Rev. A}
  \textbf{\bibinfo{volume}{68}},  \bibinfo{pages}{022312}.

\bibitem[{\citenamefont{Regal} \emph{et~al.}(2004)\citenamefont{Regal, Greiner,
  and Jin}}]{Reg04}
\bibinfo{author}{\bibnamefont{Regal}, \bibfnamefont{C.~A.}},
  \bibinfo{author}{\bibfnamefont{M.}~\bibnamefont{Greiner}}, and
  \bibinfo{author}{\bibfnamefont{D.~S.} \bibnamefont{Jin}},
  \bibinfo{year}{2004}, {``}\bibinfo{title}{Observation of resonance
  condensation of Fermionic atom pairs},{''} \bibinfo{journal}{Phys. Rev.
  Lett.} \textbf{\bibinfo{volume}{92}},  \bibinfo{pages}{040403}.

\bibitem[{\citenamefont{Retzker} \emph{et~al.}(2008)\citenamefont{Retzker,
  Thompson, Segal, and Plenio}}]{Ret08}
\bibinfo{author}{\bibnamefont{Retzker}, \bibfnamefont{A.}},
  \bibinfo{author}{\bibfnamefont{R.~C.} \bibnamefont{Thompson}},
  \bibinfo{author}{\bibfnamefont{D.~M.} \bibnamefont{Segal}}, and
  \bibinfo{author}{\bibfnamefont{M.~B.} \bibnamefont{Plenio}},
  \bibinfo{year}{2008}, {``}\bibinfo{title}{Double well potentials and quantum
  phase transitions in ion traps},{''} \bibinfo{journal}{Phys. Rev. Lett.}
  \textbf{\bibinfo{volume}{101}},  \bibinfo{pages}{260504}.

\bibitem[{\citenamefont{Roati} \emph{et~al.}(2008)\citenamefont{Roati,
  D'Errico, Fallani, Fattori, Fort, Zaccanti, Modugno, Modugno, and
  M.Inguscio}}]{Roa08}
\bibinfo{author}{\bibnamefont{Roati}, \bibfnamefont{G.}},
  \bibinfo{author}{\bibfnamefont{C.}~\bibnamefont{D'Errico}},
  \bibinfo{author}{\bibfnamefont{L.}~\bibnamefont{Fallani}},
  \bibinfo{author}{\bibfnamefont{M.}~\bibnamefont{Fattori}},
  \bibinfo{author}{\bibfnamefont{C.}~\bibnamefont{Fort}},
  \bibinfo{author}{\bibfnamefont{M.}~\bibnamefont{Zaccanti}},
  \bibinfo{author}{\bibfnamefont{G.}~\bibnamefont{Modugno}},
  \bibinfo{author}{\bibfnamefont{M.}~\bibnamefont{Modugno}}, and
  \bibinfo{author}{\bibnamefont{M.Inguscio}}, \bibinfo{year}{2008},
  {``}\bibinfo{title}{Anderson localization of a non-interacting Bose-Einstein
  condensate},{''} \bibinfo{journal}{Nature} \textbf{\bibinfo{volume}{453}},
  \bibinfo{pages}{895--898}.

\bibitem[{\citenamefont{Roumpos} \emph{et~al.}(2007)\citenamefont{Roumpos,
  Master, and Yamamoto}}]{Rou07}
\bibinfo{author}{\bibnamefont{Roumpos}, \bibfnamefont{G.}},
  \bibinfo{author}{\bibfnamefont{C.~P.} \bibnamefont{Master}}, and
  \bibinfo{author}{\bibfnamefont{Y.}~\bibnamefont{Yamamoto}},
  \bibinfo{year}{2007}, {``}\bibinfo{title}{Quantum simulation of spin ordering
  with nuclear spins in a solid-state lattice},{''} \bibinfo{journal}{Phys.
  Rev. B} \textbf{\bibinfo{volume}{75}},  \bibinfo{pages}{094415}.

\bibitem[{\citenamefont{Rusin and Zawadzki}(2010)}]{Rus10}
\bibinfo{author}{\bibnamefont{Rusin}, \bibfnamefont{T.~M.}}, and
  \bibinfo{author}{\bibfnamefont{W.}~\bibnamefont{Zawadzki}},
  \bibinfo{year}{2010}, {``}\bibinfo{title}{Zitterbewegung of relativistic
  electrons in a magnetic field and its simulation by trapped ions},{''}
  \bibinfo{journal}{Phys. Rev. D} \textbf{\bibinfo{volume}{82}},
  \bibinfo{pages}{125031}.

\bibitem[{\citenamefont{Sanpera} \emph{et~al.}(2004)\citenamefont{Sanpera,
  Kantian, Sanchez-Palencia, Zakrzewski, and Lewenstein}}]{San04}
\bibinfo{author}{\bibnamefont{Sanpera}, \bibfnamefont{A.}},
  \bibinfo{author}{\bibfnamefont{A.}~\bibnamefont{Kantian}},
  \bibinfo{author}{\bibfnamefont{L.}~\bibnamefont{Sanchez-Palencia}},
  \bibinfo{author}{\bibfnamefont{J.}~\bibnamefont{Zakrzewski}}, and
  \bibinfo{author}{\bibfnamefont{M.}~\bibnamefont{Lewenstein}},
  \bibinfo{year}{2004}, {``}\bibinfo{title}{Atomic Fermi-Bose mixtures in
  inhomogeneous and random lattices: from Fermi glass to quantum spin glass and
  quantum percolation},{''} \bibinfo{journal}{Phys. Rev. Lett.}
  \textbf{\bibinfo{volume}{93}},  \bibinfo{pages}{040401}.

\bibitem[{\citenamefont{Santini} \emph{et~al.}(2011)\citenamefont{Santini,
  Carretta, Troiani, and Amoretti}}]{San11}
\bibinfo{author}{\bibnamefont{Santini}, \bibfnamefont{P.}},
  \bibinfo{author}{\bibfnamefont{S.}~\bibnamefont{Carretta}},
  \bibinfo{author}{\bibfnamefont{F.}~\bibnamefont{Troiani}}, and
  \bibinfo{author}{\bibfnamefont{G.}~\bibnamefont{Amoretti}},
  \bibinfo{year}{2011}, {``}\bibinfo{title}{Molecular Nanomagnets as Quantum
  Simulators},{''} \bibinfo{journal}{Phys. Rev. Lett.}
  \textbf{\bibinfo{volume}{107}},  \bibinfo{pages}{230502}.

\bibitem[{\citenamefont{Schack}(1998)}]{Sch98}
\bibinfo{author}{\bibnamefont{Schack}, \bibfnamefont{R.}},
  \bibinfo{year}{1998}, {``}\bibinfo{title}{Using a quantum computer to
  investigate quantum chaos},{''} \bibinfo{journal}{Phys. Rev. A}
  \textbf{\bibinfo{volume}{57}},  \bibinfo{pages}{1634--1635}.

\bibitem[{\citenamefont{Sch\"atz} \emph{et~al.}(2007)\citenamefont{Sch\"atz,
  Friedenauer, Schmitz, Petersen, and Kahra}}]{SchFri08}
\bibinfo{author}{\bibnamefont{Sch\"atz}, \bibfnamefont{T.}},
  \bibinfo{author}{\bibfnamefont{A.}~\bibnamefont{Friedenauer}},
  \bibinfo{author}{\bibfnamefont{H.}~\bibnamefont{Schmitz}},
  \bibinfo{author}{\bibfnamefont{L.}~\bibnamefont{Petersen}}, and
  \bibinfo{author}{\bibfnamefont{S.}~\bibnamefont{Kahra}},
  \bibinfo{year}{2007}, {``}\bibinfo{title}{Towards (scalable) quantum
  simulations in ion traps},{''} \bibinfo{journal}{J. Mod. Opt.}
  \textbf{\bibinfo{volume}{54}},  \bibinfo{pages}{2317--2325}.

\bibitem[{\citenamefont{Sch\"atz} \emph{et~al.}(2004)\citenamefont{Sch\"atz,
  Leibfried, Chiaverini, Barrett, Britton, Demarco, Itano, Jost, Langer, and
  Wineland}}]{Sch04}
\bibinfo{author}{\bibnamefont{Sch\"atz}, \bibfnamefont{T.}},
  \bibinfo{author}{\bibfnamefont{D.}~\bibnamefont{Leibfried}},
  \bibinfo{author}{\bibfnamefont{J.}~\bibnamefont{Chiaverini}},
  \bibinfo{author}{\bibfnamefont{M.}~\bibnamefont{Barrett}},
  \bibinfo{author}{\bibfnamefont{J.}~\bibnamefont{Britton}},
  \bibinfo{author}{\bibfnamefont{B.}~\bibnamefont{Demarco}},
  \bibinfo{author}{\bibfnamefont{W.}~\bibnamefont{Itano}},
  \bibinfo{author}{\bibfnamefont{J.}~\bibnamefont{Jost}},
  \bibinfo{author}{\bibfnamefont{C.}~\bibnamefont{Langer}}, and
  \bibinfo{author}{\bibfnamefont{D.}~\bibnamefont{Wineland}},
  \bibinfo{year}{2004}, {``}\bibinfo{title}{Towards a scalable quantum
  computer/simulator based on trapped ions},{''} \bibinfo{journal}{Appl. Phys.
  B} \textbf{\bibinfo{volume}{79}},  \bibinfo{pages}{979--986}.

\bibitem[{\citenamefont{Schleich and Walther}(2008)}]{SW08}
\bibinfo{editor}{\bibnamefont{Schleich}, \bibfnamefont{W.~P.}}, and
  \bibinfo{editor}{\bibfnamefont{H.}~\bibnamefont{Walther}} (eds.),
  \bibinfo{year}{2008}, \emph{\bibinfo{title}{Elements of quantum information}}
  (\bibinfo{publisher}{Wiley-VCH, Weinheim}).

\bibitem[{\citenamefont{Schmied} \emph{et~al.}(2010)\citenamefont{Schmied,
  Leibfried, Spreeuw, and Whitlock}}]{Sch10}
\bibinfo{author}{\bibnamefont{Schmied}, \bibfnamefont{R.}},
  \bibinfo{author}{\bibfnamefont{D.}~\bibnamefont{Leibfried}},
  \bibinfo{author}{\bibfnamefont{R.}~\bibnamefont{Spreeuw}}, and
  \bibinfo{author}{\bibfnamefont{S.}~\bibnamefont{Whitlock}},
  \bibinfo{year}{2010}, {``}\bibinfo{title}{Optimized magnetic lattices for
  ultracold atomic ensembles},{''} \bibinfo{journal}{New J. Phys.}
  \textbf{\bibinfo{volume}{12}},  \bibinfo{pages}{103029}.

\bibitem[{\citenamefont{Schmied} \emph{et~al.}(2008)\citenamefont{Schmied,
  Roscilde, Murg, Porras, and Cirac}}]{Sch08}
\bibinfo{author}{\bibnamefont{Schmied}, \bibfnamefont{R.}},
  \bibinfo{author}{\bibfnamefont{T.}~\bibnamefont{Roscilde}},
  \bibinfo{author}{\bibfnamefont{V.}~\bibnamefont{Murg}},
  \bibinfo{author}{\bibfnamefont{D.}~\bibnamefont{Porras}}, and
  \bibinfo{author}{\bibfnamefont{J.~I.} \bibnamefont{Cirac}},
  \bibinfo{year}{2008}, {``}\bibinfo{title}{Quantum phases of trapped ions in
  an optical lattice},{''} \bibinfo{journal}{New J. Phys.}
  \textbf{\bibinfo{volume}{10}},  \bibinfo{pages}{045017}.

\bibitem[{\citenamefont{Schmied} \emph{et~al.}(2009)\citenamefont{Schmied,
  Wesenberg, and Leibfried}}]{Sch09}
\bibinfo{author}{\bibnamefont{Schmied}, \bibfnamefont{R.}},
  \bibinfo{author}{\bibfnamefont{J.~H.} \bibnamefont{Wesenberg}}, and
  \bibinfo{author}{\bibfnamefont{D.}~\bibnamefont{Leibfried}},
  \bibinfo{year}{2009}, {``}\bibinfo{title}{Optimal surface-electrode trap
  lattices for quantum simulation with trapped ions},{''}
  \bibinfo{journal}{Phys. Rev. Lett.} \textbf{\bibinfo{volume}{102}},
  \bibinfo{pages}{233002}.

\bibitem[{\citenamefont{Schmied} \emph{et~al.}(2011)\citenamefont{Schmied,
  Wesenberg, and Leibfried}}]{Schm11}
\bibinfo{author}{\bibnamefont{Schmied}, \bibfnamefont{R.}},
  \bibinfo{author}{\bibfnamefont{J.~H.} \bibnamefont{Wesenberg}}, and
  \bibinfo{author}{\bibfnamefont{D.}~\bibnamefont{Leibfried}},
  \bibinfo{year}{2011}, {``}\bibinfo{title}{Quantum simulation of the hexagonal
  Kitaev model with trapped ions},{''} \bibinfo{journal}{New J. Phys.}
  \textbf{\bibinfo{volume}{13}},  \bibinfo{pages}{115011}.

\bibitem[{\citenamefont{Schneider} \emph{et~al.}(2012)\citenamefont{Schneider,
  Porras, and Schaetz}}]{Sch12}
\bibinfo{author}{\bibnamefont{Schneider}, \bibfnamefont{C.}},
  \bibinfo{author}{\bibfnamefont{D.}~\bibnamefont{Porras}}, and
  \bibinfo{author}{\bibfnamefont{T.}~\bibnamefont{Schaetz}},
  \bibinfo{year}{2012}, {``}\bibinfo{title}{Experimental quantum simulations of
  many-body physics with trapped ions},{''} \bibinfo{journal}{Rep. Prog. Phys.}
  \textbf{\bibinfo{volume}{75}},  \bibinfo{pages}{024401}.

\bibitem[{\citenamefont{Schneider and Milburn}(2001)}]{Sch01}
\bibinfo{author}{\bibnamefont{Schneider}, \bibfnamefont{S.}}, and
  \bibinfo{author}{\bibfnamefont{G.}~\bibnamefont{Milburn}},
  \bibinfo{year}{2001}, {``}\bibinfo{title}{Entanglement in the Dicke
  model},{''} \eprint{arXiv:quant-ph/0112042}.

\bibitem[{\citenamefont{Schoelkopf and Girvin}(2008)}]{Sho08}
\bibinfo{author}{\bibnamefont{Schoelkopf}, \bibfnamefont{R.~J.}}, and
  \bibinfo{author}{\bibfnamefont{S.~M.} \bibnamefont{Girvin}},
  \bibinfo{year}{2008}, {``}\bibinfo{title}{Wiring up quantum systems},{''}
  \bibinfo{journal}{Nature} \textbf{\bibinfo{volume}{451}},
  \bibinfo{pages}{664--669}.

\bibitem[{\citenamefont{Schulte} \emph{et~al.}(2005)\citenamefont{Schulte,
  Drenkelforth, Kruse, Ertmer, Arlt, Sacha, Zakrzewski, and
  Lewenstein}}]{SchDre05}
\bibinfo{author}{\bibnamefont{Schulte}, \bibfnamefont{T.}},
  \bibinfo{author}{\bibfnamefont{S.}~\bibnamefont{Drenkelforth}},
  \bibinfo{author}{\bibfnamefont{J.}~\bibnamefont{Kruse}},
  \bibinfo{author}{\bibfnamefont{W.}~\bibnamefont{Ertmer}},
  \bibinfo{author}{\bibfnamefont{J.}~\bibnamefont{Arlt}},
  \bibinfo{author}{\bibfnamefont{K.}~\bibnamefont{Sacha}},
  \bibinfo{author}{\bibfnamefont{J.}~\bibnamefont{Zakrzewski}}, and
  \bibinfo{author}{\bibfnamefont{M.}~\bibnamefont{Lewenstein}},
  \bibinfo{year}{2005}, {``}\bibinfo{title}{Routes towards Anderson-like
  localization of Bose-Einstein condensates in disordered optical
  lattices},{''} \bibinfo{journal}{Phys. Rev. Lett.}
  \textbf{\bibinfo{volume}{95}},  \bibinfo{pages}{170411}.

\bibitem[{\citenamefont{Sch\"utzhold and Mostame}(2005)}]{Sch05}
\bibinfo{author}{\bibnamefont{Sch\"utzhold}, \bibfnamefont{R.}}, and
  \bibinfo{author}{\bibfnamefont{S.}~\bibnamefont{Mostame}},
  \bibinfo{year}{2005}, {``}\bibinfo{title}{Quantum simulator for the O(3)
  nonlinear sigma model},{''} \bibinfo{journal}{JETP Letters}
  \textbf{\bibinfo{volume}{82}},  \bibinfo{pages}{248}.

\bibitem[{\citenamefont{Sch\"utzhold}
  \emph{et~al.}(2007)\citenamefont{Sch\"utzhold, Uhlmann, Petersen, Schmitz,
  Friedenauer, and Sch\"atz}}]{Sch07}
\bibinfo{author}{\bibnamefont{Sch\"utzhold}, \bibfnamefont{R.}},
  \bibinfo{author}{\bibfnamefont{M.}~\bibnamefont{Uhlmann}},
  \bibinfo{author}{\bibfnamefont{L.}~\bibnamefont{Petersen}},
  \bibinfo{author}{\bibfnamefont{H.}~\bibnamefont{Schmitz}},
  \bibinfo{author}{\bibfnamefont{A.}~\bibnamefont{Friedenauer}}, and
  \bibinfo{author}{\bibfnamefont{T.}~\bibnamefont{Sch\"atz}},
  \bibinfo{year}{2007}, {``}\bibinfo{title}{Analogue of cosmological particle
  creation in an ion trap},{''} \bibinfo{journal}{Phys. Rev. Lett.}
  \textbf{\bibinfo{volume}{99}},  \bibinfo{pages}{201301}.

\bibitem[{\citenamefont{Semiao and Paternostro}(2012)}]{Sem10}
\bibinfo{author}{\bibnamefont{Semiao}, \bibfnamefont{F.~L.}}, and
  \bibinfo{author}{\bibfnamefont{M.}~\bibnamefont{Paternostro}},
  \bibinfo{year}{2012}, {``}\bibinfo{title}{Quantum circuits for spin and
  flavor degrees of freedom of quarks forming nucleons},{''}
  \bibinfo{journal}{Quant. Inf. Proc.} \textbf{\bibinfo{volume}{11}},
  \bibinfo{pages}{67 -- 75}.

\bibitem[{\citenamefont{Sherson} \emph{et~al.}(2010)\citenamefont{Sherson,
  Weitenberg, Endres, Cheneau, Bloch, and Kuhr}}]{She10}
\bibinfo{author}{\bibnamefont{Sherson}, \bibfnamefont{J.~F.}},
  \bibinfo{author}{\bibfnamefont{C.}~\bibnamefont{Weitenberg}},
  \bibinfo{author}{\bibfnamefont{M.}~\bibnamefont{Endres}},
  \bibinfo{author}{\bibfnamefont{M.}~\bibnamefont{Cheneau}},
  \bibinfo{author}{\bibfnamefont{I.}~\bibnamefont{Bloch}}, and
  \bibinfo{author}{\bibfnamefont{S.}~\bibnamefont{Kuhr}}, \bibinfo{year}{2010},
  {``}\bibinfo{title}{Single-atom-resolved fluorescence imaging of an atomic
  Mott insualtor},{''} \bibinfo{journal}{Nature}
  \textbf{\bibinfo{volume}{467}}, ~\bibinfo{pages}{68}.

\bibitem[{\citenamefont{Shevchenko}
  \emph{et~al.}(2010)\citenamefont{Shevchenko, Ashhab, and Nori}}]{Shev10}
\bibinfo{author}{\bibnamefont{Shevchenko}, \bibfnamefont{S.}},
  \bibinfo{author}{\bibfnamefont{S.}~\bibnamefont{Ashhab}}, and
  \bibinfo{author}{\bibfnamefont{F.}~\bibnamefont{Nori}}, \bibinfo{year}{2010},
  {``}\bibinfo{title}{Landau-Zener-Stuckelberg interferometry},{''}
  \bibinfo{journal}{Physics Reports} \textbf{\bibinfo{volume}{492}},
  ~\bibinfo{pages}{1}.

\bibitem[{\citenamefont{Simon} \emph{et~al.}(2011)\citenamefont{Simon, Bakr,
  Ma, Tai, Preiss, and Greiner}}]{Sim11}
\bibinfo{author}{\bibnamefont{Simon}, \bibfnamefont{J.}},
  \bibinfo{author}{\bibfnamefont{W.~S.} \bibnamefont{Bakr}},
  \bibinfo{author}{\bibfnamefont{R.}~\bibnamefont{Ma}},
  \bibinfo{author}{\bibfnamefont{M.~E.} \bibnamefont{Tai}},
  \bibinfo{author}{\bibfnamefont{P.~M.} \bibnamefont{Preiss}}, and
  \bibinfo{author}{\bibfnamefont{M.}~\bibnamefont{Greiner}},
  \bibinfo{year}{2011}, {``}\bibinfo{title}{Quantum simulation of
  antiferromagnetic spin chains in an optical lattice},{''}
  \bibinfo{journal}{Nature} \textbf{\bibinfo{volume}{472}},
  \bibinfo{pages}{307}.

\bibitem[{\citenamefont{Singha} \emph{et~al.}(2011)\citenamefont{Singha,
  Gibertini, Karmakar, Yuan, Polini, Vignale, Katsnelson, Pinczuk, Pfeiffer,
  West, and Pellegrini}}]{Sin11}
\bibinfo{author}{\bibnamefont{Singha}, \bibfnamefont{A.}},
  \bibinfo{author}{\bibfnamefont{M.}~\bibnamefont{Gibertini}},
  \bibinfo{author}{\bibfnamefont{B.}~\bibnamefont{Karmakar}},
  \bibinfo{author}{\bibfnamefont{S.}~\bibnamefont{Yuan}},
  \bibinfo{author}{\bibfnamefont{M.}~\bibnamefont{Polini}},
  \bibinfo{author}{\bibfnamefont{G.}~\bibnamefont{Vignale}},
  \bibinfo{author}{\bibfnamefont{M.~I.} \bibnamefont{Katsnelson}},
  \bibinfo{author}{\bibfnamefont{A.}~\bibnamefont{Pinczuk}},
  \bibinfo{author}{\bibfnamefont{L.~N.} \bibnamefont{Pfeiffer}},
  \bibinfo{author}{\bibfnamefont{K.~W.} \bibnamefont{West}}, and
  \bibinfo{author}{\bibfnamefont{V.}~\bibnamefont{Pellegrini}},
  \bibinfo{year}{2011}, {``}\bibinfo{title}{Two-Dimensional Mott-Hubbard
  Electrons in an Artificial Honeycomb Lattice},{''} \bibinfo{journal}{Science}
  \textbf{\bibinfo{volume}{332}},  \bibinfo{pages}{1176--1179}.

\bibitem[{\citenamefont{Sinha and Russer}(2010)}]{Sin10}
\bibinfo{author}{\bibnamefont{Sinha}, \bibfnamefont{S.}}, and
  \bibinfo{author}{\bibfnamefont{P.}~\bibnamefont{Russer}},
  \bibinfo{year}{2010}, {``}\bibinfo{title}{Quantum computer algorithm for
  electromagnetic field simulation},{''} \bibinfo{journal}{Quant. Inf.
  Process.} \textbf{\bibinfo{volume}{9}},  \bibinfo{pages}{385}.

\bibitem[{\citenamefont{Smirnov} \emph{et~al.}(2007)\citenamefont{Smirnov,
  Savel'ev, Mourokh, and Nori}}]{Smi07}
\bibinfo{author}{\bibnamefont{Smirnov}, \bibfnamefont{A.}},
  \bibinfo{author}{\bibfnamefont{S.}~\bibnamefont{Savel'ev}},
  \bibinfo{author}{\bibfnamefont{L.}~\bibnamefont{Mourokh}}, and
  \bibinfo{author}{\bibfnamefont{F.}~\bibnamefont{Nori}}, \bibinfo{year}{2007},
  {``}\bibinfo{title}{Modelling chemical reactions using semiconductor quantum
  dots},{''} \bibinfo{journal}{EPL} \textbf{\bibinfo{volume}{80}},
  \bibinfo{pages}{67008}.

\bibitem[{\citenamefont{Somaroo} \emph{et~al.}(1999)\citenamefont{Somaroo,
  Tseng, Havel, Laflamme, and Cory}}]{Som99}
\bibinfo{author}{\bibnamefont{Somaroo}, \bibfnamefont{S.}},
  \bibinfo{author}{\bibfnamefont{C.~H.} \bibnamefont{Tseng}},
  \bibinfo{author}{\bibfnamefont{T.~F.} \bibnamefont{Havel}},
  \bibinfo{author}{\bibfnamefont{R.}~\bibnamefont{Laflamme}}, and
  \bibinfo{author}{\bibfnamefont{D.~G.} \bibnamefont{Cory}},
  \bibinfo{year}{1999}, {``}\bibinfo{title}{Quantum simulations on a quantum
  computer},{''} \bibinfo{journal}{Phys. Rev. Lett.}
  \textbf{\bibinfo{volume}{82}},  \bibinfo{pages}{5381--5384}.

\bibitem[{\citenamefont{Somma} \emph{et~al.}(2002)\citenamefont{Somma, Ortiz,
  Gubernatis, Knill, and Laflamme}}]{Som02}
\bibinfo{author}{\bibnamefont{Somma}, \bibfnamefont{R.}},
  \bibinfo{author}{\bibfnamefont{G.}~\bibnamefont{Ortiz}},
  \bibinfo{author}{\bibfnamefont{J.~E.} \bibnamefont{Gubernatis}},
  \bibinfo{author}{\bibfnamefont{E.}~\bibnamefont{Knill}}, and
  \bibinfo{author}{\bibfnamefont{R.}~\bibnamefont{Laflamme}},
  \bibinfo{year}{2002}, {``}\bibinfo{title}{Simulating physical phenomena by
  quantum networks},{''} \bibinfo{journal}{Phys. Rev. A}
  \textbf{\bibinfo{volume}{65}},  \bibinfo{pages}{042323}.

\bibitem[{\citenamefont{Somma} \emph{et~al.}(2003)\citenamefont{Somma, Ortiz,
  Knill, and Gubernatis}}]{Som03}
\bibinfo{author}{\bibnamefont{Somma}, \bibfnamefont{R.}},
  \bibinfo{author}{\bibfnamefont{G.}~\bibnamefont{Ortiz}},
  \bibinfo{author}{\bibfnamefont{E.}~\bibnamefont{Knill}}, and
  \bibinfo{author}{\bibfnamefont{J.}~\bibnamefont{Gubernatis}},
  \bibinfo{year}{2003}, {``}\bibinfo{title}{Quantum simulations of physics
  problems},{''} \eprint{arXiv:quant-ph/0304063}.

\bibitem[{\citenamefont{Spring} \emph{et~al.}(2013)\citenamefont{Spring,
  Metcalf, Humphreys, Kolthammer, Jin, Barbieri, Datta, Thomas-Peter, Langford,
  Kundys, Gates, Smith} \emph{et~al.}}]{Spr13}
\bibinfo{author}{\bibnamefont{Spring}, \bibfnamefont{J.~B.}},
  \bibinfo{author}{\bibfnamefont{B.~J.} \bibnamefont{Metcalf}},
  \bibinfo{author}{\bibfnamefont{P.~C.} \bibnamefont{Humphreys}},
  \bibinfo{author}{\bibfnamefont{W.~S.} \bibnamefont{Kolthammer}},
  \bibinfo{author}{\bibfnamefont{X.-M.} \bibnamefont{Jin}},
  \bibinfo{author}{\bibfnamefont{M.}~\bibnamefont{Barbieri}},
  \bibinfo{author}{\bibfnamefont{A.}~\bibnamefont{Datta}},
  \bibinfo{author}{\bibfnamefont{N.}~\bibnamefont{Thomas-Peter}},
  \bibinfo{author}{\bibfnamefont{N.~K.} \bibnamefont{Langford}},
  \bibinfo{author}{\bibfnamefont{D.}~\bibnamefont{Kundys}},
  \bibinfo{author}{\bibfnamefont{J.~C.} \bibnamefont{Gates}},
  \bibinfo{author}{\bibfnamefont{B.~J.} \bibnamefont{Smith}}, \emph{et~al.},
  \bibinfo{year}{2013}, {``}\bibinfo{title}{Boson Sampling on a Photonic
  Chip},{''} \bibinfo{journal}{Science} \textbf{\bibinfo{volume}{339}},
  \bibinfo{pages}{798 -- 801}.

\bibitem[{\citenamefont{Stanescu} \emph{et~al.}(2009)\citenamefont{Stanescu,
  Galitski, Vaishnav, Clark, and Sarma}}]{Sta09}
\bibinfo{author}{\bibnamefont{Stanescu}, \bibfnamefont{T.~D.}},
  \bibinfo{author}{\bibfnamefont{V.}~\bibnamefont{Galitski}},
  \bibinfo{author}{\bibfnamefont{J.}~\bibnamefont{Vaishnav}},
  \bibinfo{author}{\bibfnamefont{C.~W.} \bibnamefont{Clark}}, and
  \bibinfo{author}{\bibfnamefont{S.~D.} \bibnamefont{Sarma}},
  \bibinfo{year}{2009}, {``}\bibinfo{title}{Topological insulators and metals
  in atomic optical lattices},{''} \bibinfo{journal}{Phys. Rev. A}
  \textbf{\bibinfo{volume}{79}},  \bibinfo{pages}{053639}.

\bibitem[{\citenamefont{Stannigel} \emph{et~al.}(2013)\citenamefont{Stannigel,
  Hauke, Marcos, Hafezi, Diehl, Dalmonte, and Zoller}}]{Sta13}
\bibinfo{author}{\bibnamefont{Stannigel}, \bibfnamefont{K.}},
  \bibinfo{author}{\bibfnamefont{P.}~\bibnamefont{Hauke}},
  \bibinfo{author}{\bibfnamefont{D.}~\bibnamefont{Marcos}},
  \bibinfo{author}{\bibfnamefont{M.}~\bibnamefont{Hafezi}},
  \bibinfo{author}{\bibfnamefont{S.}~\bibnamefont{Diehl}},
  \bibinfo{author}{\bibfnamefont{M.}~\bibnamefont{Dalmonte}}, and
  \bibinfo{author}{\bibfnamefont{P.}~\bibnamefont{Zoller}},
  \bibinfo{year}{2013}, {``}\bibinfo{title}{Constrained dynamics via the Zeno
  effect in quantum simulation: Implementing non-Abelian lattice gauge theories
  with cold atoms},{''} \eprint{arXiv:1308.0528}.

\bibitem[{\citenamefont{Stolze and Suter}(2008)}]{SS08}
\bibinfo{author}{\bibnamefont{Stolze}, \bibfnamefont{J.}}, and
  \bibinfo{author}{\bibfnamefont{D.}~\bibnamefont{Suter}},
  \bibinfo{year}{2008}, \emph{\bibinfo{title}{Quantum Computing}}
  (\bibinfo{publisher}{Wiley-VCH, Weinheim}).

\bibitem[{\citenamefont{Strini}(2002)}]{Str02}
\bibinfo{author}{\bibnamefont{Strini}, \bibfnamefont{G.}},
  \bibinfo{year}{2002}, {``}\bibinfo{title}{Error sensitivity of a quantum
  simulator. I. a first example},{''} \bibinfo{journal}{Fortsch. Phys.}
  \textbf{\bibinfo{volume}{50}},  \bibinfo{pages}{171}.

\bibitem[{\citenamefont{Struck} \emph{et~al.}(2011)\citenamefont{Struck,
  \"Olschl\"ager, Targat, Soltan-Panahi, Eckardt, Lewenstein, Windpassinger,
  and Sengstock}}]{Stu11}
\bibinfo{author}{\bibnamefont{Struck}, \bibfnamefont{J.}},
  \bibinfo{author}{\bibfnamefont{C.}~\bibnamefont{\"Olschl\"ager}},
  \bibinfo{author}{\bibfnamefont{R.~L.} \bibnamefont{Targat}},
  \bibinfo{author}{\bibfnamefont{P.}~\bibnamefont{Soltan-Panahi}},
  \bibinfo{author}{\bibfnamefont{A.}~\bibnamefont{Eckardt}},
  \bibinfo{author}{\bibfnamefont{M.}~\bibnamefont{Lewenstein}},
  \bibinfo{author}{\bibfnamefont{P.}~\bibnamefont{Windpassinger}}, and
  \bibinfo{author}{\bibfnamefont{K.}~\bibnamefont{Sengstock}},
  \bibinfo{year}{2011}, {``}\bibinfo{title}{Quantum simulation of frustrated
  magnetism in triangular optical lattices},{''} \bibinfo{journal}{Science}
  \textbf{\bibinfo{volume}{333}},  \bibinfo{pages}{996}.

\bibitem[{\citenamefont{Suzuki}(1993)}]{Suz93}
\bibinfo{editor}{\bibnamefont{Suzuki}, \bibfnamefont{M.}} (ed.),
  \bibinfo{year}{1993}, \emph{\bibinfo{title}{Quantum Monte Carlo methods in
  condensed matter physics}} (\bibinfo{publisher}{World Scientific}).

\bibitem[{\citenamefont{Szpak and Sch\"utzhold}(2011)}]{Szp11}
\bibinfo{author}{\bibnamefont{Szpak}, \bibfnamefont{N.}}, and
  \bibinfo{author}{\bibfnamefont{R.}~\bibnamefont{Sch\"utzhold}},
  \bibinfo{year}{2011}, {``}\bibinfo{title}{Quantum simulator for the Schwinger
  effect with atoms in bichromatic optical lattices},{''}
  \bibinfo{journal}{Phys. Rev. A} \textbf{\bibinfo{volume}{84}},
  \bibinfo{pages}{050101}.

\bibitem[{\citenamefont{Szpak and Sch\"utzhold}(2012)}]{SS11}
\bibinfo{author}{\bibnamefont{Szpak}, \bibfnamefont{N.}}, and
  \bibinfo{author}{\bibfnamefont{R.}~\bibnamefont{Sch\"utzhold}},
  \bibinfo{year}{2012}, {``}\bibinfo{title}{Optical lattice quantum simulator
  for QED in strong external fields: spontaneous pair creation and the
  Sauter-Schwinger effect},{''} \bibinfo{journal}{New J. Phys.}
  \textbf{\bibinfo{volume}{14}},  \bibinfo{pages}{035001}.

\bibitem[{\citenamefont{Tagliacozzo}
  \emph{et~al.}(2013{\natexlab{a}})\citenamefont{Tagliacozzo, Celi, Orland, and
  Lewenstein}}]{Tag12}
\bibinfo{author}{\bibnamefont{Tagliacozzo}, \bibfnamefont{L.}},
  \bibinfo{author}{\bibfnamefont{A.}~\bibnamefont{Celi}},
  \bibinfo{author}{\bibfnamefont{P.}~\bibnamefont{Orland}}, and
  \bibinfo{author}{\bibfnamefont{M.}~\bibnamefont{Lewenstein}},
  \bibinfo{year}{2013}{\natexlab{a}}, {``}\bibinfo{title}{Simulations of
  non-Abelian gauge theories with optical lattices},{''}
  \bibinfo{journal}{Nature Commun.} \textbf{\bibinfo{volume}{4}},
  \bibinfo{pages}{2615}.

\bibitem[{\citenamefont{Tagliacozzo}
  \emph{et~al.}(2013{\natexlab{b}})\citenamefont{Tagliacozzo, Celi, Zamora, and
  Lewenstein}}]{Tag13}
\bibinfo{author}{\bibnamefont{Tagliacozzo}, \bibfnamefont{L.}},
  \bibinfo{author}{\bibfnamefont{A.}~\bibnamefont{Celi}},
  \bibinfo{author}{\bibfnamefont{A.}~\bibnamefont{Zamora}}, and
  \bibinfo{author}{\bibfnamefont{M.}~\bibnamefont{Lewenstein}},
  \bibinfo{year}{2013}{\natexlab{b}}, {``}\bibinfo{title}{Optical Abelian
  Lattice Gauge Theories},{''} \bibinfo{journal}{Ann. Phys.}
  \textbf{\bibinfo{volume}{330}},  \bibinfo{pages}{160--191}.

\bibitem[{\citenamefont{Taylor and Calarco}(2008)}]{Tay08}
\bibinfo{author}{\bibnamefont{Taylor}, \bibfnamefont{J.~M.}}, and
  \bibinfo{author}{\bibfnamefont{T.}~\bibnamefont{Calarco}},
  \bibinfo{year}{2008}, {``}\bibinfo{title}{Wigner crystals of ions as quantum
  hard drives},{''} \bibinfo{journal}{Phys. Rev. A}
  \textbf{\bibinfo{volume}{78}},  \bibinfo{pages}{062331}.

\bibitem[{\citenamefont{Temme} \emph{et~al.}(2011)\citenamefont{Temme, Osborne,
  Vollbrecht, Poulin, and Verstraete}}]{Tem11}
\bibinfo{author}{\bibnamefont{Temme}, \bibfnamefont{K.}},
  \bibinfo{author}{\bibfnamefont{T.~J.} \bibnamefont{Osborne}},
  \bibinfo{author}{\bibfnamefont{K.~G.} \bibnamefont{Vollbrecht}},
  \bibinfo{author}{\bibfnamefont{D.}~\bibnamefont{Poulin}}, and
  \bibinfo{author}{\bibfnamefont{F.}~\bibnamefont{Verstraete}},
  \bibinfo{year}{2011}, {``}\bibinfo{title}{Quantum Metropolis sampling},{''}
  \bibinfo{journal}{Nature} \textbf{\bibinfo{volume}{471}},
  \bibinfo{pages}{87--90}.

\bibitem[{\citenamefont{Terhal and DiVincenzo}(2000)}]{Ter00}
\bibinfo{author}{\bibnamefont{Terhal}, \bibfnamefont{B.~M.}}, and
  \bibinfo{author}{\bibfnamefont{D.~P.} \bibnamefont{DiVincenzo}},
  \bibinfo{year}{2000}, {``}\bibinfo{title}{Problem of equilibration and the
  computation of correlation functions on a quantum computer},{''}
  \bibinfo{journal}{Phys. Rev. A} \textbf{\bibinfo{volume}{61}},
  \bibinfo{pages}{022301}.

\bibitem[{\citenamefont{Terraneo} \emph{et~al.}(2003)\citenamefont{Terraneo,
  Georgeot, and Shepelyansky}}]{Ter02}
\bibinfo{author}{\bibnamefont{Terraneo}, \bibfnamefont{M.}},
  \bibinfo{author}{\bibfnamefont{B.}~\bibnamefont{Georgeot}}, and
  \bibinfo{author}{\bibfnamefont{D.~L.} \bibnamefont{Shepelyansky}},
  \bibinfo{year}{2003}, {``}\bibinfo{title}{Strange attractor simulated on a
  quantum computer},{''} \bibinfo{journal}{Eur. Phys. J. D}
  \textbf{\bibinfo{volume}{22}},  \bibinfo{pages}{127}.

\bibitem[{\citenamefont{Tewari} \emph{et~al.}(2006)\citenamefont{Tewari,
  Scarola, Senthil, and Sarma}}]{Tew06}
\bibinfo{author}{\bibnamefont{Tewari}, \bibfnamefont{S.}},
  \bibinfo{author}{\bibfnamefont{V.}~\bibnamefont{Scarola}},
  \bibinfo{author}{\bibfnamefont{T.}~\bibnamefont{Senthil}}, and
  \bibinfo{author}{\bibfnamefont{S.~D.} \bibnamefont{Sarma}},
  \bibinfo{year}{2006}, {``}\bibinfo{title}{Emergence of artificial photons in
  an optical lattice},{''} \bibinfo{journal}{Phys. Rev. Lett.}
  \textbf{\bibinfo{volume}{97}},  \bibinfo{pages}{200401}.

\bibitem[{\citenamefont{Thouless}(1972)}]{Thou}
\bibinfo{author}{\bibnamefont{Thouless}, \bibfnamefont{D.~J.}},
  \bibinfo{year}{1972}, \emph{\bibinfo{title}{The quantum mechanics of
  many-body systems}} (\bibinfo{publisher}{Academic Press}).

\bibitem[{\citenamefont{Tillmann} \emph{et~al.}(2013)\citenamefont{Tillmann,
  Daki\'c, Heilmann, Nolte, Szameit, and Walther}}]{Til13}
\bibinfo{author}{\bibnamefont{Tillmann}, \bibfnamefont{M.}},
  \bibinfo{author}{\bibfnamefont{B.}~\bibnamefont{Daki\'c}},
  \bibinfo{author}{\bibfnamefont{R.}~\bibnamefont{Heilmann}},
  \bibinfo{author}{\bibfnamefont{S.}~\bibnamefont{Nolte}},
  \bibinfo{author}{\bibfnamefont{A.}~\bibnamefont{Szameit}}, and
  \bibinfo{author}{\bibfnamefont{P.}~\bibnamefont{Walther}},
  \bibinfo{year}{2013}, {``}\bibinfo{title}{Experimental boson sampling},{''}
  \bibinfo{journal}{Nature Photonics} \textbf{\bibinfo{volume}{7}},
  \bibinfo{pages}{540 -- 544}.

\bibitem[{\citenamefont{Timoney} \emph{et~al.}(2011)\citenamefont{Timoney,
  Baumgart, Johanning, Varon, Plenio, Retzker, and Wunderlich}}]{Tim11}
\bibinfo{author}{\bibnamefont{Timoney}, \bibfnamefont{N.}},
  \bibinfo{author}{\bibfnamefont{I.}~\bibnamefont{Baumgart}},
  \bibinfo{author}{\bibfnamefont{M.}~\bibnamefont{Johanning}},
  \bibinfo{author}{\bibfnamefont{A.~F.} \bibnamefont{Varon}},
  \bibinfo{author}{\bibfnamefont{M.~B.} \bibnamefont{Plenio}},
  \bibinfo{author}{\bibfnamefont{A.}~\bibnamefont{Retzker}}, and
  \bibinfo{author}{\bibfnamefont{C.}~\bibnamefont{Wunderlich}},
  \bibinfo{year}{2011}, {``}\bibinfo{title}{Quantum gates and memory using
  microwave-dressed states},{''} \bibinfo{journal}{Nature}
  \textbf{\bibinfo{volume}{476}},  \bibinfo{pages}{185--188}.

\bibitem[{\citenamefont{Torrontegui}
  \emph{et~al.}(2011)\citenamefont{Torrontegui, Ruschhaupt, Guéry-Odelin, and
  Muga}}]{Tor11}
\bibinfo{author}{\bibnamefont{Torrontegui}, \bibfnamefont{E.}},
  \bibinfo{author}{\bibfnamefont{A.}~\bibnamefont{Ruschhaupt}},
  \bibinfo{author}{\bibfnamefont{D.}~\bibnamefont{Guéry-Odelin}}, and
  \bibinfo{author}{\bibfnamefont{J.~G.} \bibnamefont{Muga}},
  \bibinfo{year}{2011}, {``}\bibinfo{title}{Simulation of quantum collinear
  chemical reactions with ultracold atoms},{''} \bibinfo{journal}{J. Phys. B}
  \textbf{\bibinfo{volume}{44}},  \bibinfo{pages}{195302}.

\bibitem[{\citenamefont{Troyer and Wiese}(2005)}]{Tro05}
\bibinfo{author}{\bibnamefont{Troyer}, \bibfnamefont{M.}}, and
  \bibinfo{author}{\bibfnamefont{U.-J.} \bibnamefont{Wiese}},
  \bibinfo{year}{2005}, {``}\bibinfo{title}{Computational complexity and
  fundamental limitations to Fermionic quantum Monte Carlo simulations},{''}
  \bibinfo{journal}{Phys. Rev. Lett.} \textbf{\bibinfo{volume}{94}},
  \bibinfo{pages}{170201}.

\bibitem[{\citenamefont{Tseng} \emph{et~al.}(2000)\citenamefont{Tseng, Somaroo,
  Sharf, Knill, Laflamme, Havel, and Cory}}]{Tse00}
\bibinfo{author}{\bibnamefont{Tseng}, \bibfnamefont{C.~H.}},
  \bibinfo{author}{\bibfnamefont{S.}~\bibnamefont{Somaroo}},
  \bibinfo{author}{\bibfnamefont{Y.}~\bibnamefont{Sharf}},
  \bibinfo{author}{\bibfnamefont{E.}~\bibnamefont{Knill}},
  \bibinfo{author}{\bibfnamefont{R.}~\bibnamefont{Laflamme}},
  \bibinfo{author}{\bibfnamefont{T.~F.} \bibnamefont{Havel}}, and
  \bibinfo{author}{\bibfnamefont{D.~G.} \bibnamefont{Cory}},
  \bibinfo{year}{2000}, {``}\bibinfo{title}{Quantum simulation with natural
  decoherence},{''} \bibinfo{journal}{Phys. Rev. A}
  \textbf{\bibinfo{volume}{62}},  \bibinfo{pages}{032309}.

\bibitem[{\citenamefont{Tsokomos} \emph{et~al.}(2010)\citenamefont{Tsokomos,
  Ashhab, and Nori}}]{Tso10}
\bibinfo{author}{\bibnamefont{Tsokomos}, \bibfnamefont{D.}},
  \bibinfo{author}{\bibfnamefont{S.}~\bibnamefont{Ashhab}}, and
  \bibinfo{author}{\bibfnamefont{F.}~\bibnamefont{Nori}}, \bibinfo{year}{2010},
  {``}\bibinfo{title}{Using superconducting circuits to engineer exotic lattice
  systems},{''} \bibinfo{journal}{Phys. Rev. A} \textbf{\bibinfo{volume}{82}},
  \bibinfo{pages}{052311}.

\bibitem[{\citenamefont{Tsomokos} \emph{et~al.}(2008)\citenamefont{Tsomokos,
  Ashhab, and Nori}}]{Tso08}
\bibinfo{author}{\bibnamefont{Tsomokos}, \bibfnamefont{D.~I.}},
  \bibinfo{author}{\bibfnamefont{S.}~\bibnamefont{Ashhab}}, and
  \bibinfo{author}{\bibfnamefont{F.}~\bibnamefont{Nori}}, \bibinfo{year}{2008},
  {``}\bibinfo{title}{Fully connected network of superconducting qubits in a
  cavity},{''} \bibinfo{journal}{New J. Phys.} \textbf{\bibinfo{volume}{10}},
  \bibinfo{pages}{113020}.

\bibitem[{\citenamefont{Uehlinger} \emph{et~al.}(2013)\citenamefont{Uehlinger,
  Jotzu, Messer, Greif, Hofstetter, Bissbort, and Esslinger}}]{Ueh13}
\bibinfo{author}{\bibnamefont{Uehlinger}, \bibfnamefont{T.}},
  \bibinfo{author}{\bibfnamefont{G.}~\bibnamefont{Jotzu}},
  \bibinfo{author}{\bibfnamefont{M.}~\bibnamefont{Messer}},
  \bibinfo{author}{\bibfnamefont{D.}~\bibnamefont{Greif}},
  \bibinfo{author}{\bibfnamefont{W.}~\bibnamefont{Hofstetter}},
  \bibinfo{author}{\bibfnamefont{U.}~\bibnamefont{Bissbort}}, and
  \bibinfo{author}{\bibfnamefont{T.}~\bibnamefont{Esslinger}},
  \bibinfo{year}{2013}, {``}\bibinfo{title}{Artificial graphene with tunable
  interactions},{''} \bibinfo{journal}{Phys. Rev. Lett.}
  \textbf{\bibinfo{volume}{111}},  \bibinfo{pages}{185307}.

\bibitem[{\citenamefont{Verstraete and Cirac}(2004)}]{Ver04}
\bibinfo{author}{\bibnamefont{Verstraete}, \bibfnamefont{F.}}, and
  \bibinfo{author}{\bibfnamefont{J.~I.} \bibnamefont{Cirac}},
  \bibinfo{year}{2004}, {``}\bibinfo{title}{Renormalization algorithms for
  quantum-many body systems in two and higher dimensions},{''}
  \eprint{arXiv:cond-mat/0407066v1}.

\bibitem[{\citenamefont{Verstraete}
  \emph{et~al.}(2009)\citenamefont{Verstraete, Cirac, and Latorre}}]{Ver08}
\bibinfo{author}{\bibnamefont{Verstraete}, \bibfnamefont{F.}},
  \bibinfo{author}{\bibfnamefont{J.~I.} \bibnamefont{Cirac}}, and
  \bibinfo{author}{\bibfnamefont{J.~I.} \bibnamefont{Latorre}},
  \bibinfo{year}{2009}, {``}\bibinfo{title}{Quantum circuits for strongly
  correlated quantum systems},{''} \bibinfo{journal}{Phys. Rev. A}
  \textbf{\bibinfo{volume}{79}},  \bibinfo{pages}{032316}.

\bibitem[{\citenamefont{Verstraete}
  \emph{et~al.}(2008)\citenamefont{Verstraete, Murg, and Cirac}}]{VMC08}
\bibinfo{author}{\bibnamefont{Verstraete}, \bibfnamefont{F.}},
  \bibinfo{author}{\bibfnamefont{V.}~\bibnamefont{Murg}}, and
  \bibinfo{author}{\bibfnamefont{J.~I.} \bibnamefont{Cirac}},
  \bibinfo{year}{2008}, {``}\bibinfo{title}{Matrix product states, projected
  entangled pair states, and variational renormalization group methods for
  quantum spin systems},{''} \bibinfo{journal}{Advances in Physics}
  \textbf{\bibinfo{volume}{57}},  \bibinfo{pages}{143--224}.

\bibitem[{\citenamefont{Verstraete}
  \emph{et~al.}(2004)\citenamefont{Verstraete, Porras, and Cirac}}]{Ver05}
\bibinfo{author}{\bibnamefont{Verstraete}, \bibfnamefont{F.}},
  \bibinfo{author}{\bibfnamefont{D.}~\bibnamefont{Porras}}, and
  \bibinfo{author}{\bibfnamefont{J.~I.} \bibnamefont{Cirac}},
  \bibinfo{year}{2004}, {``}\bibinfo{title}{Density matrix renormalization
  group and periodic boundary conditions: A quantum information
  perspective},{''} \bibinfo{journal}{Phys. Rev. Lett.}
  \textbf{\bibinfo{volume}{93}},  \bibinfo{pages}{227205}.

\bibitem[{\citenamefont{Vidal}(2008)}]{Vid08}
\bibinfo{author}{\bibnamefont{Vidal}, \bibfnamefont{G.}}, \bibinfo{year}{2008},
  {``}\bibinfo{title}{Class of quantum many-body states that can be efficiently
  simulated},{''} \bibinfo{journal}{Phys. Rev. Lett.}
  \textbf{\bibinfo{volume}{101}},  \bibinfo{pages}{110501}.

\bibitem[{\citenamefont{Vollbrecht and Cirac}(2008)}]{Vol08}
\bibinfo{author}{\bibnamefont{Vollbrecht}, \bibfnamefont{K.~G.}}, and
  \bibinfo{author}{\bibfnamefont{J.~I.} \bibnamefont{Cirac}},
  \bibinfo{year}{2008}, {``}\bibinfo{title}{Quantum simulators, continuous-time
  automata, and translationally invariant systems},{''} \bibinfo{journal}{Phys.
  Rev. Lett.} \textbf{\bibinfo{volume}{100}},  \bibinfo{pages}{010501}.

\bibitem[{\citenamefont{Vollbrecht and Cirac}(2009)}]{Voll08}
\bibinfo{author}{\bibnamefont{Vollbrecht}, \bibfnamefont{K.~G.~H.}}, and
  \bibinfo{author}{\bibfnamefont{J.~I.} \bibnamefont{Cirac}},
  \bibinfo{year}{2009}, {``}\bibinfo{title}{Quantum simulations based on
  measurements and feedback control},{''} \bibinfo{journal}{Phys. Rev. A}
  \textbf{\bibinfo{volume}{79}},  \bibinfo{pages}{042305}.

\bibitem[{\citenamefont{Volovik}(2009)}]{Vol09}
\bibinfo{author}{\bibnamefont{Volovik}, \bibfnamefont{G.~E.}},
  \bibinfo{year}{2009}, \emph{\bibinfo{title}{The Universe in a Helium
  Droplet}} (\bibinfo{publisher}{Oxford University Press}).

\bibitem[{\citenamefont{Wallraff} \emph{et~al.}(2004)\citenamefont{Wallraff,
  Schuster, Blais, Frunzio, Huang, Majer, Kumar, Girvin, and
  Schoelkopf}}]{Wal04}
\bibinfo{author}{\bibnamefont{Wallraff}, \bibfnamefont{A.}},
  \bibinfo{author}{\bibfnamefont{D.~I.} \bibnamefont{Schuster}},
  \bibinfo{author}{\bibfnamefont{A.}~\bibnamefont{Blais}},
  \bibinfo{author}{\bibfnamefont{L.}~\bibnamefont{Frunzio}},
  \bibinfo{author}{\bibfnamefont{R.-S.} \bibnamefont{Huang}},
  \bibinfo{author}{\bibfnamefont{J.}~\bibnamefont{Majer}},
  \bibinfo{author}{\bibfnamefont{S.}~\bibnamefont{Kumar}},
  \bibinfo{author}{\bibfnamefont{S.~M.} \bibnamefont{Girvin}}, and
  \bibinfo{author}{\bibfnamefont{R.~J.} \bibnamefont{Schoelkopf}},
  \bibinfo{year}{2004}, {``}\bibinfo{title}{Strong coupling of a single photon
  to a superconducting qubit using circuit quantum electrodynamics},{''}
  \bibinfo{journal}{Nature (London)} \textbf{\bibinfo{volume}{431}},
  \bibinfo{pages}{162--167}.

\bibitem[{\citenamefont{Wang and Yang}(2006)}]{Wan06}
\bibinfo{author}{\bibnamefont{Wang}, \bibfnamefont{A.~M.}}, and
  \bibinfo{author}{\bibfnamefont{X.}~\bibnamefont{Yang}}, \bibinfo{year}{2006},
  {``}\bibinfo{title}{Quantum simulation of pairing models on an NMR quantum
  computer},{''} \bibinfo{journal}{Phys. Lett. A}
  \textbf{\bibinfo{volume}{352}},  \bibinfo{pages}{304--308}.

\bibitem[{\citenamefont{Wang} \emph{et~al.}(2009)\citenamefont{Wang, Ashhab,
  and Nori}}]{Wan09}
\bibinfo{author}{\bibnamefont{Wang}, \bibfnamefont{H.}},
  \bibinfo{author}{\bibfnamefont{S.}~\bibnamefont{Ashhab}}, and
  \bibinfo{author}{\bibfnamefont{F.}~\bibnamefont{Nori}}, \bibinfo{year}{2009},
  {``}\bibinfo{title}{Efficient quantum algorithm for preparing
  molecular-system-like states on a quantum computer},{''}
  \bibinfo{journal}{Phys. Rev. A} \textbf{\bibinfo{volume}{79}},
  \bibinfo{pages}{042335}.

\bibitem[{\citenamefont{Wang} \emph{et~al.}(2011)\citenamefont{Wang, Ashhab,
  and Nori}}]{Wang11}
\bibinfo{author}{\bibnamefont{Wang}, \bibfnamefont{H.}},
  \bibinfo{author}{\bibfnamefont{S.}~\bibnamefont{Ashhab}}, and
  \bibinfo{author}{\bibfnamefont{F.}~\bibnamefont{Nori}}, \bibinfo{year}{2011},
  {``}\bibinfo{title}{Quantum algorithm for simulating the dynamics of an open
  quantum system},{''} \bibinfo{journal}{Phys. Rev. A}
  \textbf{\bibinfo{volume}{83}},  \bibinfo{pages}{062317}.

\bibitem[{\citenamefont{Wang} \emph{et~al.}(2008)\citenamefont{Wang, Kais,
  Aspuru-Guzik, and Hoffmann}}]{Wan08}
\bibinfo{author}{\bibnamefont{Wang}, \bibfnamefont{H.}},
  \bibinfo{author}{\bibfnamefont{S.}~\bibnamefont{Kais}},
  \bibinfo{author}{\bibfnamefont{A.}~\bibnamefont{Aspuru-Guzik}}, and
  \bibinfo{author}{\bibfnamefont{M.~R.} \bibnamefont{Hoffmann}},
  \bibinfo{year}{2008}, {``}\bibinfo{title}{Quantum algorithm for obtaining the
  energy spectrum of molecular systems},{''} \bibinfo{journal}{Phys. Chem.
  Chem. Phys} \textbf{\bibinfo{volume}{10}},  \bibinfo{pages}{5388--5393}.

\bibitem[{\citenamefont{Wang}
  \emph{et~al.}(2010{\natexlab{a}})\citenamefont{Wang, Wu, Liu, and
  Nori}}]{WanWu10}
\bibinfo{author}{\bibnamefont{Wang}, \bibfnamefont{H.}},
  \bibinfo{author}{\bibfnamefont{L.-A.} \bibnamefont{Wu}},
  \bibinfo{author}{\bibfnamefont{Y.-X.} \bibnamefont{Liu}}, and
  \bibinfo{author}{\bibfnamefont{F.}~\bibnamefont{Nori}},
  \bibinfo{year}{2010}{\natexlab{a}}, {``}\bibinfo{title}{Measurement-based
  quantum phase estimation algorithm for finding eigenvalues of non-unitary
  matrices},{''} \bibinfo{journal}{Phys. Rev. A} \textbf{\bibinfo{volume}{82}},
   \bibinfo{pages}{062303}.

\bibitem[{\citenamefont{Wang}
  \emph{et~al.}(2010{\natexlab{b}})\citenamefont{Wang, Liu, Feng, Yang, and
  Wang}}]{Wan10}
\bibinfo{author}{\bibnamefont{Wang}, \bibfnamefont{K.}},
  \bibinfo{author}{\bibfnamefont{T.}~\bibnamefont{Liu}},
  \bibinfo{author}{\bibfnamefont{M.}~\bibnamefont{Feng}},
  \bibinfo{author}{\bibfnamefont{W.}~\bibnamefont{Yang}}, and
  \bibinfo{author}{\bibfnamefont{K.}~\bibnamefont{Wang}},
  \bibinfo{year}{2010}{\natexlab{b}}, {``}\bibinfo{title}{Parity-relevant
  zitterbewegung and quantum simulation by a single trapped ion},{''}
  \bibinfo{journal}{Phys. Rev. A} \textbf{\bibinfo{volume}{82}},
  \bibinfo{pages}{064501}.

\bibitem[{\citenamefont{Wang and Zanardi}(2002)}]{Wan02}
\bibinfo{author}{\bibnamefont{Wang}, \bibfnamefont{X.}}, and
  \bibinfo{author}{\bibfnamefont{P.}~\bibnamefont{Zanardi}},
  \bibinfo{year}{2002}, {``}\bibinfo{title}{Simulation of many-body
  interactions by conditional geometric phases},{''} \bibinfo{journal}{Phys.
  Rev. A} \textbf{\bibinfo{volume}{65}},  \bibinfo{pages}{032327}.

\bibitem[{\citenamefont{Ward} \emph{et~al.}(2009)\citenamefont{Ward, Kassal,
  and Aspuru-Guzik}}]{War08}
\bibinfo{author}{\bibnamefont{Ward}, \bibfnamefont{N.~J.}},
  \bibinfo{author}{\bibfnamefont{I.}~\bibnamefont{Kassal}}, and
  \bibinfo{author}{\bibfnamefont{A.}~\bibnamefont{Aspuru-Guzik}},
  \bibinfo{year}{2009}, {``}\bibinfo{title}{Preparation of many-body states for
  quantum simulation},{''} \bibinfo{journal}{J. Chem. Phys.}
  \textbf{\bibinfo{volume}{130}},  \bibinfo{pages}{194105}.

\bibitem[{\citenamefont{Wei and Xue}(1997)}]{Wei97}
\bibinfo{author}{\bibnamefont{Wei}, \bibfnamefont{H.}}, and
  \bibinfo{author}{\bibfnamefont{X.}~\bibnamefont{Xue}}, \bibinfo{year}{1997},
  {``}\bibinfo{title}{Quantum isomorphic simulation},{''}
  \eprint{arXiv:quant-ph/9702050}.

\bibitem[{\citenamefont{Weimer} \emph{et~al.}(2011)\citenamefont{Weimer,
  M\"uller, B\"uchler, and Lesanovsky}}]{Weim11}
\bibinfo{author}{\bibnamefont{Weimer}, \bibfnamefont{H.}},
  \bibinfo{author}{\bibfnamefont{M.}~\bibnamefont{M\"uller}},
  \bibinfo{author}{\bibfnamefont{H.~P.} \bibnamefont{B\"uchler}}, and
  \bibinfo{author}{\bibfnamefont{I.}~\bibnamefont{Lesanovsky}},
  \bibinfo{year}{2011}, {``}\bibinfo{title}{Digital quantum simulation with
  Rydberg atoms},{''} \bibinfo{journal}{Quant. Inf. Proc.}
  \textbf{\bibinfo{volume}{10}},  \bibinfo{pages}{885--906}.

\bibitem[{\citenamefont{Weimer} \emph{et~al.}(2010)\citenamefont{Weimer,
  M\"uller, Lesanovsky, and B\"uchler}}]{Wei10}
\bibinfo{author}{\bibnamefont{Weimer}, \bibfnamefont{H.}},
  \bibinfo{author}{\bibfnamefont{M.}~\bibnamefont{M\"uller}},
  \bibinfo{author}{\bibfnamefont{Z.~P.} \bibnamefont{Lesanovsky},
  \bibfnamefont{I.}}, and
  \bibinfo{author}{\bibfnamefont{H.}~\bibnamefont{B\"uchler}},
  \bibinfo{year}{2010}, {``}\bibinfo{title}{A Rydberg quantum simulator},{''}
  \bibinfo{journal}{Nature Physics} \textbf{\bibinfo{volume}{6}},
  \bibinfo{pages}{382--388}.

\bibitem[{\citenamefont{Weimer} \emph{et~al.}(2013)\citenamefont{Weimer, Yao,
  and Lukin}}]{Wei13}
\bibinfo{author}{\bibnamefont{Weimer}, \bibfnamefont{H.}},
  \bibinfo{author}{\bibfnamefont{N.~Y.} \bibnamefont{Yao}}, and
  \bibinfo{author}{\bibfnamefont{M.~D.} \bibnamefont{Lukin}},
  \bibinfo{year}{2013}, {``}\bibinfo{title}{Collectively Enhanced Interactions
  in Solid-State Spin Qubits},{''} \bibinfo{journal}{Phys. Rev. Lett.}
  \textbf{\bibinfo{volume}{110}},  \bibinfo{pages}{067601}.

\bibitem[{\citenamefont{Weinfurtner}
  \emph{et~al.}(2011)\citenamefont{Weinfurtner, Tedford, Penrice, Unruh, and
  Lawrence}}]{Wei11}
\bibinfo{author}{\bibnamefont{Weinfurtner}, \bibfnamefont{S.}},
  \bibinfo{author}{\bibfnamefont{E.}~\bibnamefont{Tedford}},
  \bibinfo{author}{\bibfnamefont{M.}~\bibnamefont{Penrice}},
  \bibinfo{author}{\bibfnamefont{W.}~\bibnamefont{Unruh}}, and
  \bibinfo{author}{\bibfnamefont{G.}~\bibnamefont{Lawrence}},
  \bibinfo{year}{2011}, {``}\bibinfo{title}{Measurement of stimulated Hawking
  emission in an analogue system},{''} \bibinfo{journal}{Phys. Rev. Lett.}
  \textbf{\bibinfo{volume}{106}},  \bibinfo{pages}{021302}.

\bibitem[{\citenamefont{Weinstein} \emph{et~al.}(2002)\citenamefont{Weinstein,
  Lloyd, Emerson, and Cory}}]{Wei02}
\bibinfo{author}{\bibnamefont{Weinstein}, \bibfnamefont{Y.~S.}},
  \bibinfo{author}{\bibfnamefont{S.}~\bibnamefont{Lloyd}},
  \bibinfo{author}{\bibfnamefont{J.~V.} \bibnamefont{Emerson}}, and
  \bibinfo{author}{\bibfnamefont{D.~G.} \bibnamefont{Cory}},
  \bibinfo{year}{2002}, {``}\bibinfo{title}{Experimental implementation of the
  quantum baker's map},{''} \bibinfo{journal}{Phys. Rev. Lett.}
  \textbf{\bibinfo{volume}{89}},  \bibinfo{pages}{157902}.

\bibitem[{\citenamefont{Weitenberg}
  \emph{et~al.}(2011)\citenamefont{Weitenberg, Endres, Sherson, Cheneau,
  Schauss, Fukuhara, Bloch, and Kuhr}}]{Weit11}
\bibinfo{author}{\bibnamefont{Weitenberg}, \bibfnamefont{C.}},
  \bibinfo{author}{\bibfnamefont{M.}~\bibnamefont{Endres}},
  \bibinfo{author}{\bibfnamefont{J.~F.} \bibnamefont{Sherson}},
  \bibinfo{author}{\bibfnamefont{M.}~\bibnamefont{Cheneau}},
  \bibinfo{author}{\bibfnamefont{P.}~\bibnamefont{Schauss}},
  \bibinfo{author}{\bibfnamefont{T.}~\bibnamefont{Fukuhara}},
  \bibinfo{author}{\bibfnamefont{I.}~\bibnamefont{Bloch}}, and
  \bibinfo{author}{\bibfnamefont{S.}~\bibnamefont{Kuhr}}, \bibinfo{year}{2011},
  {``}\bibinfo{title}{Single-spin addressing in an atomic Mott insulator},{''}
  \bibinfo{journal}{Nature} \textbf{\bibinfo{volume}{471}},
  \bibinfo{pages}{319--324}.

\bibitem[{\citenamefont{Whitfield} \emph{et~al.}(2011)\citenamefont{Whitfield,
  Biamonte, and Aspuru-Guzik}}]{Whi10}
\bibinfo{author}{\bibnamefont{Whitfield}, \bibfnamefont{J.~D.}},
  \bibinfo{author}{\bibfnamefont{J.}~\bibnamefont{Biamonte}}, and
  \bibinfo{author}{\bibfnamefont{A.}~\bibnamefont{Aspuru-Guzik}},
  \bibinfo{year}{2011}, {``}\bibinfo{title}{Quantum computing resource estimate
  of molecular energy simulation},{''} \bibinfo{journal}{Quant. Info. Proc.}
  \textbf{\bibinfo{volume}{10}},  \bibinfo{pages}{885}.

\bibitem[{\citenamefont{Wiebe} \emph{et~al.}(2011)\citenamefont{Wiebe, Berry,
  Hoyer, and Sanders}}]{Wie11}
\bibinfo{author}{\bibnamefont{Wiebe}, \bibfnamefont{N.}},
  \bibinfo{author}{\bibfnamefont{D.~W.} \bibnamefont{Berry}},
  \bibinfo{author}{\bibfnamefont{P.}~\bibnamefont{Hoyer}}, and
  \bibinfo{author}{\bibfnamefont{B.~C.} \bibnamefont{Sanders}},
  \bibinfo{year}{2011}, {``}\bibinfo{title}{Simulating quantum dynamics on a
  quantum computer},{''} \bibinfo{journal}{J. Phys. A: Math. Theor.}
  \textbf{\bibinfo{volume}{44}},  \bibinfo{pages}{445308}.

\bibitem[{\citenamefont{Wiese}(2013)}]{Wie13}
\bibinfo{author}{\bibnamefont{Wiese}, \bibfnamefont{U.-J.}},
  \bibinfo{year}{2013}, {``}\bibinfo{title}{Ultracold Quantum Gases and Lattice
  Systems: Quantum Simulation of Lattice Gauge Theories},{''}
  \bibinfo{journal}{arXiv:1305.1602} .

\bibitem[{\citenamefont{Wiesner}(1996)}]{Wis96}
\bibinfo{author}{\bibnamefont{Wiesner}, \bibfnamefont{S.}},
  \bibinfo{year}{1996}, {``}\bibinfo{title}{Simulations of many-body quantum
  systems by a quantum computer},{''} \eprint{arXiv:quant-ph/9603028}.

\bibitem[{\citenamefont{Wilson} \emph{et~al.}(2011)\citenamefont{Wilson,
  Johansson, Pourkabirian, Johansson, Duty, Nori, and Delsing}}]{Wil11}
\bibinfo{author}{\bibnamefont{Wilson}, \bibfnamefont{C.}},
  \bibinfo{author}{\bibfnamefont{G.}~\bibnamefont{Johansson}},
  \bibinfo{author}{\bibfnamefont{A.}~\bibnamefont{Pourkabirian}},
  \bibinfo{author}{\bibfnamefont{J.}~\bibnamefont{Johansson}},
  \bibinfo{author}{\bibfnamefont{T.}~\bibnamefont{Duty}},
  \bibinfo{author}{\bibfnamefont{F.}~\bibnamefont{Nori}}, and
  \bibinfo{author}{\bibfnamefont{P.}~\bibnamefont{Delsing}},
  \bibinfo{year}{2011}, {``}\bibinfo{title}{Observation of the dynamical
  Casimir effect in a superconducting circuit},{''} \bibinfo{journal}{Nature}
  \textbf{\bibinfo{volume}{479}},  \bibinfo{pages}{376--379}.

\bibitem[{\citenamefont{Wineland} \emph{et~al.}(1998)\citenamefont{Wineland,
  Monroe, Itano, King, Leibfried, Myatt, and Wood}}]{Win98}
\bibinfo{author}{\bibnamefont{Wineland}, \bibfnamefont{D.}},
  \bibinfo{author}{\bibfnamefont{C.}~\bibnamefont{Monroe}},
  \bibinfo{author}{\bibfnamefont{W.}~\bibnamefont{Itano}},
  \bibinfo{author}{\bibfnamefont{B.}~\bibnamefont{King}},
  \bibinfo{author}{\bibfnamefont{D.}~\bibnamefont{Leibfried}},
  \bibinfo{author}{\bibfnamefont{C.}~\bibnamefont{Myatt}}, and
  \bibinfo{author}{\bibfnamefont{C.}~\bibnamefont{Wood}}, \bibinfo{year}{1998},
  {``}\bibinfo{title}{Trapped-ion quantum simulator},{''}
  \bibinfo{journal}{Phys. Scripta} \textbf{\bibinfo{volume}{T76}},
  \bibinfo{pages}{147--151}.

\bibitem[{\citenamefont{Witthaut}(2010)}]{Wit10}
\bibinfo{author}{\bibnamefont{Witthaut}, \bibfnamefont{D.}},
  \bibinfo{year}{2010}, {``}\bibinfo{title}{Quantum walks and quantum
  simulations with Bloch-oscillating spinor atoms},{''} \bibinfo{journal}{Phys.
  Rev. A} \textbf{\bibinfo{volume}{82}},  \bibinfo{pages}{033602}.

\bibitem[{\citenamefont{Wocjan}
  \emph{et~al.}(2002{\natexlab{a}})\citenamefont{Wocjan, Janzing, and
  Beth}}]{Wocj02}
\bibinfo{author}{\bibnamefont{Wocjan}, \bibfnamefont{P.}},
  \bibinfo{author}{\bibfnamefont{D.}~\bibnamefont{Janzing}}, and
  \bibinfo{author}{\bibfnamefont{T.}~\bibnamefont{Beth}},
  \bibinfo{year}{2002}{\natexlab{a}}, {``}\bibinfo{title}{Simulating arbitrary
  pair-interactions by a given Hamiltonian: graph-theoretical bounds on the
  time complexity},{''} \bibinfo{journal}{Quant. Inf. Comput.}
  \textbf{\bibinfo{volume}{2}},  \bibinfo{pages}{117--132}.

\bibitem[{\citenamefont{Wocjan}
  \emph{et~al.}(2002{\natexlab{b}})\citenamefont{Wocjan, R\"otteler, Janzing,
  and Beth}}]{Wo02}
\bibinfo{author}{\bibnamefont{Wocjan}, \bibfnamefont{P.}},
  \bibinfo{author}{\bibfnamefont{M.}~\bibnamefont{R\"otteler}},
  \bibinfo{author}{\bibfnamefont{D.}~\bibnamefont{Janzing}}, and
  \bibinfo{author}{\bibfnamefont{T.}~\bibnamefont{Beth}},
  \bibinfo{year}{2002}{\natexlab{b}}, {``}\bibinfo{title}{Simulating
  Hamiltonians in quantum networks: efficient schemes and complexity
  bounds},{''} \bibinfo{journal}{Phys. Rev. A} \textbf{\bibinfo{volume}{65}},
  \bibinfo{pages}{042309}.

\bibitem[{\citenamefont{Wocjan}
  \emph{et~al.}(2002{\natexlab{c}})\citenamefont{Wocjan, R\"otteler, Janzing,
  and Beth}}]{Woc02}
\bibinfo{author}{\bibnamefont{Wocjan}, \bibfnamefont{P.}},
  \bibinfo{author}{\bibfnamefont{M.}~\bibnamefont{R\"otteler}},
  \bibinfo{author}{\bibfnamefont{D.}~\bibnamefont{Janzing}}, and
  \bibinfo{author}{\bibfnamefont{T.}~\bibnamefont{Beth}},
  \bibinfo{year}{2002}{\natexlab{c}}, {``}\bibinfo{title}{Universal simulation
  of Hamiltonians using a finite set of control operation},{''}
  \bibinfo{journal}{Quant. Inf. Comput.} \textbf{\bibinfo{volume}{2}},
  \bibinfo{pages}{133--150}.

\bibitem[{\citenamefont{Wu} \emph{et~al.}(2002)\citenamefont{Wu, Byrd, and
  Lidar}}]{Wu02}
\bibinfo{author}{\bibnamefont{Wu}, \bibfnamefont{L.-A.}},
  \bibinfo{author}{\bibfnamefont{M.}~\bibnamefont{Byrd}}, and
  \bibinfo{author}{\bibfnamefont{D.}~\bibnamefont{Lidar}},
  \bibinfo{year}{2002}, {``}\bibinfo{title}{Polynomial-time simulation of
  pairing models on a quantum computer},{''} \bibinfo{journal}{Phys. Rev.
  Lett.} \textbf{\bibinfo{volume}{89}},  \bibinfo{pages}{057904}.

\bibitem[{\citenamefont{Wunderlich}(2009)}]{Wun09}
\bibinfo{author}{\bibnamefont{Wunderlich}, \bibfnamefont{H.~W.~C.}},
  \bibinfo{year}{2009}, {``}\bibinfo{title}{Two-dimensional cluster-state
  preparation with linear ion traps},{''} \bibinfo{journal}{Phys. Rev. A}
  \textbf{\bibinfo{volume}{79}},  \bibinfo{pages}{052324}.

\bibitem[{\citenamefont{W\"urtz} \emph{et~al.}(2009)\citenamefont{W\"urtz,
  Langen, Gericke, Koglbauer, and Ott}}]{Wur09}
\bibinfo{author}{\bibnamefont{W\"urtz}, \bibfnamefont{P.}},
  \bibinfo{author}{\bibfnamefont{T.}~\bibnamefont{Langen}},
  \bibinfo{author}{\bibfnamefont{T.}~\bibnamefont{Gericke}},
  \bibinfo{author}{\bibfnamefont{A.}~\bibnamefont{Koglbauer}}, and
  \bibinfo{author}{\bibfnamefont{H.}~\bibnamefont{Ott}}, \bibinfo{year}{2009},
  {``}\bibinfo{title}{Experimental demonstration of single-site addressability
  in a two-dimensional optical lattice},{''} \bibinfo{journal}{Phys. Rev.
  Lett.} \textbf{\bibinfo{volume}{103}},  \bibinfo{pages}{080404}.

\bibitem[{\citenamefont{Xue}(2011)}]{Xue11}
\bibinfo{author}{\bibnamefont{Xue}, \bibfnamefont{Z.-Y.}},
  \bibinfo{year}{2011}, {``}\bibinfo{title}{Simulation of anyonic fractional
  statistics of Kitaev's toric model in circuit QED},{''}
  \bibinfo{journal}{EPL} \textbf{\bibinfo{volume}{93}},
  \bibinfo{pages}{20007}.

\bibitem[{\citenamefont{Yamaguchi and Yamamoto}(2002)}]{Yam02}
\bibinfo{author}{\bibnamefont{Yamaguchi}, \bibfnamefont{F.}}, and
  \bibinfo{author}{\bibfnamefont{Y.}~\bibnamefont{Yamamoto}},
  \bibinfo{year}{2002}, {``}\bibinfo{title}{Quantum simulation of the $t$-$J$
  model},{''} \bibinfo{journal}{Superlattices and Microstructures}
  \textbf{\bibinfo{volume}{32}},  \bibinfo{pages}{343--345}.

\bibitem[{\citenamefont{Yang} \emph{et~al.}(2006)\citenamefont{Yang, Wang, Xu,
  and Du}}]{Yan06}
\bibinfo{author}{\bibnamefont{Yang}, \bibfnamefont{X.}},
  \bibinfo{author}{\bibfnamefont{A.~M.} \bibnamefont{Wang}},
  \bibinfo{author}{\bibfnamefont{F.}~\bibnamefont{Xu}}, and
  \bibinfo{author}{\bibfnamefont{J.}~\bibnamefont{Du}}, \bibinfo{year}{2006},
  {``}\bibinfo{title}{Experimental simulation of a pairing Hamiltonian on an
  NMR quantum computer},{''} \bibinfo{journal}{Chem. Phys. Lett.}
  \textbf{\bibinfo{volume}{422}},  \bibinfo{pages}{20--24}.

\bibitem[{\citenamefont{You} \emph{et~al.}(2013)\citenamefont{You, Geller, and
  Stancil}}]{You13}
\bibinfo{author}{\bibnamefont{You}, \bibfnamefont{H.}},
  \bibinfo{author}{\bibfnamefont{M.~R.} \bibnamefont{Geller}}, and
  \bibinfo{author}{\bibfnamefont{P.~C.} \bibnamefont{Stancil}},
  \bibinfo{year}{2013}, {``}\bibinfo{title}{Simulating the transverse Ising
  model on a quantum computer: Error correction with the surface code},{''}
  \bibinfo{journal}{Phys. Rev. A} \textbf{\bibinfo{volume}{87}},
  \bibinfo{pages}{032341}.

\bibitem[{\citenamefont{You} \emph{et~al.}(2007)\citenamefont{You, Liu, Sun,
  and Nori}}]{You07}
\bibinfo{author}{\bibnamefont{You}, \bibfnamefont{J.~Q.}},
  \bibinfo{author}{\bibfnamefont{Y.~X.} \bibnamefont{Liu}},
  \bibinfo{author}{\bibfnamefont{C.~P.} \bibnamefont{Sun}}, and
  \bibinfo{author}{\bibfnamefont{F.}~\bibnamefont{Nori}}, \bibinfo{year}{2007},
  {``}\bibinfo{title}{Persistent single-photon production by tunable on-chip
  micromaser with a superconducting quantum circuit},{''}
  \bibinfo{journal}{Phys. Rev. B} \textbf{\bibinfo{volume}{75}},
  \bibinfo{pages}{104516}.

\bibitem[{\citenamefont{You and Nori}(2003)}]{You03}
\bibinfo{author}{\bibnamefont{You}, \bibfnamefont{J.~Q.}}, and
  \bibinfo{author}{\bibfnamefont{F.}~\bibnamefont{Nori}}, \bibinfo{year}{2003},
  {``}\bibinfo{title}{Quantum information processing with superconducting
  qubits in a microwave field},{''} \bibinfo{journal}{Phys. Rev. B}
  \textbf{\bibinfo{volume}{68}},  \bibinfo{pages}{064509}.

\bibitem[{\citenamefont{You and Nori}(2005)}]{You05}
\bibinfo{author}{\bibnamefont{You}, \bibfnamefont{J.~Q.}}, and
  \bibinfo{author}{\bibfnamefont{F.}~\bibnamefont{Nori}}, \bibinfo{year}{2005},
  {``}\bibinfo{title}{Superconducting circuits and quantum information},{''}
  \bibinfo{journal}{Physics Today}
  \textbf{\bibinfo{volume}{58}}(\bibinfo{number}{11}),
  \bibinfo{pages}{42--47}.

\bibitem[{\citenamefont{You and Nori}(2011)}]{You11}
\bibinfo{author}{\bibnamefont{You}, \bibfnamefont{J.~Q.}}, and
  \bibinfo{author}{\bibfnamefont{F.}~\bibnamefont{Nori}}, \bibinfo{year}{2011},
  {``}\bibinfo{title}{Atomic physics and quantum optics using superconducting
  circuits},{''} \bibinfo{journal}{Nature} \textbf{\bibinfo{volume}{474}},
  \bibinfo{pages}{589--597}.

\bibitem[{\citenamefont{You} \emph{et~al.}(2010)\citenamefont{You, Shi, Hu, and
  Nori}}]{You08}
\bibinfo{author}{\bibnamefont{You}, \bibfnamefont{J.~Q.}},
  \bibinfo{author}{\bibfnamefont{X.-F.} \bibnamefont{Shi}},
  \bibinfo{author}{\bibfnamefont{X.}~\bibnamefont{Hu}}, and
  \bibinfo{author}{\bibfnamefont{F.}~\bibnamefont{Nori}}, \bibinfo{year}{2010},
  {``}\bibinfo{title}{Quantum emulation of a spin system with topologically
  protected ground states using superconducting quantum circuits},{''}
  \bibinfo{journal}{Phys. Rev. B} \textbf{\bibinfo{volume}{81}},
  \bibinfo{pages}{014505}.

\bibitem[{\citenamefont{You} \emph{et~al.}(2011)\citenamefont{You, Wang, Zhang,
  and Nori}}]{YWZ11}
\bibinfo{author}{\bibnamefont{You}, \bibfnamefont{J.~Q.}},
  \bibinfo{author}{\bibfnamefont{Z.~D.} \bibnamefont{Wang}},
  \bibinfo{author}{\bibfnamefont{W.}~\bibnamefont{Zhang}}, and
  \bibinfo{author}{\bibfnamefont{F.}~\bibnamefont{Nori}}, \bibinfo{year}{2011},
  {``}\bibinfo{title}{Manipulating and probing Majorana fermions using
  superconducting circuits},{''} \eprint{arXiv:1108.3712}.

\bibitem[{\citenamefont{Yung} \emph{et~al.}(2010)\citenamefont{Yung, Nagaj,
  Whitfield, and Aspuru-Guzik}}]{Yun10}
\bibinfo{author}{\bibnamefont{Yung}, \bibfnamefont{M.-H.}},
  \bibinfo{author}{\bibfnamefont{D.}~\bibnamefont{Nagaj}},
  \bibinfo{author}{\bibfnamefont{J.~D.} \bibnamefont{Whitfield}}, and
  \bibinfo{author}{\bibfnamefont{A.}~\bibnamefont{Aspuru-Guzik}},
  \bibinfo{year}{2010}, {``}\bibinfo{title}{Simulation of classical thermal
  states on a quantum computer: A transfer-matrix approach},{''}
  \bibinfo{journal}{Phys. Rev. A} \textbf{\bibinfo{volume}{82}},
  \bibinfo{pages}{060302}.

\bibitem[{\citenamefont{Zagoskin} \emph{et~al.}(2007)\citenamefont{Zagoskin,
  Savel'ev, and Nori}}]{Zag07}
\bibinfo{author}{\bibnamefont{Zagoskin}, \bibfnamefont{A.}},
  \bibinfo{author}{\bibfnamefont{S.}~\bibnamefont{Savel'ev}}, and
  \bibinfo{author}{\bibfnamefont{F.}~\bibnamefont{Nori}}, \bibinfo{year}{2007},
  {``}\bibinfo{title}{Modeling an adiabatic quantum computer via an exact map
  to a gas of particles},{''} \bibinfo{journal}{Phys. Rev. Lett.}
  \textbf{\bibinfo{volume}{98}},  \bibinfo{pages}{120503}.

\bibitem[{\citenamefont{Zagoskin}(1998)}]{Zago}
\bibinfo{author}{\bibnamefont{Zagoskin}, \bibfnamefont{A.~M.}},
  \bibinfo{year}{1998}, \emph{\bibinfo{title}{Quantum theory of many-body
  systems: techniques and applications}} (\bibinfo{publisher}{Springer}).

\bibitem[{\citenamefont{Zalka}(1998{\natexlab{a}})}]{Za98}
\bibinfo{author}{\bibnamefont{Zalka}, \bibfnamefont{C.}},
  \bibinfo{year}{1998}{\natexlab{a}}, {``}\bibinfo{title}{Efficient simulation
  of quantum systems by quantum computers},{''} \bibinfo{journal}{Fortschritte
  der Physik} \textbf{\bibinfo{volume}{46}},  \bibinfo{pages}{877--879}.

\bibitem[{\citenamefont{Zalka}(1998{\natexlab{b}})}]{Zal98}
\bibinfo{author}{\bibnamefont{Zalka}, \bibfnamefont{C.}},
  \bibinfo{year}{1998}{\natexlab{b}}, {``}\bibinfo{title}{Simulating quantum
  systems on a quantum computer},{''} \bibinfo{journal}{Proc. Roy. Soc. A}
  \textbf{\bibinfo{volume}{454}},  \bibinfo{pages}{313--322}.

\bibitem[{\citenamefont{Zhang} \emph{et~al.}(2009)\citenamefont{Zhang,
  Cucchietti, Chandrashekar, Laforest, Ryan, Ditty, Hubbard, Gamble, and
  Laflamme}}]{Zhan08}
\bibinfo{author}{\bibnamefont{Zhang}, \bibfnamefont{J.}},
  \bibinfo{author}{\bibfnamefont{F.~M.} \bibnamefont{Cucchietti}},
  \bibinfo{author}{\bibfnamefont{C.~M.} \bibnamefont{Chandrashekar}},
  \bibinfo{author}{\bibfnamefont{M.}~\bibnamefont{Laforest}},
  \bibinfo{author}{\bibfnamefont{C.~A.} \bibnamefont{Ryan}},
  \bibinfo{author}{\bibfnamefont{M.}~\bibnamefont{Ditty}},
  \bibinfo{author}{\bibfnamefont{A.}~\bibnamefont{Hubbard}},
  \bibinfo{author}{\bibfnamefont{J.~K.} \bibnamefont{Gamble}}, and
  \bibinfo{author}{\bibfnamefont{R.}~\bibnamefont{Laflamme}},
  \bibinfo{year}{2009}, {``}\bibinfo{title}{Direct observation of quantum
  criticality in Ising spin chains},{''} \bibinfo{journal}{Phys. Rev. A}
  \textbf{\bibinfo{volume}{79}},  \bibinfo{pages}{012305}.

\bibitem[{\citenamefont{Zhang} \emph{et~al.}(2008)\citenamefont{Zhang, Peng,
  Rajendran, and Suter}}]{Zha08}
\bibinfo{author}{\bibnamefont{Zhang}, \bibfnamefont{J.}},
  \bibinfo{author}{\bibfnamefont{X.}~\bibnamefont{Peng}},
  \bibinfo{author}{\bibfnamefont{N.}~\bibnamefont{Rajendran}}, and
  \bibinfo{author}{\bibfnamefont{D.}~\bibnamefont{Suter}},
  \bibinfo{year}{2008}, {``}\bibinfo{title}{Detection of quantum critical
  points by a probe qubit},{''} \bibinfo{journal}{Phys. Rev. Lett.}
  \textbf{\bibinfo{volume}{100}},  \bibinfo{pages}{100501}.

\bibitem[{\citenamefont{Zhang} \emph{et~al.}(2012)\citenamefont{Zhang, Yung,
  Laflamme, Aspuru-Guzik, and Baugh}}]{Zha12}
\bibinfo{author}{\bibnamefont{Zhang}, \bibfnamefont{J.}},
  \bibinfo{author}{\bibfnamefont{M.~H.} \bibnamefont{Yung}},
  \bibinfo{author}{\bibfnamefont{R.}~\bibnamefont{Laflamme}},
  \bibinfo{author}{\bibfnamefont{A.}~\bibnamefont{Aspuru-Guzik}}, and
  \bibinfo{author}{\bibfnamefont{J.}~\bibnamefont{Baugh}},
  \bibinfo{year}{2012}, {``}\bibinfo{title}{Digital quantum simulation of the
  statistical mechanics of a frustrated magnet},{''} \bibinfo{journal}{Nature
  Communications} \textbf{\bibinfo{volume}{3}},  \bibinfo{pages}{880}.

\bibitem[{\citenamefont{Zhou}
  \emph{et~al.}(2008{\natexlab{a}})\citenamefont{Zhou, Dong, Liu, Sun, and
  Nori}}]{Lan08}
\bibinfo{author}{\bibnamefont{Zhou}, \bibfnamefont{L.}},
  \bibinfo{author}{\bibfnamefont{H.}~\bibnamefont{Dong}},
  \bibinfo{author}{\bibfnamefont{Y.-X.} \bibnamefont{Liu}},
  \bibinfo{author}{\bibfnamefont{C.~P.} \bibnamefont{Sun}}, and
  \bibinfo{author}{\bibfnamefont{F.}~\bibnamefont{Nori}},
  \bibinfo{year}{2008}{\natexlab{a}}, {``}\bibinfo{title}{Quantum supercavity
  with atomic mirrors},{''} \bibinfo{journal}{Phys. Rev. A}
  \textbf{\bibinfo{volume}{78}},  \bibinfo{pages}{063827}.

\bibitem[{\citenamefont{Zhou}
  \emph{et~al.}(2008{\natexlab{b}})\citenamefont{Zhou, Gong, Liu, Sun, and
  Nori}}]{Zhou08}
\bibinfo{author}{\bibnamefont{Zhou}, \bibfnamefont{L.}},
  \bibinfo{author}{\bibfnamefont{Z.~R.} \bibnamefont{Gong}},
  \bibinfo{author}{\bibfnamefont{Y.-X.} \bibnamefont{Liu}},
  \bibinfo{author}{\bibfnamefont{C.~P.} \bibnamefont{Sun}}, and
  \bibinfo{author}{\bibfnamefont{F.}~\bibnamefont{Nori}},
  \bibinfo{year}{2008}{\natexlab{b}}, {``}\bibinfo{title}{Controllable
  scattering of a single photon inside a one-dimensional resonator
  waveguide},{''} \bibinfo{journal}{Phys. Rev. Lett.}
  \textbf{\bibinfo{volume}{101}},  \bibinfo{pages}{100501}.

\bibitem[{\citenamefont{Zinner and Jensen}(2008)}]{Zin08}
\bibinfo{author}{\bibnamefont{Zinner}, \bibfnamefont{N.~T.}}, and
  \bibinfo{author}{\bibfnamefont{A.~S.} \bibnamefont{Jensen}},
  \bibinfo{year}{2008}, {``}\bibinfo{title}{Common concepts in nuclear physics
  and ultracold atomic gasses},{''} \bibinfo{journal}{J. Phys.: Conf. Ser.}
  \textbf{\bibinfo{volume}{111}},  \bibinfo{pages}{012016}.

\bibitem[{\citenamefont{Zinner and Jensen}(2013)}]{Zin13}
\bibinfo{author}{\bibnamefont{Zinner}, \bibfnamefont{N.~T.}}, and
  \bibinfo{author}{\bibfnamefont{A.~S.} \bibnamefont{Jensen}},
  \bibinfo{year}{2013}, {``}\bibinfo{title}{Comparing and contrasting nuclei
  and cold atomic gases},{''} \bibinfo{journal}{J. Phys. G: Nucl. Part. Phys.}
  \textbf{\bibinfo{volume}{40}},  \bibinfo{pages}{053101}.

\bibitem[{\citenamefont{Zohar} \emph{et~al.}(2012)\citenamefont{Zohar, Cirac,
  and Reznik}}]{Zoh12}
\bibinfo{author}{\bibnamefont{Zohar}, \bibfnamefont{E.}},
  \bibinfo{author}{\bibfnamefont{J.~I.} \bibnamefont{Cirac}}, and
  \bibinfo{author}{\bibfnamefont{B.}~\bibnamefont{Reznik}},
  \bibinfo{year}{2012}, {``}\bibinfo{title}{Simulating Compact Quantum
  Electrodynamics with Ultracold Atoms: Probing Confinement and Nonperturbative
  Effects},{''} \bibinfo{journal}{Phys. Rev. Lett.}
  \textbf{\bibinfo{volume}{109}},  \bibinfo{pages}{125302}.

\bibitem[{\citenamefont{Zohar} \emph{et~al.}(2013)\citenamefont{Zohar, Cirac,
  and Reznik}}]{Zoh13}
\bibinfo{author}{\bibnamefont{Zohar}, \bibfnamefont{E.}},
  \bibinfo{author}{\bibfnamefont{J.~I.} \bibnamefont{Cirac}}, and
  \bibinfo{author}{\bibfnamefont{B.}~\bibnamefont{Reznik}},
  \bibinfo{year}{2013}, {``}\bibinfo{title}{Cold-Atom Quantum Simulator for
  SU(2) Yang-Mills Lattice Gauge Theory},{''} \bibinfo{journal}{Phys. Rev.
  Lett.} \textbf{\bibinfo{volume}{110}},  \bibinfo{pages}{125304}.

\bibitem[{\citenamefont{Zueco} \emph{et~al.}(2009)\citenamefont{Zueco, Galve,
  Kohler, and H\"anggi}}]{Zue09}
\bibinfo{author}{\bibnamefont{Zueco}, \bibfnamefont{D.}},
  \bibinfo{author}{\bibfnamefont{F.}~\bibnamefont{Galve}},
  \bibinfo{author}{\bibfnamefont{S.}~\bibnamefont{Kohler}}, and
  \bibinfo{author}{\bibfnamefont{P.}~\bibnamefont{H\"anggi}},
  \bibinfo{year}{2009}, {``}\bibinfo{title}{Quantum router based on ac control
  of qubit chains},{''} \bibinfo{journal}{Phys. Rev. A}
  \textbf{\bibinfo{volume}{80}},  \bibinfo{pages}{042303}.

\bibitem[{\citenamefont{Zwierlein} \emph{et~al.}(2005)\citenamefont{Zwierlein,
  Abo-Shaeer, Schirotzek, Schunck, and Ketterle}}]{Zwi05}
\bibinfo{author}{\bibnamefont{Zwierlein}, \bibfnamefont{M.~W.}},
  \bibinfo{author}{\bibfnamefont{J.~R.} \bibnamefont{Abo-Shaeer}},
  \bibinfo{author}{\bibfnamefont{A.}~\bibnamefont{Schirotzek}},
  \bibinfo{author}{\bibfnamefont{C.~H.} \bibnamefont{Schunck}}, and
  \bibinfo{author}{\bibfnamefont{W.}~\bibnamefont{Ketterle}},
  \bibinfo{year}{2005}, {``}\bibinfo{title}{Vortices and superfluidity in a
  strongly interacting Fermi gas},{''} \bibinfo{journal}{Nature}
  \textbf{\bibinfo{volume}{435}},  \bibinfo{pages}{1047--1051}.

\end{thebibliography}

\end{document}